%% file: TopRevPPNP.tex
\documentclass[twoside,12pt]{article}
\usepackage{epsfig}
\usepackage{subfigure}
\usepackage{slashed}

\usepackage{graphics}
\usepackage{amsfonts}
\usepackage{amsmath}
\usepackage{amssymb}
\usepackage{mathrsfs}
\usepackage{xspace}
\usepackage{rotating}

\def\p{\vec p}

\def\EtMiss {\slashed{E}_{T}}
\def\EtMissVec {\vec{\slashed{E}}_{T}}
\newcommand{\be}{\begin{equation}}
\newcommand{\ee}{\end{equation}}
\newcommand{\bea}{\begin{eqnarray}}
\newcommand{\eea}{\end{eqnarray}}

\def\ttbar {$t\bar{t}$\ }
\def\qbar {\bar{q}}
\def\ubar {\bar{u}}
\def\dbar {\bar{d}}

\def\bbar {\bar{b}}
\def\pbar {\bar{p}}
\def\tbar {\bar{t}}
\newcommand{\rmd}{\mathrm{d}}
\def\mtop {m_t}

\newcommand{\TeV}{\,\mbox{Te\kern-0.2exV}}
\newcommand{\GeV}{\,\mbox{Ge\kern-0.2exV}}
\newcommand{\mGeV}{\,\mathrm{Ge\kern-0.2exV}}
\newcommand{\MeV}{\,\mbox{Me\kern-0.2exV}}
\newcommand{\keV}{\,\mbox{ke\kern-0.2exV}}
\newcommand{\eV}{\,\mbox{e\kern-0.2exV}}
\newcommand{\ipb}{\,\mbox{pb}^{-1}}
\newcommand{\ifb}{\,\mbox{fb}^{-1}}
\newcommand{\pb}{\mbox{pb}}

\newcommand{\um}{\,\mbox{$\mathrm\mu$m}}
\newlength{\ziffer}

\newlength{\vorzeichen}
\newcommand{\Plus}{\settowidth{\vorzeichen}{$-$}\hspace*{\vorzeichen}}
\newcommand{\beq}{\begin{equation}}
\newcommand{\eeq}{\end{equation}}
\newcommand{\abb}{Fig.~\ref}
\newcommand{\fig}{\abb}

\newcommand{\eq}[1]{Eq.~(\ref{#1})}

\newcommand{\dzero}      {D\O\xspace}
\newcommand{\cdf}        {CDF\xspace}

\newcommand{\fermilab}  {{F{\sc ermilab}}\xspace}
\newcommand{\tevatron}  {{T{\sc evatron}}\xspace}

\newcommand{\cern}      {{C{\sc ern}}\xspace}

\newcommand{\atlas}     {{A{\sc tlas}}\xspace}
\newcommand{\lhc}       {{L{\sc hc}}\xspace}
\newcommand{\lep}       {{L{\sc ep}}\xspace}
\newcommand{\slc}       {{S{\sc lc}}\xspace}
\newcommand{\pep}       {{P{\sc ep}}\xspace}
\newcommand{\petra}     {{P{\sc etra}}\xspace}

\newcommand{\doris}     {{D{\sc oris}}\xspace}

\newcommand{\tristan}   {{T{\sc ristan}}\xspace}
\newcommand{\cms}       {{C{\sc ms}}\xspace}

\newcommand{\herwig}    {{H\sc{erwig}}\xspace}

\newcommand{\evtgen}    {{E\sc{vtgen}}\xspace}

\newcommand{\pythia}    {{P\sc{ythia}}\xspace}

\newcommand{\alpgen}    {{A\sc{lpgen}}\xspace}

\newcommand{\madgraph}  {{M\sc{adgraph}}\xspace}

\newcommand{\GEANT}     {{G\sc{eant}}\xspace}

\newcommand{\met}       {\mbox{$\not\!\!E_T$}\xspace}

\newcommand{\runi}      {Run~I\xspace}
\newcommand{\runii}     {Run~II\xspace}
\begin{document}

\title{\vspace{1cm} Status and Prospects of Top-Quark Physics} 
\author{Joseph R. Incandela$^4$, Arnulf Quadt$^1$, Wolfgang Wagner$^{2,5}$, \\
        Daniel Wicke$^{3,5}$ \\ \\
       $^1$ Universit\"at G\"ottingen, II. Physikalisches Institut, \\
            37077 G\"ottingen, Germany \\
       $^2$ Universit\"at Karlsruhe, Institut f\"ur Experimentelle Kernphysik,\\
            76128 Karlsruhe, Germany \\
       $^3$ Johannes Gutenberg Universit\"at, Institut f\"ur Physik, \\
            55099 Mainz, Germany \\
       $^4$ University of California at Santa Barbara, Santa Barbara, \\
            California 93106, Unites States of America \\
       $^5$ Bergische Universit\"at Wuppertal, Fachgruppe Physik, \\
            42119 Wuppertal, Germany}
\maketitle
\begin{abstract}
The top quark is the heaviest elementary particle observed to date. Its large 
mass of about $173\,\mathrm{GeV}/c^2$ makes the top quark act differently
than other elementary fermions, as it decays before it hadronises, passing 
its spin information on to its decay products. In addition, the top quark plays 
an important role in higher-order loop
corrections to standard model processes, which makes the top quark mass
a crucial parameter for precision tests of the electroweak theory. The top
quark is also a powerful probe for new phenomena beyond the standard
model. \\
During the time of discovery at the Tevatron in 1995 only a few
properties of the top quark could be measured. In recent years, since
the start of Tevatron Run II, the field of top-quark physics has changed
and entered a precision era. This report summarises the latest
measurements and studies of top-quark properties and gives prospects
for future measurements at the Large Hadron Collider (LHC).
\end{abstract}

\input{intro}

\input{intro_topquark}

\input{top_production}

\input{detectors}

\input{ttbar_crosssection}

\input{singletop}

\input{topmass}

\input{top_properties}

\input{new_physics}

\input{summary}

\section*{Acknowledgments}
W.W. would like to thank Jan L\"uck from Universit\"at Karlsruhe and
Peter Uwer from Humboldt Universit\"at Berlin for useful discussions
and suggestions.

%\section*{References}
\itemsep -2pt 
\bibliography{abbrevi,theory_ref,toprefs,cdfpubs,d0pubs,leprefs,lehrbuch,reviews,cmspubs,daniel}
\bibliographystyle{h-elsevier2}

\end{document}

%% file: intro.tex
%========================================================
%= Introduction =========================================
%========================================================
\section{Introduction}
\label{chap:intro_intro}

There are six known quarks in nature, the up, down, strange, charm,
bottom, and the top quark. The quarks are arranged in three pairs or
``generations''. Each member of a pair may be transformed into its
partner via the charged-current weak interaction. Together with the
six known leptons (the electron, muon, tau, and their associated
neutrinos), the six quarks constitute all of the known luminous matter
in the universe. The understanding of the properties of the quarks and
leptons and their interactions is therefore of paramount importance.

The top quark is the charge, $Q=+2/3$, and $T_3=+1/2$ member of the
weak-isospin doublet containing the bottom quark. It is the most
recently discovered quark, which was directly observed in 1995 by the
\cdf and \dzero experiments at the \fermilab\, \tevatron, a
$p\bar{p}$ collider at a centre-of-mass energy of $\sqrt{s} =
1.8\;\rm TeV$. This discovery was a great success of the Standard
Model of Elementary Particle Physics (SM), which suggested the existence of
the top quark as the weak-isospin partner of the $b$-quark already in
1977 at its discovery. Indirect evidence for the existence of the top
quark became compelling over the years and constraints on the top
quark mass, inferred from electroweak precision data, pointed exactly
at the range where the top quark was discovered. Due to its relatively
recent discovery, far less is known about the top quark than about the
other quarks and leptons.

The strong and weak interactions of the top quark are not nearly as
well studied as those of the other quarks and leptons. The strong
interaction is most directly measured in top quark pair production.
The weak interaction is measured in top quark decay and single top
quark production. There are only a few fundamental parameters
associated with the top quark in the SM: the top quark
mass and the three CKM matrix elements involving top.

Thus far, the properties of the quarks and leptons are successfully
described by the SM. However, this theory does not account
for the masses of these particles, it merely accommodates them. Due to
the mass of the top quark being by far the heaviest of all quarks, it
is often speculated that it might be special amongst all quarks and
leptons and might play a role in the mechanism of electroweak symmetry
breaking. Even if the top quark turned out to be a SM
quark, the experimental consequences of this very large mass are
interesting in their own. Many of the measurements described in this
review have no analogue for the lighter quarks. In contrast to the
lighter quarks, which are permanently confined in bound states
(hadrons) with other quarks and antiquarks, the top quark decays so
quickly that it does not have time to form such bound states. There is
also insufficient time to depolarise the spin of the top quark, in
contrast to the lighter quarks, whose spin is depolarised by
chromomagnetic interactions within the bound states.  Thus the top
quark is free of many of the complications associated with the strong
interaction. Also, top quarks are and will remain a major source of
background for almost all searches for physics beyond the Standard
Model. Precise understanding of the top signal is crucial to claim new
physics.

This review summarises the present knowledge of the properties of the
top quark such as its mass and electric charge, its production
mechanisms and rate and its decay branching ratios, {\it etc.}, and
provides a discussion of the experimental and theoretical issues
involved in their determination. Earlier reviews on top quark physics
at \runi or the earlier \runii can be found in 
\cite{wimpenny_winer,campagnari_franklin,bhat_prosper_snyder_review,tollefson_varnes,chakraborty_konigsberg_rainwater,wagner_2005,quadt:2006ma,bkehoe_2008,Pleier:2008oc,Bernreuther:2008ju,Amsler:2008zzb}. 

Since the \tevatron at \fermilab is today still the only place where
top quarks can be produced and studied directly, most of the
discussion in this article describes top quark physics at the
\tevatron. In particular, the focus is placed on the already available
wealth of results from the \runii, which started in 2001 after a five
year upgrade of the \tevatron collider and the experiments \cdf and
\dzero. However, the Large Hadron Collider, \lhc, a proton-proton
collider at a centre-of-mass energy of $\sqrt{s} = 10-14\;\rm TeV$,
commissioned with single-beam in September 2008 and planned to start
collider operation at \cern in the fall 2009, will be a prolific
source of top quarks and produce about 8 million $t\bar{t}$ events per
year (at ``low'' luminosity, $10^{33}\;\rm cm^{-2}\,s^{-1}$), a real
``top factory''.  Measurements such as the top quark mass are entirely
an issue of systematics, as the statistical uncertainty will be
negligible.

%% file: intro_topquark.tex
%\section{Introduction: The top quark in the standard model}
\section{The Top Quark in the Standard Model}
% Author: Arnulf Quadt
\label{chap:intro}
%\label{chap:top_in_SM}

\subsection{Brief Summary of the Standard Model}
\label{sec:intro_sm}

Quantum field theory combines two great achievements of physics in the
$20^{\rm th}$-century, quantum mechanics and relativity. The SM
\cite{standardmodel0,standardmodel1,standardmodel2,standardmodel3,standardmodel4,standardmodel5,standardmodel6,standardmodel7,standardmodel8,standardmodel9,standardmodel10,standardmodel11} is a
particular quantum field theory, based on the set of fields shown in
Table~\ref{tab:sm_fields}, and the gauge symmetries $SU(3)_C \times
SU(2)_L \times U(1)_Y$. There are three generations of quarks and
leptons, labelled by the index $i = 1, 2, 3$, and one Higgs field,
$\phi$.

\begin{table*}[ht]
\caption{ The fields of the SM and their gauge quantum numbers. 
$T$ and \hbox{$T_3$} are the total
weak-isospin and its third component, and $Q$ is the electric charge.}
\label{tab:sm_fields}
\centerline{
\begin{tabular}{lccc|c|c|c|c|c|c}
 & & & & $SU(3)_C$ &  $SU(2)_L$ &  $U(1)_Y$ & $T$ & $T_3$ & $Q$ \\*[1mm] \hline
$Q_L^i =$ & $\begin{pmatrix}u_L\\d_L\end{pmatrix}$ & $\begin{pmatrix}c_L\\s_L\end{pmatrix}$ & $\begin{pmatrix}t_L\\b_L\end{pmatrix}$ & 3 & 2 & $\phantom{-}1/6$ & $1/2$ & $\begin{matrix}+1/2\\-1/2\end{matrix}$ & $\begin{matrix}+2/3\\-1/3\end{matrix}$\\*[3mm]
$u_R^i =$ & $u_R$ & $c_R$ & $t_R$ & 3 & 1 & $\phantom{-}2/3$ & $0$ & $\phantom{+}0$ & $+2/3$ \\*[3mm]
$d_R^i =$ & $d_R$ & $s_R$ & $b_R$ & 3 & 1 & $-1/3$ & $0$ & $\phantom{+}0$ & $-1/3$ \\*[3mm] \hline
$L_L^i =$ & $\begin{pmatrix}{\nu_e}_L\\{e}_L\end{pmatrix}$ & $\begin{pmatrix}{\nu_\mu}_L\\{\mu}_L\end{pmatrix}$& $\begin{pmatrix}{\nu_\tau}_L\\{\tau}_L\end{pmatrix}$& 1 & 2 & $-1/2$ & $1/2$ & $\begin{matrix}+1/2\\-1/2\end{matrix}$ & $\begin{matrix}\phantom{+}0\\-1\end{matrix}$\\*[3mm]
$e_R^i =$ & $e_R$ & $\mu_R$ & $\tau_R$ & 1 & 1 & $-1$ & $0$ & $\phantom{+}0$ & $-1$\\*[3mm]
$\nu_R^i =$ & $\nu^e_R$ & $\nu^\mu_R$ & $\nu^\tau_R$ & $0$ & $0$ & $\phantom{-}0$ & $0$ & $\phantom{+}0$ & $\phantom{+}0$\\*[3mm] \hline
$\phi=$ & $\begin{pmatrix}\phi^+\\ \phi^0\end{pmatrix}$& & & 1 & 2 & $\phantom{-}1/2$ & $1/2$ & $\begin{matrix}+1/2\\-1/2\end{matrix}$ & $\begin{matrix}+1\\\phantom{+}0\end{matrix}$ \\*[3mm] \hline
\end{tabular}}
\end{table*}

Once the gauge symmetries and the fields with their (gauge) quantum
numbers are specified, the Lagrangian of the SM is fixed
by requiring it to be gauge-invariant, local, and renormalisable. The
SM Lagrangian can be divided into several pieces:
\begin{eqnarray}
{\cal{L}}_{SM} & = & {\cal{L}}_{Gauge} + {\cal{L}}_{Matter} + 
                     {\cal{L}}_{Yukawa} + {\cal{L}}_{Higgs}.
\end{eqnarray}

The first piece is the pure gauge Lagrangian, given by
\begin{eqnarray}
{\cal{L}}_{Gauge} & = & \frac{1}{2g_S^2} {\rm Tr}\; G^{\mu \nu} G_{\mu\nu} +
                        \frac{1}{2g^2}   {\rm Tr}\; W^{\mu \nu} W_{\mu\nu} -
                        \frac{1}{4{g^\prime}^2}
                                          B^{\mu \nu} B_{\mu\nu},
\end{eqnarray}
where $G^{\mu\nu}$, $W^{\mu\nu}$, and $B^{\mu\nu}$ are the gluon,
weak, and hypercharge field-strength tensors. These terms contain the
kinetic energy of the gauge fields and their self interactions. The
next piece is the matter Lagrangian, given by
\begin{eqnarray}
{\cal{L}}_{Matter} & = & i\overline{Q}_L^i\; {\not\!\!D}\; Q_L^i + 
                              i\bar{u}_R^i\; {\not\!\!D}\; u_R^i +
                              i\bar{d}_R^i\; {\not\!\!D}\; d_R^i +
                         i\overline{L}_L^i\; {\not\!\!D}\; L_L^i +
                              i\bar{e}_R^i\; {\not\!\!D}\; e_R^i.
\end{eqnarray}
This piece contains the kinetic energy of the fermions and their
interactions with the gauge fields, which are contained in the
covariant derivatives. For example, 
\begin{eqnarray}
{\not\!\!D}\; Q_L & = & \gamma^\mu
\left( \partial_\mu + i\, g_S\,G_\mu + i\, g\, W_\mu + i\,\frac{1}{6}\, 
g^\prime\, B_\mu \right) Q_L,\hspace*{7mm}
\end{eqnarray}
 since the field $Q_L$ participates in all three gauge interactions. A
 sum on the index $i$, which represents the generations, is implied in
 the Lagrangian.

These two pieces of the Lagrangian depend only
on the gauge couplings $g_S, g, g^\prime$. Their approximate values,
evaluated as $M_Z$, are
\begin{eqnarray}
g_S & \approx & 1,\hspace{1cm} g \approx 2/3,\hspace{1cm} g^\prime \approx  2/(3\sqrt{3}).
\end{eqnarray}
Mass terms for the gauge bosons and the fermions are forbidden by the
gauge symmetries.

The next piece of the Lagrangian is the Yukawa interaction of the
Higgs field with the fermions, given by
\begin{eqnarray}
{\cal{L}}_{Yukawa} & = & 
-\Gamma_u^{ij} \; \overline{Q}_L^i\;\epsilon \phi^*\;u_R^j 
-\Gamma_d^{ij} \; \overline{Q}_L^i\;         \phi  \;d_R^j 
-\Gamma_e^{ij} \; \overline{L}_L^i\;         \phi  \;e_R^j
+ h.c.,
\end{eqnarray}
where $\epsilon = i \sigma_2$ is the total antisymmetric tensor in 2
dimensions, related to the second Pauli matrix $\sigma_2$ and required
to ensure each term separately to be electrically neutral, and the
coefficients $\Gamma_u$, $\Gamma_d$, $\Gamma_e$ are $3
\times 3$ complex matrices in generation space. They need not be
diagonal, so in general there is mixing between different generations.
These matrices contain most of the parameters of the SM.

The final piece is the Higgs Lagrangian \cite{higgs1,higgs2,higgs3},
given by
\begin{eqnarray}
{\cal{L}}_{Higgs} & = & (D^\mu \phi)^\dagger\;D_\mu\phi + 
\mu^2\phi^\dagger\phi - \lambda (\phi^\dagger \phi)^2,
\end{eqnarray}
with the Higgs doublet $\phi$ as given in Table~\ref{tab:sm_fields}.
This piece contains the kinetic energy of the Higgs field, its gauge
interactions, and the Higgs potential. The coefficient of the
quadratic term, $\mu^2$, is the {\it only} dimensionful parameter in
the SM. The sign of this term is chosen such that the
Higgs field has a non-zero vacuum-expectation value on the circle of
minima in Higgs-field space given by $\langle \phi^0 \rangle =
\mu/\sqrt{2\lambda} \equiv v/\sqrt{2}$. The dimensionful parameter
$\mu$ is replaced by the dimensionful parameter $v \approx 246\;\rm
GeV$.

The acquisition of a non-zero vacuum-expectation value by the Higgs
field breaks the electroweak symmetry and generates masses for the
gauge bosons,
\begin{eqnarray}
M_W & = & \frac{1}{2}gv,\hspace{1cm}
M_Z   =   \frac{1}{2}\sqrt{g^2+g^{\prime 2}}\; v,
\hspace*{4mm} \mbox{and for the fermions,}\hspace*{2mm}
M_f   =   \Gamma_t \frac{v}{\sqrt{2}},
\end{eqnarray}
with the Yukawa coup\-ling $\Gamma_t$. Dia\-go\-na\-li\-sing the fermion mass
matrices generates the Ca\-bib\-bo-Ko\-ba\-ya\-shi-Maskawa (CKM) matrix
\cite{cabibbo,kobayashi_maskawa}, including the CP-violating phase.
\begin{figure}[!ht]
\centerline{
\includegraphics[width=0.39\textwidth,clip=]
                {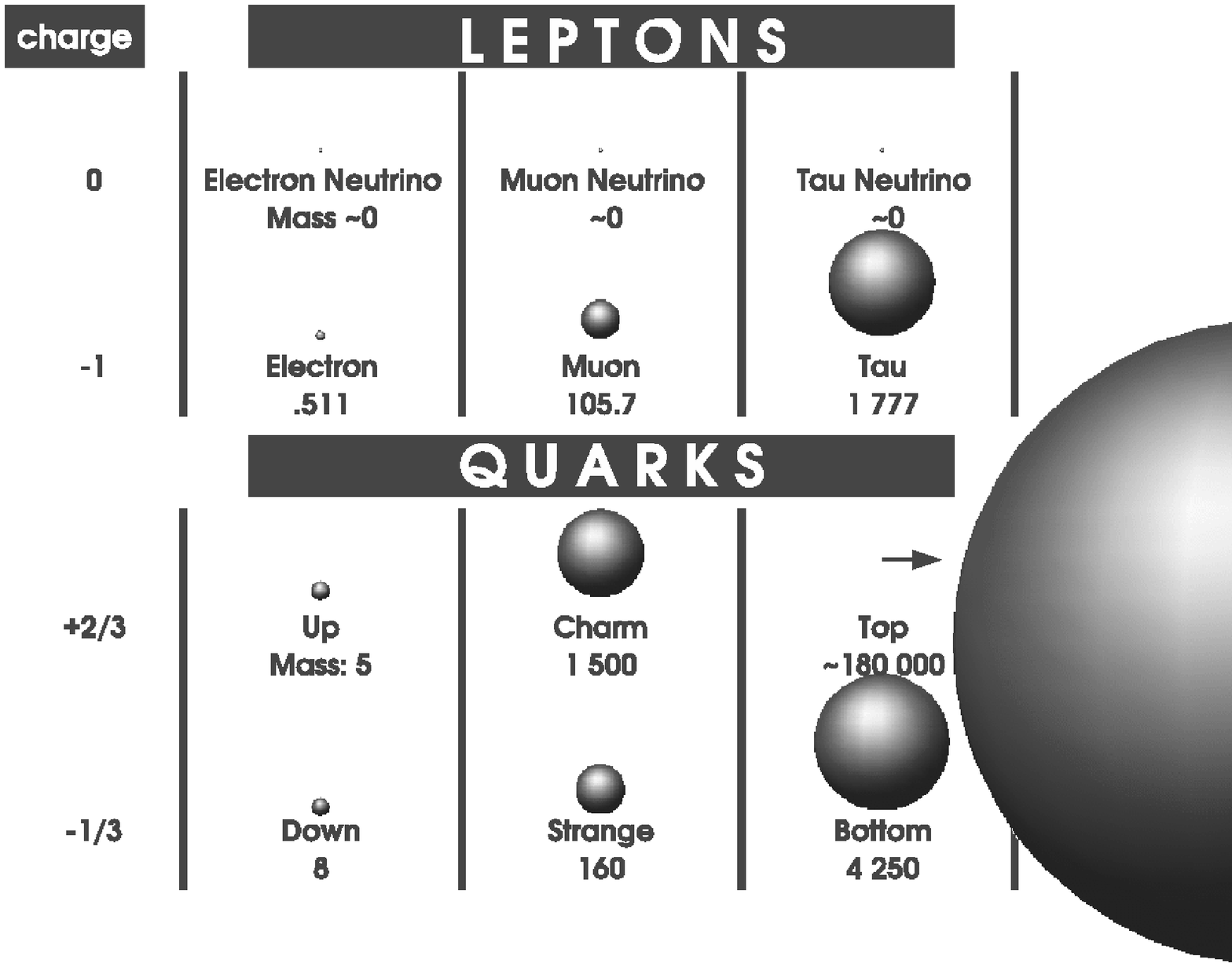} 
\hfill
\includegraphics[width=0.46\textwidth,clip=]
                {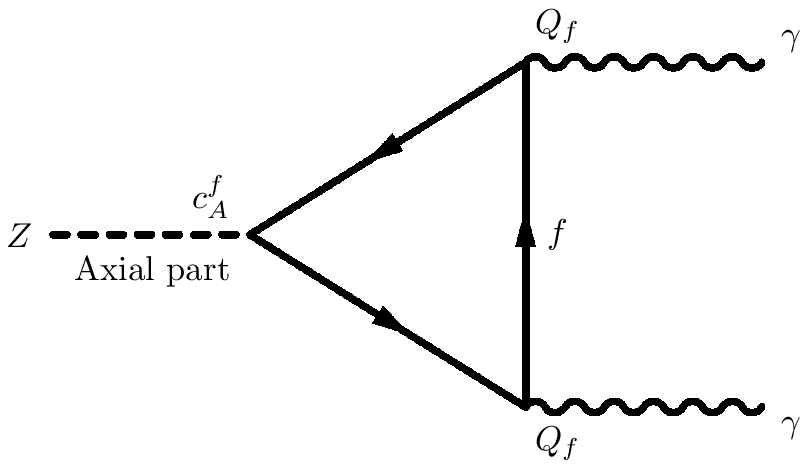} 
}
\caption{\label{fig:fermion_spectrum} Left: Table of lepton and quark
  properties such as electric charge and mass (in $\rm MeV/c^2$). The
  top quark is unique amongst all fermions due to its very large mass.
  (All elementary particles are point-like. The relative size of the 
   drawn spheres merely symbolizes the mass of 
   fermions, but does not scale linearly with the fermion mass.) 
   Right: A fermion (quark or lepton) 
  triangle diagram which potentially could cause an anomaly.}
\end{figure}

Figure~\ref{fig:fermion_spectrum} shows three lepton and quark
families with their electric charge and approximate mass. While the
neutrinos have non-zero, but very small masses, the quark masses are
much larger. The top quark, with a mass of $\approx 175\;\rm GeV/c^2$,
is by far the heaviest fermion.

\subsection{Indirect Evidence for the Existence of the Top Quark}
\label{sec:intro_topexistence}

In 1977, the $b$-quark was discovered at Fermilab \cite{herb_1977}.
The existence of a weak isospin partner of the $b$-quark, the top
quark, was anticipated for two main reasons: First it provides a
natural way to suppress the experimentally not observed
flavour-changing neutral current. The argument on which the GIM
mechanism \cite{standardmodel3} is based applies just as well for
three as for two quark doublets.

The second reason is concerned with the desire to obtain a
renormalisable gauge theory of weak interactions\footnote{The gauge
theory has to be consistent, i.e. anomaly-free, in order to be at
least unitary. The requirement of the gauge theory to be
renormalisable is stronger than to be consistent, but the former
argument is more familiar to most readers. The important consequence
of both requirements is that the gauge theory is anomaly-free.}. The
SM of electroweak interactions can be proven to be
renormalisable under the condition that the sum of the weak
hypercharges, $Y_i$, of all left-handed fermi\-ons is zero, i.e.
\begin{eqnarray} \sum_{\scriptstyle\substack{\mbox{left-handed}\\
\mbox{quarks and leptons}}} Y_i & = & 0.  \end{eqnarray} Since every
lepton multiplet contributes a value of $y = -2$ and every quark
multiplet a value of $+2/3$, the sum only vanishes if there are three
colours, i.e. every quark exists in three colour versions, and if the
number of quark flavours equals the number of lepton species. The
general proof that gauge theories can be renormalised, however, can
only be applied if the particular gauge theory is {\it anomaly
free}\footnote{A gauge theory might be renormalisable, whether or not
it is anomaly free. The general proof of renormalisability, however,
cannot be applied if it is not.}. This requires a delicate
cancellation between different diagrams, relations which can easily be
upset by ``anomalies'' due to fermion loops such as the one shown in
Figure~\ref{fig:fermion_spectrum}. The major aspect is an odd number
of axial-vector couplings. In general, anomaly freedom is guaranteed
if the coefficient\footnote{$d_{abc}$ is the coefficient in the
definition of the anomaly: $\left[ \partial_\mu J^\mu_\alpha(x)
\right]_{anom} = - \frac{1}{32 \pi^2} \cdot d_{\alpha \beta \gamma}
\epsilon^{\kappa \nu \lambda \rho}\cdot F^\beta_{\kappa\nu}(x) \cdot
F^\gamma_{\lambda\rho}(x)$, with the current $J^\mu_\alpha(x)$, the
field strength tensor $F^\beta_{\kappa \nu}$ and the total
antisymmetric tensor $\epsilon^{\kappa \nu \lambda \rho}$.}
\begin{eqnarray}\label{eqn:anomaly_free} d_{abc} & = &
\sum_{\mbox{fermions}} Tr\left[ \hat{\lambda}^a
                                       \left\lbrace \hat{\lambda}^b 
                                             ,  \hat{\lambda}^c \right\rbrace_+
                                        \right]\, = \, 0,
\end{eqnarray}
where the $\hat{\lambda^i}$ are in general the generators of the gauge
group under consideration. In the SM of electroweak
interactions, the gauge group $SU(2) \times U(1)$ is generated by the
three Pauli matrices, $\sigma_i$, and the hypercharge $Y$:
$\hat{\lambda}^i =
\sigma_i$, for $i=1, 2, 3$, and $\hat{\lambda}^4 = \hat{Y} = 2 (
\hat{Q} - \hat{T}_3)$. 
In the specific example shown in Figure~\ref{fig:fermion_spectrum},
one consequence of Equation~\ref{eqn:anomaly_free} is a relation where
each triangle is proportional to $c_A^f\, Q_f^2$, where $Q_f$ is the
charge and $c_A^f$ is the axial coupling of the weak neutral
current. Thus, for an equal number $N$ of lepton and quark doublets,
the total anomaly is proportional to:
\begin{eqnarray}
d & \propto & \sum_{i=1}^N \left( 
  \frac{1}{2}\cdot \left(0\right)^2 
 - \frac{1}{2}\cdot \left(-1\right)^2 
 +\frac{1}{2}\cdot N_c \cdot \left(+\frac{2}{3}\right)^2
 -\frac{1}{2}\cdot N_c \cdot \left(-\frac{1}{3}\right)^2\right).
\end{eqnarray}
Consequently, taking into account the three colours of each quark
($N_c = 3$), the anomalies are cancelled. Since three lepton doublets
were observed many years ago (the tau neutrino was experimentally only
observed directly in the year 2000, but the number of light neutrino
generations was known to be 3 from the \lep data on the $Z$-pole), the
lack of anomalies such as the one shown in
Figure~\ref{fig:fermion_spectrum} therefore requires the existence of
the three quark doublets.

There is a lot of indirect experimental evidence for the existence of
the top quark. The experimental limits on flavour changing neutral
current (FCNC) decays of the $b$-quark
\cite{kane_peskin_1982,bean_no_fcnc} such as $b\rightarrow s
\ell^+\ell^-$ and the absence of large tree level (lowest order)
$B_d^0 \bar{B}_d^0$ mixing at the $\Upsilon(4S)$ resonance
\cite{roy_sankar_1990,argus_1987,argus_1987_1,cleo_1993} rule out the
hypothesis of an isosinglet $b$-quark.  In other words, the $b$-quark
must be a member of a left-handed weak isospin doublet.

The most compelling argument for the existence of the top quark comes
from the wealth of data accumulated at the $e^+e^-$ colliders \lep and
\slc in recent years, particularly the detailed studies of the
$Zb\bar{b}$ vertex near the $Z$ resonance \cite{zpole_2005}. These
studies have yielded a measurement of the isospin of the $b$-quark.
The $Z$-boson is coupled to the $b$-quarks (as well as the other
quarks) through vector and axial vector charges $(v_b$ and $a_b$) with
strength (Feynman diagram vertex factor)
\begin{eqnarray}
\includegraphics[width=0.08\textwidth]
                {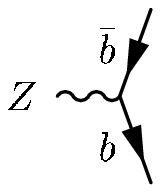} 
& = & \frac{-i\,g}{\cos\theta_W}\gamma^\mu\frac{1}{2}
                  \left(v_b - a_b\,\gamma^5 \right)\,\\
& = & \,-i\,
\sqrt{\sqrt{2}G_F M_Z^2} \; \gamma^\mu\,(v_b - a_b\,\gamma^5),
\end{eqnarray}
where $v_b$ and $a_b$ are given by
\begin{eqnarray}
v_b & = &  \left[ T_3^L(b) + T_3^R(b) \right] - 
            2 e_b \sin^2\theta_W, \;\;\;\;\mbox{and}\nonumber \\
a_b & = &  \left[ T_3^L(b) + T_3^R(b) \right].
\end{eqnarray}
Here, $T_3^L(b)$ and $T_3^R(b)$ are the third components of the weak
isospin for the left-handed and right-handed $b$-quark fields,
respectively. The electric charge of the $b$-quark, $e_b = -1/3$, has
been well established from the $\Upsilon$ leptonic width as measured
by the \doris $e^+e^-$ experiment \cite{doris_1,doris_2,doris_3}.
Therefore, measurements of the weak vector and axial-vector coupling
of the $b$-quark, $v_b$ and $a_b$, can be interpreted as measurements
of its weak isospin.

%\begin{figure}[!ht]
%\centerline{
%\includegraphics[width=0.49\textwidth,clip=]
%                {./fig_quadt/chapter1/aq_weak_b_coupling_1.eps} 
%}
%\caption{\label{fig:rb_lep2}
%  Comparison of the \lep combined
%  measurement of $R_b^0$ with the SM prediction as a
%  function of the mass of the top quark. From Reference
%  \cite{zpole_2005}.}
%\end{figure}

The (improved) Born approximation for the partial $Z$-boson decay rate
gives in the limit of a zero mass $b$-quark: \begin{eqnarray}
\Gamma_{b\bar{b}} & \equiv & \Gamma(Z\rightarrow b\bar{b})\;=\;
\frac{G_FM_Z^3}{2\sqrt{2}\pi}\,(v_b^2 + a_b^2).  \end{eqnarray} The
partial width $\Gamma_{b\bar{b}}$ is expected to be thirteen times
smaller if $T_3^L(b) = 0$. The \lep measurement of the ratio of this
partial width to the full hadronic decay width, $R_b =
\Gamma_b/\Gamma_{had} = 0.21629 \pm 0.00066$
\cite{zpole_2005}, is in excellent agreement with
the SM expectation (including the effects of the top
quark) of $0.2158$, ruling out $T_3^L(b) = 0$. The sensitivity of
$R_b$ to the mass of the top quark also shows that a top quark with a
mass around $m_t \approx 175\;\rm GeV/c^2$ is strongly favoured
\cite{zpole_2005}.

In addition, the forward-backward asymmetry in $e^+e^- \rightarrow
b\bar{b}$ below \cite{shimonaka_b_weakiso_3} and at the $Z$ pole
\cite{zpole_2005},
\begin{eqnarray}
A^0_{FB}(M_Z) & = & \frac{3}{4}\; \frac{2 v_e\,a_e}{(v_e^2+a_e^2)} \;
                    \frac{2 v_b\,a_b}{(v_b^2+a_b^2)},
\end{eqnarray}
measured to be $A_{FB}^{0,b} = 0.0992 \pm 0.0016$ is sensitive
\cite{zpole_2005,shimonaka_b_weakiso_3} to the relative size of the
vector and axial vector couplings of the $Zb\bar{b}$ vertex. The sign
ambiguity for the two contributions can be resolved by the $A_{FB}$
measurements from low energy experiments that are sensitive to the
interference between neutral current and electromagnetic amplitudes.
From earlier measurements of
$\Gamma_{b\bar{b}}$ and $A_{FB}$ at \lep,
\slc, and the low energy experiments (\pep, \petra and \tristan
\cite{elsen_b_weakiso_1,behrend_b_weakiso_2,shimonaka_b_weakiso_3}),
one obtains \cite{schaile_zerwas}
\begin{align}
T_3^L(b) & =  -0.490\;\;^{+0.015}_{-0.012} & \Rightarrow T_3^L(b) & = -1/2,\\
T_3^R(b) & =  -0.028\;\pm 0.056            & \Rightarrow T_3^R(b) & = \phantom{-}0,
\end{align}
for the third component of the isospin of the $b$-quark. This implies
that the $b$-quark must have a weak isospin partner, i.e. the top
quark with $T_3^L(t) = +1/2$ must exist.

\subsection{Indirect Constraints on the Mass of the Top Quark}
\label{sec:intro_mtopconstraints}

The precise electroweak measurements performed at \lep, \slc, NuTeV
and the $p\bar{p}$ colliders can be used to check the validity of the
SM and within its framework, to infer valuable information
about its fundamental parameters. Due to the accuracy of those
measurements, sensitivity to the mass of the top quark and the Higgs
boson through radiative corrections is gained.

\begin{figure}[!ht]
\centerline{
\includegraphics[width=0.445\textwidth,clip=]
                {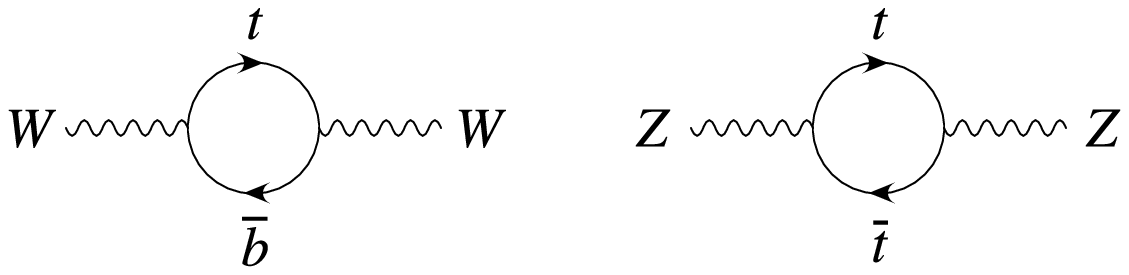} \hfill
\includegraphics[width=0.4459\textwidth,clip=]
                {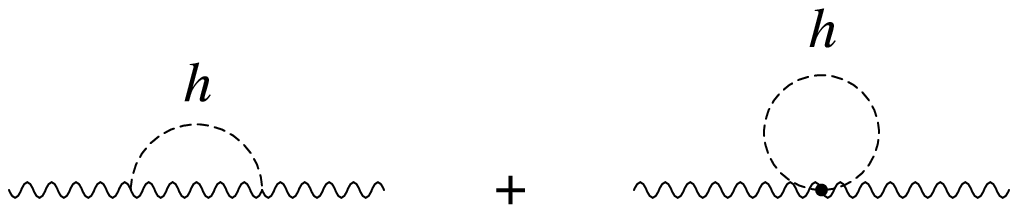} 
}
\caption{\label{fig:wz_top_loops}
  Left: Virtual top quark loops contributing to the $W$ and $Z$ boson
  masses. Right: Virtual Higgs boson loops contributing to the $W$ and
  $Z$ boson masses.}
\end{figure}

All electroweak quantities (mass, width and couplings of the $W$- and
the $Z$-boson) depend in the SM only on five
parameters. At leading, order this dependence is reduced to only three
parameters, two gauge couplings and the Higgs-field vacuum expectation
value. The three best-measured electroweak quantities can be used to
determine these three parameters: The electromagnetic coupling
constant $\alpha$, measured in low-energy experiments
\cite{alphaem_bejing}, the Fermi constant, $G_F$ determined from the
$\mu$ lifetime \cite{muon_lifetime}, and the mass of the $Z$-boson,
measured in $e^+e^-$ annihilation at \lep and \slc \cite{zpole_2005}.
By defining the electroweak mixing angle $\theta_W$ through
$\sin^2\theta_W \equiv 1 - {m_W^2}/{m_Z^2}$, the $W$-boson mass
can be expressed as: 
\begin{eqnarray} 
m_W^2 & = & \frac{\frac{\pi\alpha}{\sqrt{2} G_F}}
                 {\sin^2 \theta_W\cdot (1-\Delta r)},
\end{eqnarray}
where $\Delta r$ contains all the one-loop corrections. Contributions
to $\Delta r$ originate from the top quark by the one-loop diagrams
shown in Figure~\ref{fig:wz_top_loops} (left), which contribute to the $W$
and $Z$ masses via:
\begin{eqnarray}
(\Delta r)_{top} & \simeq & -\; \frac{3\;G_F}{8 \sqrt{2} \pi^2 \tan^2\theta_W}
                             \;\; m_t^2.
\end{eqnarray}

Also the Higgs boson contributes to $\Delta r$ via the one-loop
diagrams, shown in Figure~\ref{fig:wz_top_loops} (right):
\begin{eqnarray}
(\Delta r)_{Higgs} & \simeq & \frac{3\, G_F \, m_W^2}
                                   {8\sqrt{2} \pi^2}\;
                       \left(\ln \frac{m_H^2}{m_Z^2}-\frac{5}{6}\right).
\end{eqnarray}

%\begin{figure}[!ht]
%\centerline{
%\includegraphics[width=0.49\textwidth,clip=]
%                {./fig_quadt/chapter1/lepew_fit_1992.eps} 
%}
%\caption{\label{fig:lepew_fit_1992}
%  $\chi^2$ of the SM fit to the electroweak data as a
%  function of the top quark mass using \lep~1 data (left) and using
%  \lep~1, hadron collider and neutrino experiment data (right)
%  \cite{lepew_fit_1992}. The dependence on the Higgs boson mass, here
%  chosen to be $50, 300$ or $1000\;\rm GeV/c^2$, is weak, since $m_H$
%  enters only logarithmically in the electroweak fit, whereas $m_t$
%  enters quadratically.}
%\end{figure}

While the leading $m_t$ dependence is quadratic, i.e. very strong, the
leading $m_H$ dependence is only logarithmic, i.e. rather weak.
Therefore the inferred constraints on $m_H$ are much weaker than those
on $m_t$. This was used to successfully predict the top quark mass
from the electroweak precision data before it was discovered by \cdf
and \dzero in 1995 \cite{Abe:1995hr,Abachi:1995iq}.  Neutral
current weak interaction data, such as $e^+e^-$ annihilation near the
$Z$ pole, $\nu N$ and $e N$ deep-inelastic scattering, $\nu e$ elastic
scattering, and atomic parity violation can also be used to constrain
the top quark mass. 

The most recent indirect measurements of the top quark mass using the
$Z$-pole data together with the direct measurements of the $W$-boson
mass and total width and several other electroweak quantities yields
\cite{lepew_2008,lepew_2009}:
\begin{eqnarray}
m_{top} & = & 178.9^{+11.7}_{-\phantom{0}8.6}\;\rm GeV/c^2,
\end{eqnarray}
which is in very good agreement with the world average of the direct
measurements \cite{:2009ec}
\begin{eqnarray}
m_{top} & = & 173.1 \pm 1.3\;\rm GeV/c^2.
\end{eqnarray}
The global fit to all electroweak precision data including the world
average of the direct top quark mass measurements yields
\cite{lepew_2008,lepew_2009} the identical value as the direct
measurement dominates the fit, while a fit only to the $Z$-pole data
gives \cite{zpole_2005}:
\begin{eqnarray}
m_{top} & = & 172.6^{+13.2}_{-10.2}\;\rm GeV/c^2.
\end{eqnarray}

\begin{figure}[!ht]
\centerline{
\includegraphics[width=0.4\textwidth,,clip=]
                {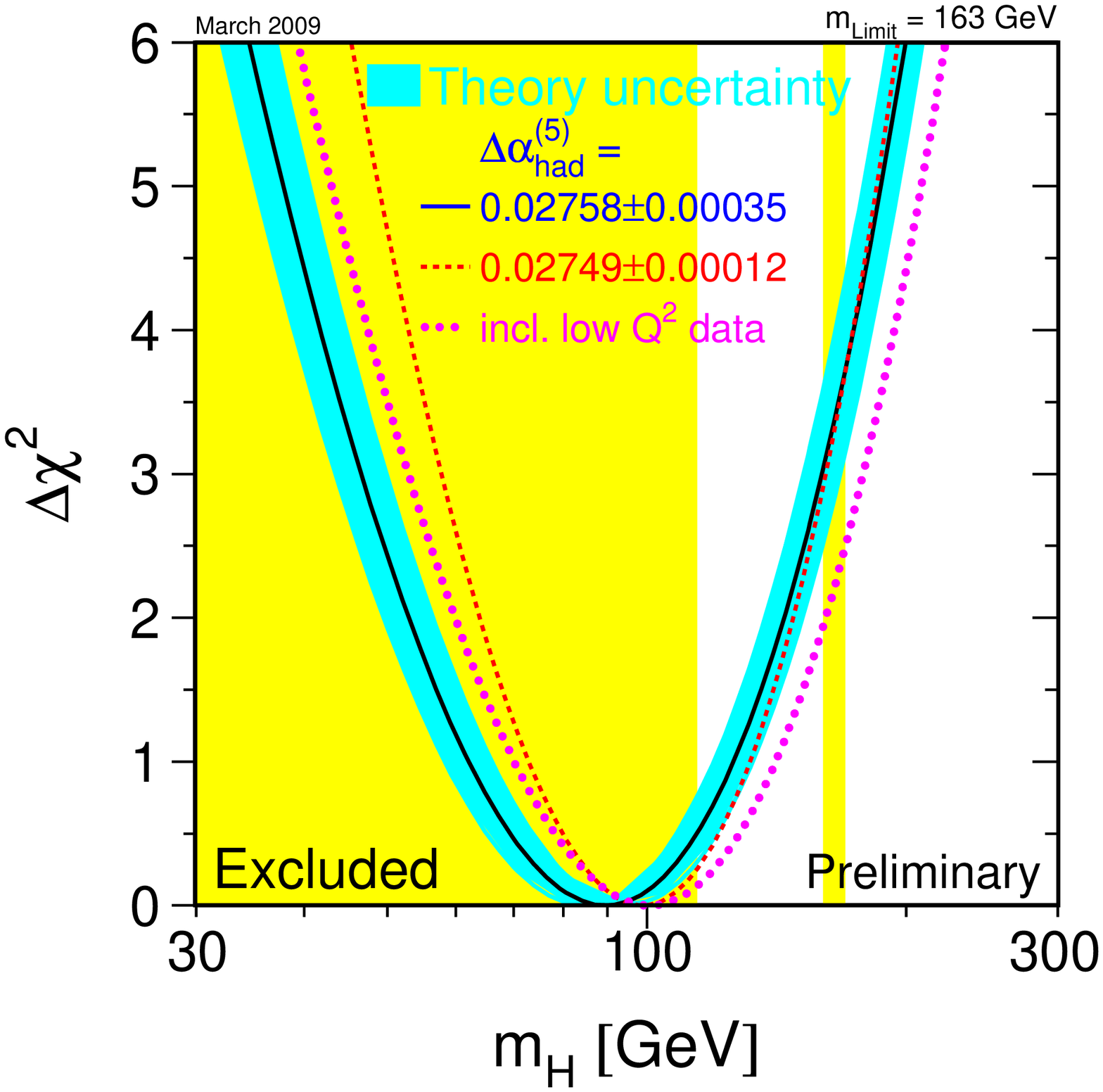} 
\hspace*{0.1\textwidth}
\includegraphics[width=0.4\textwidth,clip=]
                {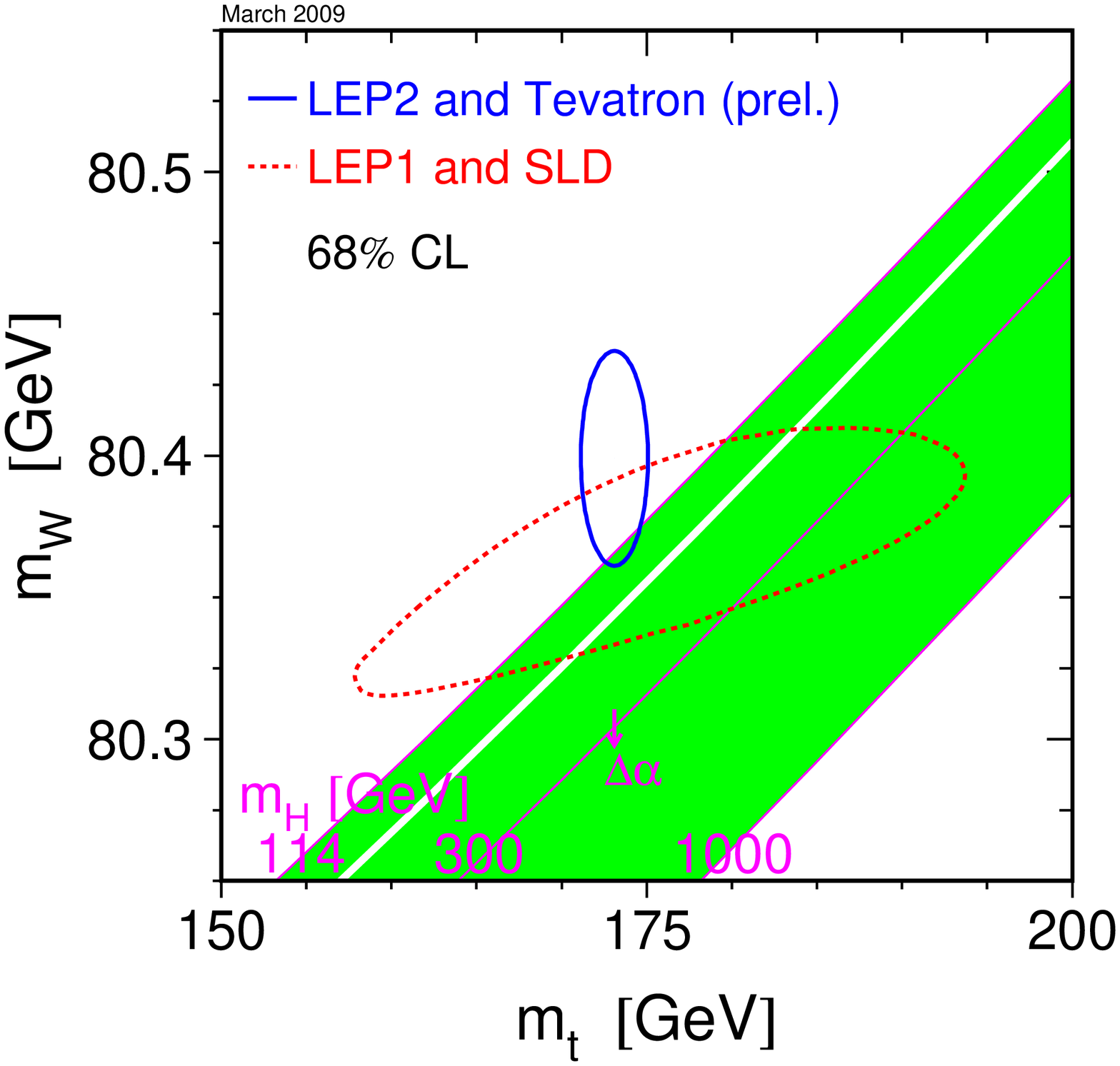} 
}
\caption{\label{fig:lepew_fit_2009} Left: Blueband plot, showing the
  indirect determination of the Higgs boson mass from all electroweak
  precision data together with the 95\% C.L. limit on the Higgs
  boson mass from the direct searches in yellow
  \cite{lep_sm_higgs,tevatron_sm_higgs}. Right: Lines of constant
  Higgs mass on a plot of $M_W$ vs. $m_t$. The dotted ellipse is the
  68\% C.L. direct measurement of $M_W$ and $m_t$. The solid ellipse is
  the 68\% C.L. indirect measurement from precision electroweak data.}
\end{figure}

The successful prediction of the mass of the top quark before its
discovery provides confidence in the precision and predictive power of
radiative corrections in the SM. Therefore, the SM
fit to the electroweak precision data including the direct
measurements of the top quark and $W$-boson mass is used to infer on
the mass of the SM Higgs boson.
Figure~\ref{fig:lepew_fit_2009} (left) shows the $\Delta \chi^2$ of
the latest fit as a function of the Higgs boson mass. The most likely
value of the Higgs mass, determined from the minimum of the $\Delta
\chi^2$ curve is $90\;^{+36}_{-27}\rm GeV/c^2$
\cite{lepew_2008,lepew_2009}, clearly indicating that the data prefers
a light Higgs boson, $m_H < 163\;\rm GeV/c^2$ at the 95\% C.L.
\cite{lepew_2008,lepew_2009}. The preferred value is slightly below
the exclusion limit of $114.4\;\rm GeV/c^2$ at the 95\% C.L. from the 
direct search for
the SM Higgs boson at \lep \cite{lep_sm_higgs} and at the
\tevatron \cite{tevatron_sm_higgs}.

Figure~\ref{fig:lepew_fit_2009} (right) shows the 68\% C.L. contour in
the $(m_t, m_W)$ plane from the global electroweak fit
\cite{lepew_2008,lepew_2009}. It shows the direct and indirect
determination of $m_t$ and $m_W$. Also displayed are the isolines of
SM Higgs boson mass between the lower limit of $114\;\rm
GeV/c^2$ and the theoretical upper limit of $1000\;\rm GeV/c^2$. As
can be seen from the figure, the direct and indirect measurements are
in good agreement, showing that the SM is not obviously
wrong. On the other hand, the fit to all data has a $\chi^2$ per
degree of freedom of 17.4/13, corresponding to a probability of
18.1\%. This is mostly due to three anomalous measurements: the $b$
forward-backward asymmetry $(A_{FB}^b)$ measured at \lep, which
deviates by $2.9\,\sigma$, the total hadronic production cross section
$(\sigma_{had}^0)$ at the $Z$-pole from \lep and the left-right cross
section asymmetry $(A_{LR})$ measured at \slc, both of which deviate
from the SM fit value by about $1.5\;\sigma$. If
$\sin^2\theta_W\,(\nu N)$, measured by the NuTeV collaboration
\cite{nutev_sin2th}, is in addition included in the fit, the measured
and fitted value of $\sin^2\theta_W\,(\nu N)$ differ by almost
$3\,\sigma$.  It seems there is some tension in the fit of the
precision electroweak data to the SM.

Measurements of $M_W$ and $m_t$ at the \tevatron could resolve or
exacerbate this tension. Improvements in the precision of the
measurement of the top quark or the $W$-boson mass at the \tevatron
translate into better indirect limits on the Higgs boson mass.  This
will also be a service to the \lhc experiments which optimise their
analysis techniques and strategies for the search for the yet elusive
SM Higgs boson in the lower mass range, preferred by the
SM electroweak fit.

\begin{figure*}[!ht]
\centerline{
\includegraphics[width=0.50\textwidth,clip=]
%                {./fig_quadt/chapter1/tmass_history_1.eps}
                {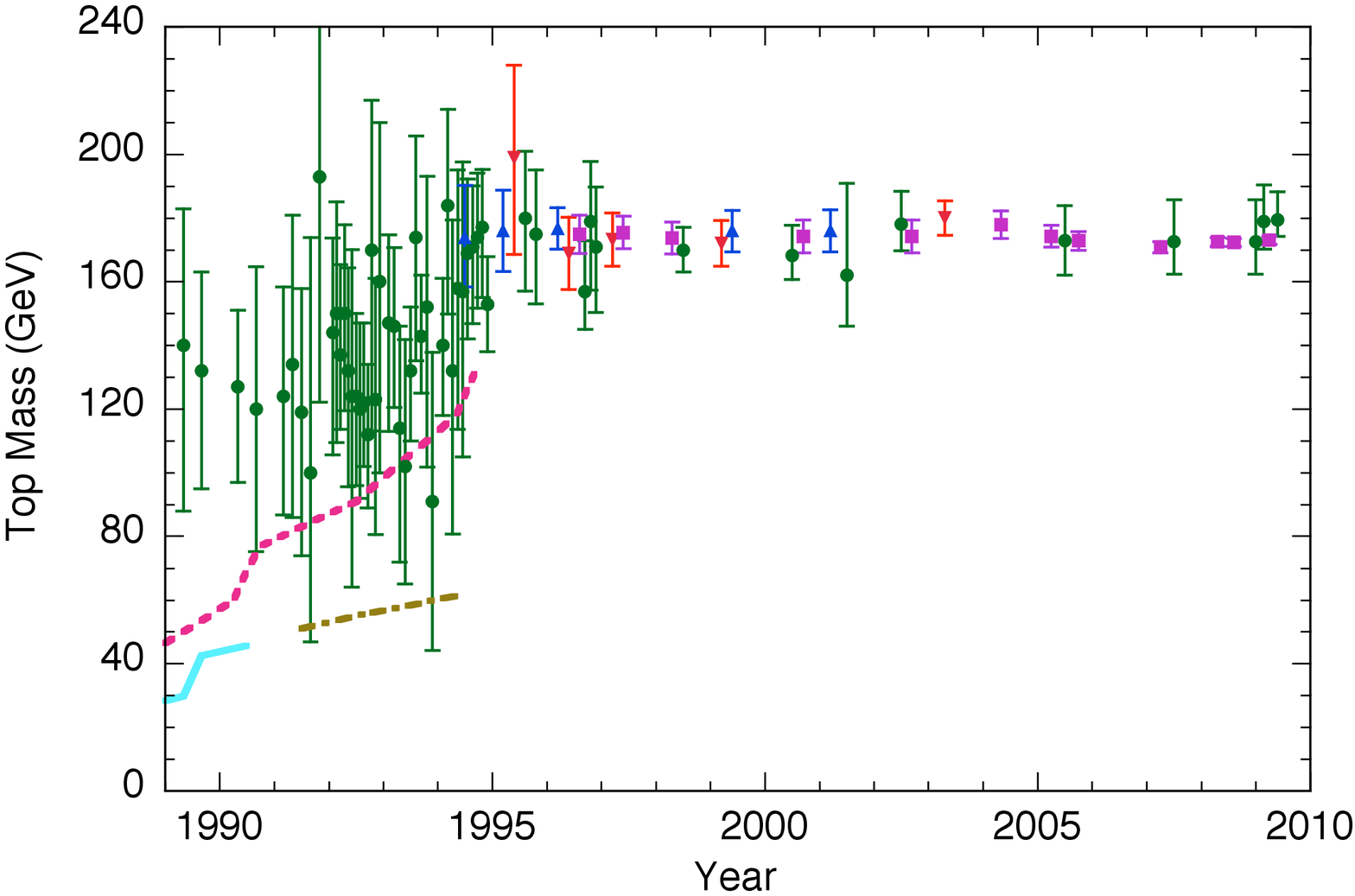}
}
\caption{\label{fig:tmass_history_1}
  History of the limits on or measurements of the top quark mass
  (updated April 2009 by C.~Quigg from \cite{priv_comm_cquigg}:
  $(\bullet)$ Indirect bounds on the top-quark mass from precision
  electroweak data; $(\blacksquare)$ World-average direct measurement
  of the top-quark mass (including preliminary results);
  $(\blacktriangle)$ published CDF and $(\blacktriangledown)$ {D\O}
  measurements; Lower bounds from $p\bar{p}$ colliders $\rm
  Sp\bar{p}S$ and the \tevatron are shown as dash-dotted and dashed
  lines, respectively, and lower bounds from $e^+e^-$ colliders
  (\petra, \tristan, \lep and \slc) are shown as a solid light grey
  line.}
\end{figure*}

Finally, in 1995, both \cdf and \dzero published the discovery of the
top quark in strong $t\bar{t}$ production
\cite{Abe:1995hr,Abachi:1995iq}, which marked the beginning of
a new era, moving on from the search for the top quark to the studies
and measurements of the properties of the top quark.
Figure~\ref{fig:tmass_history_1} shows the development of limits and
measurements on the top quark mass from indirect and direct studies at
$e^+e^-$ and hadron colliders. The top quark was discovered with a
mass of exactly the value that was predicted from global fits to
electroweak precision data.

\subsection{Top-Quark Properties}
\label{sec:intro:top_properties}

% spin -> spin correlation, ttbar production Xsec
% weak isospin -> weak coupling strength
% weak coupling (V_tb, helicity)
% strong coupling (ttbar Xsec and ttbar+jet, ttbar+jj)
% em-coupling (e+e- -> ttbar, ttbar+gamma)

The {\bf mass of the top quark} is larger than that of any other
quark.  Furthermore, the top quark mass is measured with better
relative precision (0.75\%) than any other quark \cite{:2009ec}.
Given the experimental technique used to extract the top mass, these
mass values should be taken as representing the top {\it pole mass}.
The top pole mass, like any quark mass, is defined up to an intrinsic
ambiguity of order $\Lambda_{QCD} \sim 200\;\rm MeV$ \cite{Smith97}.
The most recent combination of top quark mass measurements by the
\tevatron Electroweak/Top Working group, including preliminary \cdf
and \dzero measurements from \runii, yields $m_t = 173.1 \pm 1.3\;\rm
GeV/c^2$ \cite{:2009ec}. The prospects for the precision of 
$\mtop$ measurements at the \tevatron have recently been
revised to better than $1.5\;\rm GeV/c^2$ per experiment with the full
\runii\ data set. At the \lhc, a precision of the top-quark mass
measurement of $1-2\;\rm GeV/c^2$ is expected. At a future linear
$e^+e^-$ collider, the expected precision of a measurement of $\mtop$
from a cross section scan at the $t\bar{t}$ production
threshold is $\Delta m_{t} = 20 - 100\;\rm MeV/c^2$
\cite{martinez_miquel,hepph_0001286,hepph_0011254}.

Like most of its fundamental quantum numbers, the {\bf electric charge
of the top quark}, $q_{top}$, has not been measured so far. The
electric charge of the top quark is easily accessible in $e^+e^-$
production by measurements of the ratio $R = \frac{e^+e^-\rightarrow
  \mbox{hadrons}}{e^+e^-\rightarrow \mu^+\mu^-}$ through the top quark
production threshold. However, this region of energy is not yet
available at $e^+e^-$ colliders. Thus, alternative interpretations for
the particle that is believed to be the charge 2/3 isospin partner of
the $b$ quark are not ruled out. For example, since the correlations
of the $b$ quarks and the $W$ bosons in $p\bar{p} \rightarrow t\bar{t}
\rightarrow W^+W^-b\bar{b}$ events are not determined by \cdf or
\dzero, it is conceivable that the ``$t$ quark'' observed at the
\tevatron is an exotic quark, $Q_4$, with charge $-4/3$ with decays
via $Q_4 \rightarrow W^-b$~\cite{Chang:1998pt,Chang:1999zc,Choudhury:2001hs}. 
This interpretation is consistent with
current precision electroweak data. In order to determine the charge
of the top quark, one can either measure the charge of its decay
products, in particular of the $b$ jet via jet charge techniques, or
investigate photon radiation in $t\bar{t}$ events \cite{bauer_orr_01}.
The latter method actually measures a combination of the
electromagnetic coupling strength and the charge quantum number.
Combining the results of the two methods will thus make it possible to
determine both quantities.

At the \tevatron, with an integrated luminosity of $1-2\;\rm fb^{-1}$,
one is able to exclude at 95\% CL the possibility that an exotic quark
$Q_4$ with charge $-4/3$ and not the SM top quark was
found in \runi. At the \lhc with $10\;\rm fb^{-1}$ obtained at
$10^{33}\;\rm cm^{-2}\,s^{-1}$, it is expected to be possible to
measure the electric charge of the top quark with an accuracy of 10\%.
For comparison, at a linear collider with $\sqrt{s}=500\;\rm GeV$ and
$\int {\cal{L}} dt = 200\;\rm fb^{-1}$, one expects that $q_{top}$ can
be measured with a precision of about 10\%
\cite{tesla_physics_tdr,nlc_snowmass01}.

The SM dictates that the top quark has the same
vector-minus-axial-vector ($V$-$A$) charged-current weak interaction
$\left(-i \frac{g}{\sqrt{2}}\,V_{tb}\,\gamma^\mu
  \frac{1}{2}(1-\gamma_5)\right)$ as all the other fermions. 
Neglecting the $b$-quark mass
this implies that the $W$ boson in top-quark decay
cannot be right-handed, i.e. have positive helicity.

One of the unique features of the top quark is that on average the top
quark decays before there is time for its spin to be depolarised by
the strong interaction \cite{bigi_1986}. Thus, the top quark
polarisation\footnote{The spin of an individual top quark cannot be
  measured, only the spin polarisation of an ensemble of top quarks.}
is directly observable via the angular distribution of its decay
products. This means that it should be possible to measure observables
that are sensitive to the {\bf top quark spin} via spin correlations
\cite{tspin_kuehn,czarnecki:1990pe,Bernreuther:1994cx,Bernreuther:1995cx,Dharmaratna:1996xd,Mahlon:1995zn,stelzer_willenbrock,Bernreuther:2005is}, and
references therein.

Another interesting aspect of the production of \ttbar\ pairs via the
strong coupling is an asymmetry in the rapidity-distribution of the
$t$ and $\bar{t}$ quarks
\cite{kuehn_rodrigo_prl,Kuhn:1998kw,bowsen_ellis_rainwater}.
This effect arises at next-to-leading order, and leads to a
forward-backward asymmetry of about 5\% in $t\bar{t}$ production at
the \tevatron.

Yukawa coupling is the Higgs coupling to fermions and thus relates the
fermionic matter content of the SM to the source of mass
generation, the Higgs sector \cite{higgs1,higgs2,higgs3}. In the
SM, the Yukawa coupling to the top quark, $y_t =
\sqrt{2}\, m_t/v$, is very close to unity. This theoretically
interesting value leads to numerous speculations that new physics
might be accessed via top quark physics \cite{willenbrock_0211067}.
The Yukawa coupling will be measured in the associated $t\bar{t}H$
production at the \lhc \cite{ATLAS_TDR2,Ball:2007zza,Aad:2008zza}.

%% file: top_production.tex
\section{Top-Quark Production and Decay}
% Author: Wolfgang Wagner
\label{sec:topProd}
In this chapter, we summarize the phenomenology of top-quark production 
at hadron colliders, limiting the discussion to SM processes.
Anomalous top-quark production and non-SM decays are covered in 
chapter~\ref{sec:newPhysics}. A detailed, recent review on
top-quark phenomenology can be found in~\cite{Bernreuther:2008ju}. 
The dominating source of top quarks at the Fermilab \tevatron 
and the Large Hadron Collider (\lhc) at CERN is the
production of \ttbar pairs by the strong interaction,
while the production of single top-quarks due to charged-current
weak interactions has subleading character.

%--------------------------------------------------
\paragraph{The Factorization Ansatz}
The underlying theoretical framework for the calculation of hadronic cross
sections is the QCD-improved parton model which regards a high-energy hadron 
$A$ as a composition of quasi-free partons (quarks and gluons) sharing the 
hadron momentum $p_A$. Each parton $i$ carries a different momentum fraction 
$x_i$ and its momentum is given by $p_i = x_ip_A$. 
Based on the factorization theorem cross sections are calculated as a 
convolution of parton distribution functions (PDF) $f_{i/A}(x_i, \mu^2)$ and 
$f_{j/B}(x_j, \mu^2)$ for the colliding hadrons ($A$, $B$) and the factorized 
hard parton-parton cross section $\hat{\sigma}_{ij}$:
\begin{equation}
  \sigma (AB\rightarrow t\tbar ) = \sum_{i, j=q,\bar{q},g} \int\, 
  \rmd x_i \rmd x_j \;
  f_{i/A} (x_i, \mu^2) f_{j/B} (x_j, \mu^2) \cdot
  \hat{\sigma}_{ij}(ij \rightarrow t\tbar; \hat{s}, \mu^2)
  \label{eq:factorization}  
\end{equation}
The hadrons $AB$ are either $p\pbar$ at the \tevatron or $pp$ at the \lhc.
The variable $\hat{s}$ denotes the square of the centre-of-mass energy of 
the colliding partons: $\hat{s} = (p_i+p_j)^2 = (x_ip_A + x_jp_B)^2$.
Neglecting terms proportional to the hadron masses we have 
$\hat{s} = x_i x_j s$. The sum in (\ref{eq:factorization}) runs over all pairs 
of partons $(i, j)$ contributing to the process.
The PDF $f_{i/A}(x_i, \mu^2)$ describes the probability density for finding a 
parton  $i$ inside the hadron $A$ carrying a momentum fraction $x_i$.
The PDFs as well as $\hat{\sigma}_{ij}$ have a residual dependence on the 
factorization and renormalization scale $\mu$ due to uncalculated higher orders.
In the case of top-quark production, one typically evaluates the cross sections 
at $\mu = \mtop$.
Strictly speaking one has to distinguish between the factorization scale 
$\mu_\mathrm{F}$ introduced by the factorization ansatz and the renormalization
scale $\mu_\mathrm{R}$ due to the renormalization procedure invoked to regulate
divergent terms in the perturbation series when calculating $\hat{\sigma}_{ij}$.
However, since both scales are to some extent arbitrary parameters most authors 
have adopted the practice to use only one scale 
$\mu = \mu_\mathrm{F} = \mu_\mathrm{R}$.
If the complete perturbation series could be calculated, the result for the cross 
section would be independent of $\mu$. 
However, at finite order in perturbation theory cross-section predictions
do depend on the choice of $\mu$, and the changes when varying $\mu$ between 
$\mtop/2$ and $2\,\mtop$ are usually quoted as an indicative theoretical 
uncertainty.

\begin{figure}[t]
    \begin{center}
    \subfigure[]{
      \epsfig{file=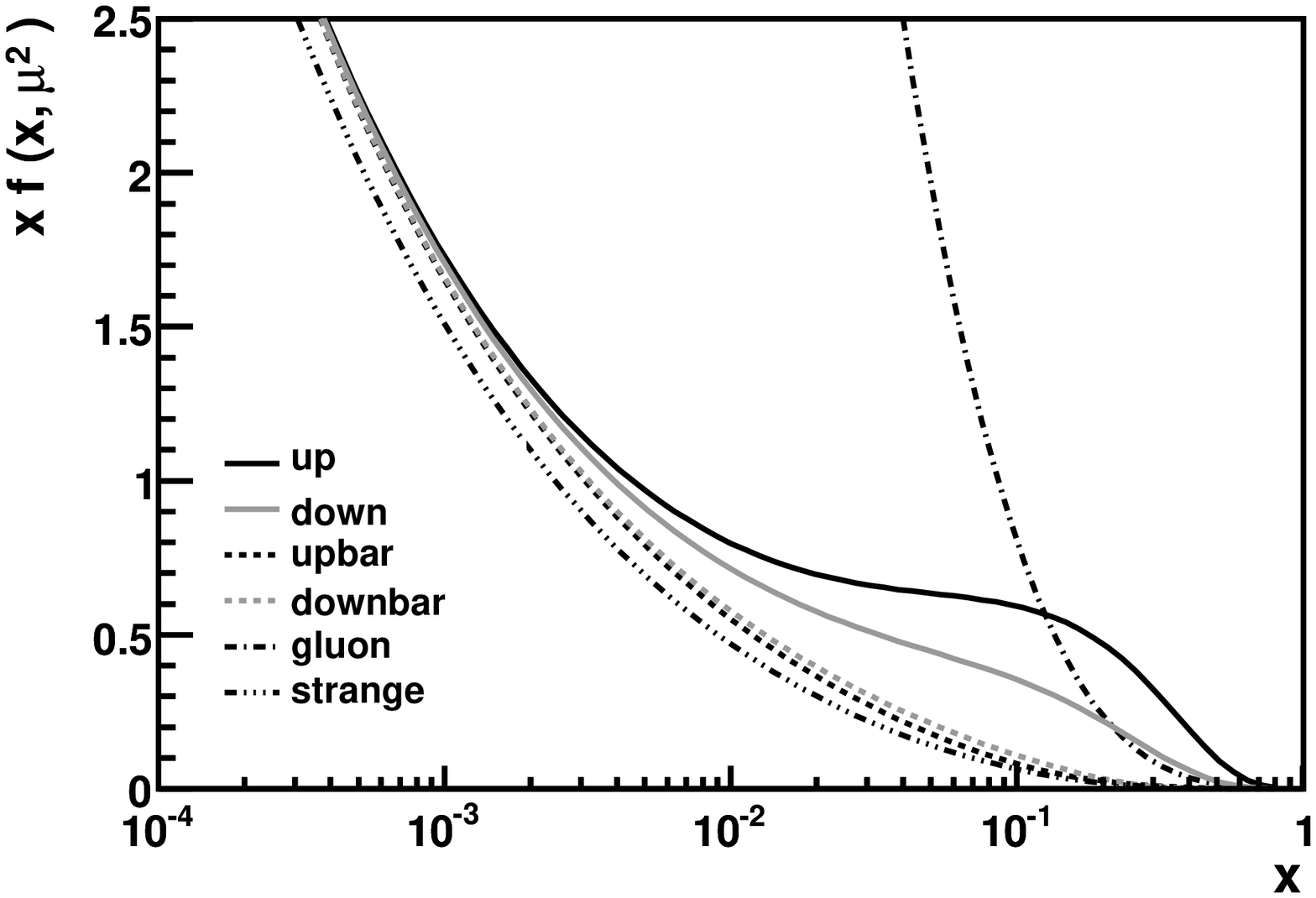, width=0.45\textwidth}
      \label{fig:pdfExample}}
    \subfigure[]{
      \epsfig{file=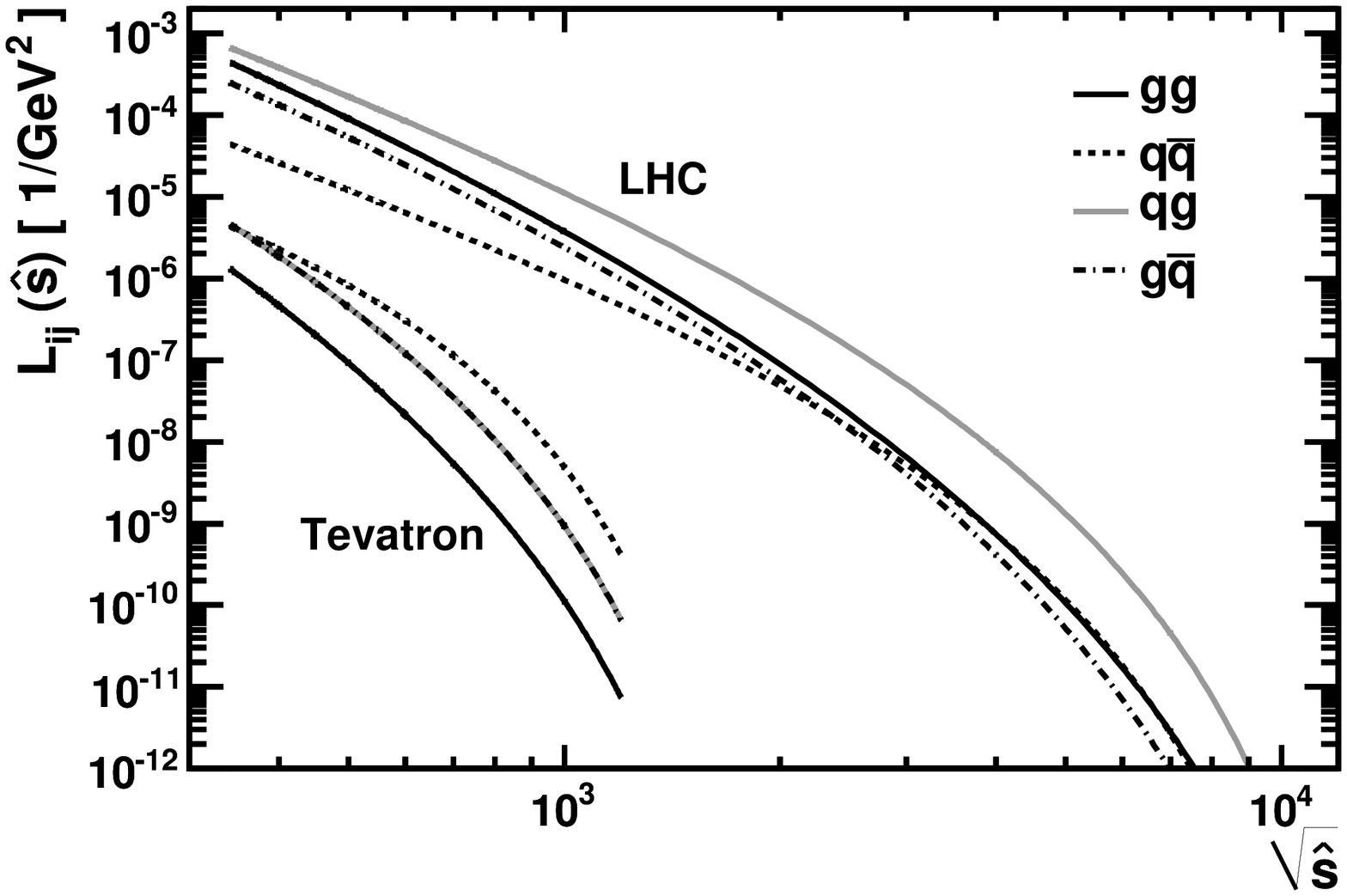, width=0.45\textwidth}
      \label{fig:partonlumi}}
    \end{center}
    \caption{\label{fig:pdf_parton}\subref{fig:pdfExample} PDFs of 
     $u$ quarks,
     $\bar{u}$ quarks, $d$ quarks, $\dbar$ quarks, $s$ quarks, and gluons 
     inside the proton. 
     The parametrization is CTEQ6.5M~\cite{Pumplin:2002vw}.
     The scale at which
     the PDFs are evaluated was chosen to be $\mu=175\;\mathrm{GeV}$ 
     ($\mu^2 = 30625\;\mathrm{GeV^2}$).
     \subref{fig:partonlumi} Parton luminosities for gluon-gluon, 
     quark-antiquark, quark-gluon, and gluon-antiquark interactions at
     the \tevatron and the \lhc~\cite{Moch:2008qy}.}
  \end{figure}

The factorization scheme serves as a method to systematically eliminate
collinear divergencies from $\hat{\sigma}_{ij}$
and absorb them into the PDFs which are extracted from global fits
to measurements of deep inelastic scattering 
experiments where either electrons, positrons or neutrinos collide with
nucleons~\cite{Bodek:1979rx,Adloff:1997mf,Chekanov:2005nn}. 
Different groups of physicists have derived parametrizations of proton PDFs 
from experimental 
data~\cite{Pumplin:2002vw,Martin:2007bv,Alekhin:2006zm,Blumlein:2006be}. 
As an example Fig.~\ref{fig:pdfExample} shows PDFs of the CTEQ 
collaboration~\cite{Pumplin:2002vw}.
For antiprotons the distributions in 
Fig.~\ref{fig:pdf_parton}\subref{fig:pdfExample} have to be reversed between
quark and antiquark.
The gluons start to dominate in the $x$ region below 0.15.
At the \tevatron, the $t\tbar$ production cross-section is driven
by the large $x$ region, since $\mtop$ is relatively large compared 
to the \tevatron beam energy ($\mtop/\sqrt{s}= 0.0875$).
At the \lhc ($\mtop/\sqrt{s}= 0.0125$) the lower $x$ region becomes 
more important.
To compare different collider setups it is instructive to compute 
parton luminosities
\begin{equation}
  L_{ij} (\hat{s}; s, \mu_F) = \frac{1}{s} \int_{\hat{s}}^{s}
  f_{i/A}\left(\frac{\tilde{s}}{s}\right) \; 
  f_{j/B}\left(\frac{\hat{s}}{\tilde{s}}\right)
  \; \frac{1}{\tilde{s}} \; \rmd\tilde{s} \ .
  \label{eq:partonlumis}
\end{equation}
The corresponding functions for the \tevatron and the \lhc are shown in 
Fig.~\ref{fig:pdf_parton}\subref{fig:partonlumi}~\cite{Moch:2008qy}. 
The increase of reach in $\sqrt{\hat{s}}$ is apparent.
In addition to that advantage, the \lhc in its final stage will have a 
luminosity which is 100 times larger than the one at the \tevatron.
It is also interesting to note that at the \lhc the $qg$ luminosity will
dominate, while at the \tevatron the $q\bar{q}$ luminosity is the largest
one.

\subsection{Top-Antitop-Quark Pair Production}
\begin{figure}[t]
    \begin{center}
    \subfigure[]{
    \includegraphics[width=0.15\textwidth]{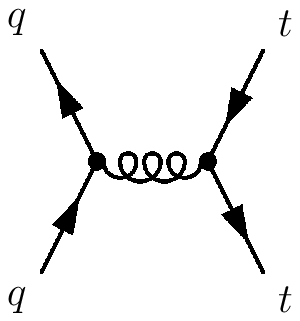}
      \label{subfig:qqtt}}  
    \hspace*{16mm}
    \subfigure[]{
    \includegraphics[width=0.15\textwidth]{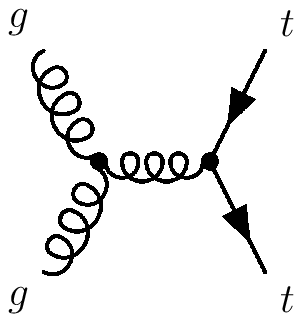}  \hspace*{4mm}
    \includegraphics[width=0.15\textwidth]{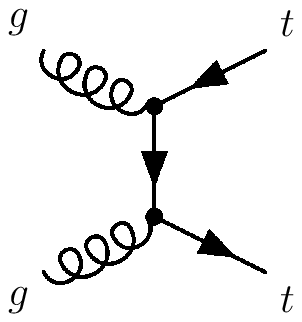}  \hspace*{4mm}
    \includegraphics[width=0.15\textwidth]{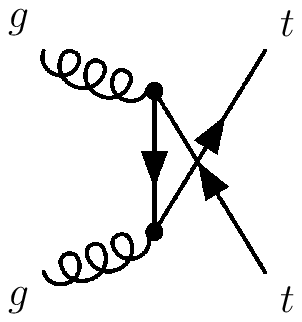}
      \label{subfig:ggtt}}
    \end{center}
    \caption{\label{fig:leadingOrderttbar} Feynman diagrams of the LO
      processes for $t\tbar$ production: 
      \subref{subfig:qqtt} quark-antiquark annihilation 
      ($q\qbar \rightarrow t\tbar$) and \subref{subfig:ggtt}
      gluon-gluon fusion ($gg \rightarrow t\tbar$).
     }
  \end{figure}

The cross section $\hat{\sigma}$ of the hard parton-parton process 
$ij \rightarrow t \tbar$ can be calculated in perturbative QCD. 
The leading order (LO) processes, contributing with $\alpha_s^2$ to the perturbation
series, are quark-antiquark annihilation, 
$q\qbar \rightarrow t\tbar$, and gluon-gluon fusion, $gg \rightarrow t\tbar$.
The corresponding Feynman diagrams for these processes are depicted
in Fig.~\ref{fig:leadingOrderttbar}.
The LO differential cross section for \ttbar
production via $q\bar{q}$ annihilation is given
by~\cite{bargerPhillips}
\begin{equation}
  \frac{\rmd\hat{\sigma}}{\rmd\hat{t}} \left( q\qbar \rightarrow t \tbar \right) =
  \frac{4\,\pi\,\alpha_s^2}{9\, \hat{s}^4} \cdot
  \left[ (m^2_t - \hat{t} )^2 + (m^2_t - \hat{u})^2 + 2m^2_t\hat{s} \right]
  \label{eq:qqttbar}
\end{equation}
where $\hat{s}$, $\hat{t}$ and $\hat{u}$ are the Lorentz-invariant 
Mandelstam 
variables of the partonic process. 
They are defined by $\hat{s} = (p_q + p_{\,\qbar})^2$,
$\hat{t} = (p_q - p_{t})^2$ and $\hat{u} = (p_q - p_{\,\tbar})^2$
with $p_i$ being the corresponding momentum four-vector of the quark $i$,
while $m_t$ denotes the top-quark mass.

A full calculation of next-to-leading order (NLO) corrections contributing
in order $\alpha_s^3$ to the inclusive parton-parton cross-section 
for heavy quark pair-production was performed in the late 1980s and
early 1990s~\cite{nason1988,beenakker1989}.
The NLO calculations involve virtual contributions to the LO processes, gluon 
bremsstrahlung processes ($q\qbar \rightarrow t\tbar + g$
and $gg \rightarrow t\tbar + g$) as well as flavor excitation processes 
like $g+q(\qbar) \rightarrow t\tbar + q(\qbar)$.
For the NLO calculation of the total hadron-hadron cross section 
$\sigma(AB \rightarrow t\tbar)$ to be consistent, one has to use NLO
determinations of the coupling constant $\alpha_s$ and the PDFs.
All quantities have to be defined in the same renormalization and factorization 
scheme as different approaches distribute radiative corrections differently 
among $\hat{\sigma}(t\bar{t})$, the PDFs and $\alpha_s$.
Most authors use the $\mathrm{\overline{MS}}$ 
(modified minimal subtraction) scheme~\cite{Bardeen:1978yd} or an extension 
of the latter. 
Corrections at NLO including full spin information are also 
available~\cite{Bernreuther:2001rq}.
Recently, analytic expressions for the NLO QCD corrections to inclusive 
$t\tbar$ production were derived~\cite{Czakon:2008ii}.

At energies close to the kinematic threshold, $\hat{s} = 4\,\mtop^2$,
$q\bar{q}$ annihilation is the dominant process, if the
incoming quarks are valence quarks, as is the case of $p\pbar$
collisions. At the \tevatron, about 85\% of $\sigma(t\tbar)$
is due to $q\bar{q}$ annihilation~\cite{Cacciari:2003fi}. 
At higher energies, the $gg$-fusion process dominates for both $p\pbar$ and $pp$ 
collisions, see also Fig.~\ref{fig:pdf_parton}\subref{fig:partonlumi}.
That is why, one can built the \lhc as a $pp$ machine without compromising  
the production cross-section.
Technically, it is of course much easier to operate a $pp$ collider,
since one spares the major challenge to produce high antiproton currents
in a storage ring. 
For the \tevatron, the ratio of NLO over LO cross-sections for $gg$ fusion 
is predicted to be 1.8 at $\mtop = 175\,\mathrm{GeV}/c^2$, for $q\bar{q}$ 
annihilation the value is only about 1.2~\cite{laenen1994}.
Since the annihilation process is dominating, the overall NLO enhancement
is about 1.25.

Contributions to $\sigma(t\tbar)$ due to radiative corrections 
are large in the region near threshold ($\hat{s} = 4\,\mtop^2$) and at high
energies ($\hat{s} > 400\;\mtop^2$). 
Near threshold the cross section is enhanced due to initial state gluon 
bremsstrahlung~\cite{laenen1992}. This effect is important at the \tevatron, 
but less relevant for the \lhc 
where gluon splitting and flavour excitation are increasingly important
effects.
The calculation at fixed NLO accuracy has been refined to systematically 
incorporate higher order corrections due to soft gluon 
radiation~\cite{Moch:2008qy,Cacciari:2008zb,Kidonakis:2008mu}. 
Technically, this is done by applying an integral transform (Mellin transform) 
to the cross section:
$\sigma_N (t\tbar)\equiv\int_0^1 \rho^{N-1}\;\sigma(\rho; t\tbar)\;\rmd\rho$,
where $\rho = 4\,\mtop^2/s$ is a dimensionless parameter.
In Mellin moment space the corrections due to soft gluon radiation
are given by a power series of logarithms $\ln N$.
For $\mu = \mtop$ the corrections are positive at all orders. 
Therefore, the resummation of the soft gluon logarithms yields an 
increase of $\sigma(t\tbar)$ with respect to the NLO value.
\renewcommand\arraystretch{1.2}
Calculations by different groups implementing the resummation 
approach~\cite{Moch:2008qy,Cacciari:2008zb,Kidonakis:2008mu} 
are in good agreement. Exemplarily, the results 
of~\cite{Cacciari:2003fi,Moch:2008ai} are quoted in Table~\ref{tab:ttbarXSection}.
\begin{table}[!t]
  \caption{\label{tab:ttbarXSection} Cross-section predictions for $t\tbar$
    production at approximate NNLO in perturbation theory including
    the resummation of initial-state gluon-bremsstrahlung.
    The cross sections are given for $p\pbar$ collisions at the \tevatron
    ($\sqrt{s} = 1.8\;\mathrm{TeV}$~\cite{Cacciari:2003fi} and 
     $\sqrt{s} = 1.96\;\mathrm{TeV}$~\cite{Moch:2008ai})
    and $pp$ collisions at the \lhc~\cite{Moch:2008ai},
    assuming $\mtop = 175\,\mathrm{GeV}/c^2$.   
    To derive the central values, the factorization and renormalization scale 
    is set to $\mu = \mtop$. The PDF parametrization CTEQ6.6 is used.}
  \begin{center}
  \begin{tabular}{ccc}
    \hline
    Initial State & $\sqrt{s}$ & $\sigma(t\tbar)$ \\
    \hline
    $p\bar{p}$ & 1.8 TeV  &  $5.19^{+0.52}_{-0.68}\,\mathrm{pb}$ \\  
    $p\bar{p}$ & 1.96 TeV &  $6.90^{+0.46}_{-0.64}\,\mathrm{pb}$ \\
    $pp$       & 10 TeV   &  $374^{+18}_{-33}\,\mathrm{pb}$ \\
    $pp$       & 14 TeV   &  $827^{+27}_{-63}\,\mathrm{pb}$ \\
    \hline
  \end{tabular}
  \end{center}
\end{table}
The quoted uncertainties include the uncertainty due to the choice of 
$\mu_\mathrm{F}$ and $\mu_\mathrm{R}$, and the uncertainty associated
with the PDF parametrization. 

\begin{figure}[t]
  \begin{center}
    \epsfig{file=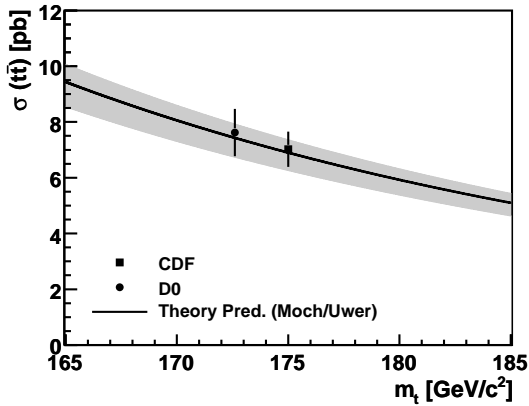, width=7.0cm}
  \end{center}
  \caption{\label{fig:sigmattbar} The $t\tbar$ cross ection in $p\bar{p}$
  collisions at $\sqrt{s} = 1.96\,\mathrm{TeV}$ as a function of $m_t$. 
  The measurements by CDF~\cite{cdfnote9448} and 
  D\O~\cite{Abazov:2008gc} are compared to the theory 
  prediction~\cite{Moch:2008ai}.}
\end{figure}

\renewcommand\arraystretch{1.0}
The cross-section predictions strongly depend on $m_t$, which is illustrated 
in Fig.~\ref{fig:sigmattbar}
that shows the predictions for the \tevatron in comparison to measurements by 
CDF~\cite{cdfnote9448} and D\O ~\cite{Abazov:2008gc}.
Within the uncertainties the measurements agree well with the theoretical
predictions.

%---------------------------------------------------------------------------    
\subsection{Single Top-Quark Production} 
\label{sec:singleTopTheo}

\begin{figure}[t]
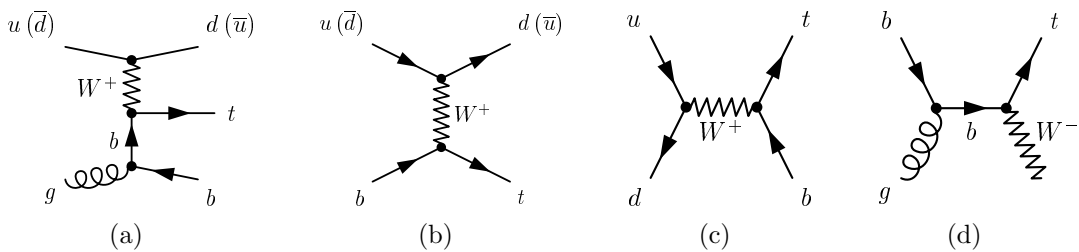

  \begin{center}    
    \subfigure[]{
      \epsfig{file=fig_wagner/singTopWg1Pict.epsi, height=26mm}
      \label{fig:WgNLO}
    } \hspace*{3mm}
    \subfigure[]{
      \epsfig{file=fig_wagner/singTopWg3Pict.epsi, height=26mm}
      \label{fig:WgLO} 
    } \hspace*{3mm}
    \subfigure[]{
      \epsfig{file=fig_wagner/singTopSChanPict.epsi, height=26mm}
      \label{fig:schan}
    }  \hspace*{3mm}
    \subfigure[]{
      \epsfig{file=fig_wagner/singTopAssocPict.epsi, height=26mm}
      \label{fig:Wt}
    }
  \end{center}
  \caption{\label{fig:singleTop} Representative Feynman diagrams for the three
    single top-quark production modes. \subref{fig:WgNLO} and \subref{fig:WgLO} 
    show $\mathit{Wg}$-fusion graphs, 
    \subref{fig:schan} the $s$-channel process, and \subref{fig:Wt} associated 
    production. 
    The graphs show single top-quark production, the diagrams for single 
    antitop-quark production 
    can be obtained by interchanging quarks and antiquarks.
   }
\end{figure}

Top quarks can be produced singly via electroweak interactions involving
the $Wtb$ vertex. There are three production modes which are distinguished by
the virtuality $Q^2$ of the $W$ boson ($Q^2 = - q^2$, where $q$ is the 
four-momentum of the $W$):  
\begin{enumerate}
\item the {\bf t channel} ($q^2 = \hat{t}\;$): 
  A virtual $W$ boson strikes a $b$ quark (a sea quark) inside the proton. 
  The $W$ boson is spacelike ($q^2  < 0$). 
  This mode is also known as $\mathit{Wg}$ fusion, since the $b$ quark 
  originates from a gluon splitting into a $b\overline{b}$ pair.
  Another name used for this process is $tq\bar{b}$ production. 
  Feynman diagrams representing this process are shown
  in Fig.~\ref{fig:singleTop}\subref{fig:WgNLO} and 
  Fig.~\ref{fig:singleTop}\subref{fig:WgLO}.
  Production in the $t$ channel is the dominant source of single top quarks 
  at the \tevatron and at the \lhc.
\item the {\bf s channel} ($q^2 = \hat{s}\;$): 
  This production mode is of Drell-Yan type and is also called $t\bbar$ 
  production. 
  A timelike $W$ boson with $q^2 \geq (\mtop + m_b)^2$ 
  is produced by the fusion of two quarks belonging to an SU(2)-isospin doublet. 
  See Fig.~\ref{fig:singleTop}\subref{fig:schan} for the Feynman diagram. 
\item {\bf associated production}: The top quark is produced in association with a
  real (or close to real) $W$ boson ($q^2 = M_W^2$). 
  The initial $b$ quark is a sea quark inside
  the proton. Fig.~\ref{fig:singleTop}\subref{fig:Wt} shows the Feynman diagram.
  The cross section is negligible at the \tevatron, but of
  considerable size at \lhc energies where associated $Wt$ production even supercedes 
  the $s$ channel. 
\end{enumerate}
In $p\pbar$ and $pp$ collisions the cross section is dominated by 
contributions from $u$ and $d$ quarks coupling to the $W$ boson on one
hand side of the Feynman diagrams. 
There is also a small contribution from the second weak-isospin 
quark doublet $(c, s)$; an effect
of about 2\% for $s$- and 6\% for $t$-channel production~\cite{heinson1997}.  
In the following, we will only consider single top-quark production via a $Wtb$ 
vertex, since the production channels involving $Wtd$ or $Wts$ vertices are 
strongly suppressed due to small CKM matrix elements.
Assuming unitarity of the three-generation CKM matrix one obtains
$|V_{td}| = 8.74^{+0.26}_{-0.37}\cdot 10^{-3}$ and 
$|V_{ts}| = (40.7\pm 1.0)\cdot 10^{-3}$~\cite{Amsler:2008zzb}.
In the following paragraphs the status of theoretical cross-section
predictions for the three single top-quark processes is reviewed.

\paragraph{t-channel Production}
\label{sec:wgfusion}
The $t$-channel process was first suggested as a potentially interesting source
of top quarks in the mid 1980s~\cite{WillenbrockDicus1986,DawsonWillenbrock1987} 
and early 1990s~\cite{Yuan1990}. If the $b$ quark is taken to be massless, 
a singularity arises, when computing the diagram in 
Fig.~\ref{fig:singleTop}\subref{fig:WgNLO} 
in case the final $\bar{b}$ quark is collinear with the incoming gluon. 
In reality, the non-zero mass of the $b$ quark regulates this collinear divergence. 
When calculating the total cross section, the collinear singularity manifests
itself as terms proportional to $\ln ((Q^2 + \mtop^2)/m_b^2)$,
causing the perturbation series to converge rather slowly. 
This difficulty can be obviated by introducing a PDF for the $b$ quark, 
$b(x, \mu^2)$, which effectively resums the logarithms to all orders of 
perturbation theory and implicitly describes the splitting of gluons into 
$b\bbar$ pairs inside the colliding hadrons~\cite{stelzerSullivan1997}.
Once a $b$-quark PDF is introduced into the calculation, the LO process is 
$q_i + b \rightarrow q_j + t$ as shown in 
Fig.~\ref{fig:singleTop}\subref{fig:WgLO}. 
In this formalism the process shown in 
Fig.~\ref{fig:singleTop}\subref{fig:WgNLO}
is a higher order correction which is already partially included
in the $b$-quark PDF.
The remaining contribution is of order $1/\ln ((Q^2 + \mtop^2)/m_b^2)$ 
with respect to the LO process.
Additionally, there are also corrections of order $\alpha_s$.
\renewcommand\arraystretch{1.2}
Cross-section calculations exist at 
NLO~\cite{harris2002,sullivan2004,Campbell:2004ch} 
and are summarised in Table~\ref{tab:singleTopCrossSection}.
The cross sections given for the \tevatron are the sum of top-quark and 
antitop-quark production. 
\begin{table}[t]
  \caption{\label{tab:singleTopCrossSection} Predicted total cross-sections 
    for single top-quark production-processes.
    The cross sections are given for $p\pbar$ collisions at the \tevatron
    ($\sqrt{s} = 1.96\;\mathrm{TeV}$) and
    $pp$ collisions at the \lhc ($\sqrt{s} = 14\;\mathrm{TeV}$).
    The cross sections of the $t$-channel process are taken 
    from~\cite{harris2002,sullivan2004} and are evaluated with 
    CTEQ5M1 PDFs.
    The values for $s$-channel and associated production are taken 
    from~\cite{Kidonakis:2006bu,Kidonakis:2007ej} using the 
    MRST2004 NNLO PDFs.
    All cross sections are evaluated at $\mtop = 175\;\mathrm{GeV}/c^2$. 
    }
  \begin{center}
  \begin{tabular}{lcccc}
  \hline
  Process & $\sqrt{s}$ & $\sigma_{tq\bar{b}}$ & 
  $\sigma_{t\bar{b}}$ & $\sigma_{Wt}$ \\ \hline
  $p\pbar \rightarrow t / \tbar$ & 1.96 TeV & $1.98^{+0.28}_{-0.22}\;\mathrm{pb}$ & 
  $1.02\pm0.08\;\mathrm{pb}$ & 
  $0.25\pm 0.03\;\mathrm{pb}$ \\ \hline
  $p p \rightarrow t$            & 14.0 TeV & $156\pm 8\;\mathrm{pb}$ & 
  $7.23^{+0.55}_{-0.47}\;\mathrm{pb}$ & $41.1\pm 4.2\;\mathrm{pb}$ \\ 
  $p p \rightarrow \tbar$        & 14.0 TeV & $91 \pm 5\;\mathrm{pb}$ & 
  $4.03^{+0.14}_{-0.16}\;\mathrm{pb}$ & $41.1\pm 4.2\;\mathrm{pb}$ \\
  \hline
  \end{tabular}
  \end{center}
\end{table}
In $pp$ collisions at the \lhc the $t$-channel cross section 
$\sigma_{tq\bar{b}}$ differs 
for top-quark and antitop-quark production, which are therefore treated
separately in Table~\ref{tab:singleTopCrossSection}. 
The contribution of soft-gluon corrections near the partonic threshold
has been investigated, but was found to be small at the 
\tevatron~\cite{Kidonakis:2006bu}. At the \lhc, the resummation approach
is not meaningful, when calculating $\sigma_{tq\bar{b}}$, since threshold 
corrections do not dominate the perturbative expansion~\cite{Kidonakis:2007ej}.

The ratio of $\sigma_{tq\bar{b}}/\sigma_{t\tbar}$ is about 30\%,
for the \tevatron as well as for the \lhc. 
The uncertainties quoted in Table~\ref{tab:singleTopCrossSection}
include the uncertainties due to the factorization scale $\mu$,
the choice of PDF parameterization, and the uncertainty in $\mtop$.
The factorization scale uncertainty is $\pm 4\%$ at the \tevatron and
$\pm 3\%$ at the \lhc.
The central value was calculated with $\mu^2 = Q^2+\mtop^2$ for the
$b$-quark PDF. 
The scale for the light-quark PDFs was set to $\mu^2 = Q^2$.
\renewcommand\arraystretch{1.0}
The dependence of the $t$-channel single top-quark cross-section is approximately 
linear in the relevant mass range, with a slope of
$-1.6\%\cdot\sigma_{tq\bar{b}}/1\,\mathrm{GeV}/c^2$ at the \tevatron and
$-0.7\%\cdot\sigma_{tq\bar{b}}/1\,\mathrm{GeV}/c^2$ at the \lhc.

An interesting feature of single top-quark production 
(in the $t$-channel and in the $s$-channel)
is that in its rest frame the top quark is 100\% polarized along
the direction of the $d$ quark 
($\dbar$ quark)~\cite{carlson1993,mahlon1997,heinson1997,stelzer1998}.
The reason for this is that the $W$ boson couples only to 
fermions with left-handed chirality.
Consequently, the ideal basis to study the top-quark spin is the one
which uses the direction of the $d$ quark as the spin axis~\cite{mahlon1997}.
In $p\pbar$ collisions at the \tevatron $t$-channel single top-quark production 
proceeds via
$ug \rightarrow td\bbar$ in about 77\% of the cases. The $d$ quark can then be
measured by the light-quark jet in the event.
The top-quark spin is best analyzed in the spectator basis for which
the spin axis is defined along the light-quark jet-direction.
However, 23\% of the events proceed via $\dbar g \rightarrow t\ubar \bbar$, 
in which case the $\dbar$ quark is moving along one of the beam directions.
For these events the spectator basis is not ideal, but since the 
light-quark jet occurs typically at high rapidity the dilution is small.
In total, the top quark has a net spin polarization of 96\% along the
direction of the light-quark jet in $tq\bbar$ production at the 
\tevatron~\cite{mahlon1997}.
In $s$-channel events the best choice is the antiproton-beam direction 
as spin basis, leading to a net polarization of 98\%~\cite{mahlon1997}. 

Since the top quark does not hadronize, its decay products
carry information about the top-quark polarization. 
A suitable variable to investigate the top-quark polarization is the 
angular distribution of electrons and muons originating
from the decay chain $t \rightarrow W^+b$, $W^+ \rightarrow \ell^+\nu_\ell$.
If $\theta_{q\ell}$ is the angle between the charged lepton and the
light-quark jet-axis in the top-quark rest-frame, 
the angular distribution is given by
$dN/d\cos \theta_{q\ell} = \frac{1}{2}\,(1+P\cos \theta_{q\ell})$~\cite{Jezabek:1994qs} with $P=1$ corresponding to fully polarized top quarks. 

\paragraph{s-channel Production}
The $s$-channel-production mode of single top-quarks ($t\bbar$ production)  
probes a complementary 
aspect of the weak charged-current interaction of the top quark, since it is
mediated by a timelike $W$ boson as opposed
to a spacelike $W$ boson in the $t$-channel process.
The leading order $s$-channel process is depicted in 
Fig.~\ref{fig:singleTop}\subref{fig:schan}.
Some first order ($\alpha_s$) corrections have the same initial and final states as
the $t$-channel diagrams shown in 
Fig.~\ref{fig:singleTop}\subref{fig:WgNLO}. 
However, these two classes of diagrams
do not interfere because they have a different colour structure.
The $t\bbar$ pair in the $t$-channel process is in a colour-octet
state, since it originates from a gluon. In $s$-channel production
the $t\bbar$ pair forms a colour-singlet because it comes from 
a $W$. The different colour structure implies that both groups of
processes must be separately gauge invariant and, therefore, they
do not interfere~\cite{WillenbrockDicus1986,Yuan1990,heinson1997}.

Cross sections for $t\bbar$ production are available at NLO 
including the resummation of soft gluon 
corrections~\cite{Kidonakis:2006bu,Kidonakis:2007ej}, see
Table~\ref{tab:singleTopCrossSection}.
The predictions for the $s$-channel cross section have a smaller uncertainty 
from the PDFs than for the $t$-channel 
because they do not
depend as strongly on the gluon PDF as the $t$-channel calculation does. 
The uncertainty in $m_t$ leads to an uncertainty in the cross
section of about 6\% ($\Delta m = 2.1\,\mathrm{GeV}/c^2$). 

The ratio of cross sections for the $t$-channel and $s$-channel mode
is 1.8 at the \tevatron ($\sqrt{s} = 1.96\;\mathrm{TeV}$) and 22
at the \lhc. In $pp$ collisions the gluon-initiated processes,
$t\tbar$ production and $t$-channel single top-quark production, 
dominate by far 
over $s$-channel single top-quark production which is 
a $q\bar{q}$-annihilation process.

\paragraph{Wt Production}
The third single top-quark production mode is characterised by the
associated production of a top quark and an on-shell (or close to
on-shell) $W$ boson.
Cross-section predictions are given in Table~\ref{tab:singleTopCrossSection}. 
It is obvious that $\mathit{Wt}$ production is
negligible at the \tevatron, but is quite important at the 
\lhc where it even exceeds the $s$-channel production-rate.
The errors quoted in Table~\ref{tab:singleTopCrossSection} include
the uncertainty due to the choice of the factorization scale
($\pm 10\%$ at the \lhc) and the PDFs ($\pm 2.5\%$ at the \lhc). 
The uncertainty in $\mtop$ 
($\Delta m = 2.1\,\mathrm{GeV}/c^2$) causes a spread of the cross section
by $\pm 3.5\%$.
At NLO, corrections to the $Wt$ cross section arise
that correspond to LO $gg\rightarrow t\bar{t}$ production with
one subsequent top-quark decay. When calculating the cross section
one has to account for that to avoid double 
counting~\cite{Tait:1999cf,Belyaev:2000me}.
First studies~\cite{Ball:2007zza,Aad:2009wy} show that $\mathit{Wt}$ 
production will be observable at the \lhc with data corresponding to 
about $10\,\mathrm{fb^{-1}}$ of integrated luminosity.

\subsection{Top-Quark Decay}
\label{sec:topdecay}
Within the SM, top quarks decay predominantly into a $b$ quark and 
a $W$ boson, while the decays $t\rightarrow d + W^+$ and $t\rightarrow s + W^+$ 
are strongly CKM-suppressed and are therefore neglected in the 
following discussion. 
The top-quark decay rate, including first order QCD corrections, is given by
\begin{equation}
  \Gamma_{t} = \frac{G_F\;\mtop^3}{8\,\pi\,\sqrt{2}}\;
  \; \left| V_{tb} \right|^2 \;
  \left( 1 - \frac{M_W^2}{\mtop^2}\right)^2\;
  \left( 1 + 2\; \frac{M_W^2}{\mtop^2}\right)\;
  \left[ 1 - \frac{2\,\alpha_s}{3\,\pi} \cdot 
  f\left(y\right)
  \right]
  \label{eq:topdecay}
\end{equation}
with $y=(M_W/\mtop)^2$ and 
$f(y) = 2\,\pi^2/3-2.5-3y+4.5y^2-3y^2\ln{y}$~\cite{Kuhn:1996ug,jezabekKuehn1989,jezabekKuehn1988}.
The QCD corrections of order $\alpha_s$ lower the LO decay rate by $-10\%$,
yielding $\Gamma_t =1.34\;\mathrm{GeV}$ at 
$\mtop = 172.6\;\mathrm{GeV}/c^2$. 
The decay width for events with hard gluon radiation ($E_g > 20\,\mathrm{GeV}$)
in the final state has been estimated to be 5 -- 10\% of $\Gamma_\mathrm{top}$,
depending on the gluon jet definition 
(cone size $\Delta R = 0.5$ to 1.0)~\cite{mrenna92}.  
Electroweak corrections to $\Gamma_\mathrm{top}$ increase the decay width by 
$\delta_\mathrm{EW}=+1.7\%$~\cite{denner1991,migneron1991}.
Taking the finite width of the $W$ boson into account leads to a
negative correction $\delta_\Gamma=-1.5\%$ such that $\delta_\mathrm{EW}$ and
$\delta_\Gamma$ almost cancel each other~\cite{jezabekKuehn1993}.

The large $\Gamma_t$ implies a very short lifetime of 
$\tau_t = 1/\Gamma_t \approx 5\cdot10^{-25}\,\mathrm{s}$ 
which is smaller than the characteristic formation time of hadrons 
$\tau_\mathrm{form} \approx 1\,\mathrm{fm}/c\approx 3\cdot 10^{-24}\;\mathrm{s}$.
In other words, top quarks decay before they can couple to light quarks and 
form hadrons. The lifetime of $t\tbar$ bound states, toponium, is too small, 
$\Gamma_{t\tbar} \sim 2\,\Gamma_{t}$, to allow for a proper
definition of a bound state with sharp binding energy, a 
feature already pointed out in the early 1980s~\cite{kuehn1981}.

The decay amplitude is dominated by the contribution from longitudinal $W$ 
bosons because the decay rate of the longitudinal component scales 
with $\mtop^3$, while the decay rate into transverse $W$ bosons increases
only linearly with $\mtop$. 
In both cases the $W^+$ couples solely to $b$ quarks of left-handed chirality,
which translates into left-handed helicity, since the $b$ quark is effectively 
massless compared to the energy scale set by $\mtop$.
If the $b$ quark is emitted antiparallel to the top-quark spin axis, 
the $W^+$ must be longitudinally polarized, $h^W$ = 0, to conserve angular
momentum.
If the $b$ quark is emitted parallel to the top-quark spin axis, the $W^+$ boson
has helicity $h^W = -1$ and is transversely polarized. 
$W$ bosons with positive helicity are thus forbidden in top-quark decays due to 
angular momentum conservation, assuming $m_b=0$. 
The ratios of decay rates into the three $W$ helicity-states are given 
by~\cite{Kuhn:1996ug}:
\begin{equation} 
  \Gamma(h^W=-1):\Gamma(h^W=0):\Gamma(h^W=+1) = 1 : 
  \frac{\mtop^2}{2\,M_W^2}:0.
\end{equation}
In the decay of antitop quarks negative helicity is forbidden.
Strong NLO corrections to the decay-rate ratios lower the fraction of 
longitudinal $W$ bosons by 1.1\% and increase the fraction of left-handed 
$W$ bosons by 2.2\%~\cite{Fischer:2000kx,Fischer:2001gp}. 
Electroweak and finite widths effects have even smaller effects on the helicity 
ratios, inducing corrections at the per-mille level~\cite{Do:2002ky}.
Taking the non-zero $b$-quark mass into account yields very small 
corrections, leading to a non-zero fraction of $W$ bosons with
positive helicity 
$\Gamma_+/\Gamma = 3.6\cdot 10^{-4}$ at Born level~\cite{Fischer:2000kx}. 

The spin orientation (helicity) of $W$ bosons from top-quark decays is 
propagated to its decay products. In case of leptonic $W$ decays the polarization 
is preserved
and can be measured~\cite{jezabekKuehn1994}.
The angular distribution of charged leptons from $W$ decays originating
from top quarks is given by
\begin{equation}
  \frac{1}{\Gamma}\frac{\rmd\Gamma}{\rmd\cos\theta_\ell} = \frac{3}{4}
  \frac{\mtop^2 \sin^2 \theta_\ell+ M_W^2 (1-\cos \theta_\ell)^2}{\mtop^2 +
  2\,M_W^2}   
\end{equation}
where $\pi-\theta_\ell$ is the angle between the $b$ quark direction and 
the charged lepton in the $W$ boson rest frame~\cite{Mahlon:1998uv}.

Given that the top quark decays almost 100\%
of the time as $t\rightarrow Wb$, typical final states for the leading
pair-production process can therefore be divided into three classes:
\smallskip

\centerline{
\begin{tabular}{llr}
{A.} & $t\overline t \to W^+\,b\,W^-\,\overline b \to
q\,\overline q'\, b\, q''\,\overline q'''\, \overline b$, & 
\hfill(46.2\%)\\*[1mm]
{B.} & $t\overline t \to W^+\,b\,W^-\,\overline b \to
q\,\overline q'\, b\, \ell\,
 \overline\nu_{\ell}\, \overline b\/ +
\overline \ell\,
 \nu_{\ell}\, b\, q\,\overline q'\, \overline b$,  & (43.5\%) \\*[1mm]
{C.} & $t\overline t \to W^+\,b\,W^-\,\overline b \to
\overline\ell\, \nu_{\ell}\, b\, \ell'\,
           \overline\nu_{\ell'}\, \overline b$, & (10.3\%)\\*[1mm]
\end{tabular}}
\smallskip

\noindent The quarks in the final state evolve into jets of hadrons.
A, B, and C are referred to as the all-jets, lepton+jets ($\ell
+$jets), and dilepton ($\ell\ell$) channels, respectively. The
relative contribution of the three channels A, B, C, including
hadronic corrections, are given in parentheses above. While $\ell$ in
the above processes refers to $e$, $\mu$, or $\tau$, most of the
results to date rely on the $e$ and $\mu$ channels. Therefore, in what
follows, $\ell$ will be used to refer to $e$ or $\mu$, unless noted
otherwise.

%% file: detectors.tex
\section{Hadron Colliders and Collider Detectors\label{sec:colliders}} 

The first hadron collider, the ISR at CERN, had 
to overcome two initial obstacles. The first was low 
luminosity, which steadily improved over time. The 
second was the broad angular spread of interesting events 
for which it was noted~\cite{Jacob:1984vc} that one needs 
`` sophisticated detectors covering at least the 
whole central region (45$^o~<\theta<~$135$^o$) 
and full azimuth''. This statement 
reflects the major revelation of the period, namely    
that hadrons have partonic substructure producing 
a strong hadronic yield at 
large transverse momentum ($p_T$). Caught unprepared, 
the ISR missed the discovery of the J/$\psi$ and 
later missed the $\Upsilon$, though it did make 
important contributions in other areas and paved the way 
for future hadron colliders. 
The partonic substructure of colliding hadrons 
leads to multiple production mechanisms, ($q\bar{q},~ qq',~qg,~gg$...) 
and a broad range of center of mass 
(CM) energies that allow one to probe mass scales
far above or below the average constituent 
CM energy. A side effect is the absence of a 
beam energy constraint so that event 
momentum and energy generally only balance 
in the plane transverse to the beam axis. 
Another aspect of hadron collisions is the fact that 
cross-sections for interesting processes are 
many orders of magnitude below the total inelastic 
cross-section ($\sigma_{tot}$). At the Fermilab Tevatron the value of 
$\sigma_{tot}$ is $\sim~60$~mb and rises to $\sim$100~mb at the LHC, 
where standard model (SM) Higgs could have a production cross-section 
of order $10^{-10}~\times~\sigma_{tot}$ or smaller, depending on $M_H$. 
High luminosities also mean multiple interactions per beam crossing. 
Hadron beams are segmented into bunches, each containing $10^{11}~-~10^{12}$
protons or anti-protons. Tevatron bunches are separated in time by 
396 ns for a bunch crossing rate of 2.5 MHz. The Tevatron currently attains 
peak luminosities in excess of $\mathcal{L}~=~3.5\times 10^{32}$~cm$^{-2}$~$s^{-1}$ 
with an average of 10 $p\bar{p}$ interactions per crossing. 
The LHC will have a bunch spacing of as little 
as 25 ns for a crossing rate of 40 MHz. 
There are expected to be an average of $\sim$25 
interactions per crossing at the design luminosity of 
$\mathcal{L}~=~10^{34}$~cm$^{-2}$~$s^{-1}$, corresponding 
to the production of as many as $10^4$ particles every 100 ns. 

In spite of a very challenging environment, hadron collider 
experiments have produced new discoveries as well as precision 
measurements. The cross-section for W production 
at the CERN SPS was $10^{-8}~\times~\sigma_{tot}$ at the time of 
its discovery~\cite{Arnison:1983rp,Banner:1983jy} and hadron collider 
experiments have since measured the W mass to high precision: 
$M_W(CDF\oplus D0)~=80.432\pm~0.039$~GeV/c$^2$~\cite{:2008ut}.
Similarly, $t\bar{t}$  production was observed at 
$10^{-10}~\times~\sigma_{tot}$ at the Fermilab 
Tevatron~\cite{Abe:1995hr,Abachi:1995iq} where 
$M_t$ has since been measured to an accuracy of better than 
1\%~\cite{:2009ec}, as discussed below.  
For the discovery of $t\bar{t}$  production, hermetic calorimetry and 
very good lepton identification enabled signal events to be efficiently 
accumulated and reconstructed and a variety of techniques 
were used to distinguish signal from backgrounds. Though there  
was considerable skepticism about the utility of silicon microstrip 
detectors in a hadron collider environment~\cite{Incandela:2007zz}, 
CDF silicon detectors~\cite{Amidei:1994iq,Cihangir:1994gx} 
dispelled all doubts, providing high precision tracking 
that made it possible to select extremely 
pure event samples by reconstructing the displaced vertices of $b$
hadron decays. 

At present, the Tevatron at FNAL and the Large Hadron Collider 
(LHC) at CERN are the only machines capable of $t\bar{t}$  production.  
The collider detectors associated with these machines are often 
compared to cameras, recording a picture each time the beams collide at 
their centers. There is some truth to this analogy, but in 
reality the collisions occur so frequently that there is initially
only enough time to draw a rough picture of what happened. If 
this picture gives an indication that something interesting 
has happened, a sequence of actions is then ``triggered'' that 
results in successively clearer pictures of the collision, each 
scrutinized to a greater extent. If a collision manages to remain 
interesting until the completion of this sequence, it is 
written to storage media. Typically only a few hundred 
collisions out of every 2 - 40 million occurring each  
second are recorded for posterity. This is a very important aspect  
of hadron colliders. The ``trigger'' system that decides
whether or not to record data is the first and most critical  
step of the analysis of data. Furthermore, the tiny fraction of all 
collisions that are recorded still represent an enormous amount 
of data. Hadron colliders can easily acquire more than a petabyte 
($10^{15}$ bytes) of raw data per year.

Though different in their details, the hadron collider detectors at the 
Tevatron and LHC are similar in most respects. Primary  $p\bar{p}$ ($pp$)
interactions have a roughly Gaussian probability distribution 
along the beam-line, peaking at the detector center   
with $\sigma~\sim~25$ cm ($\sim~5$ cm) at the Tevatron (LHC).
In all cases, the central detector regions are cylindrical 
with their axes oriented along the horizontal direction. The 
cylinders are sandwiched between endcap regions arrayed as planar 
disks. Though it was not the case in previous generations of 
collider detectors at the Tevatron and SPS collider, all detectors 
now include a magnetic spectrometer, with the tracking system surrounded 
by calorimetry and muon chambers. Electrons, photons, and hadronic 
jets are identified using calorimeters and tracking while 
muons are identified by muon stations and tracking. The 
data are collected using a multi-level trigger system.
The coordinate system used with hadron collider detectors is 
right-handed with the z axis along the direction of one of the beams, 
the y axis vertical and the x axis in the plane of the accelerator. 
The coordinates are also expressed in relation to particle 
trajectories via azimuthal angle $\phi$, rapidity $y$, and 
pseudorapidity $\eta$. The latter are functions of the polar 
angle $\theta$ and particle velocity $\beta$: 
$y(\theta,\beta)\equiv \frac{1}{2} ln[(1+\beta cos\theta)/(1-\beta cos\theta)]$, 
and 
$\eta(\theta)\equiv y(\theta,\beta=1)$. 

In the remainder of this section, essential features of the 
Tevatron and LHC, the general purpose detectors that 
record the remnants of hadron collisions from these machines 
and the basic methods used to select and reconstruct $t\bar{t}$  
events are presented.

\subsection{Collider Experiments at the Tevatron in Run 2 
(2001-present)\label{sec:tevatron}}

Upgrades to the FNAL accelerator complex, including 
the addition of the new Main Injector, electron cooling, and 
the anti-proton recycler have enabled a luminosity increase  
of more than one order of magnitude relative to Run 1 (1992-1996). 
In addition, the superconducting magnets used in 
Run 2 are operated at a lower temperature allowing margin 
for higher currents to produce higher fields, supporting 
an increase in CM energy from 1.8 TeV to 1.96 TeV. 
In Run 1, luminosity peaked at 
$\mathcal{L}\sim 3\times 10^{31}$~cm$^{-2}$~s$^{-1}$ 
with bunch spacing of $\sim$3.5~$\mu s$. In Run 2 
peak luminosities are currently 
within about 35\% of the Run 2 target of  
$\mathcal{L}\sim 5\times 10^{32}$~cm$^{-2}$~s$^{-1}$. To date, the Tevatron  
has delivered over 6 fb$^{-1}$  to the CDF and D0 
experiments and is adding new data at the rate 
of $\sim 2.5~-~3.0$~fb$^{-1}$/year. To handle  
the higher collision rates, the  
detectors were upgraded with new front-end and trigger 
electronics. Both CDF and D0 also installed completely 
new and more powerful tracking systems for Run 2 that include
large-area, multi-layer silicon detectors.

%\subsubsection{The Collider Detector at Fermilab (CDF) \label{sec:cdf}}
%
The CDF Run 2 upgrade~\cite{Blair:1996kx} includes silicon tracking, a wire 
drift chamber, a time-of-flight (TOF) detector, 
new endplug calorimeters, extended muon 
systems, and new front-end electronics, trigger, and 
data acquisition (DAQ). The goals of the 722,000 channel silicon system 
were to increase tracking coverage to $\vert\eta\vert \le 2$, provide 
precise 2 dimensional (2D) and reasonable 3 dimensional (3D) 
impact parameter and vertex resolution and  
dead-time-less triggering on displaced tracks. 
The core of the system is the SVX II detector~\cite{Blair:1996kx,Sill:2000zz} 
which has 5 double-sided layers extending 
to $\pm$45 cm from the center of the detector as 
necessitated by the broad span of interactions 
along the beam axis. The SVX II hit information is used in the  
online trigger~\cite{Bardi:2001uv} which makes it possible 
to collect large samples of hadronic 
charm and bottom particle decays as well as to 
tag long-lived particles in high 
$p_T$ events. Each of the 5 layers of SVX II contains double-sided 
Silicon that combines axially oriented strips 
with strips oriented either at 90$^o$ or 1.2$^o$ to the axial direction. 
The 1.2$^o$ stereo angle provides less ambiguous matching to
the axial strips but this is achieved at the cost of  
hit position resolution in the direction parallel to the 
strips. Typical hit resolutions in the z coordinate 
for correctly matched axial-stereo hit pairs are  
$\sim 1$~mm ($\sim 0.05$~mm) for 1.2$^o$ (90$^o$) stereo angle.
Inside the inner bore of the SVX II detector 
and mounted directly on the beam pipe in the radial range 
1.35~-~1.70 cm is the Layer 00 silicon 
detector~\cite{Hill:2004qb}. This layer uses single-sided 
25~$\mu m$ pitch detectors with 50~$\mu m$ readout pitch. 
The readout electronics are mounted outside of the tracking 
volume. The silicon is cooled to $\sim$~0$^o$C 
and can operate at high bias voltages as necessitated by radiation 
damage. Layer 00 reduces uncertainty in the transverse impact
parameter of tracks which in turn improves high $p_T$ 
$b$ jet identification ($b$ tagging) and proper lifetime
resolution in low $p_T$ $b$ hadron decays. The SVT trigger 
and Layer 00 silicon were critical ingredients in the 
CDF observation of $B_s$ mixing~\cite{Abulencia:2006ze}. Just
outside the SVX II is the ISL~\cite{Affolder:2000tj} 
which is comprised of two layers of double-sided 
silicon (1.2$^o$ stereo) at radii of 20 and 28~cm. The ISL 
extends $\pm$95~cm along the beam axis. In the central 
region,  $\vert\eta\vert \le 1$, the ISL 
is surrounded by a wire drift chamber~\cite{Affolder:2003ep}
with 96 wire planes grouped into eight superlayers, alternating  
between axial and $\pm$3$^o$ stereo angle. Groups of 12 sense wires 
and 17 potential wires are contained in cells separated by Au-plated 
mylar sheets. Front-end electronics can measure charge deposition 
for particle identification. The tracking system has 
transverse momentum resolution $\sigma_{p_T}/{p_T}^2 \sim 0.0017$ 
(GeV/c)$^{-1}$ and $r~-~\phi$ impact parameter resolution of $\sim 40~ \mu m$ 
which includes a 30 $\mu m$ contribution from the beamspot.
Further particle identification 
is made possible by the TOF system~\cite{Acosta:2004kc} installed 
just beyond the drift chamber and constructed as a cylindrical array  
of scintillator bars.  Scintillation light produced when a
charged particle traverses the TOF is collected at both ends of the bars  
by fine-mesh photomultipliers. The time resolution is of order 120 ps
and enables kaon identification for $b$ flavor tagging. For improved 
electron and jet finding in the forward region, as well as 
improved missing transverse energy ($\EtMiss$) resolution, CDF installed new 
scintillator-tile endplug calorimeters~\cite{deBarbaro:1998sc}. 
These have resolution comparable to the central calorimeters, namely 
$\sigma_E/E\sim 15/\sqrt{E} \oplus 7$\% 
in the electromagnetic section which is designed to interact with 
and stop photons and electrons, and $\sigma_E/E\sim 70/\sqrt{E} \oplus 4$\% 
in the hadronic section which is designed to interact with and stop hadrons. 
Outside the calorimeters, muon systems were upgraded to fill gaps and 
extend coverage. 
The CDF trigger system has 3 levels. The Level 1 hardware trigger 
has a pipeline of 42 cells allowing a 50 kHz accept rate. 
At Level 2, there is a combination of hardware and firmware with 
a two stage asynchronous pipeline comprised 
of four buffer cells to allow an accept rate of $\sim$~1 kHz. At Level 3 a 
Linux processor farm fully reconstructs events and selects $\sim$100  
events per second for storage.

%\subsubsection{D0 \label{sec:d0}}
%
The D0 Run 2 detector~\cite{Abazov:2005pn} includes a 793,000
channel Silicon Microstrip Tracker (SMT)~\cite{Cooper:2008zzb} 
and a Central Fiber Tracker (CFT) inside a new 2 T solenoidal  
magnet with a radius of 60 cm. The Run 2 upgrade also includes 
enhanced muon systems, and all new front-end electronics, trigger, 
and DAQ. The D0 silicon includes 'F disks' which reside between and just 
beyond the barrels, and 'H disks' which are installed 
in the far forward regions. The barrels consist of 
axial and either 90$^o$ or 2$^o$ stereo layers while the 
F and H disks contain $\pm$15$^o$ and  $\pm$7.5$^o$  
symmetric u-v doublets, respectively. 
Surrounding the silicon is the 2.6 m 
long CFT made up of eight layers of axial and 
$\pm$~2$^o$ stereo doublets of 1 mm diameter scintillating
fibers. The 77,000 fibers are connected to visible light
photon counters (VLPC) operated at $\sim10^o$C with
$\le 0.1$\% noise occupancy. The axial layers are 
used in a Level 1 track trigger. The Silicon and Fibers are
surrounded by the 2 T superconducting solenoidal magnet.
The tracker has good momentum resolution and 
$r~-~\phi$ impact parameter resolution of $\sim 40 \mu m$ 
which includes a 30 $\mu m$ contribution from the beamspot.
As for other experiments, the magnetic spectrometer can be used
to improve the calibration of the calorimeter by comparing the 
transverse momenta of isolated tracks, (e.g. from electron candidates) 
with the corresponding transverse energy measurements. 

In Run 2 D0 has retained its existing hermetic
LAr calorimeter which has resolution 
$\sigma_E/E\sim 15\%/\sqrt{E}$ for electrons 
and $\sigma_E/E\sim 50\%/\sqrt{E}$ for pions. 
The total jet energy resolution is $\sim$15\% at 30 GeV 
falling to $\sim$10\% at 100 GeV and to $\sim$5-7\% above 200 GeV. 
There are also improved forward and central preshower detectors.
The central preshower uses 7 mm prismatic scintillator strips 
to reduce the electron trigger rate by a factor of 5. The D0 muon
system is the outermost part of the detector and consists of 
a system of proportional drift tubes in the region 
$\vert\eta\vert~<~1$ and mini drift tubes that extend the coverage
to $\vert\eta\vert~\sim~2$. Scintillation counters are used for 
triggering and also for cosmic ray and beam-halo $\mu$ rejection. 
The system is completed by Toroidal magnets and special shielding.
Each subsystem has 3 layers with the innermost layer in between
the calorimeter and iron of the toroid. The average energy loss of 
1.6 GeV in the calorimeter and 1.7 GeV in the iron is taken into 
account in determining the momenta of muons. The $p_T$ resolution 
of tracks from the SMT and CFT matched to $\mu$ segments is 
$\sigma_{p_T}/p_T~=~0.02~\oplus~0.002~p_T$ GeV/c~\cite{Abazov:2006ka}. 
The calorimeters use switched capacitor arrays for pipelining data. 
The silicon and fiber readout use the SVX2 chip which has a 32 
cell pipeline. The multi-level trigger system focuses on regions of 
interest within events to avoid full event reconstruction. 
Level 0 is a beam crossing trigger (``minimum bias''). 
Level 1 is based upon calorimetry, preshower, fiber tracker, 
and muon scintillators as well as some information on 
displaced tracks from the silicon system and has a 2 kHz 
accept rate. The level 2 hardware trigger has 1 kHz accept rate.
Level 3 is a processor farm with 50 Hz accept rate. 

\subsection{Collider Detectors at the CERN LHC \label{sec:lhc}}

The LHC~\cite{Evans:2008zz} will collide protons at $\sqrt{s}=$~10 TeV
in 2009 and 2010 with a bunch spacing of 50~ns. 
The luminosity will increase in tandem with increased  
understanding of the safe operation modes of the machine. Physics 
running will start at $\mathcal{L}\sim 10^{31}$~cm$^{-2}$~s$^{-1}$ 
and then increase in several steps to a target of  
$\mathcal{L}\sim 10^{32}$~cm$^{-2}$~s$^{-1}$. Of order 100~pb$^{-1}$ 
of data are planned to be delivered to each of the 
experiments. During a shutdown that is currently planned 
for 2011, additional protection systems will be installed 
in the LHC and the 1232 superconducting dipole magnets  
will be trained to operate at, or near to, the design 
current of $\sim$12 kA. Each dipole is 40m long and operates 
at 1.9$^o$K. Subsequent running will be at or near 
$\sqrt{s}=$~14 TeV with 25 ns bunch spacing and luminosity 
of $\mathcal{L}=\sim 10^{33}$~cm$^{-2}$~s$^{-1}$ for several years. 
The luminosity will then rise to $\mathcal{L}\sim 10^{34}$~cm$^{-2}$~s$^{-1}$ 
after exchanging Copper collimators for new ones made from Carbon. 

%\subsubsection{A Toroidal LHC ApparatuS (ATLAS)\label{sec:atlas}}

ATLAS is the largest collider experiment ever  
built~\cite{:2008zzm}. Its size is determined by the extent of 
its muon system which uses 20~m diameter air-core toroids and 
endcap planes separated by 46~m. The ATLAS inner detector
starts with pixel layers with 
$50\,\mathrm{\mu m}\times\,400\,\mathrm{\mu m}$ pixels.
There are three cylindrical layers 
at radii of 5.1, 8.9, and 12.3~cm and three end-cap disks per 
side. In total, the system contains 80$\,$M  pixels. 
Outside the pixels there are four double-sided, 
shallow stereo barrel layers of silicon microstrips 
at radii of 30 to 51~cm and nine endcap disks per end. 
The total area of silicon is $\sim$~61~m$^2$. The Silicon is
operated at -5 to -10~$^o$C to avoid destructive annealing that 
occurs after type inversion due to radiation exposure and 
which is accelerated at higher temperatures~\cite{Lindstrom:2001ww}.
Beyond the silicon is a  transition-radiation, straw-tube tracker 
with 73 and 160 planes in the barrel and each end cap, respectively.
Single hit resolution is $\sim$130 $\mu m$. The tracker will 
have momentum resolution of 
$\sigma_{p_T}/p_T\sim 5\times\,10^{-5}~p_T\;\oplus\;0.01$ ($p_T$ in GeV/c)
with coverage for $\vert\eta\vert\le 2.5$.
The ATLAS tracker is situated in a 2T
solenoidal magnetic field. The calorimeter includes a Pb-LAr accordion
electromagnetic system in the central region
surrounded by a scintillator-tile hadronic
section. In the forward regions the electromagnetic
and hadronic systems both use LAr. Preshower
detectors are installed ahead of the electromagnetic
calorimeter inside the solenoid cryostat wall and
employ narrow strips to allow pointing to the interaction
vertex with resolution of $\sim 50\,\mathrm{mrad}/\sqrt{E}$. Overall
the energy resolutions expected to be achieved are
$\sigma(E)/E=10/\sqrt{E}\oplus0.2$\% with energy in GeV 
in the electromagnetic and $\sigma(E)/E\sim 50~(80)/\sqrt{E}\oplus~3~(6)$\% 
in the central (forward) tile hadronic system with coverage of 
$\vert\eta\vert\le 3.2 ~(4.9)$
The muon system will provide momentum resolution
of order 2-3\% for intermediate momenta and
7-10\% at 1 TeV with coverage of $\vert\eta\vert\le 2.7$. 
The ATLAS trigger has three levels. Level 1 is made
up of custom electronics and uses coarse resolution data from 
the calorimeters and muon system. Level 2 is a software 
trigger run on a processor farm with full resolution data  
for regions associated with Level 1 trigger objects. 
The highest level is the Event Filter (EF) which has full resolution 
data for the full detector and runs full offline reconstruction and 
selection algorithms. 

%
%\subsubsection{Compact Muon Solenoid (CMS) \label{sec:cms}}

The CMS experiment~\cite{:2008zzk} has an 
inner detector that includes a pixel detector with three 
barrel layers at radii between 4.4, 7.3 and 10.2 cm and a silicon 
strip tracker with 10 barrel layers extending from a radius of 20 cm
to a radius of 1.1 m. Layers 1,2,5, and 6 of the silicon strip tracker 
are double-sided with 100 mrad stereo angle.  The size of the pixels is 
$100\times150~\mu m^2$. Endcap systems consist of two disks in
the pixel detector and three small inner disks plus nine large outer 
disks in the strip tracker on each side of the barrel. The silicon 
strip disks are composed of concentric rings of modules with ring 1
being closest to the beam. The innermost two of three rings in 
the small disks, and rings 1,2 and 5 of the large disks are double-sided.
The disks extend tracking acceptance to $\vert\eta\vert \le 2.5$. 
The CMS tracker is the 
largest silicon tracker ever built with more than  200 m$^2$ 
of active silicon area. All told, there are 66 M pixels and 
9.6$\,$M silicon strips. The momentum resolution of the tracker is 
$\sigma_{p_T}/p_T\sim1.0~(2.0)\times 10^{-4}~p_T \oplus 0.008~(0.02)$ 
with $p_T$ in GeV/c in the central (forward) regions.
For muons, the additional lever arm of the muon system 
allows an improvement in resolution to 
$\sim4.5\%\times\sqrt{p_T ~(TeV)}$. The asymptotic transverse 
(longitudinal) impact parameter resolution is expected to 
be better than 35 (75) $\mu m$. The CMS calorimeters  
include a PbWO$_4$ crystal electromagnetic calorimeter with 
Si avalanche photodiode readout surrounded by a Cu and 
scintillator tile hadronic calorimeter. The calorimeters 
and tracking elements are inside a 4T solenoidal magnetic field. 
The electromagnetic calorimeter is expected to achieve resolution of 
$\sigma(E)/\sqrt{E~(GeV)}\sim2.7 ~(5.7)/E\oplus 0.55$~\% in the central 
(forward) region. The hadronic calorimeter 
will have resolution of $\sigma(E)/\sqrt{E~(GeV)}\sim 100/E\oplus 5$~\%.

\subsection{Event Reconstruction and Pre-selection for $t\bar{t}$ \label{sec:event}} 

Although $t\bar{t}$ production is a rare process, it is at a sufficiently high 
scale to generate observable quantities that can be 
triggered online and selected offline with high efficiency. However, the ability to 
discriminate $t\bar{t}$ from background is another story and depends to a 
greater extent on details of particular final state topologies and detector 
capabilities. As discussed earlier, top quarks within the SM are predicted to decay 
into a $W$ boson and a $b$ quark with a branching fraction of nearly 
100\%~\cite{Amsler:2008zzb}. The presence or absence of leptonic $W$ decays 
is then the primary factor in shaping final state signatures. One-third
of $W$ bosons decay to leptons, including 
$\tau\nu_{\tau}$ with the $\tau$ decaying about one-third of the time 
to an $e$ or $\mu$ plus two neutrinos. The remaining two-thirds of $\tau$ 
decays involve hadrons and one neutrino. In view of these 
considerations, $t\bar{t}$ final states
with 0, 1, or 2 potentially well-isolated electrons or muons from $W$ decays  
will occur at rates of roughly 54.9\%, 38.4\% and 6.7\%, respectively. 
These will be respectively referred to as the ``all-hadronic'', ``lepton plus jets'', 
and ``dilepton'' modes or channels of top events, where it is 
generally understood that ``lepton'' in this context refers to an $e$ or $\mu$.  
Whenever $\tau$ leptons are used, they are called out explicitly, as for example
is done for the $\ell\tau$ dilepton channels. 
All $t\bar{t}$ final states include two high energy jets 
arising from fragmentation of $b$ quarks. The leptonic modes 
(including hadronic $\tau$ decays) have an imbalance in transverse momentum 
($\EtMiss$) due to the presence of undetected neutrinos. There are
two additional jets for each hadronic $W$ decay and there are sometimes additional  
jets from QCD radiation.  Existing collider experiments can identify isolated 
leptons from $W$ decays with very high efficiency and they can also 
measure their momenta with very good resolution. Jets and $\EtMiss$ in 
$t\bar{t}$ events are also easily detected but are not as  
well-resolved. It is worth noting that almost all high energy events at hadron 
colliders contain only jets because energetic isolated 
electrons and muons are due to rare weak 
decays involving real or virtual intermediate vector bosons. 
Lepton multiplicity is therefore highly correlated (anti-correlated) with signal 
purity (abundance) in event samples selected for $t\bar{t}$ studies.  

Most measurements of top quark properties begin with 
the determination of the production 
cross section simply because it  
is almost a direct by-product of the  
selection of an event sample, including optimization of 
the discrimination of
$t\bar{t}$ from backgrounds and establishing the detailed 
make-up of the final sample.  Event sample selection can be 
divided into three stages; (i) online object 
(e.g. $e,~\mu,~$jet, $\EtMiss$) reconstruction and triggering,
(ii) offline event reconstruction and pre-selection and 
(iii) the use of specialized information and techniques 
to better distinguish $t\bar{t}$ events from backgrounds 
and to understand as well as possible the contributions 
and characteristics of $t\bar{t}$ and various background 
processes that enter into the final sample. 

This section provides an overview of object reconstruction, triggering, 
event selection and some of the methods used to 
discriminate signal from background in $t\bar{t}$ studies  
performed by the four major hadron colliders discussed 
sections~\ref{sec:tevatron} and~\ref{sec:lhc}. In the interest of space and 
pedagogy, no attempt is made to provide every detail of  
object and event reconstruction for all published $t\bar{t}$ studies.
Rather the emphasis here is on the general ideas behind, and motives for,  
using some of the more important methods. Many more details can be found in 
the references cited in the text. A review of $t\bar{t}$ cross section measurements 
is then presented in section~\ref{sec:xsection}.

%\subsubsection{Reconstruction and object identification\label{sec:reco}}
%
The basic components of $t\bar{t}$ events that need to 
be identified and reconstructed are; (i) the primary interaction 
vertex, (ii) electrons, (iii) muons, (iv) jets and (v) $\EtMiss$. 
Additionally, for better discrimination of $t\bar{t}$ from backgrounds, 
special methods are used to tag jets resulting from $b$ quarks and 
discriminating variables are constructed based on event topology and the 
kinematic properties of the objects in the event. 
Discriminating variables are used either for simple cut-based selection 
algorithms or as inputs to likelihood functions and neural networks. 
For any given object, all of the hadron collider experiments 
typically employ similar reconstruction and identification techniques,
though possibly in different order or  
with a different emphasis, depending upon the particular strengths of the apparatus. 

The primary vertex for a given event is found by associating 
well-measured tracks to a common point of origin using, for instance,
a Kalman filter~\cite{kalman}.  Interaction 
vertices are found event-by-event and their locations for a large 
number of events can be used to define the beamline in the luminous
region, defined as the region along the z axis where the probability 
for interactions is high. The probability of an interaction as a 
function of z in this region has an approximately Gaussian distribution 
with $\sigma\sim{30}\,(\sim{5})\,\mathrm{cm}$ at each Tevatron (LHC) interaction region
with peak at, or near to, the nominal geometric center of the detector. 
Typically there is more than one interaction per bunch crossing and more 
than one interaction vertex may be reconstructed. Provided that the 
multiplicity is not too high, it is generally true that only one of 
the interactions involves a hard-scatter with high constituent CM energy. 
The others are typically soft-scatters known collectively as 
``minimum bias'' events. The choice of a primary vertex to associate 
with $t\bar{t}$ production can thus be done in
several ways. One can choose the vertex with the highest $p_T$ sum of 
associated tracks. In leptonic modes, the primary vertex can be chosen to be
the one to which the lepton is associated. In D0 a ``minimum 
bias probability''~\cite{Abazov:2007kg} is computed for each reconstructed 
vertex based on the transverse momenta and total number of associated tracks.
The primary vertex with the lowest minimum bias probability can then 
be selected as the hard-scatter vertex. 

Unlike primary vertex reconstruction, lepton reconstruction involves not only 
the tracking system but also additional detector subsystems specifically 
designed for this purpose. Electron reconstruction in its most basic form 
is the simple matching of a localized deposit of energy in the electromagnetic (EM) 
calorimeter with a track reconstructed in the tracking system. 
The tracks are required to be above some $p_T$ threshold (typically no
higher than 10-20 GeV/c) and must pass quality cuts which are usually 
based on the number of hits found on tracking elements and the quality of 
the track fit as measured by the $\chi^2$ per degree of freedom ($\chi^2/N_{dof}$),
for example. The energy deposited by an electron 
should be well-contained in the electromagnetic section which 
is designed and built with a large number of interaction lengths. The energy is
also expected to have a relatively small lateral spread in $\eta~-~\phi$ in 
the absence of bremsstrahlung. The EM cluster is obtained by combining signals from 
neighboring cells using a variety of methods from
simple seed-based, fixed proximity groupings or patterns, 
to algorithms using a small radius cone  in $\eta~-~\phi$ such as 
$R=\sqrt{\Delta\eta ^2 +\Delta\phi^2}~\sim 0.2$. More 
sophisticated algorithms that take into account bremsstrahlung 
photons by recapturing isolated EM clusters separated in azimuth 
from the main cluster are also employed in some cases.
Electron discriminating variables include the relative proportions of hadronic  
and electromagnetic energy such as the ratio $E_{HAD}/E_{EM}$ or 
the EM fraction ($f_{EM}=E_{EM}/E$) of the cluster energy. In 
D0~\cite{Abazov:2007kg} a fixed cut, $f_{EM}~>~0.9$,  is used while in 
CDF~\cite{Abulencia:2006kv} it is energy dependent; 
$E_{HAD}/E_{EM} \le 0.055+0.00045 E(GeV)$.  
More fine-grained detector elements in ATLAS~\cite{Aad:2009wy}, 
CDF and D0, provide additional shower shape information that is 
used to build likelihood discriminants. 

It is interesting to note that the high radiation environment 
of the LHC has led to a greater 
use of silicon detectors for tracking as discussed in 
sections~\ref{sec:tevatron} and~\ref{sec:lhc}. 
This has resulted in an increase in the amount of material which  
leads to a substantial probability for bremsstrahlung  
radiation for low $E_T$ electrons. This has important implications 
for $e$ and $\gamma$ reconstruction. In CMS~\cite{Bayatian:2006zz} for instance, 
the Kalman filter track algorithm valid for $p_T\ge 30$~GeV 
is replaced by a non-linear Gaussian-Sum Filter (GSF)~\cite{Fruhwirth:2004ma} 
in which state vectors and errors are Gaussian
mixtures of components whose weights depend on the measurements. 
Bethe-Heitler~\cite{Bethe:1934za} modeling of energy loss in tracking material
is also performed. The track parameters that are obtained are more meaningful 
along the entire track trajectory which allows more powerful matching 
discriminants to be employed because the difference between the estimated 
$p_T$ of the electron at the primary vertex and that at the face 
of the EM calorimeter is a good measure of the fraction of energy 
lost to bremsstrahlung. 

Muon reconstruction in its simplest form involves a match between a track in the 
tracking system and a track, segment or simply some number of hits 
in the outer muon stations. In CDF and D0 isolated high $p_T$ tracks 
in the tracker that pass 
standard quality cuts are extrapolated to the muon chambers situated beyond 
the calorimeter in search of a matched muon track segment.  In D0 segments 
are defined when at least 2 (1) wire (scintillator) hits found in the three stations 
are matched. In ATLAS and CMS, the muon systems are used to obtain standalone
muon reconstruction and $p_T$ measurements. Global muons in CMS correspond to
muon tracks that are matched to tracker tracks. In all cases, the tracker
provides most of the muon track parameter information. However, at very 
high momentum, the large lever arms of the CMS and ATLAS muon systems 
improve the $p_T$ measurements substantially. 
All experiments apply additional requirements to minimize contamination 
from a number of sources including hadron punch-through, decay in 
flight (DIF) hadrons, beam-halo and Cosmic ray muons. Punch-through hadrons are those 
rare cases in which hadrons from hadronic showers manage to escape the 
calorimeter and generate signals in the muon stations. Decay
in flight hadrons involve real muons as in $\pi^+\rightarrow\mu\nu_{\mu}$ 
and $K^+\rightarrow\mu\nu_{\mu}$. These are dominant decay 
modes at nearly 100\% and 63.5\%, respectively, but the lifetimes 
are long ($c\tau_{\pi}=$~7.8~m, $c\tau_{K}=$~3.7~m) and they are 
generally highly boosted so that the decays occur relatively rarely inside
the detectors. Also, when they occur they are generally inside jets and 
can be suppressed, but not entirely eliminated, 
by a variety of clean-up and isolation requirements. 
In the case of $t\bar{t}$ the candidate muons are from 
$W$ decays and so they are prompt (i.e. originating from the primary vertex) 
and generally well-isolated. Useful requirements are that the calorimeter energy 
should be consistent with what is expected for a minimum ionizing particle (MIP)
such as a muon. The distance of closest approach $d$ to the
primary vertex in the transverse plane divided by its uncertainty, 
also known as impact parameter significance $S_d$,  
should be consistent with what is expected for a prompt track 
(e.g. $S_{d}\equiv d/\sigma_d\le 3$ in CDF). The track must also match  
to the appropriate primary vertex in z. 

For both electrons and muons, there are different classes 
of quality that are defined, called ``loose'', ``medium'' 
and ``tight'' etc., for which progressively stronger 
requirements are used, and lower efficiencies are obtained. The key
element in this progression is lepton isolation. 
Calorimeter and track isolation require that the fraction of summed 
$E_T$ of calorimeter towers or the summed $p_T$ of tracks in a cone of radius 
$R\sim~0.4-0.5$ about the centroid of the calorimeter cluster or 
the lepton flight path, excluding the contribution 
of the lepton itself, should be small, or a small fraction of the energy of
the lepton. Requirements are usually on the order of a maximum of 1 to 6 
GeV or a fraction not to exceed $\sim 0.1$.

In all cases, lepton triggering and reconstruction efficiencies are studied 
via $Z\rightarrow \ell\ell$ decays. The basic goal is to obtain a large, 
pure and $unbiased$ sample of high $p_T$ isolated leptons by using 
as few lepton discriminants as possible. The procedure, which dates 
back (at least) to the CERN SPS collider experiments is now known as ``Tag and Probe'' and 
typically begins with a sample of single lepton triggers that   
each contain a corresponding high $p_T$ tight lepton. The next step is to 
identify a potential second high $p_T$  lepton in the event using 
fewer lepton discriminants. If the two objects form an 
invariant mass that is comparable to the $Z$ mass, then it is very possible that 
the second candidate is a true lepton. In practice, the invariant mass 
distribution for many such events will show a peak at $M_Z \sim 91$~GeV over 
a background that can be measured in sidebands. The tight and loose  
selection criteria for the two lepton candidates can be
varied to obtain an acceptable signal-to-background ratio. At this point, one
has an unbiased sample of real leptons. These represent
the denominator in subsequent measurements of efficiency. For instance one can
select the second lepton without requiring it 
to have fired a lepton trigger and then see how often it has done so to extract 
a relative trigger efficiency. More generally, one can measure the
relative efficiency of any criteria that was not used to select the 
probe lepton. 

Like electrons, jets are constructed from calorimeter towers but are generally 
broader and involve substantial energy deposits in both the EM and the 
hadronic layers. They are frequently constructed from calorimeter cells 
and towers using iterative cone algorithms of various types. CDF and 
ATLAS use a default cone size of $R~=~0.4$ while CMS and D0 use  
$R~=~0.5$. In some cases there are multiple algorithms run all of whose 
results are available for use in an analysis. These may include infrared 
safe cone and $K_t$  algorithms~\cite{Seymour:2006vv}. Once candidate 
jets are formed, selections may be applied to avoid problems associated with 
noisy towers and to remove electrons and photons. The EM fraction, 
for instance is required to be neither too small nor too large 
(e.g. $0.05\le f_{EM} \le 0.95$). In addition there can 
be requirements on the longitudinal and lateral energy distributions. 
D0 for instance 
requires that the ratio of leading to next-to-leading $E_T$ cells be less
than 10, that no tower contain more than 90\% of the total jet energy 
and that the fraction of energy in the outermost hadronic layer be less
than 40\%. In all experiments, towers or cells associated with an 
identified $e$ or $\gamma$ are removed and one applies  
corrections to take into account response imperfections and  
other effects. The latter may include a correction for the energy offset 
due to contributions from the underlying event, pile-up of 
signal from successive bunch crossings or multiple interactions within a
single bunch crossing, and noise in the calorimeter electronics.
There is an additional correction for the $\eta$ dependence of 
calorimeter response and possibly for time dependence of the response.
The latter is particularly important in calorimeters that use scintillators 
as active elements since the light yield diminishes with age and 
radiation dose. The jet energy scale (JES) is obtained from  
precisely measured EM objects by assuming $E_T$ conservation in 
$\gamma~+~jet$ events in order to obtain an average jet correction as a 
function of $E_T$. At $E_T\approx 10$~GeV jet corrections can be as large 
as 100\% but are generally no more than $\sim 5-10$\% at $E_T\ge 100$.

The presence of a $\nu$ is inferred from energy imbalance in the transverse 
plane, or $\EtMiss$. The $x$ and $y$ components of raw $\EtMiss$ are calculated simply 
as the negative sum over the corresponding components of the $E_T$ in calorimeter
cells. The sum  includes all calorimeter 
cells that remain after noise suppression algorithms and corrections are applied. 
Several corrections are then made for cells associated with objects for 
which energy corrections are known. Thus the jet energy corrections that are 
applied to jets are subtracted vectorially. 
Similarly, the generally smaller energy corrections applied to clusters associated 
with electrons and photons are vectorially subtracted and the small vector $E_T$  
contribution from a muon in the calorimeter is added while the muon 
$p_T$ vector is subtracted. Finally, additional corrections are applied for 
$E_T$ losses due to gaps in coverage and dead material such as occurs in 
the walls of LAr cryostats. 

\begin{figure}[tb]
\begin{center}
\includegraphics[width=0.95\textwidth]{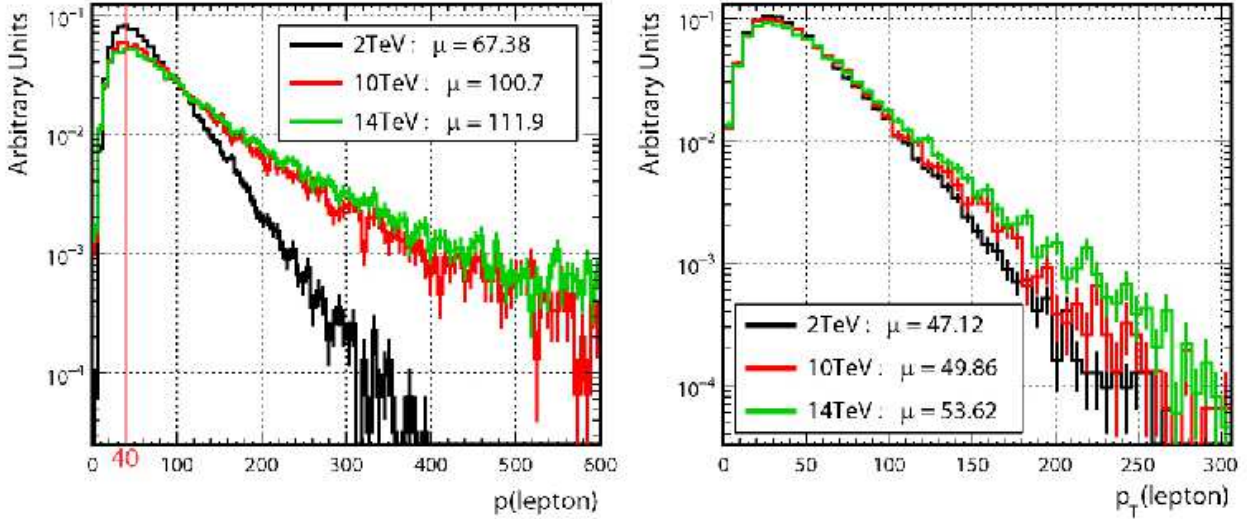}
%\begin{minipage}[t]{8 cm}
%\epsfig{file=cdf_dil_2800pb.eps,scale=0.5}
%\end{minipage}
%\begin{minipage}[t]{16.5 cm}
\caption{Lepton $p$ (left) and $p_T$ (right) for $e$'s and $\mu$'s from
$W$ decays in $t\bar{t}$ at the Tevatron and LHC.
\label{fig:top_leptons}}
%\end{minipage}
\end{center}
\end{figure}

%\subsubsection{Triggering and sample selection \label{sec:selection}} 
%
Triggering and preliminary event selection of leptonic modes are based 
upon the presence of at least one high $E_T$, isolated $e$ or $\mu$ 
and/or large $\EtMiss$ as defined in the preceeding section. 
To avoid various backgrounds, as discussed below, 
thresholds for these quantities are set as high as possible
without incurring substantial loss of $t\bar{t}$ signal.  
It turns out that the optimal threshold for leptons in $t\bar{t}$ events 
is roughly the same at the LHC as at the Tevatron. 
In fact, the threshold is typically around $20\,\mathrm{GeV}$ which is  
the level where one would successfully retain most leptons 
coming from $W$'s produced at rest.  
This is the result of several factors.  
First, top quarks are produced with very little excess
momentum because they are produced near threshold as a result of  
steeply falling parton distributions for $q\bar{q}$ annihilation at the 
Tevatron and $gg$ fusion at the LHC. Secondly the boost imparted to the 
$W$ from the decay of the top quark has little effect on the position of the
peak in the lepton energy distribution for the decay $W\rightarrow \ell\nu$. 
This is because of the polarization of the $W$; $\sim$70\% longitudinal and 
$\sim$30\% left-handed polarization, which means that 
the $\ell$ is on average emitted in a direction slightly 
opposed to the direction of flight of the $W$, neatly   
canceling out the effect of the boost. 
The most probable value of the energy of the $\ell$ stays quite near 
to $M_W/2\sim 40\,\mathrm{GeV}$ as seen in the generator level distributions shown in
Figure \ref{fig:top_leptons}.
%%% Figure %%%
A cut on $p_T~>~20$~GeV thus retains the peak of the distribution 
of the leptons over the entire central region of the detector, ($\vert\eta\vert ~\le~1.3$). 
Of course a threshold of $20\,\mathrm{GeV}$ is less efficient for an $e$ 
or $\mu$ coming from the decay chain $W\rightarrow \tau\nu$, 
with $\tau\rightarrow\ell\nu_{\ell}\bar{\nu}_{\tau}$ due to the energy 
carried by the two additional neutrinos. 

The $t\bar{t}$ all-hadronic mode is selected by requiring multijet final states. 
For triggering purposes the calorimeter granularity can be simplified as in the 
case of CDF for which it is a 24$\times$24 grid of ``trigger towers'' 
in $\eta~-~\phi$ with each tower spanning 
$\Delta\phi\times\Delta\eta~=~15^o\times~0.2$. 
In the most recent study of all-hadronic $t\bar{t}$ events 
by CDF~\cite{Aaltonen:2007qf} with $1.02\,\mathrm{fb^{-1}}$ of data, the
triggering at each level proceeds as follows. Level 1 requires a single
tower with $E_T~\ge~10$~GeV. Level 2 requires  $\sum_j E_{T,j}\ge~175$~GeV
where the sum is over all trigger towers. Using more accurate
energy estimates, Level 3 requires at least four jets with $E_T~\ge~10$~GeV. 
It is in fact relatively typical to find that the trigger and event 
pre-selection require at least four jets above an $E_T$ threshold of 
order 10-30~GeV. This is a particularly difficult channel 
because of the extremely large multijet backgrounds. After the final stage of 
triggering the CDF event sample contains 4.3 M events with a Monte Carlo (MC) 
estimated efficiency of 80\% for all-hadronic $t\bar{t}$ events, corresponding 
to a $S/B$ of roughly 1/1100 where the background is essentially all multijet 
events. Further improvements in event selection for all $t\bar{t}$ modes and the
all-hadronic mode in particular requires more discriminating power.

%\subsubsection{Discriminating $t\bar{t}$ signal from backgrounds\label{sec:signal}} 

It is very important to note that in the preceeding discussion 
the triggers and 
event selections do not make use of all of the components of the corresponding 
$t\bar{t}$ signatures. In particular there are always two  
additional b jets and two additional light quark or charm jets per hadronic $W$ decay. 
Indeed, for 0, 1 and 2 lepton modes, the main concentrations of signal appear  
in samples of events with $\ge~2$, $\ge~4$ or $\ge~6$ jets, respectively. 
Similarly, the purity of the selected samples increases with each additional 
$b$ jet that's identified. 
The reason that these event characteristics are not used in the standard event selection
(often called ``pre-selection'') is that it is important to have access to events 
from signal-depleted regions which are similar in many respects to events in 
signal regions. These provide access to control samples that can be used to test one's 
understanding of important SM backgrounds via data-driven and/or 
MC-based methods. The ability to 
extrapolate an understanding of backgrounds in control regions 
to the signal region can range from very direct and simple to very 
tenuous and speculative. In general, it is somewhere between
these extremes. It is however one of the most important stages 
of the analysis for the experimentalist because one cannot generally rely 
upon simulations to provide an accurate characterization of the behavior of 
many backgrounds or of the signal. Many of the more important   
and creative contributions to experimental hadron collider 
physics and $t\bar{t}$ studies in particular have been associated with this 
stage of analysis. 

Once the events have been triggered online and pre-selected offline, the next step 
is generally focused on methods for enhancing the $S/B$ ratio while 
retaining a substantial signal component. Ideally the methods employed do 
not bias the kinematic properties that are likely to be studied after 
a cross section is measured. A good example
is the use of $b$ tagging. The most powerful techniques for identification of 
$b$ jets, or ``$b$ tagging'' are based upon the fact that $b$ hadrons have 
long lifetimes ($\tau\sim$~1.5~ps) so that with typical transverse energies of 
50 to 100 GeV in $t\bar{t}$ events they have a mean travel path of 
order 5 mm. If the $b$ hadron 
decays to two or more charged stable particles, it can often be detected via 
a well-reconstructed vertex that is significantly displaced from the primary 
vertex~\cite{Abe:1995hr,Abulencia:2006in,Abazov:2006ka}.
Since a $b$ hadron generally decays sequentially to a
$c$ meson which also has a fairly long lifetime, it is possible for 
the visible charged tracks to originate from two different displaced vertices.
In some cases, this spoils the ability to isolate a well-reconstructed 
vertex. Nevertheless, the $b$ jet can be tagged by algorithms that rely 
upon the low probability for jets of lighter flavor partons to produce tracks 
with large impact parameters ($d$) relative to the primary 
vertex~\cite{Buskulic:1993ka,Abulencia:2006kv} or beamline. Alternatively a $b$ tag 
can rely upon the appearance of an $e$ or $\mu$ that is closely associated 
with a hadronic jet~\cite{Abe:1994st,Acosta:2005zd}. 
In the latter case, the lepton is assumed to result from the 
semileptonic decay of the $b$ or subsequent $c$ meson. Such leptons  
are more difficult to identify than those coming from $W$ decays because 
they are not isolated. However, they do retain many of the characteristics
of leptons and often have a substantial component of  
their momentum transverse to the jet axis. 
They also often have large positive 
impact parameters but this information is not always used in 
order to retain high efficiency and avoid correlation with the  
lifetime tagging techniques.

Secondary vertex tagging in CDF operates on a per-jet basis,
where only tracks within the jet cone are considered for
each jet in the event. Tracks are 
selected based on $p_T$, track fit quality ($\chi^2/N_{dof}$) 
and the quality of associated Silicon detector hits because
only tracks whose impact parameters are well-resolved will 
contribute usefully to secondary vertex reconstruction.
Jets with at least two such tracks can produce
a displaced vertex and so are called ``taggable''. 
%Next, tracks in the jet are selected on the basis of 
%their impact parameter significance 
%($S_{d}\equiv d/\sigma_{d}$) with
%respect to the primary vertex. 
The CDF SecVtx algorithm~\cite{Acosta:2004hw} developed in Run 1
uses a two-pass approach to find secondary vertices. In the first pass, relaxed 
track requirements are used ($p_T > 0.5$~GeV/c $S_{d} >2.5$), 
but high purity is achieved by requiring at least three tracks 
of which one has $p_T > 1.0$~GeV/c. If the first pass is 
unsuccessful, track requirements are tightened 
($p_T > 1.0$~GeV/c $S_{d} >3$) and an attempt is made to 
find a vertex with two tracks of which one has $p_T > 1.5$~GeV/c.
Once a vertex is found, the two dimensional (2D)
decay length ($L_{2D}$) is calculated as the projection of 
the distance from the primary to the secondary vertex projected 
onto the jet axis in the transverse plane. The sign of
$L_{2D}$ is positive (negative) if the angle between the jet 
axis and the vector from the primary to secondary vertex in
the transverse plane is less (more) than 90$^o$. The 
latter corresponds to the unphysical situation in which the 
secondary vertex is behind the primary vertex relative to 
the jet direction. Secondary vertices corresponding to the 
decays of $b$ and $c$ hadrons generally have large positive 
values of $L_{2D}$.  Those from mismeasured tracks 
(``mistags'') will have
smaller, possibly negative values. The latter are suppressed by 
requiring ``positive tags'' with 
$S_{L_{2D}}\equiv L_{2D}/\sigma_{L_{2D}}~>~3$.
Nevertheless, ``negative tags'' with $S_{L_{2D}}~<~-3$ 
are used to estimate the false positive tag rate.
Note that values of $\sigma_{L_{2D}}\sim~$100~-~200~$\mu m$ 
are typical. Other experiments use similar but 
different algorithms based on secondary vertex finding. 
In D0 for instance~\cite{Abazov:2006ka}, jets are formed 
as clusters of tracks and secondary vertices prior to being 
associated to calorimeter jets. As for CDF, 
track selection relies upon the quality of Silicon detector 
information, and impact parameter significance and
a calorimeter jet is $b$ tagged if it has significant 
positive 2D decay length. In all instances, care is
taken to reject tags consisting of pairs of tracks 
that construct an invariant mass comparable to 
known long-lived light flavor mesons such as 
$K_S^o\rightarrow\pi^+\pi^-$ and 
$\Lambda^o\rightarrow p^+\pi^-$ or photon conversions 
$\gamma \rightarrow e^+e^-$.

In the absence of vertexing, there are several commonly used
algorithms for $b$ tagging. The simplest is the ``track counting'' 
method in which one counts tracks associated with a particular jet 
that have positive impact parameter significance above some 
threshold and declare a candidate $b$ tag when there are some 
minimum number of these tracks. In some cases~\cite{Abazov:2005ey} 
a jet is $b$-tagged if there are $N\ge 2~(\ge~3)$ tracks
with $S_{d}>~3~(>~2)$. The $b$-tag efficiency in this case is 
slightly higher than that for secondary vertexing but 
a drawback is that the mistag rate is doubled. Another drawback of 
this algorithm is that it does not take into account the fact that 
the probability for a jet of light flavor origin ($uds$ or $g$) to 
pass the tagging requirements increases with the $E_T$ of the jet as 
a result of increasing numbers of tracks inside the jet cone. This
means that the mistag rate rises with $E_T$. 
The problem is overcome in a more sophisticated version of the 
track counting algorithm known as the ``jet probability'' 
algorithm~\cite{Buskulic:1993ka,Abulencia:2006kv}.  For this
algorithm, the negative side of the track  
impact parameter distribution is mirrored about zero 
to obtain a data-driven definition of positive impact parameter 
resolution. This is used to determine a per-track probability that 
a track with a given significance $S_{d}$ is in fact consistent 
with having come from the primary vertex. The products of such probabilities for
all tracks in the jet is then used to define a jet probability $P_J$ 
which is the probability that an ensemble of $N$ tracks in the jet cone 
with impact parameter significances  $S_{d}^j$ for $j~=~1,...,N$ could 
be produced by fragmentation and hadronization of a light flavor parton with 
no lifetime. The negative impact parameter 
distributions are obtained from large samples of jet events for which the 
contamination of heavy flavor is small. To take into account 
possible variations in track parameter resolutions, CDF defines 
72 different track categories corresponding to different values of 
$\p_T$, $\eta$ and the number of silicon detector hits on the track. 
Resolution functions for each category are a convolution of 
four Gaussian distributions. 
The jet probabilities constructed from the per-track probabilities
are properly normalized to take into 
account the number of tracks contributing to the  
calculation of $P_J$. As a result, a sample of jets originating from 
light flavor partons will have a $P_J$ distribution which is flat 
from 0 to 1 while the distribution for 
a sample of jets originating from heavy flavor 
($b$ and $c$) partons will tend to be concentrated at very low 
probabilities. The algorithm has a stable mistag rate as a function of jet $E_T$.
It also has the attractive feature of providing a 
continuous variable, the value of $P_J$, for use in selecting 
events. 

The algorithms described above have the potential for efficient 
identification of $b$ jets with fairly low mis-identification rates. 
Purity can be increased at the cost of efficiency, 
but it is difficult to extend their performance 
beyond efficiencies of order 50-60\%, and so more 
sophisticated approaches are taken for additional gains. 
D0 for example uses a neural-network $b$ tagger~\cite{Shary:2008pr} 
and the LHC experiments have developed a variety of tagging algorithms 
including neural networks and methods that combine vertex, impact parameter 
and jet shape information into multivariate discriminants~\cite{Bayatian:2006zz}. 

Measurements of $t\bar{t}$ production cross sections that rely upon $b$-tagging 
assume the SM branching ratio $Br(t\rightarrow Wb)\sim$~100\%. There 
are however physics models in which this is not true and so 
an interesting approach to $t\bar{t}$ signal discrimination 
is the use of event topology information. D0 has 
used a set of 13 variables to build a discriminant 
function~\cite{Abazov:2007kg} in the lepton plus jets channel. 
Compared to  $t\bar{t}$ events in which the final 
state objects are distributed somewhat isotropically, 
background events from hard-scatter processes 
have more of a ``back-to-back'' structure.  
These observations can be quantified using 
$aplanarity$ $\mathcal{A}~=~\frac{3}{2}\lambda_3$ and 
$sphericity$ $\mathcal{S}~=~\frac{3}{2}(\lambda_2 + \lambda_3)$, defined by 
the normalized 2D momentum tensor $\mathcal{M}_{ij}~=~\sum p_i p_j/\vert \vec{p}^2\vert$
of the jets in the event and the $\lambda_k$
are ordered eigenvalues of the $\mathcal{M}_{ij}$ (with $\lambda_1$ the largest). In 
top quark events, these quantities are larger than for background.
In addition,  $t\bar{t}$ events are produced close to threshold and so have 
less longitudinal boost, resulting in a larger proportion of energy in 
the central rapidities. One thus obtains discrimination power 
from  $centrality$ $\mathcal{C}$, defined as the ratio of
the scalar sum of $p_T$ of the jets to the scalar sum of the energy 
of the jets. Other useful variables include $H_T\equiv\sum p_T$ and 
$M_T$ which are the scalar sum $p_T$ and transverse mass, respectively, of the 
four leading jets in the event.

%% file: ttbar_crosssection.tex
\section{ Measurement of the $t\bar{t}$ Production Cross Section \label{sec:xsection}} 

Measurements of the top pair production cross section ($\sigma_{t\bar{t}}$) 
can be sensitive to new particles which come from or decay to top quarks. 
Processes that could enhance $\sigma_{t\bar{t}}$ include Little Higgs 
\cite{Schmaltz:2005ky}, and strong dynamics \cite{Hill:1993hs}. Supersymmetry 
\cite{Nilles:1983ge} could produce multi-lepton, multijet signatures with 
significant $\EtMiss$  akin to $t\bar{t}$ signatures and possibly also the cascade 
$t\rightarrow H^{+}b\rightarrow\tau\nu b$ which would alter observed top quark 
branching ratios. Conversely $t\bar{t}$ production must be well 
understood at the LHC where it is a background for most new physics models. Of 
course, $\sigma_{t\bar{t}}$ is also of interest in its own 
right along with other properties of this conspicuously heavy quark. 
In Run 2, CDF and D0 have accumulated $\sim$50 times the integrated luminosity 
of Run 1. Current theoretical uncertainties on top production at the Tevatron 
have $\sim$10\% uncertainties. At the LHC the product of cross-section and 
luminosity will be roughly three orders of magnitude higher, making the LHC
a true top quark factory.

The cross section is determined by the number of observed candidate events 
$N_{obs}$, the estimated number of background events $B$, the integrated 
luminosity $\int \mathcal{L} dt$, and the $t\bar{t}$ acceptance times efficiency 
$\mathcal{A}\times\epsilon$. The latter is defined as the fraction of 
simulated $t\bar{t}$ events that pass all selection criteria after correcting for 
known differences between real and simulated data as estimated with real and 
simulated control samples. With regard to the latter, there are a variety of 
programs regularly used. Exact leading-order calculations such 
as \alpgen~\cite{Mangano:2002ea} and \madgraph~\cite{Maltoni:2002qb} provide the 
partonic final states underlying some important background processes such 
as $W/Z~+~n~jets$ with $n~=1,2,3,~\ge 4$. These programs are coupled with showering MC 
such as \pythia~\cite{Sjostrand:2000wi} and \herwig~\cite{Corcella:2000bw} 
which are often used for $t\bar{t}$ signal samples 
and rare electroweak backgrounds such as di-bosons. 
Heavy flavor decays are handled by programs such as 
\evtgen~\cite{Lange:2001uf}. In all cases, the 
\GEANT~\cite{Agostinelli:2002hh,Allison:2006ve} program is 
used to simulate the response of the detectors.  More details on
Monte Carlo routines used in specific measurements of the $t\bar{t}$ production 
cross sections are available in the cited references.   

%% could discuss simulations here, Alpgen etc. 
The final cross section can be calculated by maximizing the likelihood of 
obtaining $N_{obs}$ events given the number expected $N_{exp}~=~S~+~B$, where  
$S~=~\sigma_{t\bar{t}}\mathcal{A}\epsilon \int\mathcal{L} dt$. Uncertainties 
are then taken from the cross section values where the logarithm of the 
likelihood decreases by 0.5, and systematic uncertainties are included in 
the likelihood as nuisance parameters with Gaussian probability distributions.
The result of the likelihood maximization is equal to that obtained from the 
familiar formula 
$\sigma_{t\bar{t}}=(N_{obs}-B)/(\mathcal{A}\epsilon\int\mathcal{L}dt)$,
and yields statistical uncertainties for a Poisson probability distribution 
while also having the capacity to extract a single cross section from multiple 
data samples.

%In this section the current understanding of $t\bar{t}$ production at the 
%Tevatron is reviewed, followed by a brief look at plans for $t\bar{t}$ 
%production studies at the LHC. 
What follows is a broad-brushed presentation, emphasizing 
important methods and results. Details are available in the references. Unless
otherwise specified, the cross section values presented in this section assume  
$M_t~=~175$~GeV/c$^2$.

\subsection{Measurements of $\sigma_{t\bar{t}}$ in the Dilepton Channel \label{sec:dilepton}}

At hadron colliders the main SM processes with true high-$p_T$ $e\mu$ dilepton 
final states are $t\bar{t}$, $W^{\pm}W^{\mp}$ and 
$Z/\gamma^*\rightarrow \tau\tau\rightarrow e\mu\nu_e\nu_{\mu}\nu_{\tau}\nu_{\tau}$. 
These processes are distinguishable from one another by jet counts and $\EtMiss$.  
There is typically less $\EtMiss$ in $Z\rightarrow \tau\tau$ events due to the 
many neutrinos which tend to wash out their collective impact and steal  
energy from the other leptons ($\ell ~=~ e,\mu$). The $W/Z$ processes 
are generally produced with fewer and softer jets than $t\bar{t}$. The 
hadronic tau decay channels, labeled as $\ell\tau$ and $\tau\tau$, 
arise from these same processes but their experimental signatures are not 
as distinctive. Nevertheless, they can be statistically separated from jets to some 
extent, as discussed below. The $ee$ and $\mu\mu$ final states again involve 
the same SM processes but must also contend with Drell-Yan (DY) 
$Z/\gamma^*\rightarrow \ell^+\ell^-$. To combat this, one raises 
the $\EtMiss$  threshold for all opposite-sign same-flavor (OSSF) dileptons and 
either vetoes events in the ``Z mass window'',  
($\vert M_{\ell\ell}-M_Z\vert \le 15$ GeV), or raises the $\EtMiss$  threshold 
further there. 

In addition to sources of real lepton pairs, there are ``instrumental backgrounds'' 
in which leptons are faked by jets or other types of leptons. An isolated $\mu$ 
could be faked by a jet that fragments primarily to a single leading $p_T$ charged 
hadron that decays to a $\mu$, or by a hadron that punches through the hadron 
calorimeter to leave a signal in the muon system. A jet containing mainly $\pi^o$'s 
can fake an isolated $e$ if a high $p_T$ electron from $\gamma\rightarrow ee$  
provides a matched track.  Hadronic $\tau$ decays have 
signatures similar to very narrow jets 
with low multiplicities of charged hadrons. Thus for $ee,~e\mu,~\mu\mu$, and 
$\ell\tau$ samples a significant background is $W~+~jets$ with 
$W\rightarrow e\nu,~\mu\nu$~or~$\tau\nu$ and a fake $e$ or $\mu$. 
A smaller background arises from multijet events requiring two fake leptons 
and fake $\EtMiss$. Though rare, these ``QCD'' events do occur because 
the QCD production cross section is many orders of magnitude higher than those   
for processes with real high $p_T$ isolated leptons.

Of course $t\bar{t}$ events always contain a pair of bottom quarks while the 
other SM sources of real or fake dileptons rarely do. Thus $b$ tagging can be 
used to improve $S/B$ in dilepton events. This was rarely done in earlier 
Tevatron measurements because of a shortage of events and the inherent  
purity of the $t\bar{t}$ dilepton signature. Larger dilepton samples are now 
available and analyses include $b$ tagging to produce ultra-pure $t\bar{t}$ samples.

%best dilepton result, CDF conf. 2.8/pb

\begin{figure}[tb]
\begin{center}
\includegraphics[width=0.4\textwidth]{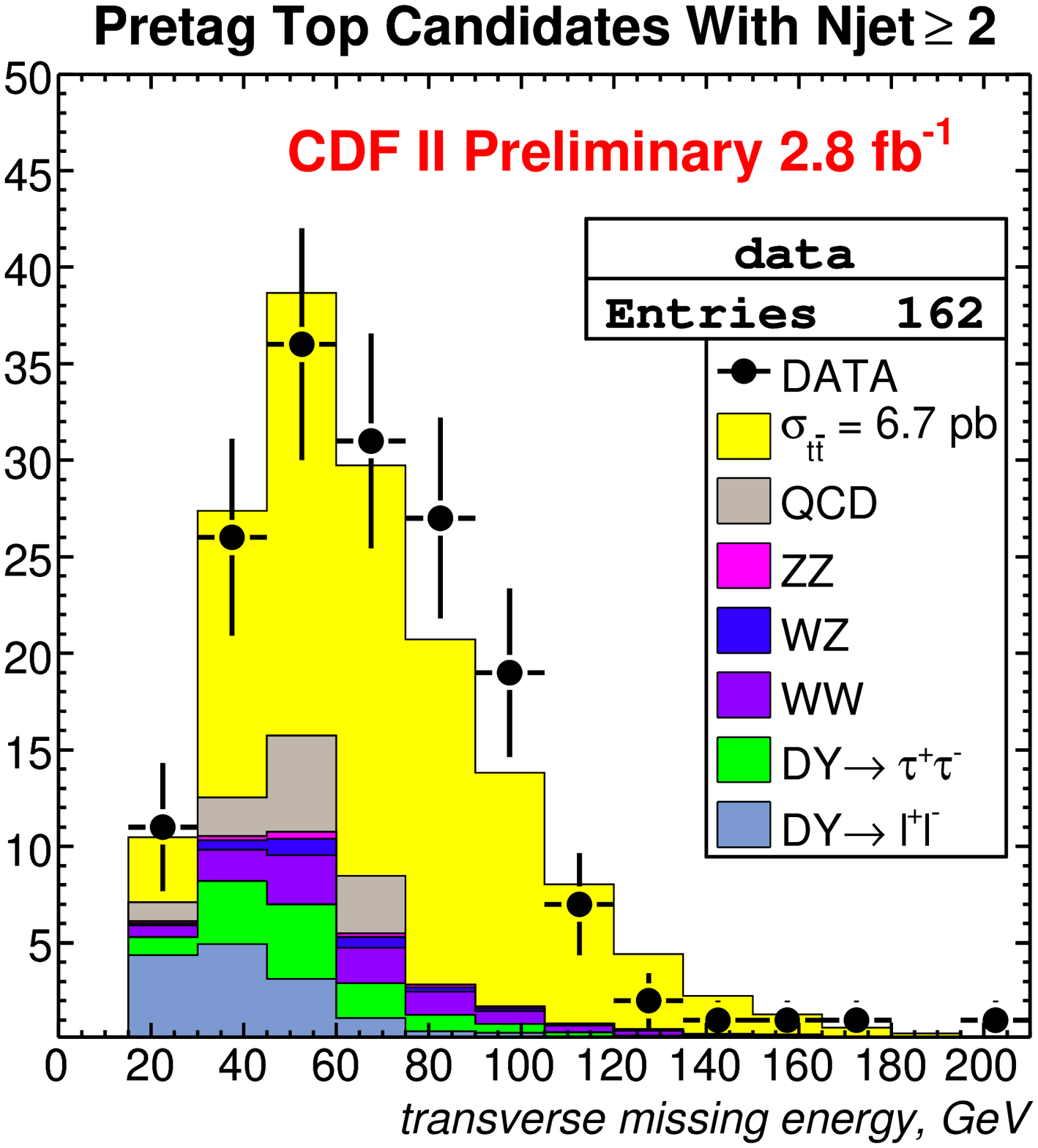}
\includegraphics[width=0.4\textwidth]{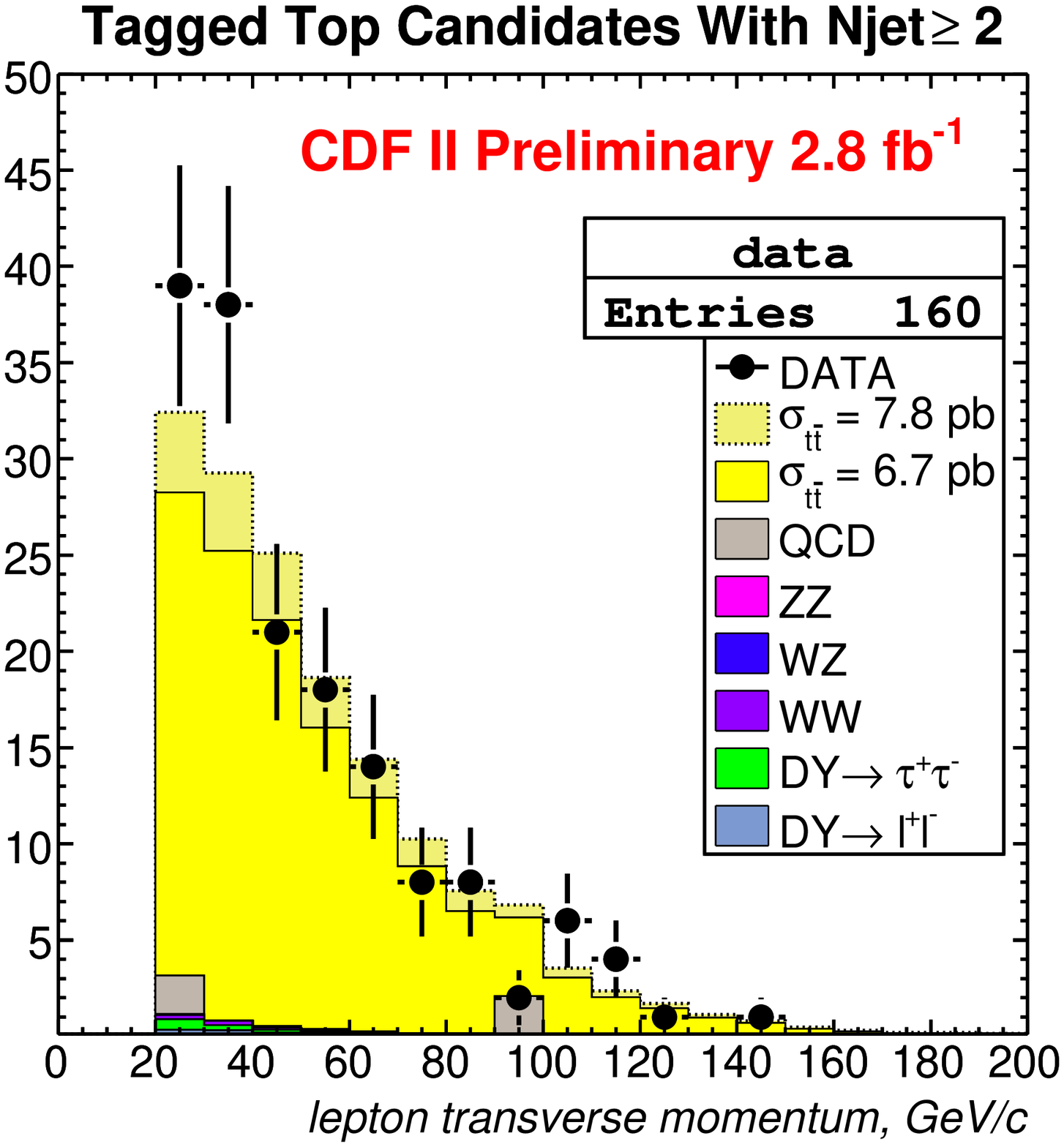}
%\begin{minipage}[t]{8 cm}
%\epsfig{file=cdf_dil_2800pb.eps,scale=0.5}
%\end{minipage}
%\begin{minipage}[t]{16.5 cm}
\caption{Comparison of data to expectations for missing $E_T$ for 
pre-tag (left) and lepton $E_T$ for tagged (right) CDF dilepton events \cite{cdfnote9399}. \label{fig:cdf_dil_2800pb}}
%\end{minipage}
\end{center}
\end{figure}

The most recent CDF measurement of $\sigma_{t\bar{t}}$ in the dilepton channel 
\cite{cdfnote9399} makes use of $2.8\,\mathrm{fb^{-1}}$ of data. 
The basic selection and treatment of backgrounds is typical of  
most such measurements. Event selection includes two opposite sign (OS) leptons, 
large $\EtMiss$ and two jets with $E_T~>~$30 and $E_T~>15\,\mathrm{GeV}$. 
Remaining backgrounds 
fall into two categories as discussed above in which both leptons are real or at 
least one is fake. Diboson 
($WW,~WZ,~ZZ$) and $Z/\gamma^*$ are estimated from MC samples 
taking into account acceptances, efficiencies, theoretical cross sections, and 
the appropriate integrated luminosity. Differences between 
simulation and data are corrected as discussed in section \ref{sec:event}. 
Jet multiplicity scale factors for processes involving multiple jets from 
QCD radiation such as $WW$ and $Z/\gamma^*$ are also implemented. The contribution 
of events with a fake lepton is estimated with a sample of same sign (SS) dileptons 
having the same kinematic selection as the OS sample. It is assumed that 
for the $W~+~jets$ background a fake lepton will pair with a real lepton with the 
same or opposite sign at the same rate. This is verified in the background dominated
$e\mu~+~\le 1$~jet control sample. Kinematic and geometric acceptance $\mathcal{A}$ 
is determined in simulation. Prior to $b$ tagging, the ``pre-tag'' sample has a 
total of 162 events and an estimated background of $51.9\pm4.5$ events. The 
dominant uncertainties are in the estimated background from fake leptons and the 
jet energy scale (JES). Requiring at least one $b$ tag leaves 80 events with a
background $4.0\pm1.7$ events.  The cross sections obtained without and with 
a $b$-tag requirement are:
$\sigma_{t\bar{t}}^{0b}=6.67\pm0.77~\rm{(stat)}~\pm0.43~\rm{(syst)}~\pm0.39~
\rm{(lumi)~pb}$, and $\sigma_{t\bar{t}}^{1b}=7.81\pm0.92~\rm{(stat)}~\pm0.68~
\rm{(syst)}~\pm0.45~\rm{(lumi)~pb}$ where the systematic uncertainty is a 
convolution of acceptance and background uncertainties. The $\EtMiss$ distribution 
for events in the pre-tag sample and the lepton $E_T$ distribution in the tagged  
sample are shown in Fig. \ref{fig:cdf_dil_2800pb}.

%Previous CDF dilepton measurements \cite{Abe:1997iz,Acosta:2004uw} did not 
%use $b$ tagging. 
%The former was similar to that described above in that it required 
%two well-identified leptons ($\ell ~\equiv~ e,\mu$). 
%It was performed in Tevatron Run 1 and found 9 
%candidates with an estimated background of 2.4$\pm$0.5 events 
%in a sample corresponding to an integrated luminosity of $109\pm7$~pb$^{-1}$ 
%to yield $8.2^{+4.4}_{-3.4}$~pb to be compared to the 
%expected value of $\sim 5$~pb at $\sqrt{s}~=~1.8$ TeV and $M_t~=~175$ GeV.  
%Ref.  

The first CDF dilepton measurement in Run 2 \cite{Acosta:2004uw} did not 
use $b$ tagging but combined two measurements, one of which required the 
standard OS leptons while the other required only one $e$ or 
$\mu$ paired with an OS, isolated high $p_T$ track (referred to as a ``track-lepton''). 
For a data sample of $197\pm12$~pb$^{-1}$ they reported a cross section 
of $7.0^{+4.4}_{-3.4}$~pb. The track-lepton approach has 
the potential to increase the signal 
acceptance, particularly for $W\rightarrow\tau\nu$ decays, and so improves  
sensitivity to $t\rightarrow H^{+}b\rightarrow\tau\nu b$. 
The track-lepton approach has been refined and used with 1.1~fb$^{-1}$ by 
CDF \cite{Aaltonen:2009ve} and 0.4~fb$^{-1}$ by D0 \cite{Abazov:2007bu}. 
Note that the probability for a jet to fake a track-lepton is more than an 
order of magnitude higher than that for a jet to fake an $e$ or $\mu$, making 
the $W~+~jets$ background more significant. Indeed, with higher 
statistics samples the standard way of using jet data to parameterize 
the per-jet fake rates versus $E_T$ and $\eta$ begins to 
show inadequacies when applied to control samples like the $W~+~1j$ sample. 
The CDF analysis used jets in $\gamma~+~jets$ and $Z~+~jets$ data to 
parameterize the per-jet fake rates applied to $W~+~jets$ events. The 
method takes into account the fact that quark jets have roughly an 
order of magnitude higher probability to fake a track-lepton 
than do gluon jets, and $W/Z/\gamma~+~jets$ events often contain a leading quark jet in the 
final state whereas dijets more often contain gluons. The new approach 
reduces the relative uncertainty on the fake lepton background to $<$~20\% from
30\%~-~50\% in prior Run 2 studies. Other sources of backgrounds for 
this analysis are $Z/\gamma^*\rightarrow ee,~\mu\mu,~\tau\tau$ in association 
with jets. The $\tau\tau$ contribution is estimated with MC events for which  
jet multiplicity is weighted to match data (e.g. $\ell\ell~+~jets$ events
in the $Z$ mass window). The $ee$ and $\mu\mu$ contributions outside 
the $Z$ mass window that pass $\EtMiss$ cuts are estimated by multiplying the 
ratio of simulated DY events passing cuts outside and inside the $Z$ window 
times the observed number of events inside the $Z$ window. The final 
cross section combining tagged and untagged sub-samples is 
$\sigma_{t\bar{t}}=9.6\pm 1.2~\rm{(stat)}~^{+0.6}_{-0.5}~
\rm{(syst)}~\pm0.6~\rm{(lumi)~pb}$.
%The resulting cross sections in the pre-tag and tagged sample (requiring at least 1 b
%tag per event) are:
%$${\sigma_t\bar{t}}^{pre}=8.3\pm1.3~\rm{(stat)}~\pm0.8~\rm{(syst)}~\pm0.5~\rm{(lumi)}~pb$$ 
%$${\sigma_t\bar{t}}^{b}=10.5^{+1.4}_{-1.3}~\rm{(stat)}~^{+0.8}_{-0.7}~\rm{(syst)}~\pm 0.6~\rm{(lumi)~pb}$$
%while the combined cross section of the tagged and untagged sub-samples is:
%$${\sigma_t\bar{t}}=9.6\pm1.2~\rm{(stat)}~^{+0.6}_{-0.}~\rm{(syst)}~\pm0.6~\rm{(lumi)~pb}$$

In the D0 analysis \cite{Abazov:2007bu} the track-lepton category requires a $b$-tag. 
To extract the background with fake isolated muons D0 employs a ``tight'' sample 
containing $N_T$ events passing all $\mu\mu$ selection criteria and a ``loose'' sample of 
$N_L$ events with only one $\mu$ required to be isolated. The loose and
tight event counts are then related to the signal-like events ($N_{sl}$) and 
background-like events ($N_{bl}$) in the ``loose'' sample where signal-like events 
contain muons from $W/Z$ decays and background-like events contain at least one 
muon candidate from sources associated with a jet. Thus  
$N_L~=~N_{sl}~+~N_{bl}$ and $N_T~=~\epsilon N_{sl}~+~f_{\mu} N_{bl}$  
where $\epsilon$ ($f_{\mu}$) is the probability for the second muon in 
signal-like (background-like) events to pass the isolation requirement. 
The same method expanded to four equations and four unknowns corresponding to  
pairings of real or fake leptons or track-leptons is used to estimate 
$W~+~jets$ and multijet backgrounds from special control samples. The 
efficiencies for the signal-like loose events to pass tight isolation 
criteria are obtained from simulated DY events while the corresponding efficiencies for 
background-like loose events were measured in multijet data with very low $\EtMiss$. 
The cross section obtained by combining all channels is
$\sigma_{t\bar{t}}=7.4\pm1.4~\rm{(stat)}~\pm0.9~\rm{(syst)}~\pm0.5~\rm{(lumi)}~pb$.
 
More recently D0 carried out a study with a dataset of 
$1\,\mathrm{fb^{-1}}$~\cite{Abazov:2009si} 
using a more sophisticated approach to $\tau$ identification. Three types of $\tau$ 
decays are defined and neural networks ($NN_{\tau}$) for each type are developed 
using discriminating variables based on longitudinal and transverse 
shower profiles and calorimeter isolation. The three types 
of $\tau$ decay are characterized by a single track with energy deposited in
the hadronic calorimeter (``$\pi^{\pm}$-like''), a single track with energy 
in both the EM and hadronic calorimeters (``$\rho^{\pm}$-like'') and 
2 or 3 tracks with invariant mass below 1.1 or 1.7 GeV, respectively.
%The main source of background to the $\ell\ell$ channels with 
%$\ell~=~e,\mu$ are $Z/\gamma^*\rightarrow\ell\ell$ and 
%$Z/\gamma^*\rightarrow\tau\tau$ with both $\tau$'s decaying leptonically, 
%$\tau\rightarrow\ell\nu_{\ell}\nu_{\tau}$. Similar approaches are used to treat
%these backgrounds. 
The cross section obtained for the combination of all channels is 
$\sigma_{t\bar{t}}=7.5\pm~1.0~\rm{(stat)}~^{+0.7}_{-0.6}~\rm{(syst)}~^{+0.6}_{-0.5}~\rm{(lumi)~pb}$. The $\ell\tau$ channel was used to obtain a first
measurement of the cross-section times branching ratio:
${\sigma_{t\bar{t}}}\times Br(t\bar{t}\rightarrow \ell\tau b\bar{b}) = 0.13^{+0.09}_{-0.08}~\rm{(stat)}~^{+0.06}_{-0.06}~\rm{(syst)}~^{+0.02}_{-0.02}~\rm{(lumi)~pb}$
for $M_t=170\,\mathrm{GeV}$, in agreement with the SM expectation of $0.14\pm0.02$ pb.
%$${\sigma_{t\bar{t}}}=7.6*{+4.9}_{-4.3}~\rm{(stat)}~^{+3.5}_{-3.4}~
%\rm{(syst)}~^{+1.4}_{-0.9}~\rm{(lumi)~pb}$$
In the $\ell\tau$ channel, QCD multijet backgrounds are partially reduced by requiring 
$15<\EtMiss<200\,\mathrm{GeV}$ and a significant azimuthal separation between the $p_T$ of the
$\ell$ and 
$\EtMiss$. The final cleanup of the $\ell\tau$ channel is provided by the requirement 
of at least one $b$-tag. The simulated background from $W~+~\ge~2$~jet events is 
normalized by fitting the transverse mass distribution of the isolated lepton and $\EtMiss$~ to
data. The multijet background is estimated with $SS$ events, taking into 
account $SS$ contributions from other backgrounds and $t\bar{t}$ with MC. 

A final CDF dilepton measurement of note in recent years is one that employs
a global fitting method \cite{Abulencia:2006mf} to 
simultaneously extract the production cross sections for $t\bar{t},~W^+W^-$
and $Z\rightarrow\tau\tau$ events. These processes populate regions of 
the $\EtMiss$-$N_{j}$ plane to different degrees, where $N_j$ is the number of 
jets in the event. A study of the broader 
$\EtMiss$-$N_{j}$ plane has the potential to improve our understanding of all   
three processes and at the same time provide greater 
sensitivity to the appearance of discrepancies with the SM. 
The results for all three processes are consistent with SM 
expectations and the $t\bar{t}$ result is $\sigma_{t\bar{t}}=8.5^{+2.7}_{-2.2}$~pb. 

\subsection{Measurements of $\sigma_{t\bar{t}}$ in the Lepton+Jets Channel \label{sec:leptonjets}}

The $\ell~+jets$ channel ($\ell~=~e,~\mu$) is not intrinsically as pure as the dilepton 
channel. On the other hand, it has a much higher branching fraction and so 
can afford more stringent selection criteria to yield large event samples with good 
$S/B$. This is why the $\ell~+jets$ channel played the largest role in the 
discovery of the top quark \cite{Abe:1995hr,Abachi:1995iq} and carries the largest 
weight in the measurement of $M_t$ and many other top quark properties. The two main 
ways to obtain pure samples of $t\bar{t}$ events in this channel are 
the use of $b$ tags and event topology information. 
The main backgrounds are $W~+~jets$ events with $W\rightarrow\ell\nu_{\ell}$ 
and multijet events where a jet is misidentified as a lepton (as discussed in 
section \ref{sec:dilepton}) and energy is mis-measured resulting in  
substantial $\EtMiss$. The $t\bar{t}$ signal tends to occur with jet multiplicities 
$N_j~\ge 4$ but there is also a non-negligible fraction with $N_j~=3$.  
By contrast, the main  backgrounds decrease rapidly with increasing jet multiplicities.

%D0 L+jets Kin and B tag

In the kinematic likelihood analyses of Ref.~\cite{Abazov:2007kg}, D0 
selects events with $N_j~\ge 4$. The multijet background contribution 
is determined with data samples selected with loose and tight lepton 
identification as was discussed for the dilepton mode.  
The ``loose - tight'' sample, containing those events passing the 
loose, but not the tight criteria is used to model distributions of the 
variables used in the kinematic likelihood for multijets after correcting 
for residual $t\bar{t}$ and $W/Z~+~jets$ events. The variables, 
discussed in Section \ref{sec:event}, exploit the topological differences 
between $t\bar{t}$ and backgrounds. They include aplanarity $\mathcal{A}$ and 
centrality $\mathcal{C}$, the $\sum {p_T}^2$ and $\sum {\eta}^2$ for the four 
leading jets, and many others. Templates for the distributions of these variables 
in $t\bar{t}$ and $W/Z~+~jets$ are based on simulated MC samples. The validity 
of the templates is tested on the $N_j~=3$ sample. The templates are used 
in a maximum likelihood fit to the data. The multijet normalization is 
constrained to its estimated contribution within uncertainties. The $t\bar{t}$ 
cross section is measured separately for the different lepton types and all types 
combined. The combination result for 425 pb$^{-1}$ is: 
${\sigma_{t\bar{t}}}=6.4^{+1.3}_{-1.2}~\rm{(stat)}~\pm0.7~\rm{(syst)}~\pm 0.4~\rm{(lumi)~pb}$.
%These results are considerably more precise than the previous published results
%\cite{Abazov:2005ex} obtained with 230 pb$^{-1}$. 
D0 has recently published a result based on 0.9 
fb$^{-1}$ \cite{Abazov:2008gc} which alters somewhat the choice of 
discriminating variables that are used to include measures of the 
separation in $\phi$ and $\eta~-~\phi$ between pairs of objects. 
The analysis also includes the $N_j~=3$ sample in the measurement, requiring that 
$\sum_{i=1}^3 {p_T}^i~>~120$ GeV for these events.  They obtain 
${\sigma_{t\bar{t}}}=6.62\pm0.78~\rm{(stat)}~\pm0.36~\rm{(syst)}~\pm 0.40~\rm{(lumi)~pb}$. 
This is combined with an analysis in which at least one $b$ tag is required in place of 
the kinematic likelihood. The $b$ tag algorithm uses a neural network   
which combines parameters sensitive to the displaced decay vertices of the 
B hadrons. A typical operating point in D0 top analyses
achieves a $b$ jet efficiency of $\sim$54\% and light flavor mistag rate of $\sim$1\%.
The multijet background is determined as described above and 
contributions from other background sources in the inclusive sample
are estimated by multiplying events by their probability to be
$b$-tagged. The latter is parameterized for jet $E_T$ and $\eta$ using 
MC samples with corrections for object reconstruction differences between data and MC
and for the estimated $t\bar{t}$ content of the sample. The latter depends on 
$\sigma_{t\bar{t}}$ and so the process is repeated iteratively allowing the 
$t\bar{t}$ cross section estimate to vary until the result stabilizes. The 
result obtained is
$\sigma_{t\bar{t}}=8.05\pm0.54~\rm{(stat)}~\pm0.7~\rm{(syst)}~\pm 0.49~\rm{(lumi)~pb}$. The combined result is: $\sigma_{t\bar{t}}=7.42\pm0.53~\rm{(stat)}~\pm0.46~\rm{(syst)}~\pm 0.45~\rm{(lumi)~pb}$. 
The main sources of systematic uncertainty are the selection efficiency, b-tagging, 
JES and MC modeling which each contribute $\sim 0.2$ pb. 

%CDF L+jets Kin and B tag

CDF has used secondary vertex ($SVX$), Jet Probability ($JPB$), and Soft Lepton 
($SLT_{\mu}$) $b$-tag algorithms as well as kinematic likelihoods in Run 2 measurements of 
$\sigma_{t\bar{t}}$.  The $JPB$ algorithm was used with a 318 pb$^{-1}$ dataset 
\cite{Abulencia:2006kv} requiring at least one tight 
($P_J~\le~$1\%) $b$-tag to obtain 
${\sigma_{t\bar{t}}}=8.9\pm~1.0~\rm{(stat)}~^{+1.1}_{-1.0}\rm{(syst)}$~pb for
$M_t~=~178$~GeV/c$^2$. The $SVX$ algorithm was used with a comparable dataset 
\cite{Abulencia:2006yk} to obtain ${\sigma_{t\bar{t}}}=5.8\pm~1.2~\rm{(stat)}
~^{+0.9}_{-0.7}\rm{(syst)}$~pb. The $SLT_{\mu}$ algorithm described briefly in Section 
\ref{sec:event} was recently applied to 2 fb$^{-1}$ of data \cite{:2009ax}. 
This analysis makes use of a new way of measuring the dominant 
background of fake muons in $W~+~jets$ from decay in flight and punch-through hadrons. 
First, kaons, pions and protons are identified 
by reconstructing $D^{*+}\rightarrow D^o\pi^+\rightarrow K^-\pi^+\pi^+$, and 
$\Lambda\rightarrow p\pi^-$ decays (and their conjugates) using data 
collected with the SVT trigger. The muon fake rate is then measured for each 
particle type in bins of $p_T$ and $\eta$ and convoluted with the corresponding
probabilities in these bins for the various particle types to appear in 
$W~+~jets$ events as taken from simulation. They obtain 
$\sigma_{t\bar{t}}~=~9.1~\pm ~1.6$~pb.

The most recent and the most precise $t\bar{t}$ production cross section measurements are
performed by CDF with nearly 2.8 fb$^{-1}$ of data where the uncertainty is 
limited by the systematic uncertainty which in turn is dominated by the 5.8\%
uncertainty on the integrated luminosity. To overcome this hurdle the 
ratio $\mathcal{R} \equiv \sigma_{t\bar{t}}/\sigma_Z$ is measured using 
$\ell~+~jets$ events selected in a fairly standard way. 
%leptons and jets with $E_T~>~20$~GeV, leading jet $E_T~>~30$ and $\EtMiss$~$>$~35 GeV. 
The $t\bar{t}$ and $Z$ cross sections are then measured using events 
collected with identical triggers. The $t\bar{t}$ events are 
discriminated from backgrounds using either an artificial neural network 
($ANN$) \cite{cdfnote9474} based on kinematic and topological variables like 
those used by D0, or by means of the secondary vertex $b$-tag algorithm 
\cite{cdfnote9616}.  The dominant $W~+~jets$ background is modeled 
in the $ANN$ study using simulated samples of $W~+~n~jets$ with $n~=~0,1,...,4$ 
partons. In the $SVX$ study the $W~+~jets$ background has to be broken into 
separate heavy flavor ($HF$) and light flavor ($LF$) jet categories. Because $W~+~HF$  
events are poorly modeled in Monte Carlo, a data-driven correction factor 
is calculated in a non-signal region and used to estimate the amount 
of $W~+~HF$ in the signal region. $W~+~LF$ events only enter the final sample when 
a light flavor jet is mistagged. This background is estimated by applying a 
parameterization of the mistag rate from jet data to events that pass selections 
except for a secondary vertex tag. 
For the non-$W$ background in both analyses, 
the $\EtMiss$ distribution in data is fit to templates for non-$W$ events (taken 
for instance from dijet triggers passing all selection requirements except  
the $\EtMiss$ requirement) and a template for $t\bar{t}$ signal from simulation.  
The result for the $SVX$ analysis is 
${\sigma_{t\bar{t}}}=7.1\pm0.4~\rm{(stat)}~\pm0.6~\rm{(syst)}~\pm 0.4~\rm{(lumi)~pb}$.
The inclusive $Z/\gamma^*$ cross section is then measured in the range 
$66~\le ~M_{\ell\ell}~\le~ 116$~GeV using consistent triggers and definitions of leptons as
for the $t\bar{t}$ cross section to obtain:
$Br(Z/\gamma^*\rightarrow \ell\ell)\times \sigma_{(Z/\gamma^*\rightarrow \ell\ell)}
=253.5\pm 1.1  ~\rm{(stat)}~\pm4.5~\rm{(syst)}~\pm 14.9~\rm{(lumi)~pb}$
Together with the $t\bar{t}$ result one obtains 
$1/\mathcal{R}~=~35.7~\pm 2.0 \rm{(stat)}~\pm 3.2~\rm{(syst)}$, where $\mathcal{R}$ is
the ratio of $t\bar{t}$ to $Z$ production. 
Multiplying $\mathcal{R}$ by the current theoretical value for the $Z$ cross-section 
($\sigma_z~=~251.3\pm 5.0$pb) one obtains
${\sigma_{t\bar{t}}}=7.0\pm0.4~\rm{(stat)}~\pm0.6~\rm{(syst)}~\pm 0.1~\rm{(theory)~pb}$.
A similar procedure was carried out for the $ANN$ analysis to obtain
$1/\mathcal{R}~=~36.5~^{+2.1}_{-2.3}~\rm{(stat)}~^{+1.9}_{-2.0}~\rm{(syst)}$ and 
${\sigma_{t\bar{t}}}=6.89\pm0.41~\rm{(stat)}~^{+0.41}_{-0.37}~\rm{(syst)}~\pm 0.14~\rm{(theory)~pb}$. 

\subsection{Measurements of $\sigma_{t\bar{t}}$ in the All-Hadronic Channel \label{sec:allhadronic}}

Despite having the largest branching ratio, the absence of leptons in the
all-hadronic channel makes it the most difficult of all channels in which to isolate
$t\bar{t}$. Since the first observations of $t\bar{t}$ in the all-hadronic channel in 
Run 1 by CDF \cite{Abe:1997rh} and D0 \cite{Abbott:1999mr}. both experiments have 
continued to improve their strategies for approaching this challenging measurement.
D0 measured $\sigma_{t\bar{t}}$ with 405~pb$^{-1}$ in Run 2, selecting 
events with $\ge$6 jets, employing ``loose'' and ``tight'' secondary vertex 
$b$-tagging corresponding to requirements on decay length significance 
$L_{2D}/\sigma_{L_{2d}}~>$ 5 or 7, respectively. Events were required to have 
at least one tight or two loose tags. Single and double tagged events were treated
separately. A neural network was used to estimate the $t\bar{t}$ content of the 
samples in order to extract a cross section. The neural network used a variety of
variables to distinguish $t\bar{t}$ from QCD multijets which is the only relevant 
background. The 6 variables used include ones that are similar to those 
discussed previously in the context of other kinematic discriminant analyses but 
also included a mass $\chi^2$ variable $\mathcal{M}$ which utilized the $W$ 
and top quark mass constraints associated with $t\bar{t}$ events. The combined 
cross section is found to be  ${\sigma_{t\bar{t}}}=4.5^{+2.0}_{-1.9}~\rm{(stat)}~^{+1.4}_{-1.1}~\rm{(syst)}~\pm 0.3~\rm{(lumi)~pb}$.

More recently CDF has performed a measurement with 1.02~fb$^{-1}$ of data using 
a basic event selection of 6 or more jets, of which at least one is $b$-tagged, 
and kinematic requirements imposed by means 
of a neural network \cite{Aaltonen:2007qf}. The neural network improves the 
$S/B$ of the pre-tagged event sample by 60\% to 1/12  as 
compared to the previous cut-based measurement 
\cite{Abulencia:2006in}. The multijet background is
estimated from data using $t\bar{t}$-depleted control samples such as
multijet events with exactly four jets with an estimated  
$t\bar{t}$ fraction of roughly 1/3600. These events are used to parameterize the
per-jet $b$-tag rate in multijet events, comprising both real heavy flavor tags and 
mistags. The parameterization is then validated on multijet events with higher numbers of  
jets prior to selection with the neural-network, which is at a stage 
where the samples are dominated by 
background.  After the neural-network and $b$-tag
requirements, 926 tags are observed in 772 events for which the background is estimated
to account for 567~$\pm$~28 of the tags. The cross section obtained with the selected
sample is ${\sigma_{t\bar{t}}}=8.3\pm1.0~\rm{(stat)}~^{+2.0}_{-1.5}~\rm{(syst)}~\pm 0.5~\rm{(lumi)~pb}$

\subsection{Combined Cross Sections from the Tevatron and Prospects for the LHC  \label{sec:xsectionlhc}}

CDF and D0 have both recently released $t\bar{t}$ production cross section values 
obtained by combining results from various analyses. D0 combines results from 
the $\ell~+~jets$, $\ell\ell$ and $\ell\tau$ final states ($\ell~=~e,\mu$) measured with 
1~fb$^{-1}$ of Run 2 data \cite{Abazov:2009ae} with an assumed top mass value 
of $M_t~=~170$~GeV/c$^2$.
The combined cross section is calculated with a joint likelihood function that is 
the product of Poisson probabilities for 14 non-overlapping subsamples of events. 
Additional Poisson terms representing backgrounds estimated 
separately in each subsample are also included. Systematic uncertainties are included
as nuisance parameters with Gaussian distributions centered at zero and widths 
corresponding to a single standard deviation in the parameter uncertainty.
The final result is:
$${\sigma_{t\bar{t}}}=8.18^{+0.98}_{-0.87}~{\rm pb} ~~~~~~~~~~{\rm D0~~1~fb}^{-1}$$
CDF combines five preliminary measurements performed with 
2.8 fb$^{-1}$ of data \cite{cdfnote9448}.
One of the measurements is performed in the dilepton channel while the other 
four are in the $\ell~+~jets$ channel using different methods to discriminate 
against background including $ANN$, $SVX$ and $SLT_{\mu}$ already discussed 
earlier. The fourth measurement uses a Soft Electron $b$-tag ($SLT_e$). The
combination uses the BLUE technique \cite{Lyons:1989gh} taking into account 
statistical and systematic correlations. The resulting cross section is:
$${\sigma_{t\bar{t}}}=7.02\pm0.63~{\rm pb}  ~~~~~~~~~~~~{\rm CDF~2.8~fb}^{-1}$$

%$${\rm CDF~2.8~fb^{-1}:{\sigma_{t\bar{t}}}=7.02\pm0.30{\rm (stat)}
%~\pm0.38{\rm (syst)}~\pm0.41{\rm (syst)}~{\rm pb}$$

These measurements are in good agreement with theoretical expectations and 
have uncertainties comparable to those associated with the theoretical calculations.
Reaching this level of accuracy has required many years during which statistically 
significant signal samples were accumulated and improved methods of analysis 
were developed and exploited. The ATLAS and CMS experiments have paid close 
attention to the Tevatron $t\bar{t}$ studies over the years and now contain 
many of the seasoned Tevatron experts.  
As discussed in Section \ref{sec:lhc} the LHC 
will be a top factory, producing huge numbers of $t\bar{t}$ events. The ATLAS and
CMS collaborations are currently preparing to study  events in early 
data taking. These studies have concentrated on the leptonic final states
with an initial goal of assessing the $t\bar{t}$ production cross section at 10~TeV as 
a background to new physics searches. ATLAS has studied the 
discrimination afforded by reconstruction of the hadronically decaying 
top quark in the $\ell~+~jets$ channel with and without $b$ tagging \cite{Aad:2009wy} 
which may not be available immediately at the start of data-taking. The 
dilepton channel without $b$ tagging has also been studied by ATLAS as well as 
CMS \cite{Ball:2007zza}. Once detectors and software are well understood, and large 
event samples are collected, the LHC experiments will begin to produce results
comparable in quality to the Tevatron results but the environment of the LHC is
different from that of the Tevatron and several new elements will require 
new adaptations. For instance, there will be substantially more jets in 
$t\bar{t}$ and $W/Z~+~jets$ events. For $E_T~\ge$~15 GeV, the cross section
for $t\bar{t}~+~\ge$~1 jet saturates the total $t\bar{t}$ production cross section. 
Additionally, while it is true that $\sigma_{t\bar{t}}$ increases 
100-fold relative to the Tevatron while the inclusive $W$ and $Z$ cross sections
increase by roughly a factor of 5, the cross sections for 
$W/Z$ production in association with four or more jets \cite{Mangano:2008ag}
increase as much or more than $\sigma_{t\bar{t}}$ production 
and it is these events that are the backgrounds to $t\bar{t}$ in the  $l~+~jets$ channels, 
for instance. All in all, the study of
$t\bar{t}$ production will continue to be challenging at the LHC, but it will also
continue to be rewarding and potentially crucial to the discovery of new physics.

%% file: singletop.tex
\section{Measurement of Single Top-Quark Production}
% Author: Wolfgang Wagner

While $t\bar{t}$-pair production via the strong interaction is the main source 
of top quarks at the \tevatron and the \lhc, the SM also predicts the production 
of single top-quarks via charged-current weak interactions, in which a virtual 
$W$ boson is either exchanged in the $t$-channel ($tq\bar{b}$ production) or in the 
$s$-channel ($t\bar{b}$ production), see
section~\ref{sec:singleTopTheo}. At the \lhc associated $Wt$ production will 
also be important.
While Run I and early Run II searches for single top-quarks at 
CDF~\cite{CDFsingleTopRun1,CDFsingleTopANNRun1,CDFsingleTop2005} and 
D\O~\cite{d0SingleTopCutsRun1,d0SingleTopNNRun1,Abazov:2005zz,Abazov:2006uq} 
could only set upper limits on the production cross section, recent analyses, 
using more data and advanced analysis techniques, have
found first evidence for singly produced top 
quarks~\cite{Abazov:2006gd,Abazov:2008kt,Aaltonen:2008sy}.
Updates of theses analyses to larger data sets lead to the observation of single
top-quark production at the level of five standard deviations in 
March 2009~\cite{Aaltonen:2009jj,Abazov:2009ii}.

\subsection{Observation of Single Top Quarks at CDF and D\O}
\label{subsec:stObservation}
The main thrust of these analyses was to establish evidence for single
top-quarks, considering both production modes relevant at the \tevatron, 
$t$-channel and $s$-channel, as one single-top signal. 
The ratio of $t$-channel to $s$-channel events is assumed to be given by
the SM.
That is why, this search strategy is often
referred to as {\em combined search}. By measuring the inclusive 
single top-quark cross section and using theory predictions one can
deduce the absolute value of the CKM matrix element $|V_{tb}|$, without
assuming there are only three generations of quarks. If one would 
measure $|V_{tb}|$ to be significantly smaller than one this would 
be a strong indication for a fourth generation of quarks or other
effects beyond the SM~\cite{Bobrowski:2009ng,Alwall:2006bx}.

Single top-quark events feature a $Wb\bar{b}$ ($s$-channel) or $Wbq\bar{b}$ 
($t$-channel) partonic final state.
The $W$ boson originating from the top-quark decay is reconstructed in the leptonic decay
modes $e\nu_e$ or $\mu\nu_\mu$, while hadronic $W$ decays and decays to 
$\tau\nu_\tau$ are not considered because of large backgrounds from 
QCD-induced jet production. 
The quarks from the hard scattering process manifest themselves as 
hadronic jets with transverse momentum. Additional jets may arise from hard gluon radiation
in the initial or final state. 
The experimental signature of SM single top-quarks is therefore given
by one isolated high-$p_\mathrm{T}$ charged lepton, large missing transverse 
energy ($\EtMiss$), and two or three high-$E_T$ jets.

At D\O, various trigger algorithms have been used requiring an electron or
muon candidate and one or two jets. The backbone of the CDF analyses are
collision data triggered by central electron or muon candidates. The lepton
coverage is extended by one additional trigger path which requires large $\EtMiss$
and an energetic electromagnetic cluster in the forward calorimeter,
and one path asking for large $\EtMiss$ and two jets. The latter one targets
muon events in which a muon candidate could not be established at trigger 
level, but is reconstructable offline.

\paragraph{Event Selection} 
At CDF, the offline event selection requires exactly one 
isolated electron with $E_T>20\,\mathrm{GeV}$ or one isolated muon with
$p_T > 20\,\mathrm{GeV}/c$. Electrons and muons are reconstructed up to
$|\eta|<1.6$. D\O \ applies slightly lower $p_T$ thresholds and asks for
electrons with $E_T>15\,\mathrm{GeV}$ and muons with 
$p_T > 18\,\mathrm{GeV}/c$. Electrons are identified in the central
region up to $|\eta|<1.1$ and muons up to $|\eta|<2.0$.
In order to reduce the $Z$+jets, $t\bar{t}$, and diboson backgrounds,
events with a second lepton candidate are rejected by both collaborations. 
Cosmic ray and photon conversion events are identified and removed.
Neutrinos from $W$-boson decay remain undetected and cause an imbalance
in the transverse momentum sum. CDF selects events with 
$\EtMiss > 25\;\mathrm{GeV}$, while 
D\O \ applies the cut $15\;\mathrm{GeV} < \EtMiss < 200\;\mathrm{GeV}$.
QCD-multijet background without a leptonic $W$ decay is further minimized 
with additional requirements on, e.g., the angle between the direction 
of $\EtMiss$ and the lepton candidate, thereby removing events in which the 
lepton and the jet are in a back-to-back configuration.
Reducing QCD-multijet events to a level of just a few percent is crucial 
because it is very difficult to model the full event kinematics of
these misidentified events.

Hadronic jets are identified at CDF by a fixed-cone algorithm with
radius $\Delta R \equiv \sqrt{(\Delta \eta)^2+(\Delta \phi)^2} = 0.4$,
while D\O \ uses a midpoint-cone algorithm with $\Delta R = 0.5$.
Jet-energy corrections are applied to account for instrumental 
effects~\cite{Bhatti:2005ai} and convert reconstructed jet energies to
particle-level energies. 
Jets with the same $\eta$ and $\phi$ as a reconstructed electron are
removed from the list of jets to avoid double-counting physics objects.
CDF asks for two or three jets with $|\eta|< 2.8$ and
$E_T >20\,\rm{GeV}$. The D\O \ analyses use events with two, three, or
four jets with $E_T >15\,\rm{GeV}$ and $|\eta|< 3.4$, but there are
additional requirements on the leading and subleading jet.
The leading jet has to have $E_T >25\,\rm{GeV}$ and $|\eta|< 2.5$,
the second leading one $E_T >20\,\rm{GeV}$.
Both collaborations further require at least one of the jets to be
identified as originating from a $b$ quark.
For a jet to be $b$ tagged at CDF it must contain a reconstructed 
secondary vertex consistent with the decay of a $b$ 
hadron~\cite{Acosta:2004hw}. D\O \ utilizes an advanced $b$-tagging 
algorithm based on neural networks (NN)~\cite{Scanlon:2006wc}. 
To be subjected to the tagging 
algorithm, the jets have to contain at least two tracks that fulfill
some minimum quality criteria and point to a common origin. Those
jets are considered to be {\it taggable}.
The NN uses various input variables to discriminate $b$-quark jets 
from other jets, such as the impact parameter 
significances of the tracks associated to the jet, the mass and the
decay length significance of the secondary vertex, and the number
of tracks used to reconstruct the secondary vertex. For a jet to be
considered a $b$-quark jet the output of the NN has to be
above a certain threshold. The average probability for a light-quark
jet to be misidentified as a $b$-quark jet at this operating point 
is 0.5\%, while the average tagging efficiency for true $b$-quark 
jets with $|\eta|<2.4$ is 47\%. 

\paragraph{Background Estimation}
The expected background event rates are obtained from a compound model
comprising simulated events, theoretical cross sections, and 
normalizations in background-dominated control samples.
The basic strategy is to normalize the main component, the $W$+jets
rate, to the observed yield before applying the $b$-tagging algorithm.
This sample of events is the so-called {\it pretag} data set.
The idea behind this is that the samples of simulated $W$+jets events
describe the event kinematics well, but do not predict the event
rate correctly, and therefore it is sufficient to scale up the
event rate by a constant factor. For the normalization in the pretag
data to work the contributions to this data set which are not due to $W+$ jets
have to be determined. The $t\bar{t}$ contribution is estimated by
normalizing samples of simulated events -- CDF 
uses {\sc pythia}~\cite{Sjostrand:2006za},
D\O \ {\sc alpgen}~\cite{Mangano:2002ea} -- to NLO cross sections. CDF estimates 
small contributions of a few electroweak processes, diboson 
production ($WW$, $WZ$, $ZZ$) and $Z+$ jets in the same way, while
D\O \ incorporates those in the $W+$ jets estimate.
However, the main challenge in understanding the pretag data set is to estimate
the fraction of QCD-multijet events. CDF tackles this challenge by
removing the cut on $\EtMiss$ and fitting the $\EtMiss$ 
distribution. The region of $\EtMiss < 25\,\mathrm{GeV}$ is dominated
by QCD-multijet events and provides a normalization for this
event class in the signal region. D\O \ uses the so-called
{\it matrix method} to determine the contribution of QCD-multijet
events. This method links the number of events with real and
misidentified leptons in the pretag data set to the numbers in a 
control sample with relaxed lepton identification cuts by applying 
cut efficiencies that are derived in specific calibration data 
sets, i.e. $Z\rightarrow \ell\ell$ data or events with
$\EtMiss < 10\,\mathrm{GeV}$. 
The factors to normalize the $W+$ jets background model to
the observed pretag data are derived separately for each
$W+n$ jets sample and for the different lepton categories.
At D\O \ these scale factors vary across the samples from
$0.9\pm0.2$ to $1.6\pm 0.4$. Both collaborations use the LO matrix-element
generator {\sc alpgen}~\cite{Mangano:2002ea} combined with parton showering 
and the underlying event model by {\sc pythia}~\cite{Sjostrand:2006za}.
To properly describe the flavor composition in $W$+jets data,
CDF scales up the fraction of $Wb\bar{b}$ and $Wc\bar{c}$ events
predicted by {\sc alpgen} by a factor of $1.4\pm0.4$,
as demonstrated by studies in the $W+1$ jet sample where one
jet is identified as a $b$-quark jet.
D\O \ uses a correction factor of 1.47 for $Wb\bar{b}$ and $Wc\bar{c}$,
and 1.38 for the $Wcj$ contribution.

D\O \ derives the event yield in the $b$-tagged data sets from 
samples of simulated pretagged events after proper normalization 
by applying tag-rate functions derived from control samples. 
While D\O \ uses the tag-rate parameterization for all flavors,
CDF applies the $b$-tagging algorithm to $b$-quark jets and 
$c$-quark jets in simulated events. The rates of tagged heavy 
flavor jets are subsequently corrected by a constant factor of $0.95 \pm 0.05$.
The rate of misidentified light-quark jets at CDF is obtained
from mistag-rate functions applied to the observed pretag 
data subtracting all contributions which are not due to $W+$ light jets.  

CDF simulates single-top events using the LO matrix-element
generator~{\sc madevent}~\cite{Alwall:2007st}. The two $t$-channel processes
in Fig.~\ref{fig:singleTop}\subref{fig:WgLO} and
in Fig.~\ref{fig:singleTop}\subref{fig:WgNLO} are produced and
combined~\cite{Lueck:2006hz} to one sample to match the event kinematics as predicted
by a fully differential NLO calculation~\cite{sullivan2004}.
D\O \ uses the Monte Carlo generator 
{\sc comphep-singletop}~\cite{Boos:2006af,Boos:2004kh}
which has the matching between the two $t$-channel processes
built in.

The resulting prediction and the observed event yields are given in 
Table~\ref{tab:stop_eventyield}. 
\begin{table}[!t]
  \caption{\label{tab:stop_eventyield} Numbers of expected and observed 
    lepton+jets events in the CDF and D\O \ single top-quark analyses using
    $3.2\,\mathrm{fb^{-1}}$ or $2.3\,\mathrm{fb^{-1}}$, respectively.
    The total prediction is in some cases not equal to the sum of 
    the processes because of round-off effects.}
  \begin{center}
  \begin{tabular}{lcccccc}
    \hline
            & \multicolumn{2}{c}{CDF} & & \multicolumn{3}{c}{D\O} \\
    Process & $W + 2$jets  & $W + 3$jets & & $W + 2$jets & $W + 3$jets &  $W + 4$jets \\
    \hline
    $tq\bar{b}$+$t\bar{b}$ & $146 \pm 21$ & $45 \pm 7\ \,$ &  & $139\pm 18$ & $63\pm 10$ & $21\pm 5$ \\
    $W$+jets   & $1701 \pm 372$ & $503\pm 109$ & & $1829\pm 161$ & $637\pm 61\ \,$ &
                 $180\pm 18$ \\
    $Z$+jets / dibosons & $125\pm 9\ \,$ & $46\pm 3\ \,$ & & $229\pm 38$ & $85\pm 17$ & $26\pm 7$ \\
    $t\bar{t}$ & $204\pm 30$ & $482\pm 70\ $ & & $222\pm 35$ & $436\pm 66\ \,$ & $484\pm 71$ \\

    QCD multijets & $\ \,90\pm 36$ & $35\pm 14$ & & $196\pm 50$ & $73\pm 17$ & 
            $30\pm 6$ \\
    \hline
    Total prediction & $2265\pm 375$ & $1112\pm 130\ \,$ & & $2615\pm 192$ & 
            $1294\pm 107\ $ & $742\pm 80$ \\
    Observation in data & 2229 & 1086 & & 2579 & 1216 & 724 \\
    \hline  
  \end{tabular}
  \end{center}
\end{table}
D\O \ finds a total 
acceptance of $(2.1\pm 0.3)\%$ for the $t$-channel process
and $(3.2\pm 0.4)\%$ for the $s$-channel. The CDF analyses
use an acceptance of $(1.8\pm 0.3)\%$ for the $t$-channel
and $(2.7\pm 0.4)\%$ for the $s$-channel. 
In the lepton+jets data set corresponding to $3.2\,\mathrm{fb^{-1}}$ CDF
expects $191\pm 28$ single top-quark events, while the D\O \ 
expectation is $223\pm 32$ signal events in 
$2.3\,\mathrm{fb^{-1}}$.
The dominating background process is $Wb\bar{b}$, followed
by misidentified $W+$ light-quark jet events, $Wc\bar{c}$,
and $Wcj$. In the $W+3$ jets data set $t\bar{t}$ is the
most important background.

At CDF, an additional analysis investigates the $\EtMiss$+jets data set which is
disjoint to the lepton+jets data set mentioned above. The analysis selects events
in which the $W$ boson decays into $\tau$ leptons and those in which the electron
or muon are not identified. 
The main background in the $\EtMiss$+jets channel is QCD-multijet production
which is reduced by means of a NN discriminant that is computed using
15 kinematic variables. A cut on this discriminant reduces the background by
77\%, while the single top-quark signal is only diminished by 9\%.
In a data set corresponding to 
$2.1\,\mathrm{fb^{-1}}$  CDF selects 1411 candidate events. The expectation value
of single top-quark events is $64\pm 10$.

\begin{figure}[t]
  \begin{center}    
    \subfigure[]{
      \epsfig{file=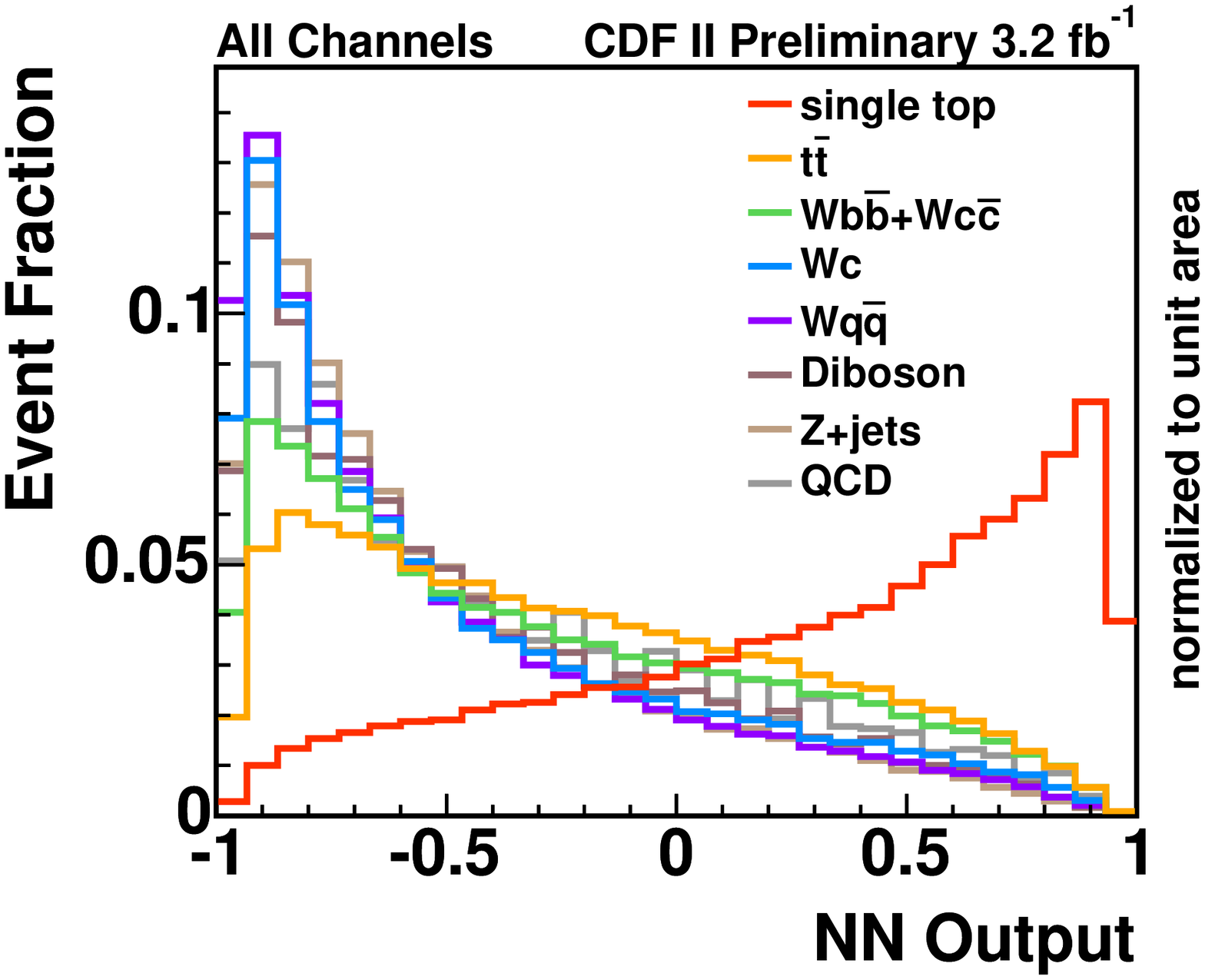, width=0.31\textwidth}
      \label{fig:CDF_NN_Temp}
    } 
    \subfigure[]{
      \epsfig{file=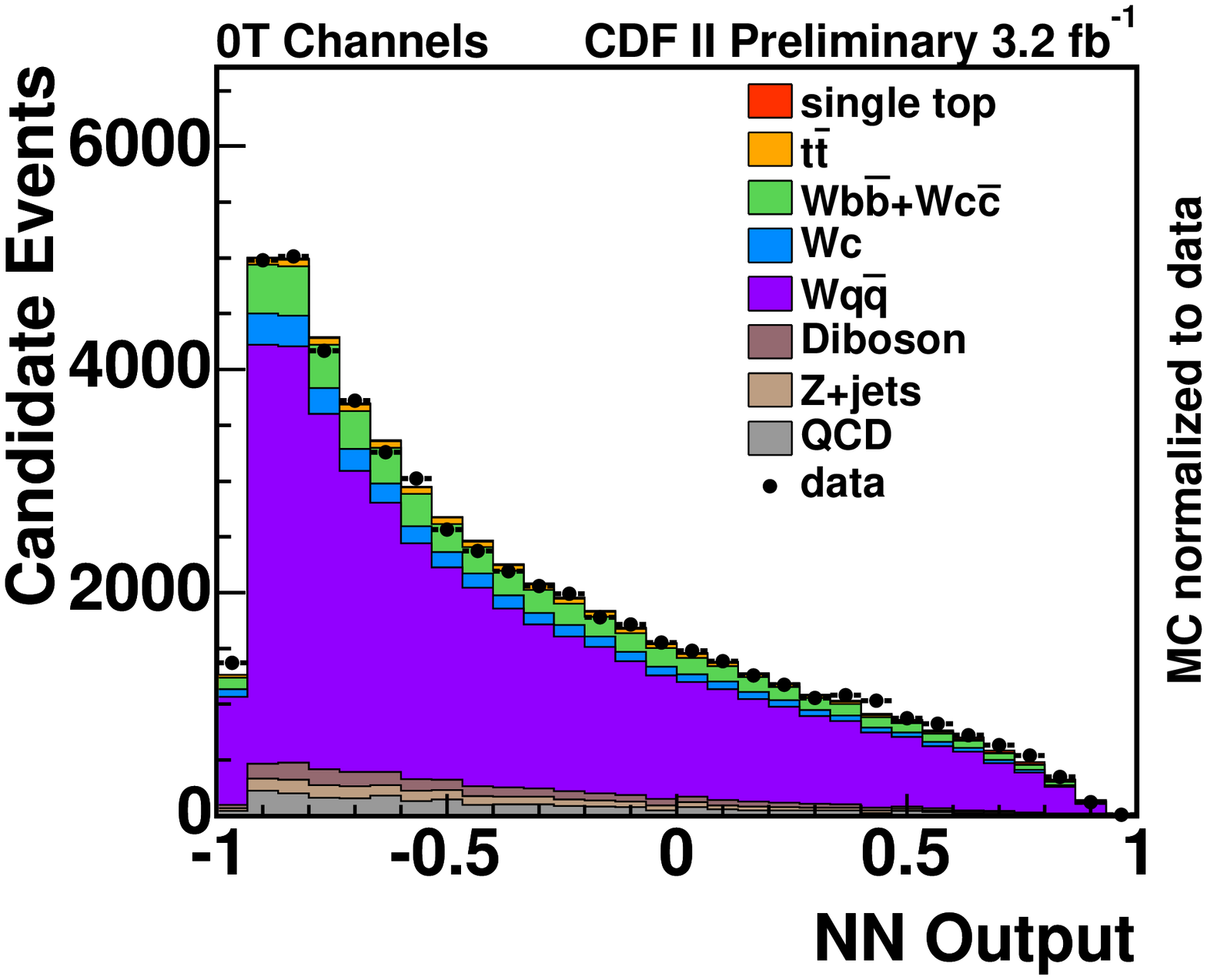, width=0.31\textwidth}
      \label{fig:CDF_NN_Pretag} 
    } 
    \subfigure[]{
      \epsfig{file=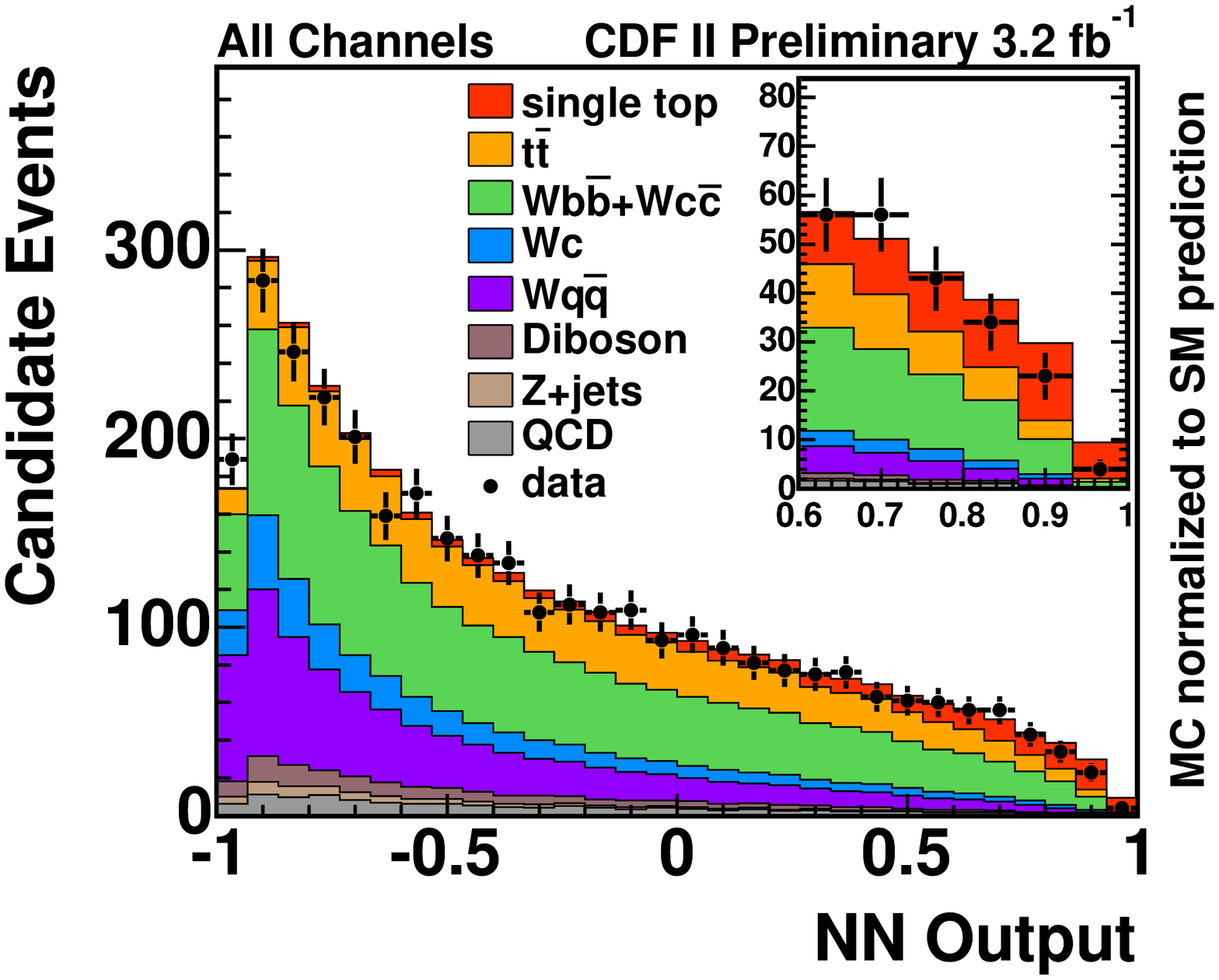, 
              width=0.31\textwidth}
      \label{fig:CDF_NN_Data}
    }  
  \end{center}
  \caption{\label{fig:CDF_NN_Analysis} CDF NN analysis.
    \subref{fig:CDF_NN_Temp} Normalized shapes of the NN output.
    \subref{fig:CDF_NN_Pretag} NN output on the untagged data set for validation. 
    \subref{fig:CDF_NN_Data} NN output on the selected data set compared to 
    the SM prediction.}
\end{figure}

\paragraph{Multivariate Analyses}
Even though the single top-quark production cross section is 
predicted~\cite{harris2002,Kidonakis:2006bu}
to amount to about 40\% of the $t\bar{t}$ cross section,
the signal has been obscured for a long time due to the very challenging 
background situation. 
After the cut-based event selection sketched above 
the signal-to-background ratio is only about 5 to 6\%. 
% S/B: Zahl gemittelt ueber W+2 und W+3 Jets Bin. CDF: 6.1%. D0: 4.9%. 
Further kinematic cuts on the event topology
proofed to be prohibitive, since the number of signal events in the
resulting data set would become too small. Given this challenging situation
the analysis groups in both \tevatron collaborations turned towards
the use of multivariate techniques, in order to exploit as much information
about the observed events as possible. The explored techniques comprise
artificial neural networks (NN), LO matrix elements (ME), 
boosted decision trees (BDT), and likelihood ratios.
All these multivariate techniques aim at maximizing the separation between 
the single top-quark signal and the backgrounds by combining the information 
contained in several
discriminating variables to one powerful discriminant. An example of these 
discriminants is shown in Fig.~\ref{fig:CDF_NN_Analysis}\subref{fig:CDF_NN_Temp}.
To ascertain that these advanced methods work reliably, extensive checks 
have been undertaken. 
In a first step, it is important to check the modeling of input variables. 
This can be done on the set of selected events, but these checks are 
statistically limited. It is therefore more meaningful to check the modeling 
of the event kinematics in the untagged data set which comprises twenty to fifty 
times more events and consists of events in which at least one jet is
taggable, but no jet is actually identified as a $b$-quark jet. 
A very important further check is to compute the discriminant on the
untagged data set, which is completely dominated by background, and compare the
observed distribution to the one obtained from the background model, see 
Fig.~\ref{fig:CDF_NN_Analysis}\subref{fig:CDF_NN_Pretag}. 
Finally, after additional tests of robustness of the method and further
checks of the background model have been passed, 
the analysis techniques are applied to the signal
sample. Fig.~\ref{fig:CDF_NN_Analysis}\subref{fig:CDF_NN_Data} shows the
NN discriminant at CDF in the lepton+jets data set. The observed distribution
is compared to the expectation based on the background model and the SM 
prediction for the rate of single top-quark events.

\begin{figure}[!t]
  \begin{center}    
    \subfigure[]{
      \epsfig{file=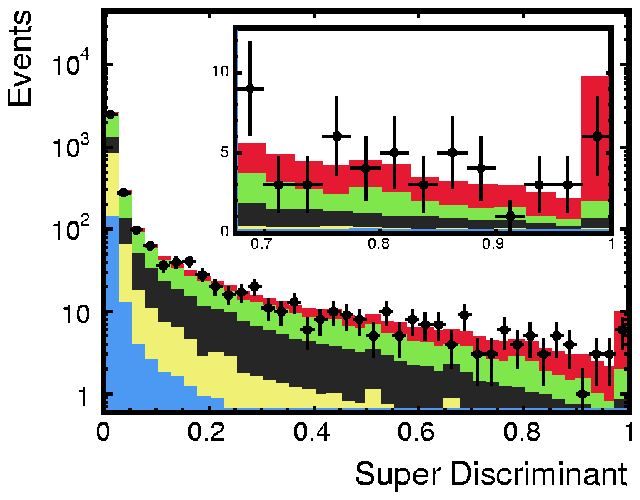, width=0.38\textwidth}
      \epsfig{file=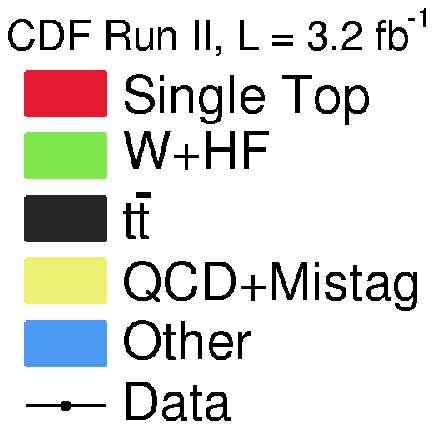, width=0.16\textwidth}
      \label{fig:CDF_Combo_Discrim}
    } \hspace{5mm}
    \subfigure[]{
      \epsfig{file=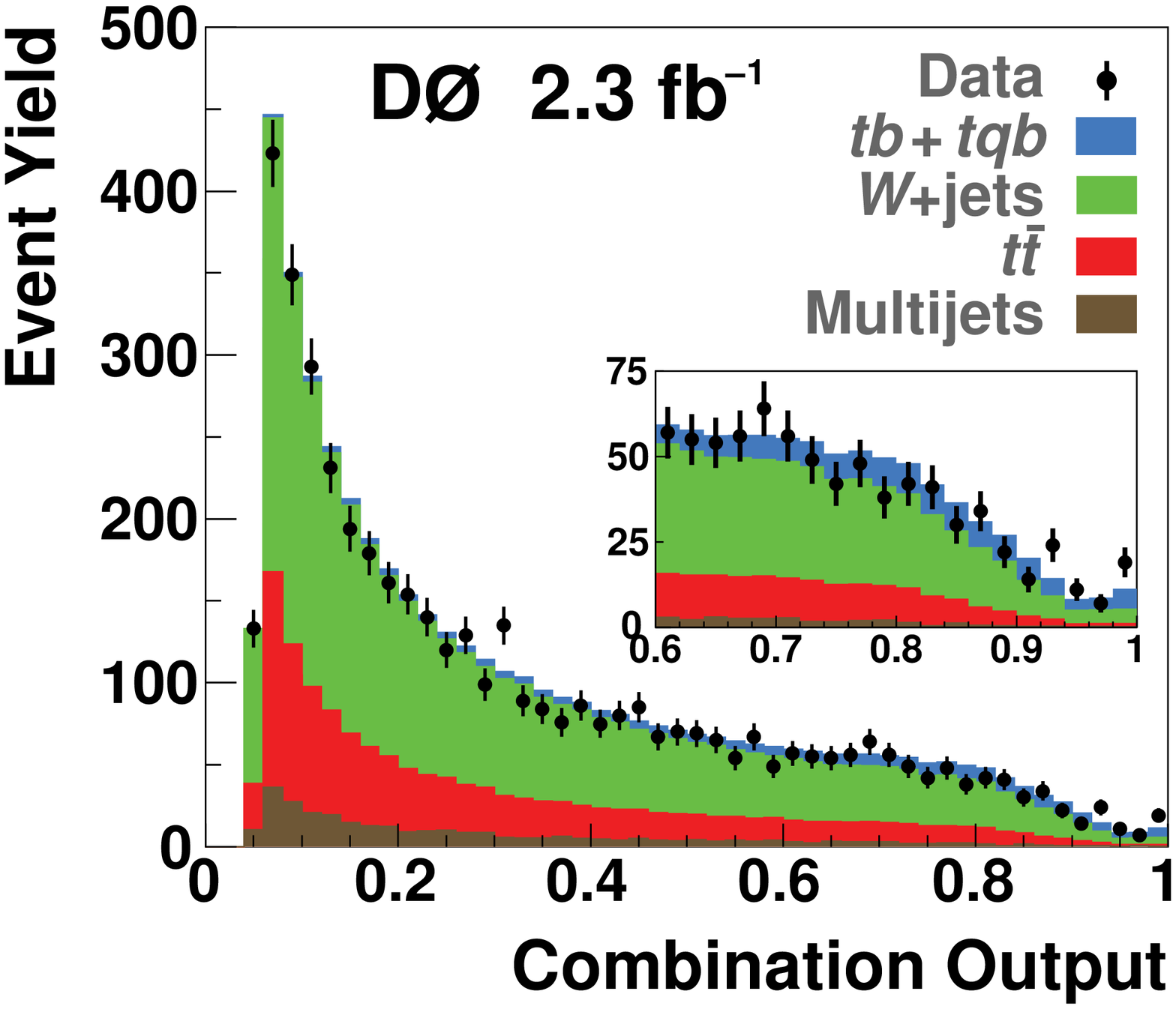, 
	width=0.33\textwidth}
      \label{fig:D0_Combo_Discrim} 
    } 
  \end{center}
  \caption{\label{fig:Combo} Distributions of super-discriminants,
    \subref{fig:CDF_Combo_Discrim} at CDF. 
    \subref{fig:D0_Combo_Discrim} at D\O.
    The observed data are compared to the SM prediction. 
   }
\end{figure}

Even though all analyses run on the lepton+jets data set share the same events
they are not fully correlated, the typical correlation being about 70\%. It is
therefore worthwhile to combine the analyses. Both collaborations do this by
computing a super-discriminant which takes the individual discriminants as 
input and combines them based on NN techniques. The resulting distributions
are displayed in Fig.~\ref{fig:Combo}.

The templates of the discriminant distributions for the different 
processes, see e.g. Fig.~\ref{fig:CDF_NN_Analysis}\subref{fig:CDF_NN_Temp}, 
are compared to the observed data distribution using a Bayesian maximum 
likelihood technique. 
Systematic uncertainties arise from uncertainties in the jet-energy corrections,
the modeling of initial and final state gluon radiation, the
factorization scale used to generate the simulated events, the 
modeling of the underlying event and the hadronization process,
and efficiencies for $b$-tagging, lepton identification and trigger.
The effects of these uncertainties on the background rates, 
the signal acceptance and detection efficiency and the shape of the
discriminant templates are investigated with dedicated Monte Carlo samples
or by studying control samples of collision data, and then incorporated
in the likelihood function in a parameterized form. 
The posterior probability density of the single top-quark cross section is
then obtained by integrating over the parameters associated with 
the systematic uncertainties using Gaussian prior distributions and by
applying a flat prior distribution to the signal cross section
which is zero for negative values and one for positive ones.
\renewcommand\arraystretch{1.2}
The cross sections measured by the various analyses and the combination 
are summarized in Table~\ref{tab:stop_measured_cross_sections}.
\begin{table}[!t]
  \caption{\label{tab:stop_measured_cross_sections} Cross sections measured
    by the multivariate analyses searching for single top-quarks.
    The quoted cross sections are the sum of $t$-channel and $s$-channel
    cross sections for top quark and antitop quark production. 
    Additionally, the expected and observed significance 
    in standard deviations (SD) are given. The theoretical prediction at NLO is
    quoted for $m_t=175\,\mathrm{GeV}/c^2$.}
  \begin{center}
  \begin{tabular}{lccc}
    \hline
    Analysis & Cross Section & Expected Significance & Observed Significance \\
             &  [pb]         & [standard deviations] & [standard deviations] \\ 
    \hline
    \multicolumn{4}{c}{CDF} \\
    Neural Networks (NN) & $1.8^{+0.6}_{-0.6}$ & 5.2 & 3.5 \\
    Boosted Decision Tree (BDT) & $2.1^{+0.7}_{-0.6}$ & 5.2 & 3.5 \\
    Matrix Elements (ME) & $2.5^{+0.7}_{-0.6}$ & 4.9 & 4.3 \\ 
    Likelihood Function & $1.6^{+0.8}_{-0.7}$ & 4.0 & 2.4 \\
    $s$-Channel Likelihood & $1.5^{+0.9}_{+0.8}$ & 1.1 & 2.0 \\ \hline
    Super-Discriminant & $2.1^{+0.6}_{-0.5}$ & $> 5.9\ $ & 4.8 \\
    $\EtMiss$+Jets & $4.9^{+2.5}_{-2.2}$ & 1.4 & 2.1 \\ \hline
    Combined & $2.3^{+0.6}_{-0.5}$ & $> 5.9\ $ & 5.0 \\
    \hline
    \multicolumn{4}{c}{D\O} \\
    Boosted Decision Tree (BDT) & $3.7^{+1.0}_{-0.8}$ & 4.3 & 4.6 \\
    Neural Networks (NN) & $4.7^{+1.2}_{-0.9}$ & 4.1 & 5.2 \\
    Matrix Elements (ME) & $4.3^{+1.0}_{-1.2}$ & 4.1 & 4.9 \\ \hline
    Combined & $3.9\pm 0.9$ & 4.5 & 5.0 \\
    \hline
    Theory (NLO, $m_t=175\,\mathrm{GeV}/c^2$) & $2.9\pm 0.4$ & -- & -- \\
    \hline
  \end{tabular}
  \end{center}
\end{table}
CDF quotes the measured cross sections at $m_t = 175\;\mathrm{GeV}/c^2$,
D\O at $m_t = 170\;\mathrm{GeV}/c^2$. However, the acceptance
correction due to $m_t$ are relatively small, namely 
$+0.02\,\mathrm{pb/(GeV}/c^2)$.

% Significance
The significance of the expected and observed single-top signal is determined
using a frequentist approach based on hundreds of millions of ensemble tests. 
While CDF uses the so-called $Q$-value as a test statistic, D\O \ uses the measured 
cross section. The $Q$-value is defined as the ratio of the posterior probability 
density for the observed data assuming the predicted single-top cross section
over the posterior probability density assuming no single top-quarks to be present.
Based on the chosen test statistic the collaborations calculate the
probability ($p$-value) to obtain the measured or a higher cross section 
under the assumption the observed data contain only background. 
The observed and expected significances are expressed in terms of standard 
deviations of a Gaussian
distribution and given in Table~\ref{tab:stop_measured_cross_sections}.
It is interesting to note that the CDF lepton+jets analyses are all below the 
theoretically expected cross section of $2.9\pm 0.4\,\mathrm{pb}$, while
all D\O \ measurements indicate a value above the prediction. 
Since the analyses in each collaboration use the same events, they are
highly correlated with a correlation coefficient between 60\% or 70\%.
In summary, both collaborations have observed single top-quark production via the
weak interaction at the level of five standard deviations.

\paragraph{Determination of $\mathbf{|V_{tb}|}$}
\label{sec:singletopvtb}
The measured single top-quark production cross sections can be used to 
determine the absolute value of the CKM-matrix element $V_{tb}$, if one assumes
$V_{tb} \gg V_{ts}$, $V_{tb} \gg V_{td}$, and a SM-like left-handed
coupling at the $Wtb$ vertex. Contrary to indirect determinations
in the framework of flavor physics~\cite{Amsler:2008zzb} the extraction
of $V_{tb}$ via single top-quark production does not assume unitarity
of the CKM matrix and is thereby sensitive to a fourth generation
of quarks. The assumption of $V_{ts}$ and $V_{td}$ being small compared
to $V_{tb}$ enters on the production side, since top quarks can also be
produced by $Wts$ and $Wtd$ vertices, and in top-quark decay.
The phenomenological analysis of~\cite{Bobrowski:2009ng} as well as the 
measurement of $R_b$, see section~\ref{sec:top_prop_vtb}, indicate
that this assumption is well justified.

To determine $|V_{tb}|$ the analysts divide the measured single top-quark
cross section by the predicted value, which assumes $|V_{tb}|=1$, and take 
the square root. CDF obtains 
$|V_{tb}| = 0.91\pm 0.11 (\mathrm{stat + syst}) \pm 0.07 (\mathrm{theory})$
and $|V_{tb}| > 0.71$ at the 95\% C.L.
D\O \ finds $|V_{tb}| = 1.07 \pm 0.12$ 
and $|V_{tb}| > 0.78$ at the 95\% C.L.
The lower limits assume a flat prior for $|V_{tb}|^2$ in the interval
$[0,1]$.

\subsection{Separation of t-channel and s-channel Events}
\label{subsec:sep_search}
\begin{figure}[!t]
  \begin{center}    
    \subfigure[]{
      \epsfig{file=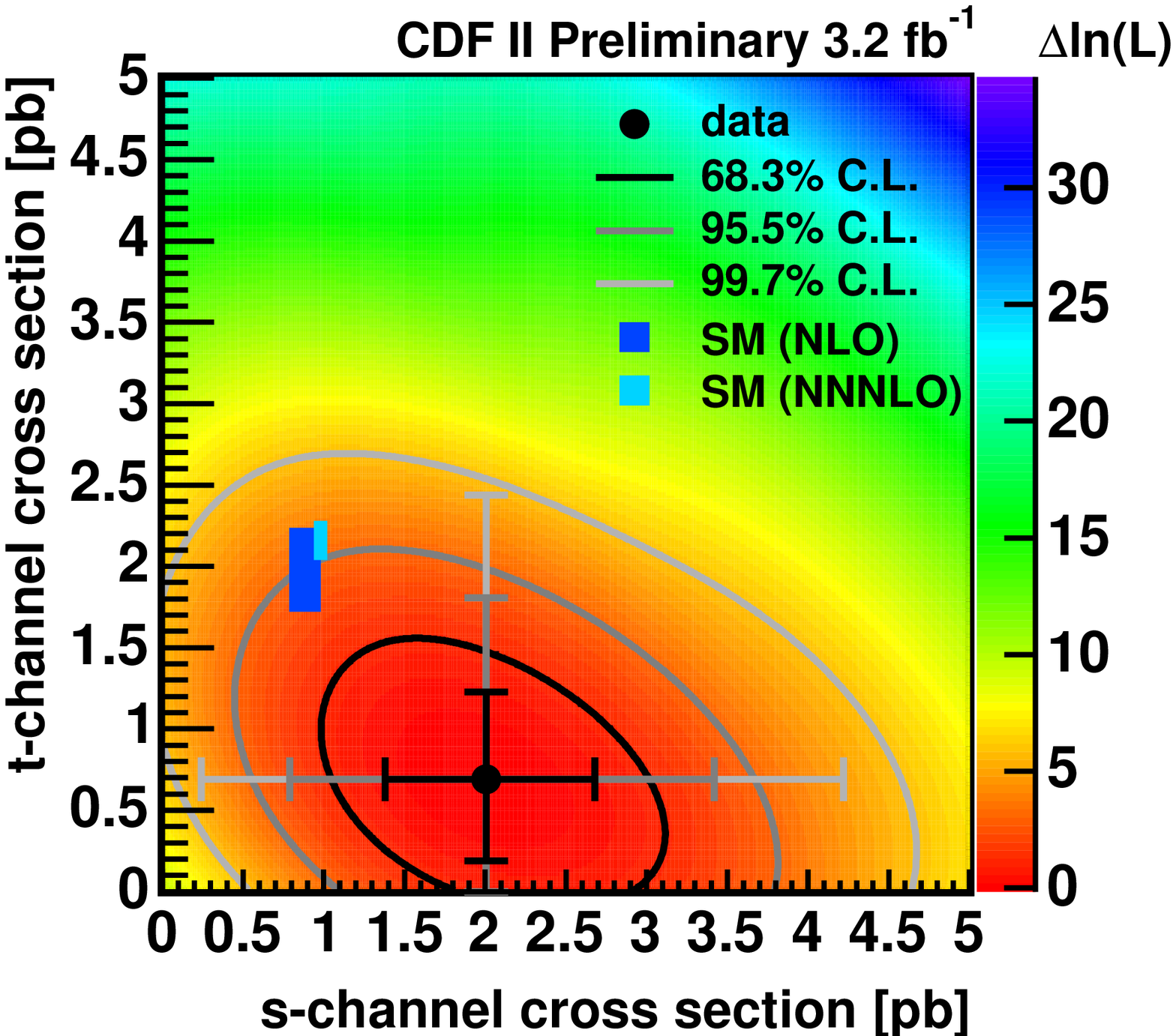, width=0.39\textwidth}
      \label{fig:CDF_Sep}
    } \hspace{0.15\textwidth}
    \subfigure[]{
      \epsfig{file=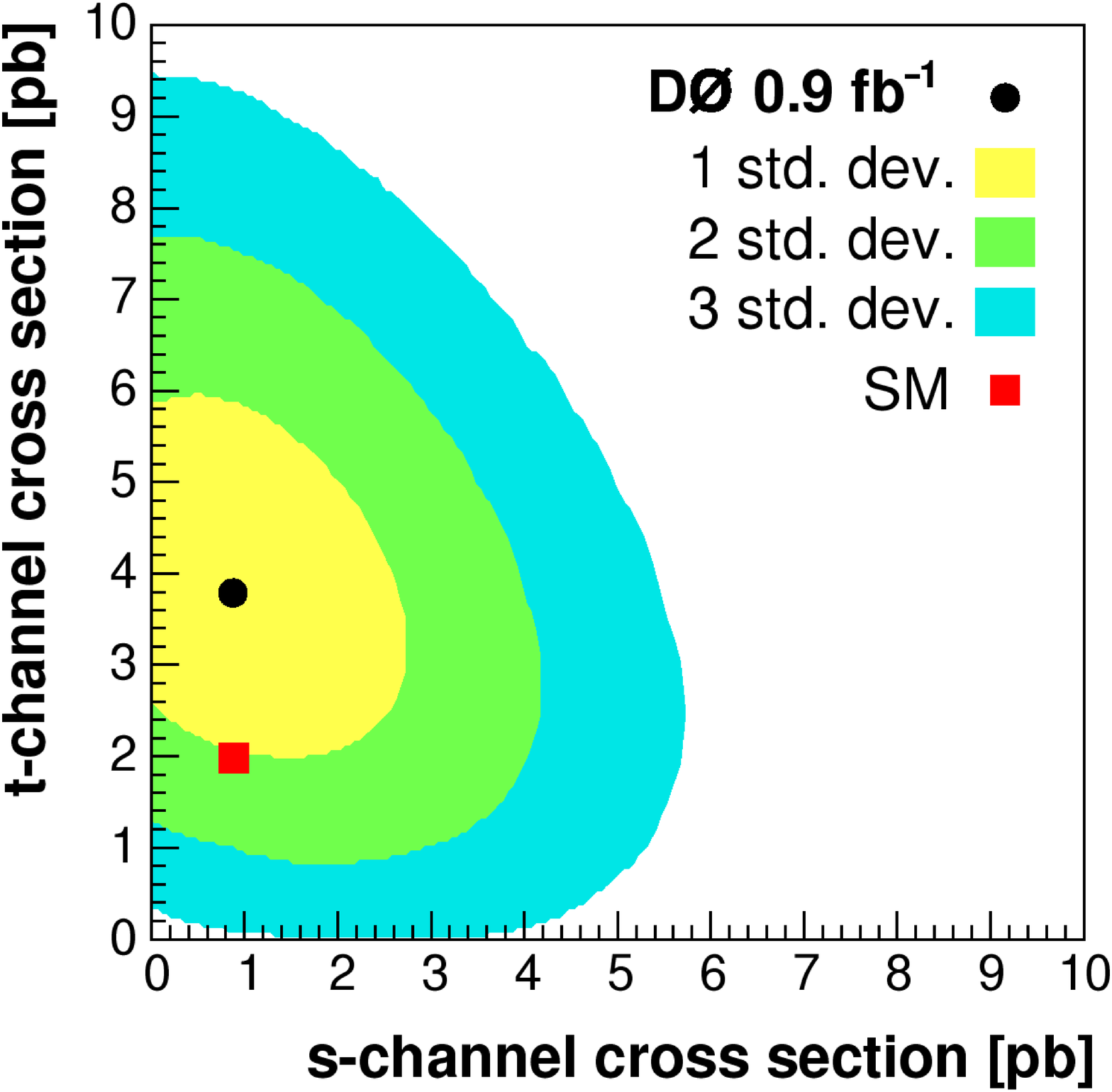, 
	width=0.32\textwidth}
      \label{fig:D0_Sep} 
    } 
  \end{center}
  \caption{\label{fig:Sep_Search} Results of the analyses separating
    $t$-channel ($tq\bar{b}$) and $s$-channel ($t\bar{b}$) events. 
    The figures show the
    contours of $t$-channel versus the $s$-channel cross section,
    \subref{fig:CDF_Sep} obtained from the NN separate-search analysis at CDF,
    \subref{fig:D0_Sep} obtained from the BDT analysis at D\O.
    The separate search of D\O \ uses only $0.9\,\mathrm{fb^{-1}}$.
   }
\end{figure}

In addition to the combined-search analyses described in the previous section
CDF and D\O \ have also attempted to separate $t$-channel and $s$-channel
single top-quark events. Both processes are treated as independent, i.e.,
the assumption of the ratio of the two cross sections to be given by the
SM is dropped. The separation of the two channels is important because they
are subject to different systematic uncertainties. 
The $t$-channel cross section, for example, is quite sensitive to the $b$-quark 
PDF, respectively, the gluon PDF. In addition, the two channels also exhibit
a different sensitivity to physics beyond the SM~\cite{tait2001}.
In Fig.~\ref{fig:Sep_Search} the resulting contours in cross section 
space are shown.
CDF measures $\sigma_{tq\bar{b}} = 0.7^{+0.5}_{-0.5}\,\mathrm{pb}$
and $\sigma_{t\bar{b}} = 2.0^{+0.7}_{-0.6}\,\mathrm{pb}$. The quoted uncertainties
include statistical and systematic components.
The most probable
values of the D\O \ analysis are at
$\sigma_{tq\bar{b}} = 3.8\,\mathrm{pb}$ and 
$\sigma_{t\bar{b}} = 0.9\,\mathrm{pb}$~\cite{Abazov:2008kt}.
The values for the $t$-channel cross section are pointing in different
directions. While CDF measures about $2\sigma$ lower than the predicted
value 
($\sigma_{tq\bar{b}}=1.98^{+0.28}_{-0.22}\;\mathrm{pb}$ at NLO
and $m_t=175\,\mathrm{GeV}/c^2$),
D\O \ measures about $1\sigma$ higher.

\subsection{Prospects for Single Top-Quark Measurements at the LHC} 
\label{subsec:singletopLHC}
Collisions at the \lhc will be a copious source of top-quarks. Not only
the $t\bar{t}$ cross section, but also the cross sections for 
single top-quark production will be a factor of 100 higher than at the
\tevatron. It will therefore be possible to not only measure cross sections,
but also carefully study properties of single top-quarks in detail, like for 
example the 
polarization. The \lhc experiments ATLAS and CMS have thoroughly studied
the prospects of single top-quark measurements at the 
\lhc~\cite{Ball:2007zza,Aad:2009wy}, which we will briefly summarize here.

Single top-quark analyses at ATLAS are based on a common preselection of
events, which asks for an isolated electron or muon with 
$|\eta| \leq 2.5$ and $p_T > 30\,\mathrm{GeV}/c$, 
$\EtMiss > 20\,\mathrm{GeV}$, and
at least two jets with $|\eta| \leq 5.0$ and 
$p_T > 30\,\mathrm{GeV}/c$, one of which has to be identified to 
originate from a $b$ quark~\cite{Aad:2009wy}. Events with an additional isolated
lepton with $p_T > 10\,\mathrm{GeV}/c$ are vetoed to reduce
the number of $t\bar{t}$ dilepton events.
Events with three or more additional jets with 
$p_T > 15\,\mathrm{GeV}/c$ are also removed.

After preselection, the analyses optimized for the different single top-quark
channels follow separate branches. The selection of single-top $t$-channel 
candidate events dwells on the signature of a high-$p_T$ light-quark jet
in the forward direction, requiring $|\eta| \geq 2.5$. 
The $b$-tagged jet has to have $p_T > 50\,\mathrm{GeV}/c$. Based on these
cuts a signal-to-background ratio of 37\% will be reached.
With collision data corresponding to $1\,\mathrm{fb^{-1}}$ ATLAS expects
to measure $|V_{tb}|$ with a relative uncertainty of $\pm12\%$.

The discrimination of $s$-channel single top-quark events and $Wt$ events
will be much more challenging than measuring the $t$-channel cross section.
ATLAS is therefore studying multivariate techniques to isolate
those events. Evidence for $s$-channel production appears to be
achievable at the $3\sigma$ level with $30\,\mathrm{fb^{-1}}$ of
collision data, while evidence for $Wt$ production may be reached
with $10\,\mathrm{fb^{-1}}$ using a BDT analysis.

Performance studies at CMS are based on sets of simulated events 
corresponding to $10\,\mathrm{fb^{-1}}$~\cite{Ball:2007zza}. 
The $t$-channel analysis asks for one isolated muon with 
$|\eta| \leq 2.1$ and $p_T > 19\,\mathrm{GeV}/c$, 
$\EtMiss > 40\,\mathrm{GeV}$, and two jets, one of which is identified
as a $b$-quark jet. This $b$-tagged jet is required to have
$|\eta| < 2.5$ and $p_T > 35\,\mathrm{GeV}/c$, the light-quark jet
must be in the forward direction with $|\eta| > 2.5$ and
$p_T > 40\,\mathrm{GeV}/c$. Further cuts are placed on the
transverse mass of the reconstructed $W$ boson,
$50 < M_T(W) < 120\,\mathrm{GeV}/c^2$, and the invariant mass of 
the reconstructed top-quark candidate,
$110 < M_{\ell\nu b} < 210\,\mathrm{GeV}/c^2$. With these selection
cuts CMS reaches a signal-to-background ratio of 1.34. The 
relative uncertainty on $|V_{tb}|$ is expected to be 5\%.
For the $Wt$-production mode the dilepton and semileptonic channels
have been studied at CMS. The signal-to-background ratio is found to be
0.37 and 0.18, respectively.
The $s$-channel will be even more difficult yielding a
signal-to-background ratio of 0.13.

%% file: topmass.tex
\section{Top-Quark Mass Measurements}
% Author: Arnulf Quadt
\label{sec:topmass}

The direct observation of the top quark in 1995
\cite{Abe:1995hr,Abachi:1995iq} was anticipated since the
$b$-quark was expected to have an isospin partner to ensure the
viability of the Standard Model, and therefore not a big surprise.
What was a surprise is the very large mass of the top-quark, almost 35
times the $b$-quark mass. The top-quark mass is a fundamental
parameter in the Standard Model, and plays an important role in
electroweak radiative corrections, and therefore in constraining the
mass of the Higgs boson. A large value of the top-quark mass
\cite{:2009ec} indicates a strong Yukawa coupling to the Higgs,
and could provide special insights in our understanding of electroweak
symmetry breaking \cite{Simmons:1998my}. The top-quark mass could
have a different origin than the masses of the other light quarks.
Thus, precise measurements of the top-quark mass provide a crucial
test of the consistency of the Standard Model and could indicate a
strong hint for physics beyond the Standard Model.  In doing that, it
is important to measure and compare the top-quark mass in the
different decay channels. Since all top mass measurements assume a
sample composition of $t\bar{t}$ and Standard Model background events,
any discrepancy among the measured top masses could indicate the
presence of non-Standard Model events in the samples.

\subsection{Lepton+Jets Channel}
\begin{figure*}[!ht]
\centerline{
\includegraphics[width=0.49\textwidth,clip=]
                {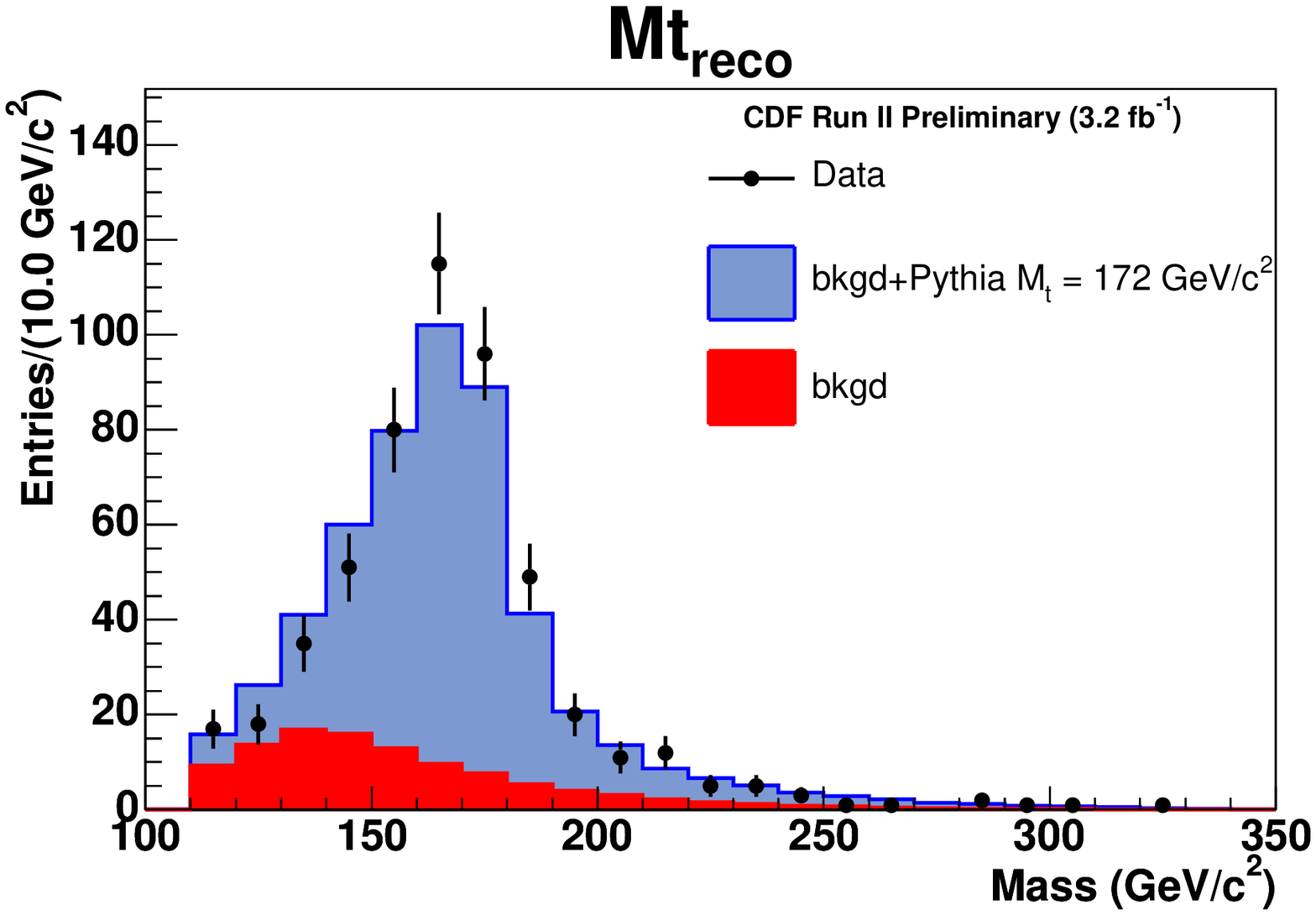}
\hfill
\includegraphics[width=0.49\textwidth,clip=]
                {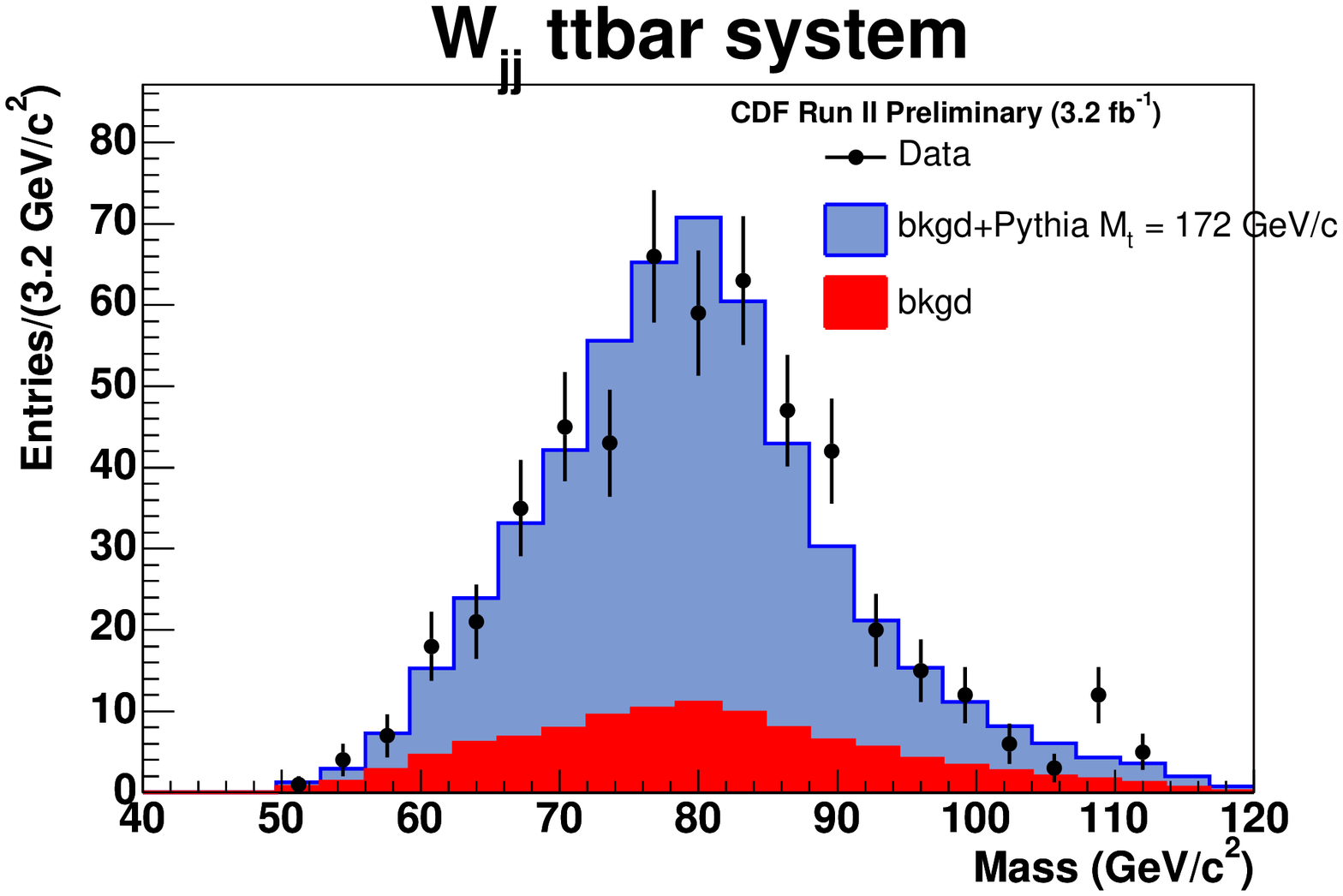}
}
\caption{\label{fig:tmass_ljets1} Top-quark mass analyses in the
  lepton+jets channel. The comparison between data and estimation in
  the lepton jet channel in the reconstructed top-quark (left) and the
  $W$-boson (right) mass using a template method with {\it in situ}
  calibration by \cdf \cite{cdf_9679}.}
\end{figure*}

The top mass has been measured in the lepton+jets, dilepton and the
all-jets channel by both \cdf and \dzero. At present, the most precise
measurements come from the lepton+jets channel containing four or more
jets and large missing $E_T$. The samples for the mass measurement are
selected using topological (topo) or $b$-tagging methods. In this
channel, four basic techniques are employed to extract the top mass.
In the first, the so-called {\bf ``template method'' (TM)}
\cite{abe_prd50_1994,Abe:1995hr,Abachi:1995iq}, an
over-constrained (2C) kinematic fit is performed to the hypothesis $t
\overline t \rightarrow W^+\, b\, W^-\, \overline b \rightarrow \ell\,
\nu_\ell\, b\, q\, \overline q'\, \overline b$ for each event,
assuming that the four jets of highest $E_T$ originate from the four
quarks in $t \overline t$ decay. There are 24 possible solutions
reflecting the allowed assignment of the final-state quarks to jets
and two possible solutions for the longitudinal momentum, $p_z$, of
the neutrino when the $W$ mass constraint is imposed on the leptonic
$W$ decay. The number of solutions is reduced to 12 when a jet with an
identified secondary vertex is assigned as one of the $b$ quarks, and
to 4 when the event has two such secondary vertices. A $\chi^2$
variable describes the agreement of the measurement with each possible
solution under the \ttbar hypothesis given jet-energy resolutions. The
solution with the lowest $\chi^2$ is defined as the best choice,
resulting in one value for the reconstructed top quark mass per event.
The distribution of reconstructed top-quark mass from the data events
is then compared to templates modeled from a mixture of signal and
background distributions for a series of assumed top masses, see 
Fig.~\ref{fig:tmass_ljets1}. The best
fit value for $\mtop$ and its uncertainty are obtained from
a maximum-likelihood fit.

\begin{figure*}[!ht]
\centerline{
\includegraphics[width=0.39\textwidth,clip=]
                {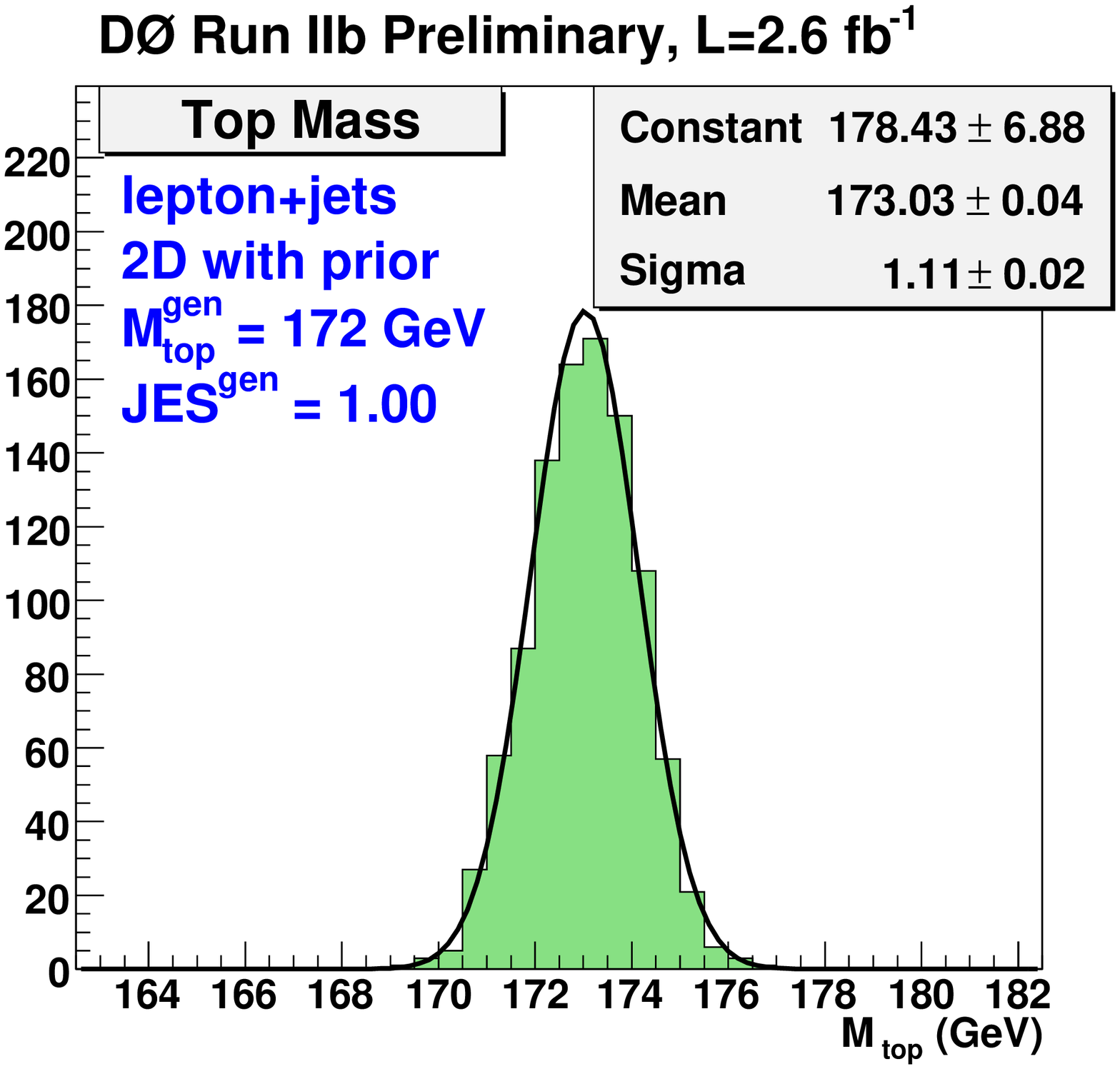}
\hfill
\includegraphics[width=0.59\textwidth,height=6.5cm,clip=]
                {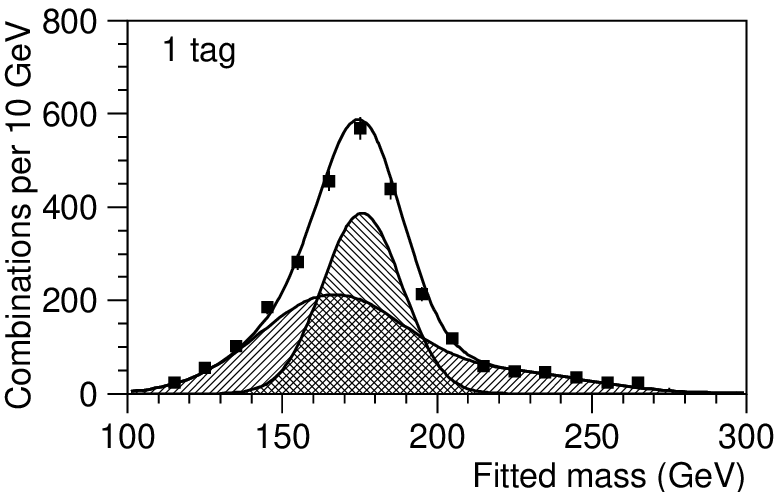}
}
\caption{\label{fig:tmass_ljets2} Top-quark mass analyses in the
  lepton+jets channel. Left(\dzero): Distributions of fitted top quark
  masses from ensemble tests performed on $m_t^{gen} = 172.5\;\rm
  GeV/c^2$ and $JES_{gen}=1.00$ MC samples using the Matrix Element
  Method in lepton+jets event with neural network $b$-tagging and {\it
    in situ} calibration is shown \cite{d0_5877}.  Right(\dzero):
  Prediction of the shapes of the fitted mass distribution for the
  wrong and the correct permutations (hatched) and the sum of the two
  (black line) using the fitted parameters of an Ideogram method. The
  sum of the two is compared to the simulated data containing a
  weighted sum of all solutions (correct and wrong), for the default
  jet energy scale and a generated top quark mass of $175\;\rm
  GeV/c^2$ for events with one $b$-tag. \cite{Abazov:2007rk}}
\end{figure*}

In the second method, the {\bf ``Matrix Element/Dynamic Likelihood
  Method'' (ME/DLM)}, similar to that originally suggested by Kondo et
al.  \cite{dlm1,dlm2,dlm3,dlm5} and Dalitz and Goldstein
\cite{dalitz_goldstein_1,dalitz_goldstein_2,dalitz_goldstein_3}, a
probability for each event is calculated as a function of $\mtop$, 
using a LO matrix element for the production and
decay of \ttbar pairs. All possible assignments of reconstructed jets
to final-state quarks are used, each weighted by a probability
determined from the matrix element. The correspondence between
measured four-vectors and parton-level four-vectors is taken into
account using probabilistic transfer functions. The results of
ensemble tests using the ME method at D\O \ are shown in 
Fig.~\ref{fig:tmass_ljets2}.

In a third method, the {\bf ``Ideogram Method''}
\cite{delphi_ideogram,mulders_ideogram}, which combines some of the
features of the above two techniques, each event is compared to the
signal and background mass spectrum, weighted by the $\chi^2$
probability of the kinematic fit for all 24 jet-quark combinations and
an event probability. The latter is determined from the signal
fraction in the sample and the event-by-event purity, as determined
from a topological discriminant in Monte Carlo events. An additional
variation on these techniques is the ``Multivariate Likelihood'' (ML)
technique, where an integral over the matrix element is performed for
each permutation, and then summed with weights determined by the
$b$-tagging information on each jet. Backgrounds are handled in the ML
technique by ``deweighting'' events according to a background
probability calculated using variables based on the topology of the
event.

\begin{figure*}[!ht]
\centerline{
\includegraphics[width=0.49\textwidth,clip=]
                {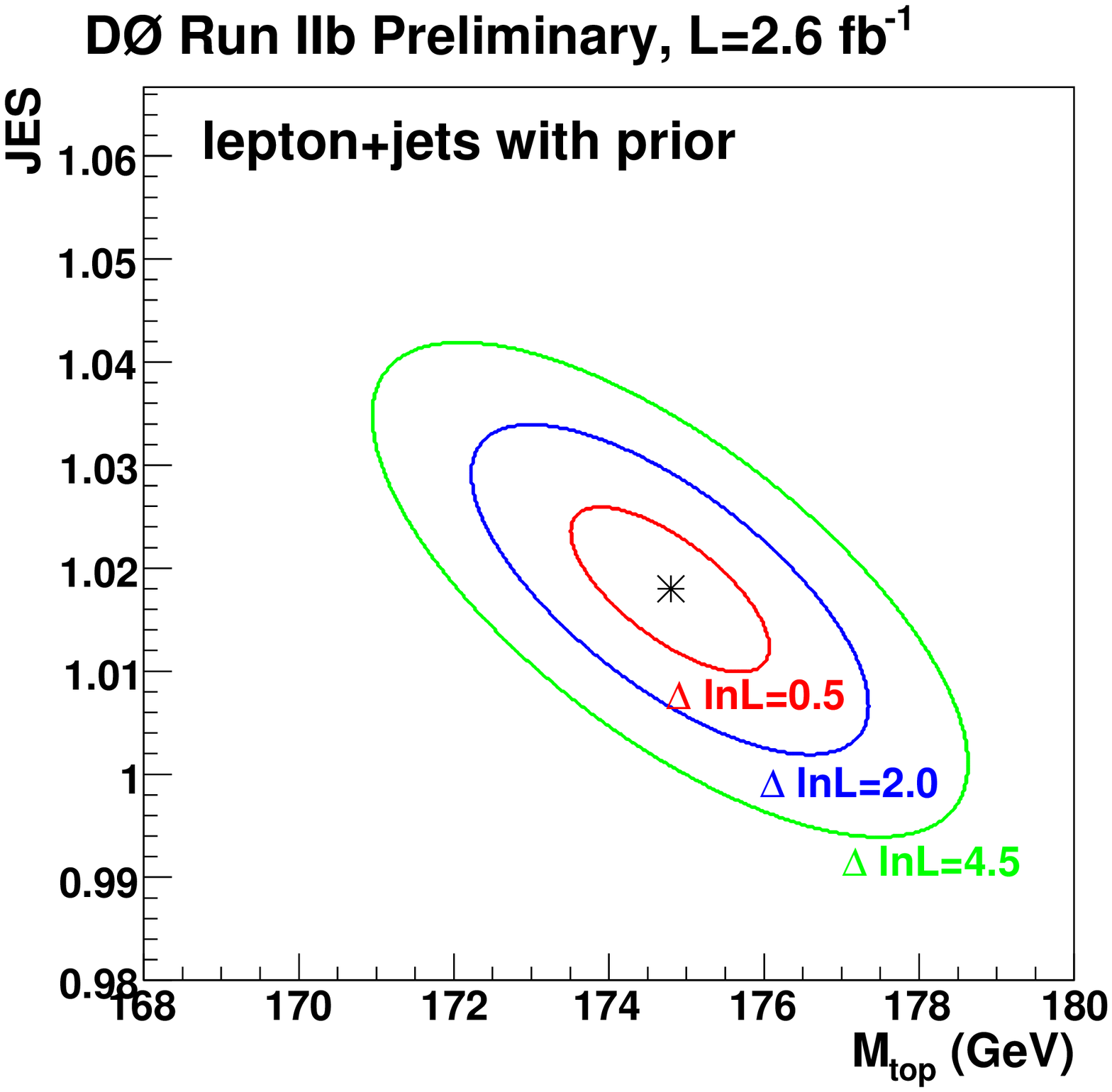}
\hfill
\includegraphics[width=0.49\textwidth,height=8.5cm,clip=]
                {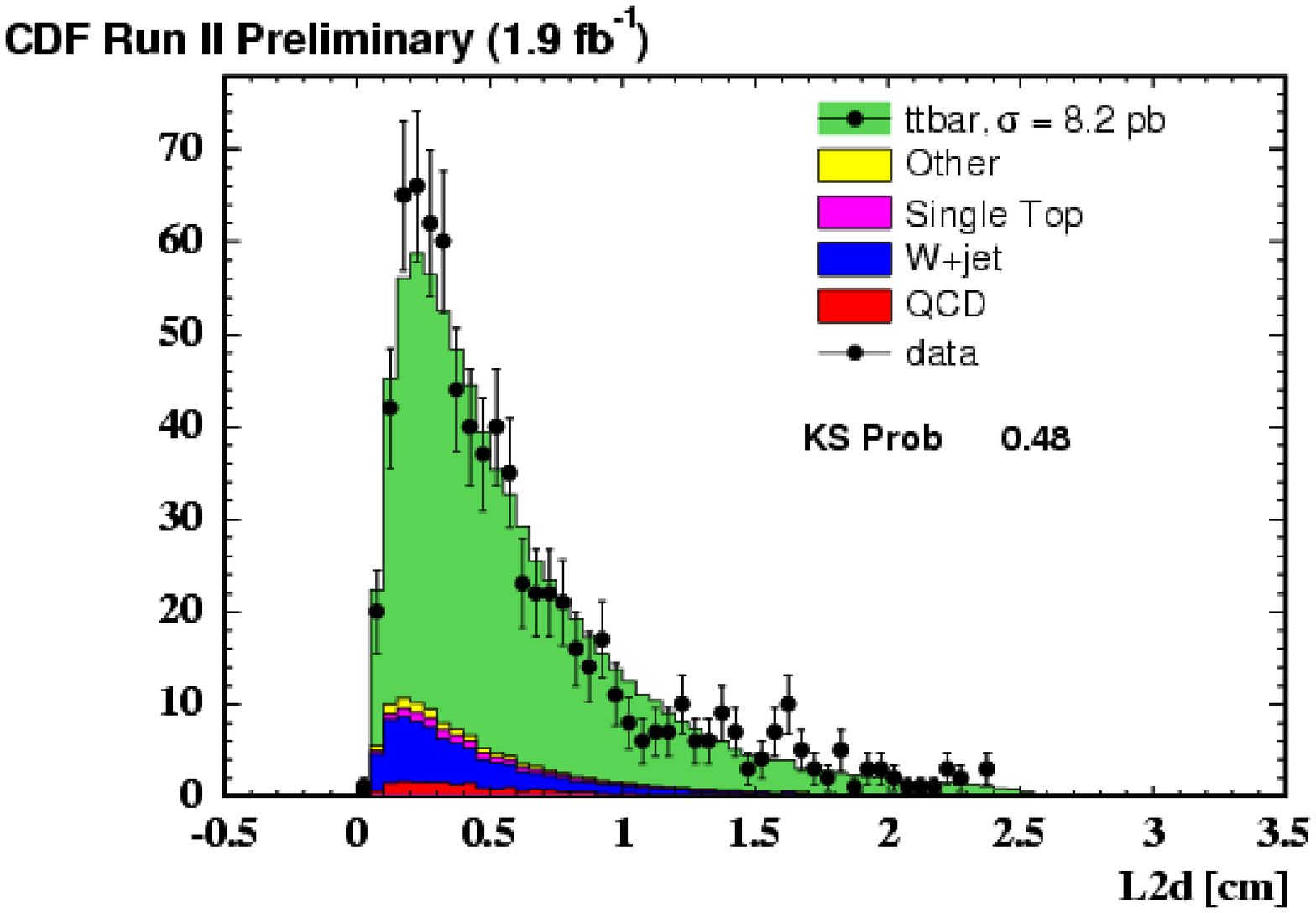}
}
\caption{\label{fig:tmass_ljets3} Top-quark mass analyses in the
  lepton+jets channel. Left(\dzero): Calibrated result of the 2D (JES
  vs. $m_t$) analysis using the Matrix Element Method in lepton+jets
  event with neural network $b$-tagging and {\it in situ} calibration
  is shown \cite{d0_5877}. Right (CDF): For each lepton+jets event
  passing the selection, the decay length, L2d, of the two leading
  SecVtx tagged jets is recorded.  Signal and background distributions
  for top mass hypotheses similar to the measured results ($m_t =
  178\;\rm GeV/c^2$) area shown \cite{cdf_9414}.}
\end{figure*}

With at least four jets in the final state, the dominant systematic
uncertainty on $\mtop$ is from the uncertainty on the
jet-energy scale. \cdf (TM, ME, ML) and \dzero (ME) have reduceed the
jet energy scale uncertainty by performing a simultaneous, {\it in
  situ} fit to the $W\rightarrow jj$ hypothesis using the jets without
identified secondary vertices, see Fig.~\ref{fig:tmass_ljets1} (right) 
and Fig.~\ref{fig:tmass_ljets3} (left). Also simultaneous measurements of the
top-quark mass, the light-quark and the $b$-quark jet energy scale are
proposed \cite{Fiedler:2007tf}.

The fourth technique, the {\bf Decay Length Method''}
\cite{Hill:2005zy,Abulencia:2006rz} relies solely on tracking, and
thus avoids the jet-energy scale uncertainty. This methods exploits
the fact that, in the rest frame of the top quark, the boost given to
the bottom quark has a Lorentz factor $\gamma_b \approx 0.4\, m_t /
m_b$. The measurement of the transverse decay length $L_{xy}$ of the
$b$-hadrons from the top quark decay is therefore sensitive to $\mtop$.  
Fig.~\ref{fig:tmass_ljets3} (right) displays the distribution of the
two-dimensional decay length in \ttbar candidate events at CDF.

\subsection{Di-Lepton Channel}
Additional determinations of the top mass come from the dilepton
channel with two or more jets and large missing $E_T$, and from the
all-jets channel. The dilepton channel, with two unmeasured neutrinos,
is under-constrained by one measurement. It is not possible to extract
a value for the top-quark mass from direct reconstruction without
adding additional information. Assuming a value for $m_t$, the
$t\overline t$ system can be reconstructed up to an eight-fold
ambiguity from the choice of associating leptons and quarks to jets
and due to the two solutions to the $p_z$ of each neutrino. Recently,
an analytic solution to the problem has been proposed
\cite{Sonnenschein:2006ud}. At the \tevatron, two basic techniques are
employed: one based on templates and one using matrix elements. 

\begin{figure*}[!th]
\centerline{
\includegraphics[width=0.45\textwidth,clip=]
                {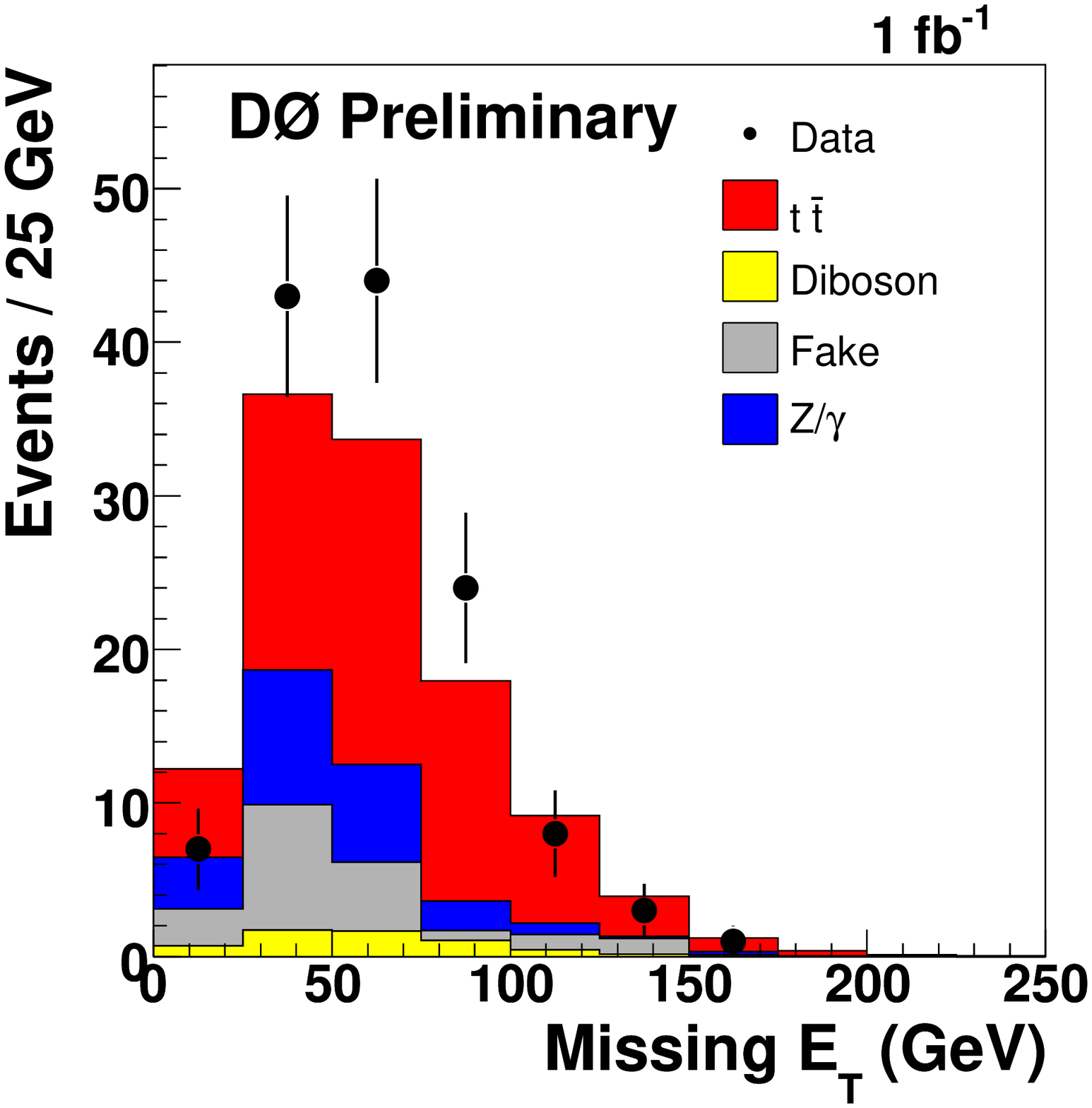}
\hfill
\includegraphics[width=0.54\textwidth,height=8cm,clip=]
                {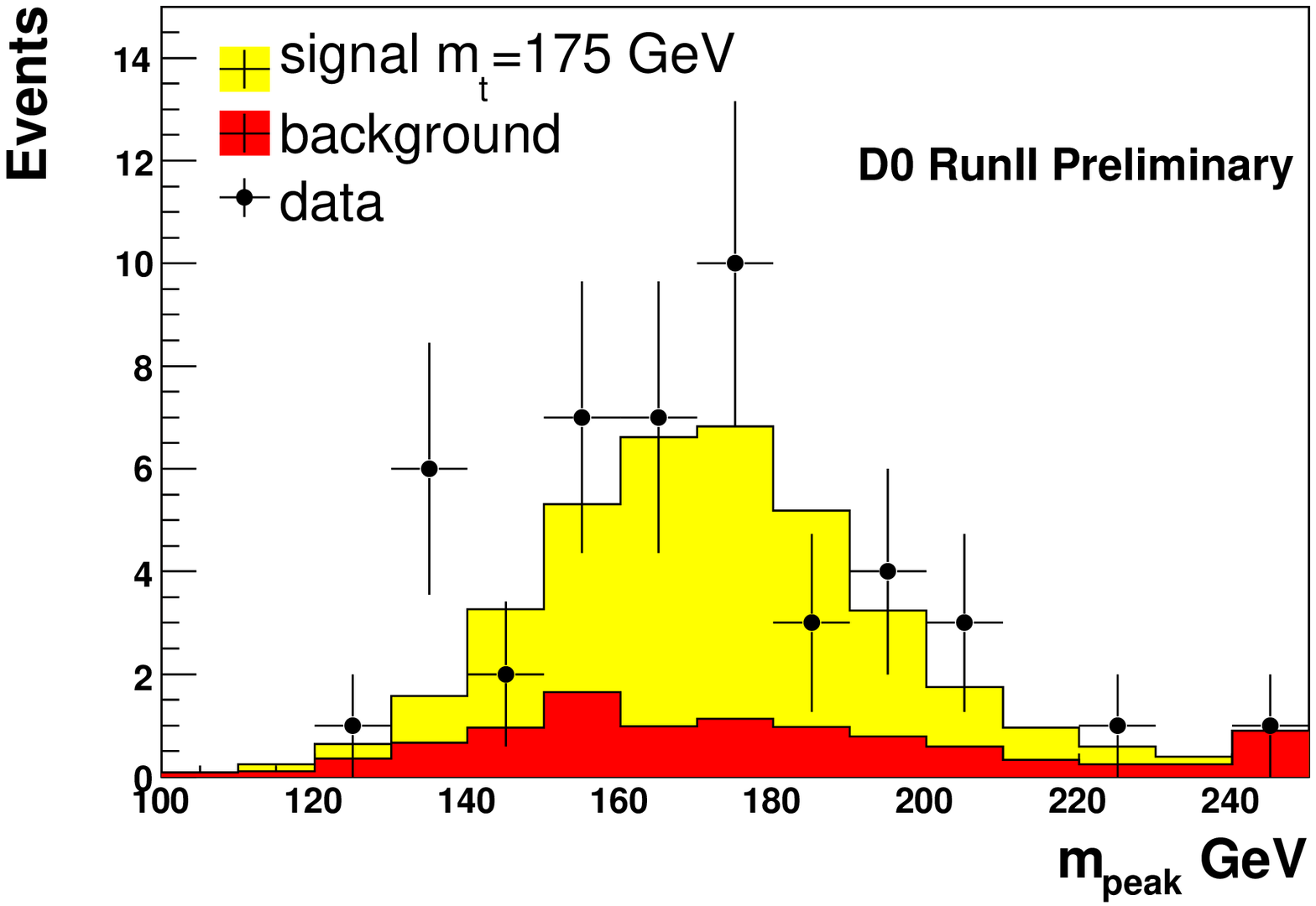}
}
\caption{\label{fig:tmass_dilepton1} Top-quark mass analyses in the
  di-lepton channel. Left(\dzero): DATA/MC comparison for the
  combination of dilepton channels for \met as used in the
  neutrino-weighting method with neural network $b$-tagging and a
  topological selection \cite{d0_5746}. Right(\dzero): Comparison of
  peak masses in data and Monte Carlo using a matrix weighting method
  with a topological selection \cite{d0_5463}.}
\end{figure*}

The first class of techniques incorporates additional information to
render the kinematic system solvable. In this class, there are two
techniques that assign a weight as a function of $\mtop$ for each
event based on solving for either the azimuth, $\phi$, of each
neutrino given an assumed $\eta$, {\boldmath ($\eta (\nu)$)}
\cite{Abbott:1997fv,Abbott:1998dn,Abe:1998bf}, or for $\eta$
of each neutrino given an assumed $\phi$, {\boldmath ($\phi (\nu)$)}
\cite{Abulencia:2005uq}. An alternative approach, {\bf (${\cal{M}}WT$)}
\cite{Abbott:1997fv,Abbott:1998dn}, solves for $\eta$ of each
neutrino requiring the sum of the neutrino $\vec{p}_T$'s to equal the
measured missing $E_T$ vector. In another technique, {\boldmath
  ($p_z(t\overline t)$)} \cite{cdf_7797}, the kinematic system is
rendered solvable by the addition of the requirement that the $p_z$ of
the $t\overline t$ system, equal to the sum of the $p_z$ of the $t$
and $\overline t$, be zero within a Gaussian uncertainty of $180\;\rm
GeV/c$. In a variation of the $p_z(t\bar{t})$ technique, the
theoretical relation between the top mass and its production cross
section is used as an additional constraint. In most of the techniques
in this class, a single mass per event is extracted and a top-mass
value found using a Monte Carlo template fit to the single-event
masses in a manner similar to that employed in the lepton+jets TM
technique. The \dzero $(\eta(\nu))$ analysis uses the shape of the
weight distribution as a function of $m_t$ in the template fit. 
The $\EtMiss$ distribution for the events of this analysis is shown
in Fig.~\ref{fig:tmass_dilepton1} (left).

The second class, {\bf ME/DLM}, uses weights based on the SM LO
matrix element for an assumed mass given the measured
four-vectors (and integrating over the unknowns) to form a joint
likelihood as a function of $\mtop$ for the ensemble of fitted
events.
\begin{figure*}[!th]
\centerline{
\includegraphics[width=0.38\textwidth,clip=]
                {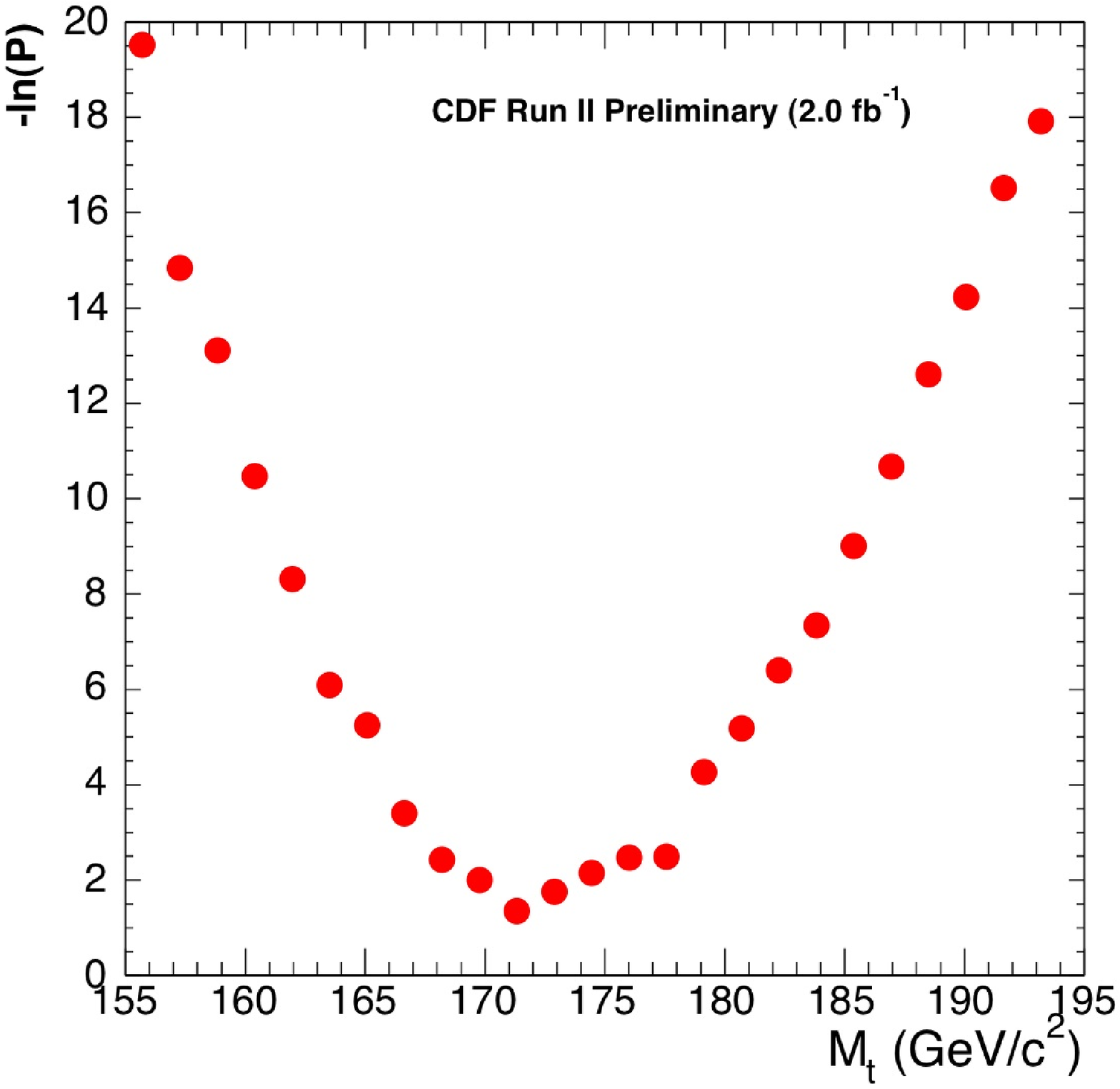}
\hfill
\includegraphics[width=0.52\textwidth,clip=]
                {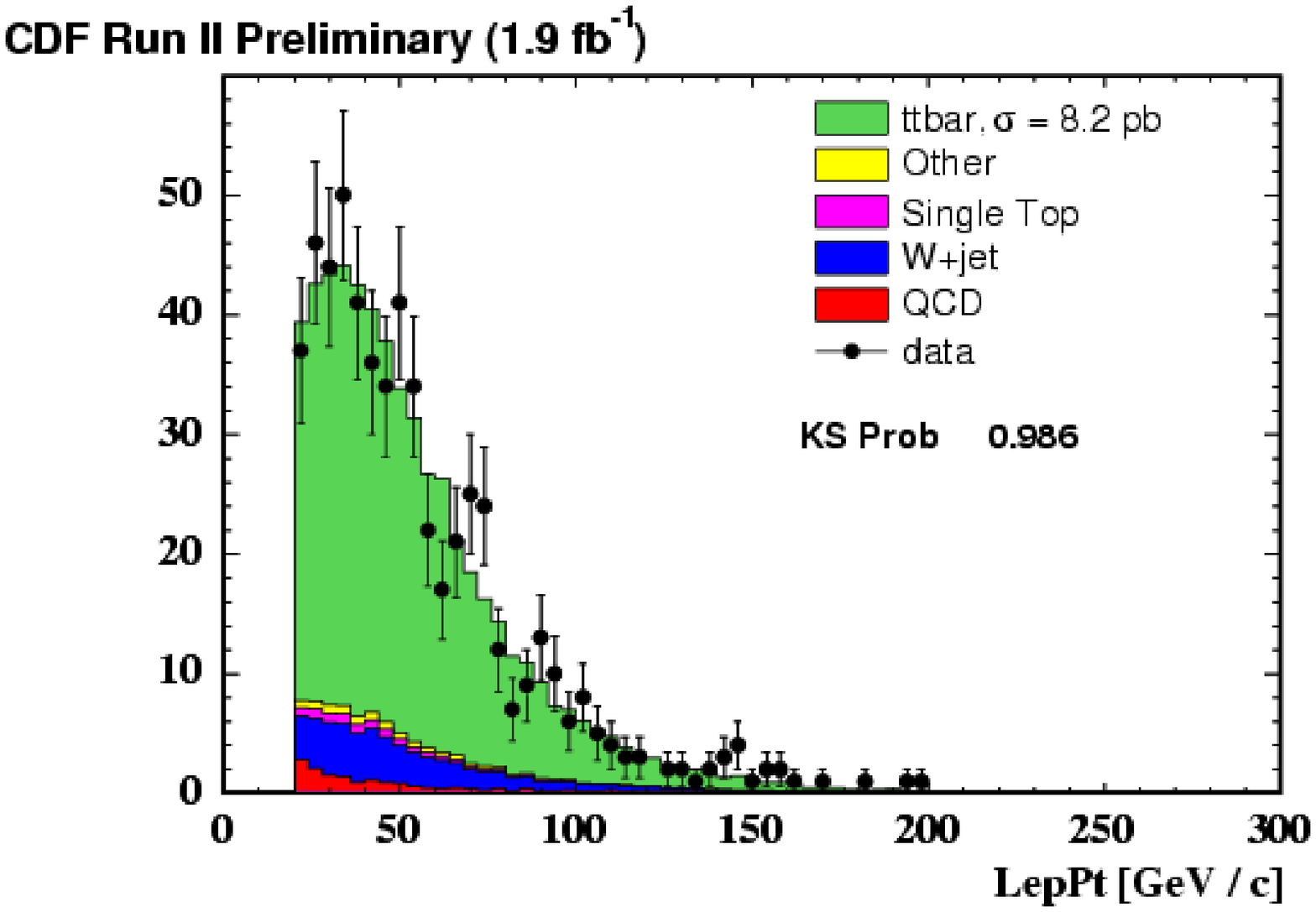}
}
\caption{\label{fig:tmass_dilepton2} Top-quark mass analyses in the
  di-lepton channel. Left(\cdf): Final posterior probability density
  as a function of top pole mass for the 344 candidate events in the
  \cdf data using a matrix element method \cite{Aaltonen:2008bd}.
  Right(\cdf): Signal, background, and data for the lepton $p_T$
  distribution, using a hypothesis top mass $m_t = 173\;\rm GeV/c^2$
  in a lepton-$p_T$ analysis \cite{cdf_9414}.}
\end{figure*}
Examples of this technique are shown in Fig.~\ref{fig:tmass_dilepton1} (right)
and Fig.~\ref{fig:tmass_dilepton2} (left).

The $p_T$ spectrum of the leptons in the dilepton channel has also
been used to extract a top mass measurement \cite{cdf_8959}, see 
Fig.~\ref{fig:tmass_dilepton2} (right). The
resulting statistical uncertainty of the measurement is large, but as
with the $L_{xy}$ technique, it is free of the systematic uncertainty
due to the jet-energy scale.

In the most recent set of CDF results, a measurement has been done
using the lepton+jets and dilepton channels simultaneously. In the
lepton+jets channel, the TM is used together with an {\it in situ} $W
\rightarrow jj$ fit. In the dilepton channel, $\eta(\nu)$ is used plus
a fit to the scalar sum of transverse energies $(H_T)$, which is
sensitive to the top mass.

\subsection{All-Jets Channel}
\begin{figure*}[!ht]
\centerline{
\includegraphics[width=0.49\textwidth,clip=]
                {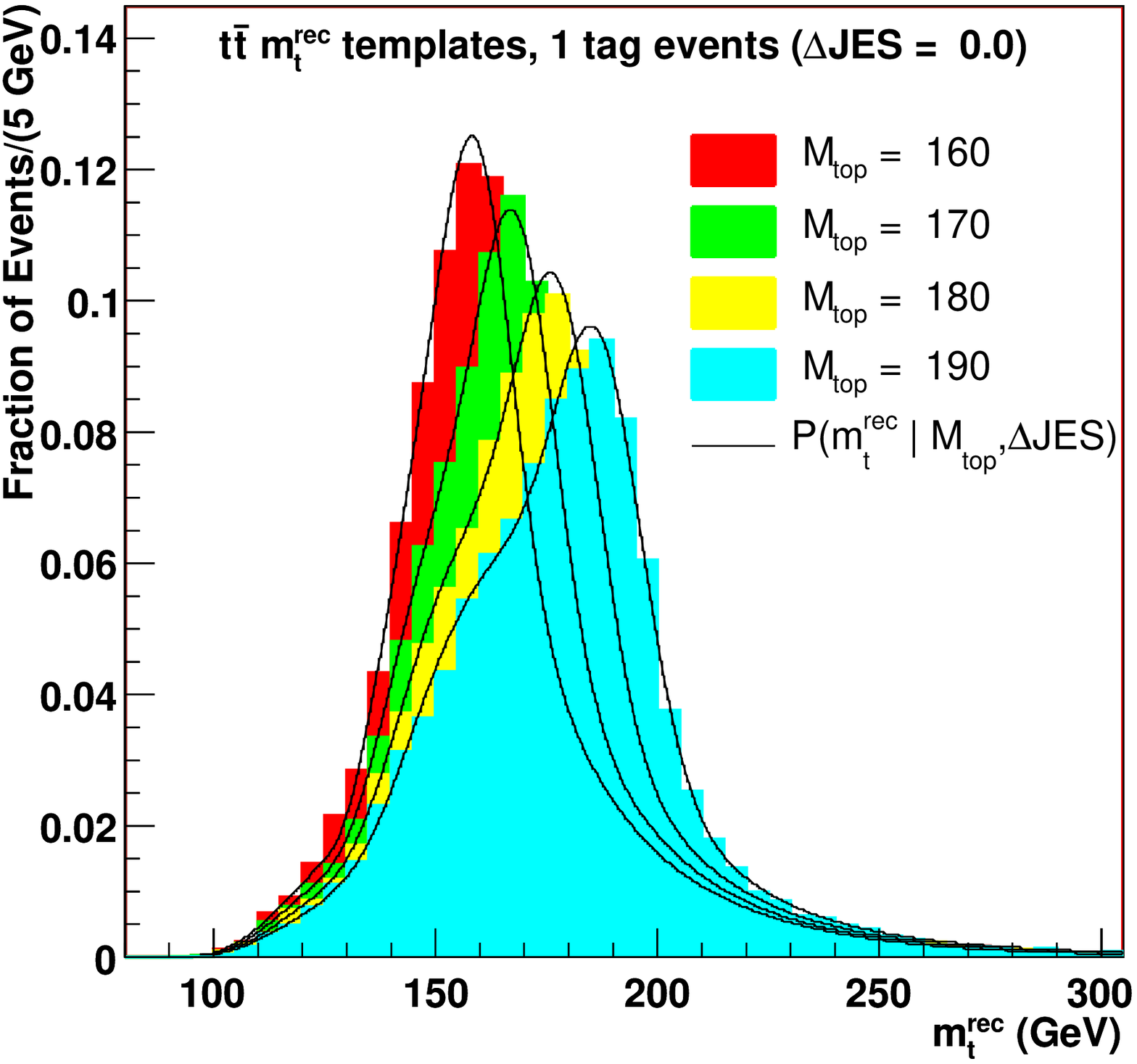}
\hfill
\includegraphics[width=0.49\textwidth,clip=]
                {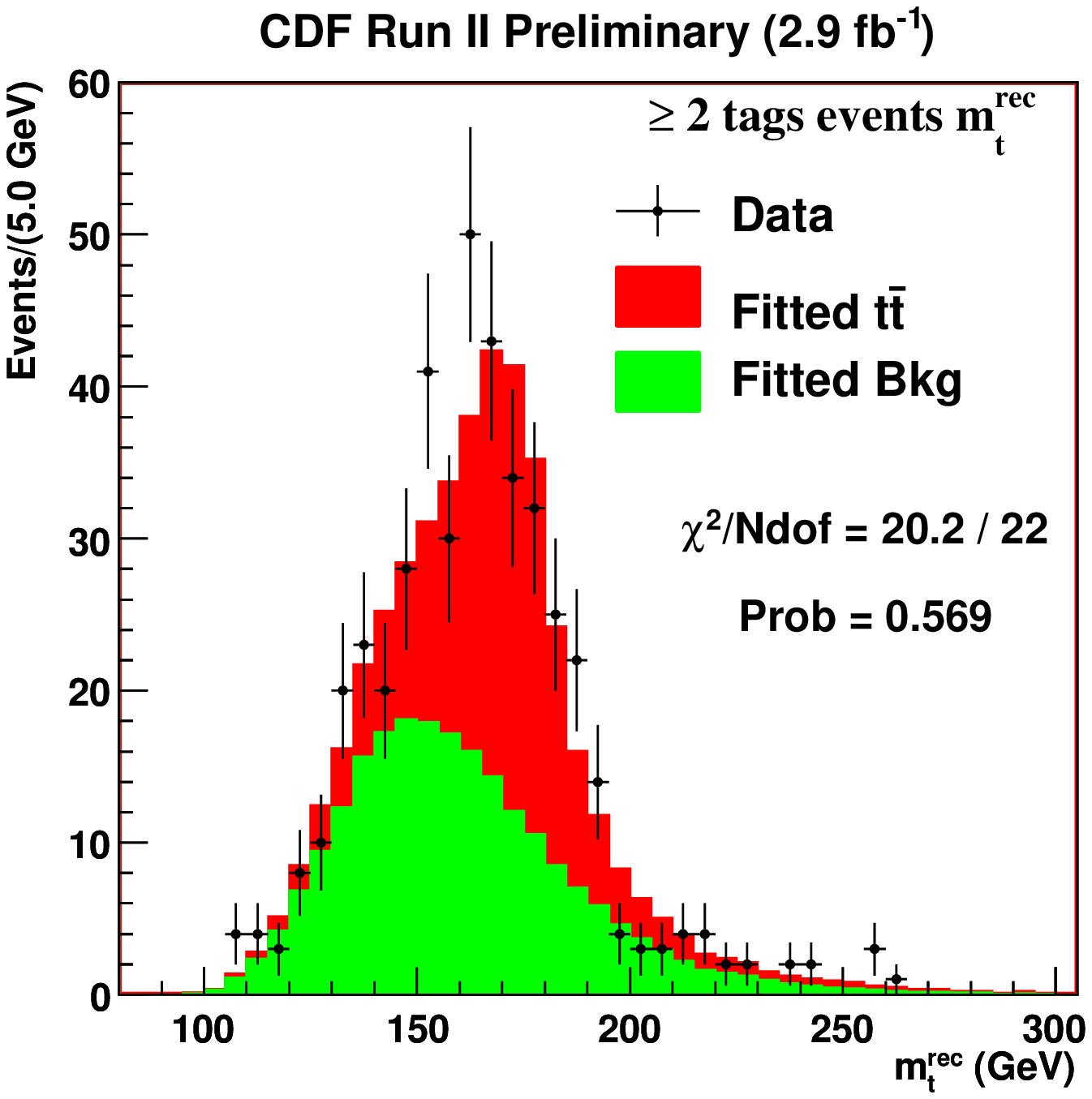}
}
\caption{\label{fig:tmass_alljets1} Top-quark mass analyses in the
  all-jets channel. Left(\cdf): Reconstructed signal $m_t$
  distribution, as a function of the input top mass for events with
  exactly 1 $b$-tag, and their respective parametrization \cite{cdf_9694}.
  Right(\cdf): Top reconstructed invariant masses for signal events
  with at least 2 b-tags \cite{cdf_9694}.}
\end{figure*}

In the all-jets channel there is no unknown neutrino momentum to deal
with, but the S/B is the poorest. Both, \cdf and D\O, use events with
6 or more jets, of which at least one is $b$-tagged. In addition, both
experiments have employed a neural network selection based on an array
of kinematic variables to improve the S/B. At \dzero, a top-quark mass
is reconstructed from the jet-quark combination that best fits the
hadronic $W$-mass constraint and the equal-mass constraint for the two
top quarks. At \cdf, the top-quark mass for each event was
reconstructed applying the same fitting technique used in the
$\ell$+jets mode. In the most recent analysis, the {\it in situ}
jet-energy scale calibration from the $W \rightarrow jj$ fit is also
used. At both, \cdf and D\O, the resulting mass distribution is
compared to Monte Carlo templates for various top quark masses and the
background distribution, see Fig.~\ref{fig:tmass_alljets1} for an 
example, and a maximum-likelihood technique is used to
extract the final measured value of $m_t$ and its uncertainty.

\dzero also measures the top-quark mass via comparison of the \ttbar
production cross section with the Standard Model expectation
\cite{d0_5459}. This method has the advantage that it is very
simple and sensitive to the top-quark pole mass, which is a very well
defined concept. The fully-inclusive cross-section calculation, used
for comparison, contains current best theoretical knowledge with
reduced-scheme or scale-dependence.

It should be noted that the different techniques make assumptions
about the SM-like production and decay of the top quark at
different levels. In general, methods which make stronger assumptions
about the top quark production and decay mechanism, such as the Matrix
Element/Dynamic Likelihood Method, have the highest sensitivity to the
top quark mass. Simple template methods in the lepton+jets channel,
which only rely on energy and momentum conservation in the kinematic
reconstruction and detector resolution functions, are on the one hand
less sensitive to $\mtop$, but on the other hand more stable
with respect to possible modification in the details of the production
and decay mechanisms involved.

\begin{table*}[p]
\caption{Measurements of $\mtop$ by \dzero and \cdf and their 
    average~\cite{:2009ec}. It is a combination of Run~I and Run~II 
    measurements (labelled with $\star$), yielding a $\chi^2$ of 6.3 for 
    10 degrees of freedom.}

\begin{footnotesize}
\label{tab:tmass_summary}
\begin{center}
\begin{tabular}{|l|l|l|l|l|} \hline
\multicolumn{1}{|c|}{$m_t\;\rm (GeV/c^2)$} & 
\multicolumn{1}{c|}{Source} & 
\multicolumn{1}{c|}{$\int {\cal{L}} dt$} & 
\multicolumn{1}{c|}{Ref.} & 
\multicolumn{1}{c|}{Method}\\ \hline
$173.3 \pm 5.6 \pm 5.5$ &D\O\phantom{F}~Run~I& 125 &~\cite{Abbott:1998dc,Abachi:1997jv}  &$\ell$+jets, TM\\
$180.1 \pm 3.6 \pm 3.9$ &D\O\phantom{F}~Run~I& 125 &~\cite{Abazov:2004cs,d0_0407005}{\;$\star$}&$\ell$+jets, ME\\
$168.4 \pm 12.3\pm 3.6$ &D\O\phantom{F}~Run~I& 125 &~\cite{Abbott:1998dn}{\;$\star$}&$\ell \ell$, $\eta (\nu)$/${\cal{M}}WT$\\
$178.5 \pm 13.7\pm 7.7$ &D\O\phantom{F}~Run~I& 110 &~\cite{d01_mt_allj_plb}    &all jets\\
\hline
$179.0 \pm 3.5 \pm 3.8$ &D\O\phantom{F}~Run~I& 110-125 &~\cite{Abazov:2004cs,d0_0407005}& D\O\ comb.\\
\hline
%
% from here on D0-Run-II l+jets
%
$169.1^{+5.9}_{-5.2}$   &D\O\phantom{F}~Run~II&1000    &~\cite{d0_5907} & $\ell\ell$,$\ell$ +jets, $\sigma_{t\bar{t}}$\\
$173.7 \pm 0.8 \pm 1.6$ &D\O\phantom{F}~Run~II&3600    &~\cite{d0_5877}{\;$\star$} & $\ell$+jets/$b$-tag, ME with $W \rightarrow jj$\\
$171.5 \pm 1.8 \pm 1.1$ &D\O\phantom{F}~Run~II&1000    &~\cite{Abazov:2008ds} & $\ell$+jets/$b$-tag, ME with $W \rightarrow jj$\\
$170   \pm 7$           &D\O\phantom{F}~Run~II&1000    &~\cite{Abazov:2008gc} & $\ell$+jets, $\sigma_{t\bar{t}}$\\
$173.7 \pm 4.4{^{+2.1}_{-2.0}}$&D\O\phantom{F}~Run~II& 420&~\cite{Abazov:2007rk}& $\ell$+jets/$b$-tag, Ideogram\\
%$170.3{^{+4.1}_{-4.5}}{^{+1.2}_{-1.8}}$&D\O\phantom{F}~Run~II& 370 &~\cite{Abazov:2006bd} & $\ell$+jets/$b$-tag, ME with $W \rightarrow jj$\\
%$169.2{^{+5.0}_{-7.4}}{^{+1.5}_{-1.4}}$&D\O\phantom{F}~Run~II& 370 &~\cite{Abazov:2006bd} & $\ell$+jets/topo,    ME with $W \rightarrow jj$\\
%
% from here on D0-Run-II dilepton
%
$174.8 \pm 3.3 \pm 2.6$ &D\O\phantom{F}~Run~II&3600    &~\cite{d0_5897}{\;$\star$} & $e\mu$, ME\\
$176.0 \pm 5.3 \pm 2.0$ &D\O\phantom{F}~Run~II&1000    &~\cite{d0_5746} & $\ell \ell$/$b$-tag, $\eta (\nu)$\\
$175.2 \pm 6.1 \pm 3.4$ &D\O\phantom{F}~Run~II&1000    &~\cite{d0_5463} & $\ell \ell$/topo, ${\cal{M}}WT$\\
$171.5^{+9.9}_{-8.8}$   &D\O\phantom{F}~Run~II&1000    &~\cite{d0_09012137}   &$\ell \ell$, $\tau$+jets, $\sigma_{t\bar{t}}$\\
%$178.1 \pm 6.7 \pm 4.8$ &D\O\phantom{F}~Run~II& 370    &~\cite{Abazov:2006bg} &$\ell \ell$, $\eta (\nu)$/${\cal{M}}WT$\\
\hline
$174.2\pm 0.9\pm 1.5$   &D\O\phantom{F}~Run~II&3600 &~\cite{d0_5900}{\;$\star$}&  D\O\ comb.\\
\hline
$176.1 \pm 5.1 \pm 5.3$ &\cdf~Run~I& 110 &~\cite{Abe:1998bf,Abe:1997vq,Affolder:2000vy}{\;$\star$}&$\ell$ + jets\\
$167.4 \pm10.3 \pm 4.8$ &\cdf~Run~I& 110 &~\cite{Abe:1998bf}{\;$\star$}& $\ell \ell$\\
$186.0 \pm10.0 \pm 5.7$ &\cdf~Run~I& 110 &~\cite{cdf1_alljets_prl}{\;$\star$}& all jets\\
\hline
$176.1 \pm 6.6        $ &\cdf~Run~I& 110 &~\cite{Abe:1998bf}& \cdf comb. \\
\hline
%
% from here on CDF-Run-II l+jets
%
$181.3 \pm 12.4 \pm 3.5$   &\cdf~Run~II&2000 &~\cite{cdf_9518}& $\ell$ + jets/soft-$\mu$, TM\\
$170.9 \pm 2.2 \pm 1.4$    &\cdf~Run~II&1000 &~\cite{Abulencia:2007br}& $\ell$ + jets, ME with $W\rightarrow jj$\\
$172.1 \pm 1.1 \pm 1.1$    &\cdf~Run~II&3200 &~\cite{cdf_9692}{\;$\star$}& $\ell$ + jets, multivariat with $W\rightarrow jj$\\
$168.9 \pm 2.2 \pm 4.2$    &\cdf~Run~II&1000 &~\cite{cdf_8669}& $\ell$ + jets, TM\\
$171.7 {^{+1.4}_{-1.5}} \pm 1.1$ &\cdf~Run~II&3200 &~\cite{cdf_9679}& $\ell\ell$,$\ell$ + jets, TM with $W\rightarrow jj$\\
$175.3 \pm 6.2 \pm 3.0$    &\cdf~Run~II&1900 &~\cite{cdf_9414}{\;$\star$}& $\ell$ + jets, Decay Length\\
$172.1 \pm 7.9 \pm 3.0$    &\cdf~Run~II&2700 &~\cite{cdf_9683}& $\ell$ + jets, Lepton $p_T$\\
$171.6 \pm 2.0 \pm 1.3$    &\cdf~Run~II&1700 &~\cite{cdf_9135}& $\ell$ + jets, DLM with $W\rightarrow jj$\\
%
% from here on CDF-Run-II dilepton
%
$171.2 \pm 2.7 \pm 2.9$    &\cdf~Run~II&1900 &~\cite{Aaltonen:2008bd}{\;$\star$}& $\ell\ell$, ME\\
$170.4 \pm 3.1 \pm 3.0$    &\cdf~Run~II&1800 &~\cite{cdf_8951}& $\ell\ell$, ME\\
$167.3 \pm 4.6 \pm 3.8$    &\cdf~Run~II&1000 &~\cite{Abulencia:2006ry}& $\ell\ell$/$b$-tag, ME\\
$172.0 {^{+5.0}_{-4.9}} \pm 3.6$ &\cdf~Run~II&1800 &~\cite{cdf_8955}& $\ell \ell$/$b$-tag, $\eta (\nu)$\\
$156   \pm 20  \pm 4.6$    &\cdf~Run~II&1800 &~\cite{cdf_8959}& $\ell\ell$, Lepton $p_T$\\
%$167.9 \pm 5.2 \pm 3.7$    &\cdf~Run~II& 340 &~\cite{Abulencia:2005uq}& $\ell\ell$, combination\\
$165.1 {^{+3.3}_{-3.2}} \pm 3.1$ &\cdf~Run~II&2800 &~\cite{cdf_09013773}& $\ell \ell$/$b$-tag, $\phi (\nu)$\\
$169.7 {^{+5.2}_{-4.9}} \pm 3.1$ &\cdf~Run~II&1200 &~\cite{:2007jw}& $\ell \ell$, $p_z(t\bar{t})$\\
$170.7 {^{+4.2}_{-3.9}} \pm 2.6$ &\cdf~Run~II&1200 &~\cite{:2007jw}& $\ell \ell$, $p_z(t\bar{t})$+$\sigma_{t\bar{t}}$\\
%$169.7 {^{+8.9}_{-9.0}} \pm 4.0$ &\cdf~Run~II& 340 &~\cite{Abulencia:2005uq,Abulencia:2006js}& $\ell \ell$, $\phi (\nu)$\\
%
% from here on CDF-Run-II alljets
%
$165.2 \pm 4.4 \pm 1.9$    &\cdf~Run~II&1900 &~\cite{cdf_9265}& all jets, Ideogram\\
$174.0 \pm 2.2 \pm 4.8$    &\cdf~Run~II&1000 &~\cite{Aaltonen:2007qf}& all jets, TM\\
$174.8 \pm 2.4 {^{+1.2}_{-1.0}}$ &\cdf~Run~II&2900 &~\cite{cdf_9694}{\;$\star$}& all jets, TM\\
$171.1 \pm 3.7 \pm 2.1$,   &\cdf~Run~II&1000 &~\cite{cdf_08111062}& all jets, ME+TM\\
\hline
$172.6 \pm 0.9 \pm 1.2$      &\cdf~Run~II&3200 &~\cite{cdf_9714}&\cdf comb. \\
\hline
$178.0 \pm 4.3     $& \cdf \& D\O\ & 110-125 &~\cite{tevewwg_1} & Run-I combination  \\
$173.1 \pm 1.3{~^*}$& \cdf \& D\O\ & 110-340 &~\cite{:2009ec} & world-average (2009)\\
\hline
\end{tabular}
\end{center}
\end{footnotesize}
\end{table*}

The different measurements of $\mtop$, described in this
section, are summarised in Table~\ref{tab:tmass_summary}. The
systematic uncertainty (second uncertainty shown) is comparable to or
larger than the statistical uncertainty, and is primarily due to the
uncertainties in the jet-energy scale and in the Monte Carlo modeling.
In the \runii analyses, \cdf and \dzero have controlled the jet-energy
scale uncertainty via {\it in situ} $W \rightarrow jj$ calibration
using the same \ttbar events, as mentioned above, and via new
techniques such as the decay-length technique.

The Tevatron Electroweak working Group (TevEWWG), responsible for the
combined \cdf/\dzero average top mass in
Table~\ref{tab:tmass_summary}, took account of correlations between
systematic uncertainties in the different measurements in a
sophisticated manner \cite{:2009ec}. The latest TevEWWG world
average \cite{:2009ec}, including published and some preliminary
\runii results, yields $m_t = 173.1 \pm 1.3\;\rm GeV/c^2$ (statistical
and systematic uncertainties combined in quadrature).

\subsection{Measurement of the Top-Quark Mass at the LHC}
At the LHC there will be 8 million \ttbar pairs produced per year at a 
luminosity of $10^{33}\;\rm cm^{-2}s^{-1}$. Such large event samples will 
permit precision measurements
of the top-quark parameters. The statistical uncertainties on $m_t$
will become negligible, and, using essentially the methodes developed
at the \tevatron, systematic uncertainties, better than $\pm 2\;\rm
GeV/c^2$ per channel are anticipated
\cite{ATLAS_TDR2,Aad:2008zza,Ball:2007zza}.

For example, \cms measures $\mtop$ in the lepton+jets channel
using a full kinematic fit to the events, and using the result of this
fit in an event-by-event likelihood as a function of the top mass. In
a data sample of $10\;\rm fb^{-1}$, \cms expects a statistical
uncertainty on the top-quark mass of only $200\;\rm MeV/c^2$, and a
systematic uncertainty of $1.1\;\rm GeV/c^2$ if the dominating
uncertainty, the $b$-jet energy scale, is known to 1.5\%. In the
dilepton channel already in $1\;\rm fb^{-1}$ a measurement with a
precision of $4.5\;\rm GeV/c^2$ can be made, improving to $0.5 (\rm stat.)
\pm 1.1 (\rm syst.)\;\rm GeV/c^2$ in $10\;\rm fb^{-1}$.

Similarly, \atlas studies measures the top-quark mass in the
lepton+jets channel using kinematic fits. In the final expected result
for $1\;\rm fb^{-1}$, the systematic uncertainty dominates over the
statistical one, and amounts to $0.7\,\rm GeV/c^2$ per \% of $b$-jet
energy scale uncertainty, $0.2\,\rm GeV/c^2$ per \% of light-quark jet
energy scale uncertainty, and $\sim 0.3\;\rm GeV/c^2$ due to
uncertainties related to initial or final state radiation.

An interesting alternative method to measure the top mass has been
studied by \cms, and involves the selection of events where a
$b$-quark hadronizes into a $J/\Psi (+X)$ and the $J/\Psi$ into two
leptons. The $W$-boson from the same top quark also decays
leptonically. The invariant mass of the three leptons is sensitive to
the top mass; the systematic uncertainties of this method include
$b$-decay modelling and the lepton energy scale, but not the $b$-jet
energy scale, and it is thus almost orthogonal to the standard
methods. In $20\;\rm fb^{-1}$ the statistical error could reach $\sim
1\;\rm GeV/c^2$ and the systematic error $\sim 1.5\;\rm GeV/c^2$,
dominated by the theory systematic that may be further reduced by new
calculations.

Given the experimental techniques used to extract the top mass, these
mass values should be taken as representing the top {\it pole mass}.
The top pole mass, like any quark mass, is defined up to an intrinsic
ambiguity of order $\Lambda_{QCD} \sim 200\;\rm MeV$ \cite{Smith97}.
For further discussion see, for example, \cite{Hoang:2008xm} and
references therein. High energy physicists around the world have
started planning for a future $e^+e^-$ linear collider, which may
become operational in the next decade. Such a machine will offer new
means for precision studies of the top quark properties and dynamics.
For example, the top quark mass could be measured with a precision of
$\approx 20\;\rm MeV/c^2$ from a threshold scan
\cite{tesla_physics_tdr,nlc_snowmass01,nlc_physics_tdr}.

%\begin{figure*}[p]
%\centerline{
%\includegraphics[width=0.85\textwidth,clip=]
%                {./fig_quadt/chapter2/topmass_tev0309.eps}
%}
%\caption{\label{fig:tmass_summary_1} Summary of the top-quark mass
%  measurements by \dzero and \cdf in Run-I or Run-II. This plot only
%  shows results which are used in the combination to the world average
%  top-quark mass as of March 2009 by the Tevatron Electroweak/Top
%  Working Group \cite{:2009ec}.}
%\end{figure*}
%
%\newpage
%\pagebreak

%% file: top_properties.tex
%
\section{Top-Quark Production and Decay Properties}
% Author: Daniel Wicke
% $Id: top_properties.tex,v 1.17 2009/07/10 18:29:31 wgw Exp $
%

% Plan: 7p
% = EW
% $W$ Helicity w=3 2Seiten
% FCNC    w=1  0.67
% Vtb
% = 
% Charge w=1   0.67
% = Strong
% Afb w=1 0.67
% Gluon vs. Quark production w=1 0.67
% Differential Xsec w=1 0.67
% =
% Width and Lifetime w=1 0.67

In Section~\ref{sec:topProd} we have described the phenomenology of the top
quark expected within the SM. To establish the top quark
discovered at the \tevatron as the SM top quark it is important to verify these
properties experimentally. % and to set limits on possible deviations. 
This section will summarize measurements of interaction properties of the top
quark and the corresponding limits on  possible
deviations from the SM that do not assume explicit presence of
non-SM particles. 
First measurements of the properties of electroweak interactions of the top
quark shall be summarized. Then verifications of electrical properties will 
be described, followed by measurements of the strong interaction properties of the
top quark. Finally, measurements of the top quarks width and lifetime are covered.
Measurements that involve  particles beyond the SM  or the Higgs boson are covered in
Section~\ref{sec:newPhysics}.

\subsection{Properties of the Electroweak Interaction of the Top Quark}
\begin{figure}[!tb]
  \centering
\includegraphics[width=0.30\textwidth]{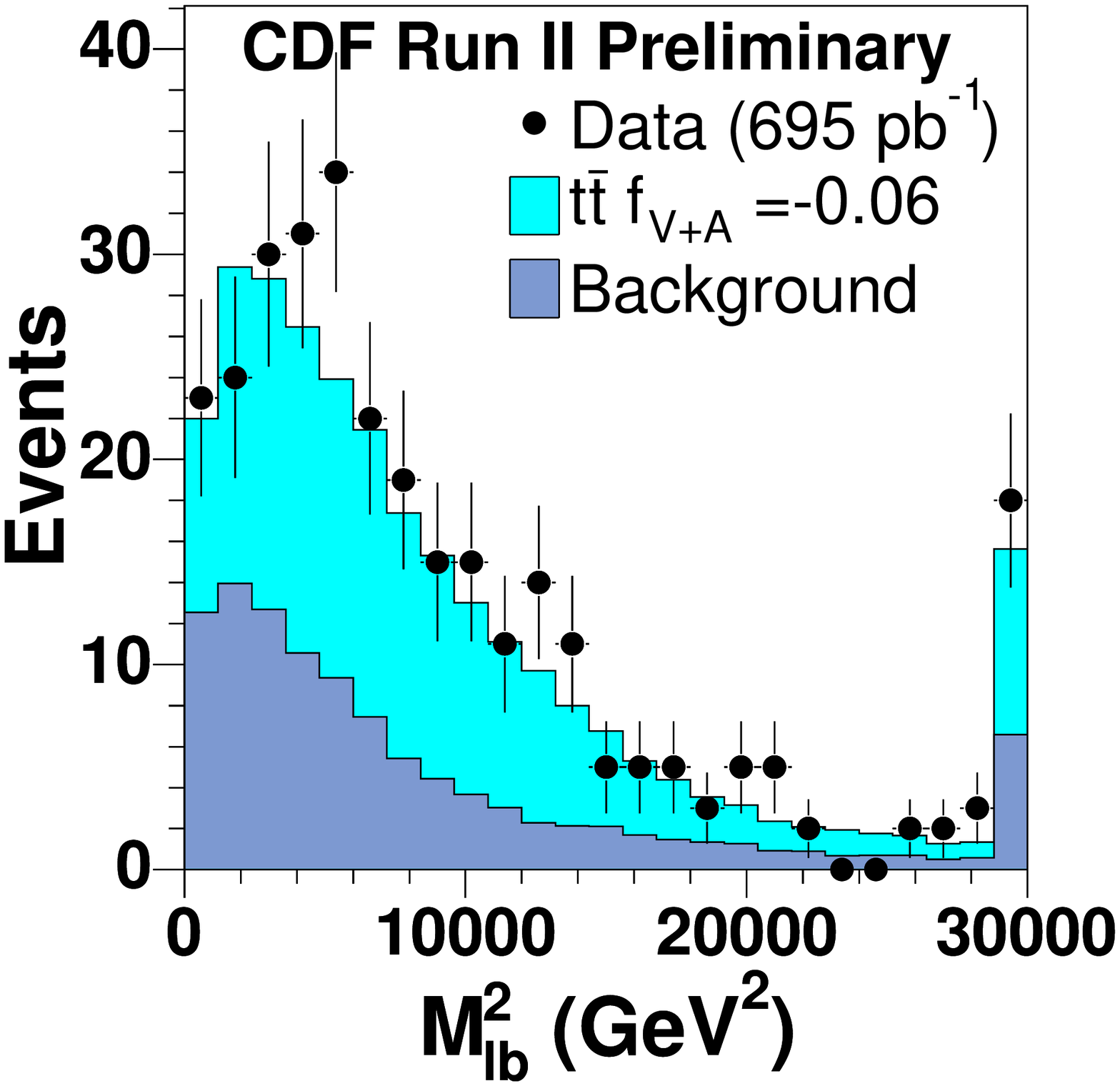}
\hspace*{0.05\textwidth} 
\includegraphics[width=0.48\textwidth]{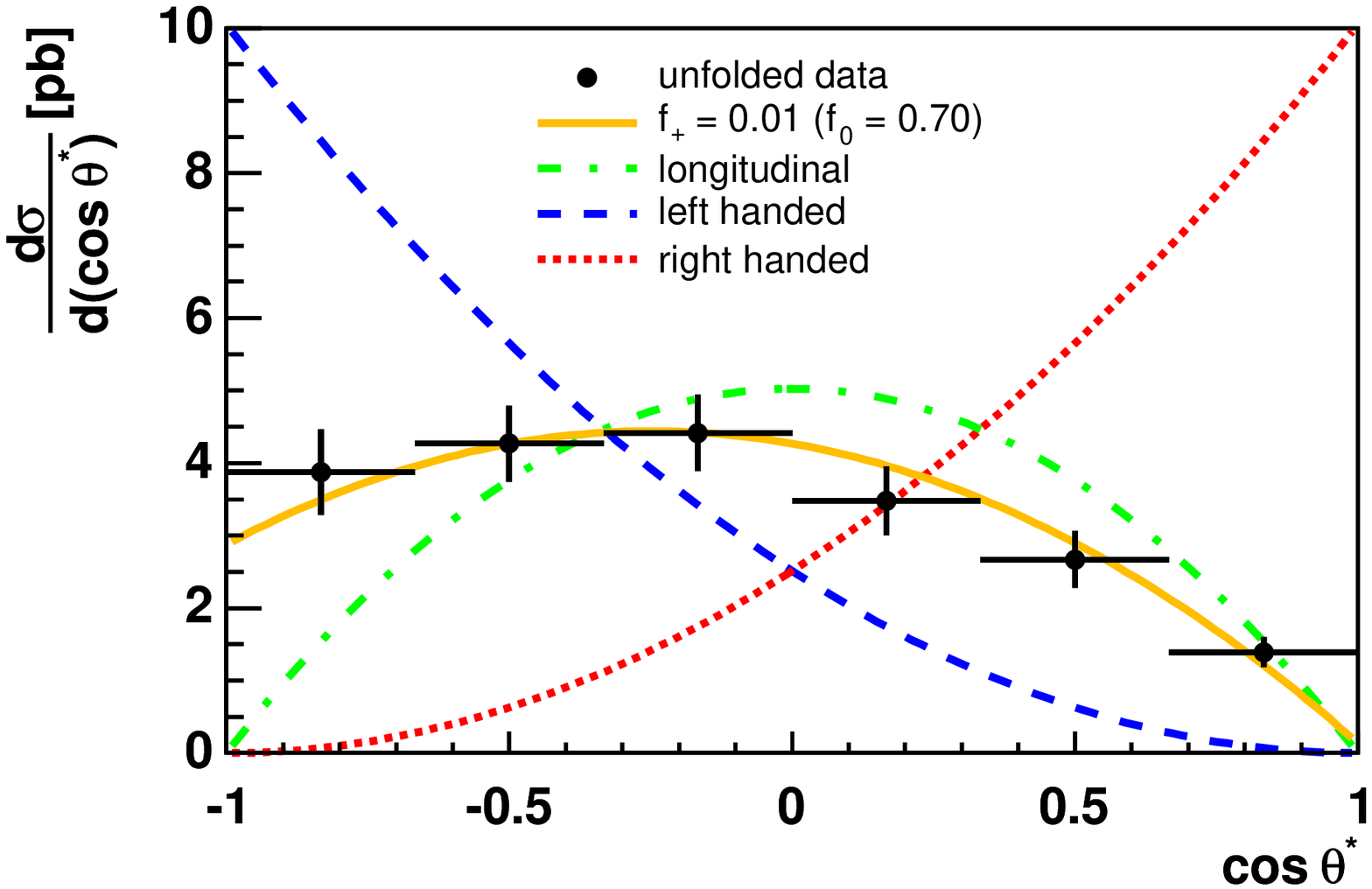}  

  \caption{Distribution of the lepton-$b$-quark invariant mass squared with
    measured by CDF~\cite{Abulencia:2006iy}. Observed $\cos\theta^*$
    distribution after deconvolution of measurement effects, compared to
    expectations of the individual polarizations and the best fit
    combination~\cite{Aaltonen:2008ei}.} 
  \label{fig:cdf:mlb-1}
\end{figure}

\paragraph{\boldmath $W$ boson helicity}
One of the first properties of the electroweak interaction with top quarks is
that of the helicity states of the $W$ boson occurring in top-quark decays.
The SM expects the top quark to decay to a $b$ quark and a $W$ boson. The
$V-A$ structure of the charged weak current restricts the polarization of the
$W$ boson in the SM. At the known values of $m_t$ and $m_b$ 
the respective fractions are
about $30\%$ left handed~($-$) and about $70\%$ longitudinal~($0$). Only a negligible
fraction of right handed~($+$) $W$ bosons is expected in the SM. Depending on the
$W$ helicity ($-,0,+$) the charged lepton in the $W$ decay prefers
to align with the $b$ quark direction, stay orthogonal or escape in the direction
opposite to the $b$ quark. Several observables are sensitive to these differences:
 the  transverse
momentum of the lepton, $p_T^\mathrm{lept}$, the lepton-$b$-quark invariant mass, 
$M_{lb}$ (\ref{fig:cdf:mlb-1} left),
and the angle between the lepton and the $b$-quark direction, $\cos\theta^*$. For best
sensitivity at \tevatron energies $\cos\theta^*$ is measured in the $W$ rest
frame.

The decay angle $\cos\theta^*$ has been used by both experiments in several
analyses with increasing
luminosity~\cite{Abulencia:2006ei,Aaltonen:2008ei,Abazov:2006hb,Abazov:2007ve}. 
This observable requires the reconstruction of the full top pair decay
kinematics to find the rest frame of the $W$ boson that is needed to calculate
$\cos\theta^*$. In the most recent results use datasets corresponding to
luminosities between $1.9$ and $2.7\ifb$ \cite{Aaltonen:2008ei,Abazov:2007ve}. 

The CDF analyses concentrate on top-pair events in the lepton+jets
channel requiring an isolated lepton, $\EtMiss$ and at least
four jets, one of which is required to  be identified as $b$-quark jet. 
Two methods to reconstruct the top-pair event kinematics are used. One method
recovers the unmeasured neutrino momentum from the $\EtMissVec$
and from solving the quadratic equation that connects the neutrino and lepton
momenta with $m_W$. 
The other method %(which will be called ``template'' method) 
uses a constrained kinematic fit to determine the lepton and
parton momenta, where these momenta are allowed to float within the
experimental uncertainties of the measured quantities. Constraints are built
from the reconstructed $W$-boson mass and the equality of the top and anti-top masses
constructed from the fitted lepton and parton momenta.
Both methods require the association of the measured jets to the partons of the
top-pair topology and use the quality of the constrained fit to select this
assignment. 

The first analysis with full reconstruction uses the signal simulation to derive acceptance
functions which are then convoluted with the theoretical predicted number of
events in each bin of $\cos\theta^*$. The helicity fractions are then taken from  a
binned likelihood fit after subtracting the background estimation from data, 
see Fig.~\ref{fig:cdf:mlb-1} (right).
In the second analysis signal templates for the three different helicity
states are combined the background expectations. The helicity
fractions are taken from an unbinned likelihood fit with proper correction for
acceptance effects. 
The results of both analyses agree in the analyzed $1.9\ifb$ of data. They
are combined with the BLUE method~\cite{Lyons:1988rp}:
\bea
f_0 & =& \Plus 0.66\pm 0.16 \mathrm{(stat)}\pm 0.05  \mathrm{(syst)} \qquad %\nonumber\\
f_+  =        -0.03\pm 0.06 \mathrm{(stat)}\pm 0.03  \mathrm{(syst)}
\eea
with a correlation of about $-90\%$ between $f_0$ and $f_+$~\cite{Aaltonen:2008ei}.

CDF also performed an analysis using the Matrix Element technique with
$1.9\ifb$ and finds slightly smaller statistical uncertainties at the cost of
slightly increased systematics~\cite{CdfNote9144}.

D\O\ has reconstructed the decay angle in both the lepton+jets and the
dilepton channel of top pair decays. Events are selected by requiring an isolated lepton, $\EtMiss$
and at least four jets. No second lepton is allowed in the event. Dilepton events
are selected with two isolated charged leptons with opposite charge, large
$\EtMiss$ and at least two jets.  A veto on $Z$ boson events
is applied. 

The top-pair decay kinematics in lepton+jets events is reconstructed using
a constraint fit similar to the CDF analysis. To select the optimal jet parton
assignment in addition to the fit quality, the probability to find the
observed $b$ tags in the chosen assignment is considered.
The decay angle, $\cos\theta^*$, is computed from the leptonic side and
a second measurement of the absolute value is taken from the hadronic side.
The kinematics of dilepton events can be solved assuming $m_t$ with a
fourfold ambiguity. In addition, two possible assignments of the two leading
jets in $p_T$ to the $b$ quarks are considered. To explore the full phase space
consistent with the measurements, the measured jet and charged lepton momenta
are fluctuated according to the detector resolution and $\cos\theta^*$ is
computed for each fluctuation. The average  over all solutions
and all fluctuations is computed for each jet to find two $\cos\theta^*$
values per event. The resulting distribution for  RunIIb data is shown in
Fig.~\ref{fig:d0:whel-costheta}. 
\begin{figure}
  \centering
  \includegraphics[width=0.332\textwidth]{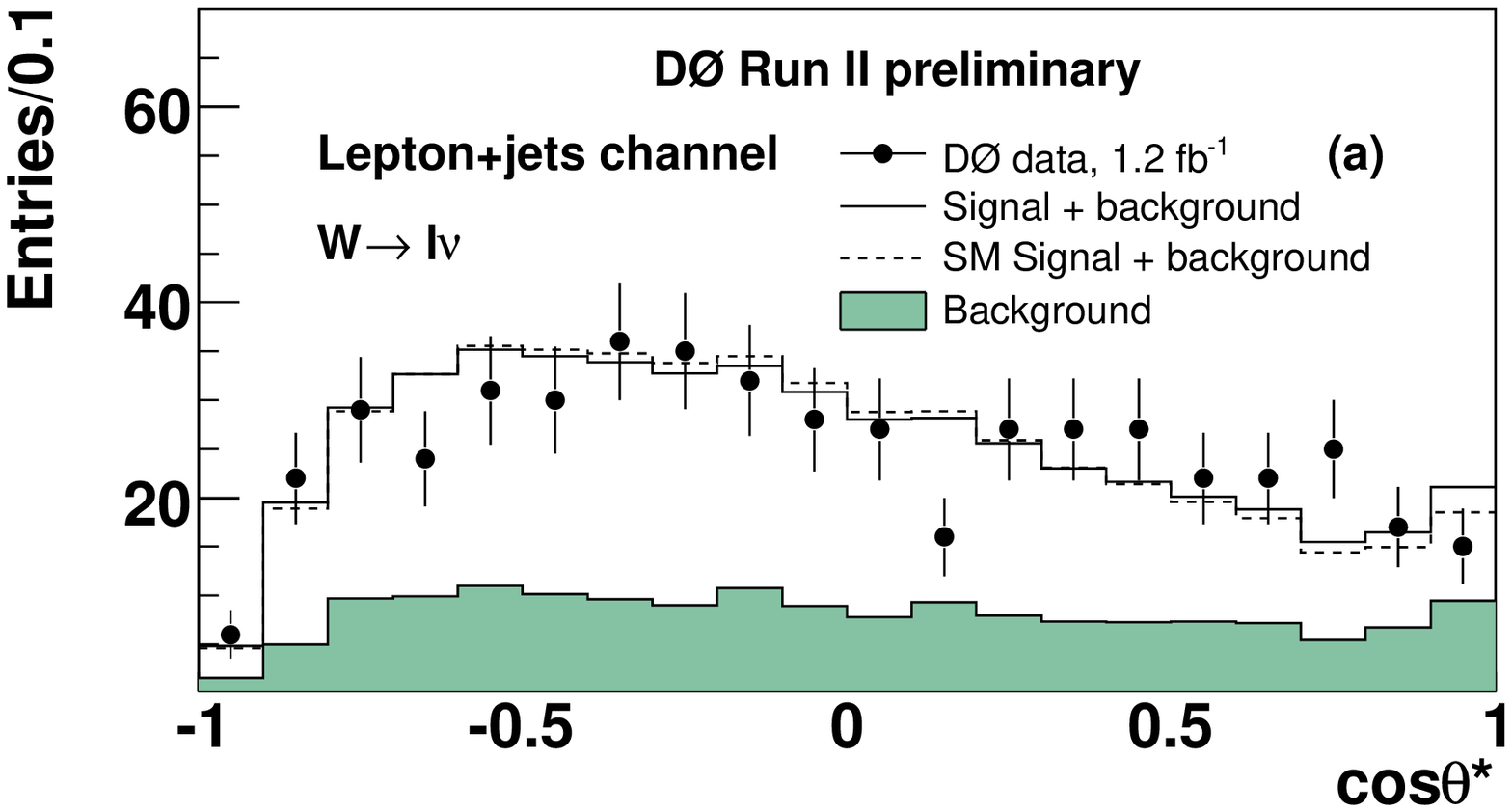}%
  \includegraphics[width=0.332\textwidth]{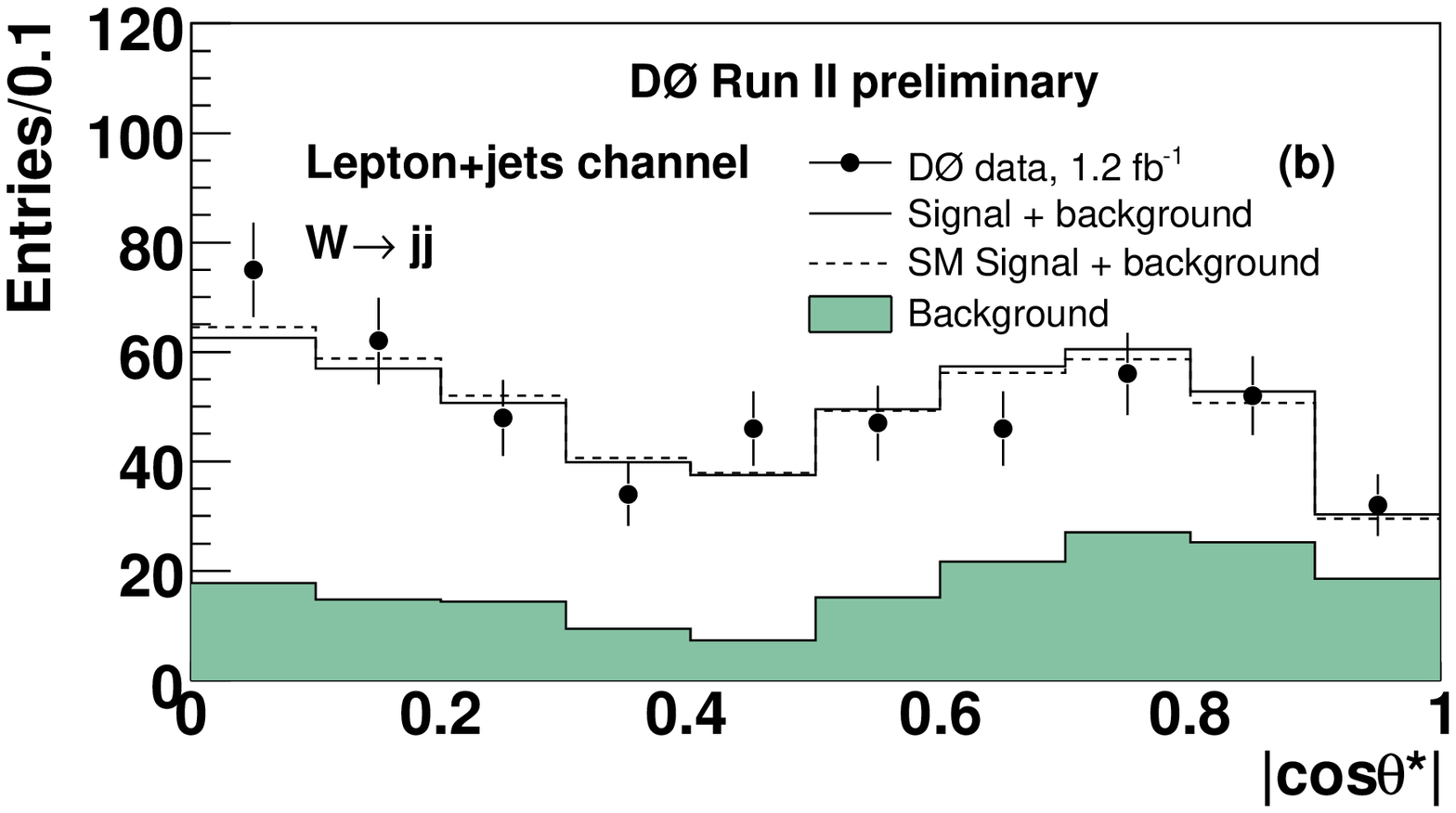}%
  \includegraphics[width=0.332\textwidth]{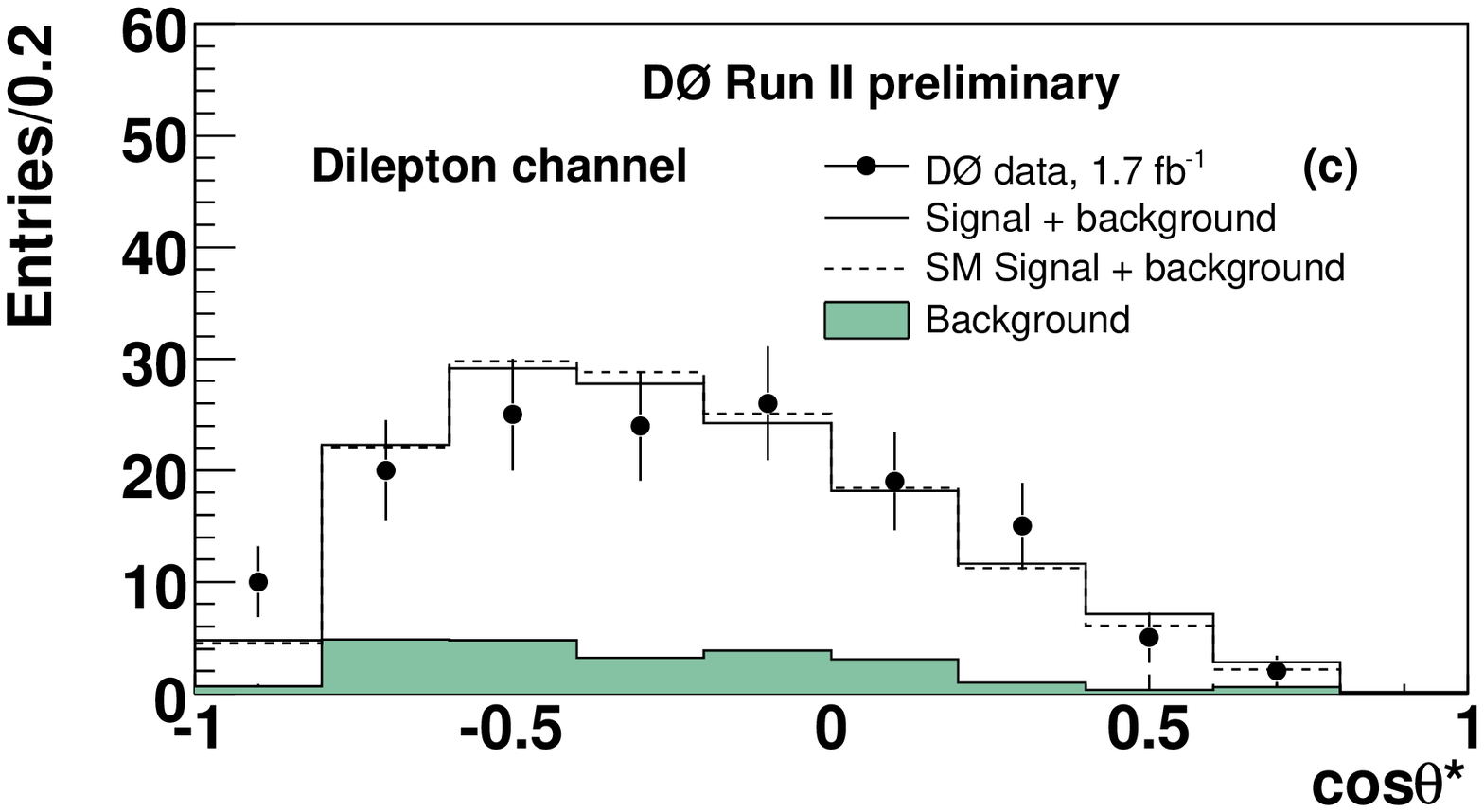}%
  \caption{Distributions of $\cos\theta^*$ as measured in % $1.2-1.7\ifb$ of
  D\O\
  RunIIb data. The left plot shows the results from the leptonic decay
  in lepton+jet events, the middle plot the one from the hadronic
  decay. The right plot shows the distribution obtained from dilepton events.
  In addition to the data shown with error bars all plots contain 
  the best fit prediction is shown as full line, the SM as a dashed line 
  histogram. The shaded area represents the background
  contribution~\cite{Abazov:2007ve}.}
  \label{fig:d0:whel-costheta}
\end{figure}

Simulations with specific $W$ boson helicities are used to construct a binned
likelihood function, which is optimized to find the helicity fractions most
consistent with the observed data.
Combining the datasets of RunIIa and RunIIb %, which are analyses separately,
to a total luminosity of $2.2\!-\!2.7\ifb$ D\O\ finds
\bea
f_0&=&0.490\pm0.106\mathrm{(stat)}\pm0.085\mathrm{(syst)}\qquad %\nonumber\\
f_+=0.110\pm0.059\mathrm{(stat)}\pm0.052\mathrm{(syst)}\quad\mbox.
\eea
The correlation between the two numbers is $-80\%$~\cite{Abazov:2007ve}.

The lepton-$b$-quark invariant mass $M_{lb}$ has been used by CDF 
in lepton+jets events with one or two identified $b$-jets and 
and dilepton events with two identified $b$ jets.
For each selected event $M^2_{lb}$ is computed. 
In lepton+jets events with a single identified $b$ jet the lepton the
computation is unambiguous, though only correct in half the cases.
For lepton+jets events with two identified $b$-jets a two dimensional
distribution of $M^2_{lb}$ is constructed. One dimension being $M^2_{lb}$
computed with the higher energetic $b$-jet, the other with the lower energetic
$b$-jet. 
In the dilepton events the two dimensional histogram is filled twice,
i.e. using each of the leptons, c.f.~Fig.~\ref{fig:cdf:mlb-1}.
The distributions obtained in data are compared to simulation. Background is
estimated from a combination of simulated events and data. Signal
samples are simulated with left and right handed couplings. The fraction of
left-handed $W$ bosons is obtained from a binned likelihood with nuisance
parameters to describe the systematics: $f_+=-0.02\pm 0.07$ or  $f_+<-0.09$ at
the 95\% C.L.  The longitudinal fraction is kept at
its SM value of about $70\%$.
All helicity measurements are in good agreement with the SM expectations
of a pure $V-A$ coupling.

\paragraph{\boldmath The CKM element $V_{tb}$}
\label{sec:top_prop_vtb}
The strength of the charged weak interaction of the top quark is defined by
the product of the weak coupling and the $tb$ element of the CKM matrix,
$V_{tb}$. In principle this is a free parameter of the SM, but as the CKM
matrix needs to be unitary, the knowledge about the other elements constrains
the absolute value within the SM to 
$0.9990<\left|V_{tb}\right|<0.09992$~\cite{Amsler:2008zz,Alwall:2006bx}.
Physics beyond the SM invalidates these assumptions and relaxes constraints
on $\left|V_{tb}\right|$. 

Deviations from the SM value modifies both the top quark decay and the
weak production of single top quarks.  
As described in Section~\ref{sec:singletopvtb} the measured
cross-section for single top-quark production in both experiments yields
a lower limit of $\left|V_{tb}\right|>0.78$~\cite{Abazov:2009ii,Aaltonen:2009jj}.
The analyses assume that top-quark decay is dominated by $t\rightarrow
Wb$.

In top-pair events lower values of $\left|V_{tb}\right|$ reduce the amount of
$b$ quarks in events with top pair signature. Both experiments have used the
ratio of top-quark pair events with zero, one or two identified $b$ jets to
extract the fraction of top quarks decaying to $Wb$: 
$R_b=\frac{\left|V_{tb}\right|^2}{\left|V_{td}\right|^2+\left|V_{ts}\right|^2+\left|V_{tb}\right|^2}$.
CDF has investigated lepton+jets and dilepton events and find $R_b$ being
consistent with one~\cite{Acosta:2005hr}. D\O\ has investigated lepton+jets events with an
integrated luminosity of  $0.9\ifb$. Again $R_b$ is found to be consistent with
one and sets the currently best limit on $R_b>0.79$ at the $95\%$~C.L. This limit is 
converted to a limit on the ratio of $\left|V_{tb}\right|^2$ to the
off-diagonal elements:
$\frac{\left|V_{tb}\right|^2}{\left|V_{td}\right|^2+\left|V_{ts}\right|^2}>3.8$
at the $95\%$~C.L.~\cite{Abazov:2008yn}. The only assumption entering this limit is that top quarks
cannot decay to quarks other that the known SM quarks, which is
valid even in presence of an additional generation of quarks as long as the
$b'$ quark is heavy enough. 
Both, the results from single top quark and those from top-quark pair events
are consistent with the SM expectation.
%The direct result confirms the indirect results from $B\rightarrow
%X_s\gamma$~\cite{Bobrowski:2009ng,Alwall:2006bx}. 

\paragraph{Flavor-Changing Neutral Currents} 
Flavor-changing neutral currents (FCNC) do not appear in the SM
at tree level and are suppressed in quantum loops. 
Anomalous couplings could lead to enhancements of FCNC in
the top sector and their observation would be a clear sign of new 
physics~\cite{Fritzsch:1989qd,AguilarSaavedra:2004wm}.
The \tevatron experiments have looked for FCNC both in the (singly) production
of top quarks~\cite{Aaltonen:2008qr,Abazov:2007ev} and in top-quark
decays~\cite{Abe:1997fz,Aaltonen:2008aaa}. 
Limits on the single-top production through anomalous couplings were also 
set with \lep and HERA 
data~\cite{Heister:2002xv,Abdallah:2003wf,Achard:2002vv,Abbiendi:2001wk,Chekanov:2003yt,Aktas:2003yd,H1:FCNC2007}.

The top decay through a $Z$ boson was searched for in  $1.9\ifb$ of CDF
data~\cite{Aaltonen:2008aaa}. The event selection aims to identify events in which the $Z$ boson decays
leptonically and the second top decay through a $W$ boson into hadrons. Two
leptons of the same type and opposite charge are required with an invariant
mass that lies within $15\GeV$ of the $Z$ boson mass.  

To separate signal from background the mass of the $W$ boson is reconstructed
from two jets, the top quark is reconstructed by adding a third jet and a second
top quark is reconstructed from the $Z$ boson with the fourth jet. 
A $\chi^2$ variable is built from the differences of the reconstructed masses
to $m_W$ and $m_t$, respectively.
The $\chi^2$ of the jet parton assignment that yields the lowest $\chi^2$ is used to build a
distribution of $\chi^2$ values.
The distribution observed in data is compared to 
signal and background templates derived from simulation and from control
samples, c.f.~Fig.~\ref{fig:cdf-fcnc-toppairs}. Data agree well with the SM expectation and thus limits  on the
branching fraction $t\rightarrow Zq$ are set. For this the
Feldman-Cousins method is applied and yields  \mbox{${\cal B}(t\rightarrow
Zq)<3.7\%$} at the $95\%$~C.L.~\cite{Aaltonen:2008aaa}.
The Run~I result in addition set limits on  a flavor-changing currents in the
photon plus jet mode of \mbox{${\cal B}(t\rightarrow
\gamma u)+{\cal B}(t\rightarrow
\gamma c)<3.2\%$}~\cite{Abe:1997fz}.
\begin{figure}
  \centering
\includegraphics[width=0.5\textwidth]{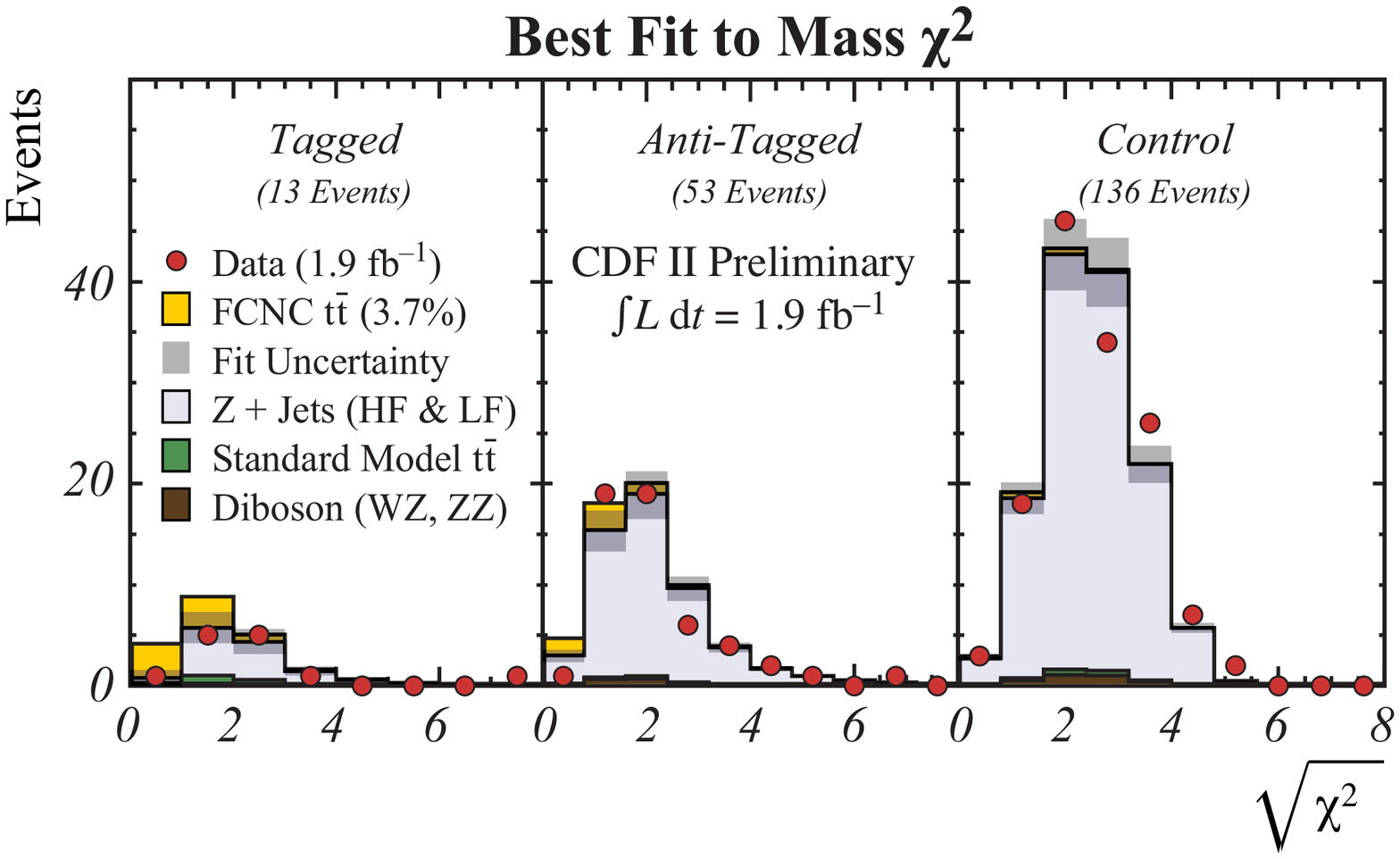}  
\hspace*{0.03\textwidth}
\includegraphics[width=0.35\textwidth]{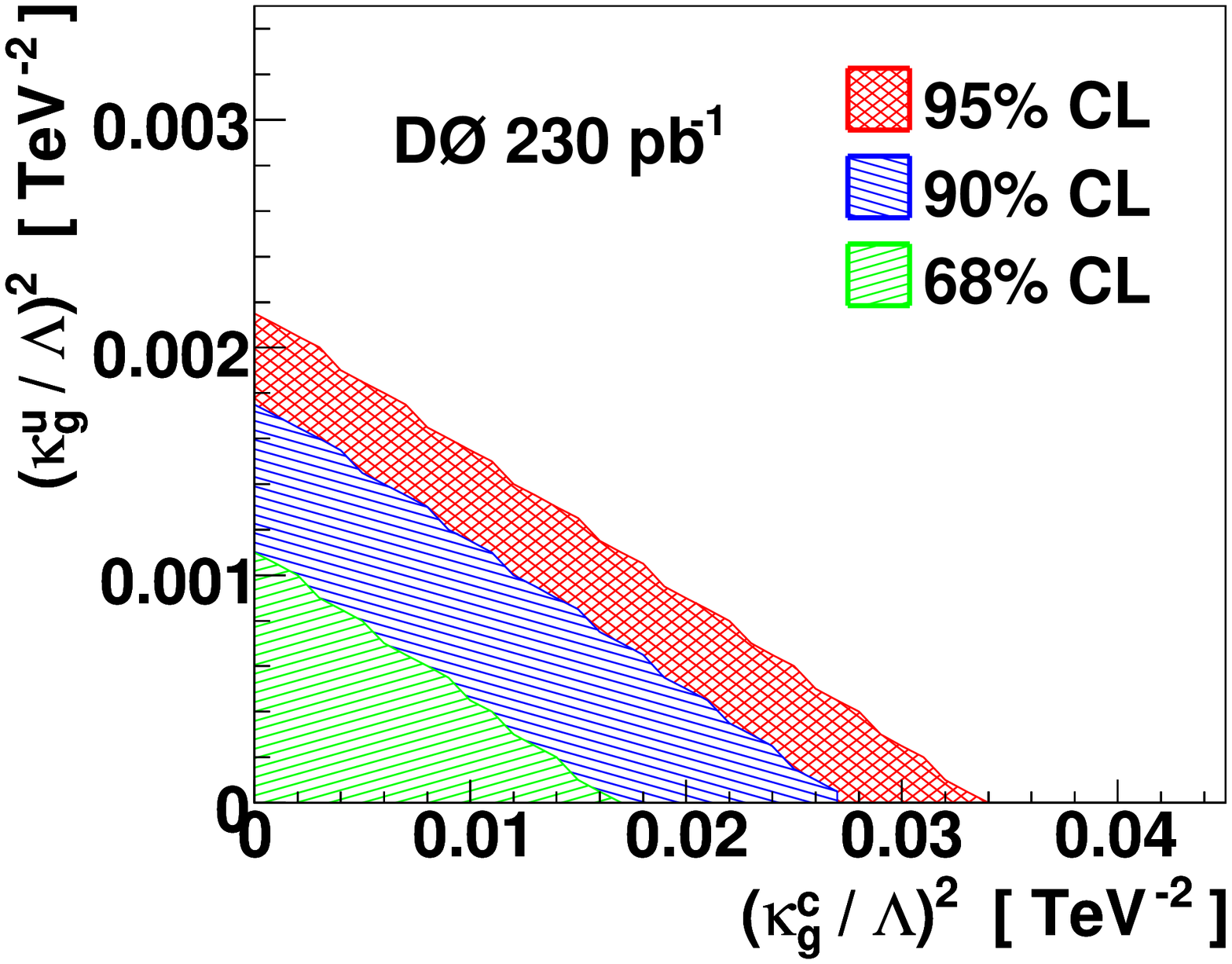}
  \caption{Left: Mass $\chi^2$ distribution observed by CDF in the two signal and one
    control sample compared the the best fit expectation plus a signal of
    $3.7\%$ branching to $Z$ boson~\cite{Aaltonen:2008aaa}. Right: Exclusion contour obtained by D\O\
    in a two dimensional approach~\cite{Abazov:2007ev}. }
  \label{fig:cdf-fcnc-toppairs} \label{fig:d0-fcnc-singletop}
\end{figure}

While the above study of top-quark decays addresses FCNC through the 
$Z$ boson, investigations of the production of single top
events can be used to restrict anomalous gluon couplings.
CDF looks for the production of single top quarks
without additional jets, $u(c)+g\rightarrow t$~\cite{Aaltonen:2008qr}. 
To select such events with a
leptonic top-quark decay, one isolated lepton, $\EtMiss$ and
exactly one hadronic jet are required. The jet must be identified
as $b$-quark jet. Additional cuts are used to reduce the non-$W$ backgrounds as in
the single top analyses, see~Section~\ref{subsec:stObservation}.
D\O\ investigated a sample of single top quark production one additional jet~\cite{Abazov:2007ev}. 
Both experiments train neural networks to distinguish the FCNC from SM
production. The neural network outputs are used to measure the 
cross-section of FCNC production of single top quarks. Both experiments find
no evidence for such production and set limits. 
CDF concludes $\sigma_t^{\mathrm{FCNC}}<1.8\pb$ at the 
$95\%$~C.L.~\cite{Aaltonen:2008qr}. 
This cross-section limit is converted to limits on FCNC top-gluon coupling
constants following~\cite{Liu:2005dp,Yang:2006gs}. 
Under the conservative assumption that only one of the couplings differs from zero,
CDF finds 
$\kappa_{gtu}/\Lambda<0.018\TeV^{-1}$ or $\kappa_{gtc}/\Lambda<0.069\TeV^{-1}$.
Expressed in terms of the top branching fraction through these processes these limits
correspond to ${\cal B}(t\rightarrow u+g)<3.9\cdot 10^{-4}$ and 
${\cal B}(t\rightarrow c+g)<5.7\cdot 10^{-3}$~\cite{Aaltonen:2008qr,Zhang:2008yn}.

D\O\ computes the limits directly in terms of the  anomalous
gluon couplings $\kappa_{gtu}/\Lambda$ and  $\kappa_{gtc}/\Lambda$.
A likelihood is folded with a prior flat in the FCNC cross sections and
exclusion contours are computed as contours of equal probability that contain
$95\%$ of the volume.  These two-dimensional limits on the squared couplings 
are shown in \fig{fig:d0-fcnc-singletop}~(right).

\subsection{Electrical Properties of  Top Quark}
The top quark's electrical properties are fixed by its charge.
However, in reconstructing top quarks the charges of the objects
usually aren't checked. Thus an exotic charge
value of $\left|q_t\right|=4e/3$ isn't excluded by standard analyses.
To distinguish between the SM and the exotic top charge it is
necessary to reconstruct the charges of the top-quark decay products,
the $W$ boson and the $b$ quark.  

CDF and D\O\ have investigated lepton+jets events with at least two
identified $b$ jets. CDF also includes dilepton events with one and two
identified jets. 
The $W$ boson charge is taken from the charge
of the reconstructed lepton. As a measure of the $b$ quark charge a jet charge
is defined as a weighted sum of the charges of the tracks inside the $b$ jet
$Q_\mathrm{jet}=\sum{w_i q_i}/\sum w_i $. As a weight CDF uses the component
of the track momenta longitudinal to the jet
$w_i^\mathrm{CDF}= (\vec{p_i}\cdot\vec{p}_{\mathrm{jet}})^{\kappa}$ with
$\kappa=0.5$; D\O\ uses the component transverse to the jet momentum
$w_i^\mathrm{D\O}=p_{\perp,i}^\kappa$ with $\kappa=0.5$.
These jet charges will have significant event-to-event fluctuations, but are
good for a statistical analysis. For the statistical analysis it is critical
to understand the purity of the $b$-tagged jets and the amount of
contributions from $c$-quark jets and from light-quark jets. Both experiments use dijet
data to calibrate their jet-charge observable. Finally, one of the identified $b$
jets has to be assigned to the lepton, putting the other to hadronic top-quark
decay. That assignment is taken which is best described by a kinematic fit with
constraints on  $m_t$ and $m_W$. 
For the dilepton events CDF bases the corresponding decision 
on the invariant mass values between the associated lepton and $b$ quark.

Likelihood methods are used to test the consistency of the data
with the SM and the exotic scenario. Both experiments confirm the SM. The
strongest results stems from CDF. Using data corresponding to $1.5\ifb$, 
a Bayes factor of $2\log B_f=12$ is found.

\subsection{Properties of the Strong Interaction of the Top Quark}
\paragraph{Forward Backward Asymmetry}
At the \tevatron the initial state of proton antiproton is not an eigenstate
under charge conjugation. Thus in principle 
also the final state may change under this operation. 
In QCD, however, such a charge asymmetry appears only at NLO and arises mainly 
from interference between contributions symmetric and
anti-symmetric under the exchange of top and anti-top
quarks~\cite{Kuhn:1998kw}. 
Experimentally, CDF and D0 investigated forward-backward asymmetries~\cite{Aaltonen:2008hc,CdfNote9724,d0:2007qb},
$
A_{\mathrm{FB}}=({N_{\mathrm F}-N_{\mathrm B}})/({N_{\mathrm F}+N_{\mathrm
  B}})
$,
where $N_{\mathrm F}$ and $N_{\mathrm B}$ are the number of events observed in
forward and backward direction, respectively.
The forward and backward directions are either defined in the
laboratory frame, i.e. according to the sign of the rapidity of the top quark,
$y_t$, or can be defined in the frame where the top pair system rests along
the beam axis, i.e., according to the sign of the rapidity difference between
top and anti-top quark, $\Delta y=y_t-y_{\bar t}$. The two different
definitions of forward and backward yield two different asymmetries that are
labeled $A_{\mathrm{FB}}^{p\bar p}$ and $A_{\mathrm{FB}}^{t\bar t}$ according to their rest frame of definition.
In the SM at NLO, asymmetries are expected to be  
$0.05$ and $0.08$, respectively~\cite{Antunano:2007da}, but at NNLO 
significant corrections for the contributions from $t\bar t+X$ are 
predicted~\cite{Dittmaier:2007wz}.
The smallness of the asymmetries expected within the SM make them a
sensitive probe for new physics.

The measurements by CDF and D\O\ use lepton+jets events with at least one
identified $b$ jet. For each event the kinematics  is reconstructed with a
kinematic fit that uses constraints on the reconstructed top quark and $W$
boson mass. %The details of this fit differed between the analyses. 
The jet-to-parton assignment with the best fit is used. The rapidity of the hadronically decayed
(anti-)top quark, $y_h$ is multiplied by the minus the charge, $Q_\ell$, of the
lepton to obtain the top quark rapidity, $y_t=-Q_\ell\, y_h$. Correspondingly
the rapidity difference $\Delta y=y_t-y_{\bar t}$ is computed as $\Delta y= Q_\ell\,(y_\ell-y_h) $,

\begin{figure}
  \centering
  \includegraphics[width=0.25\textwidth,clip,trim=0mm 0mm 0mm 19mm]{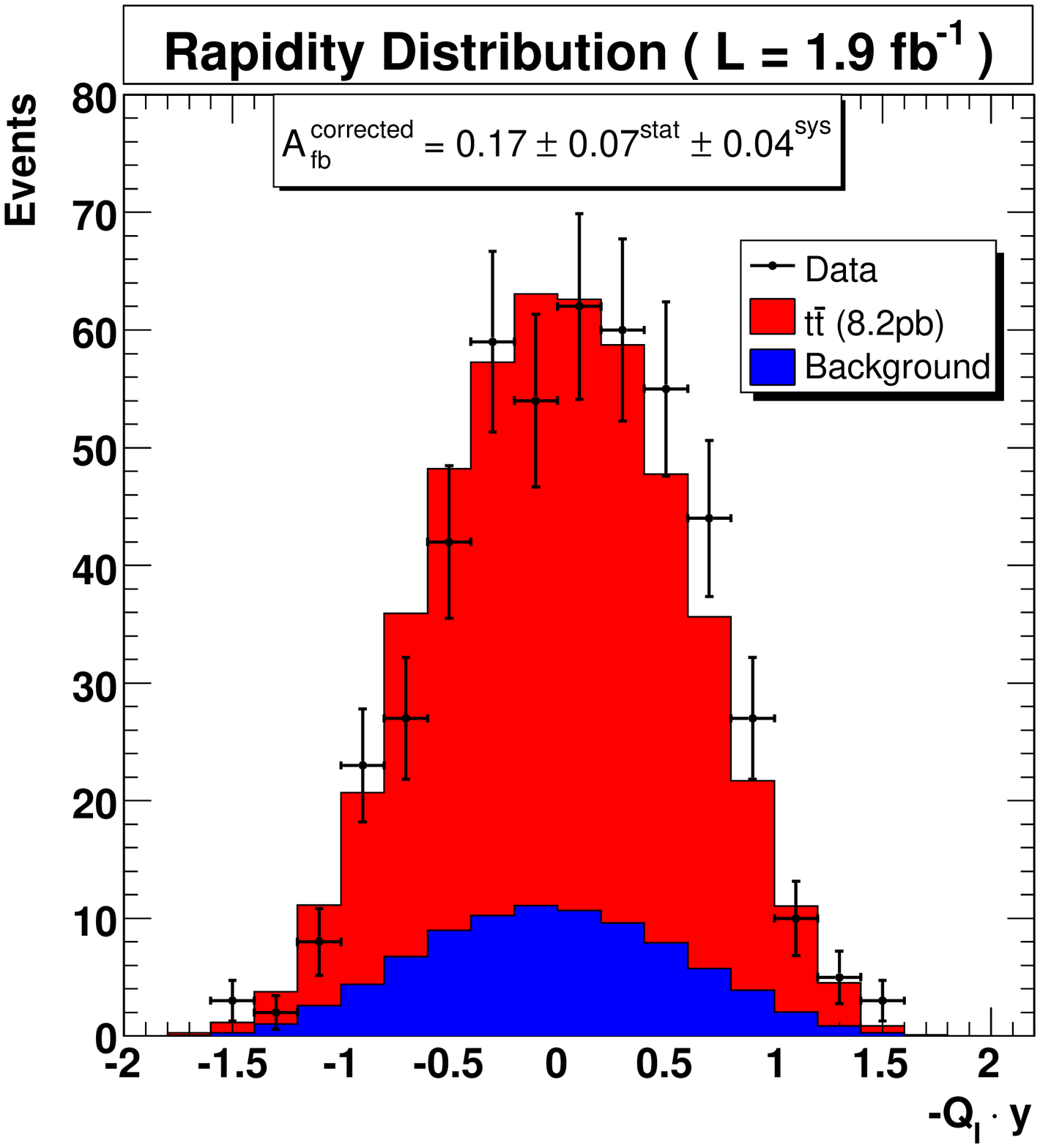}
  \quad
  \includegraphics[width=0.25\textwidth]{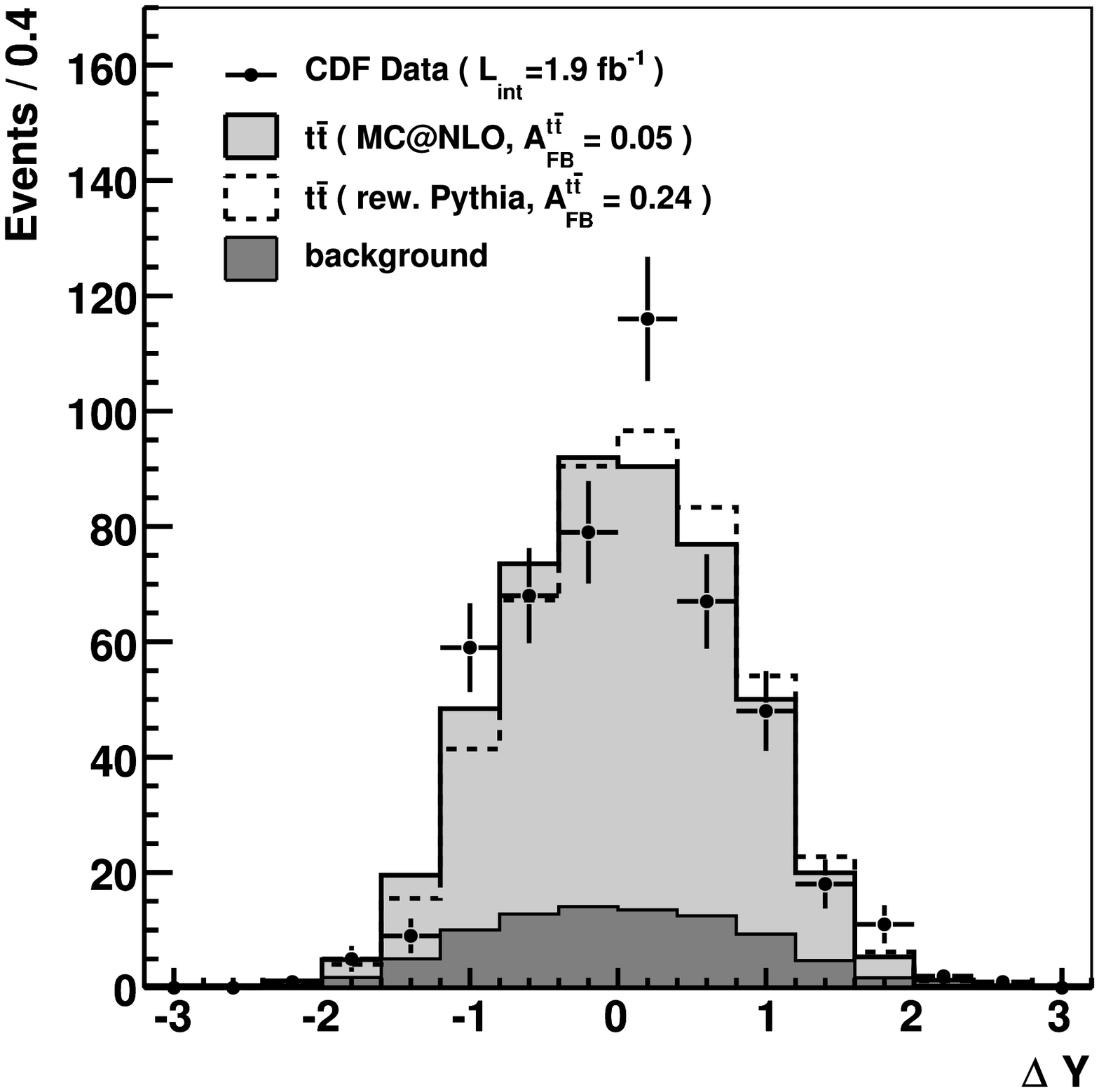}
  \quad
  \includegraphics[width=0.42\textwidth]{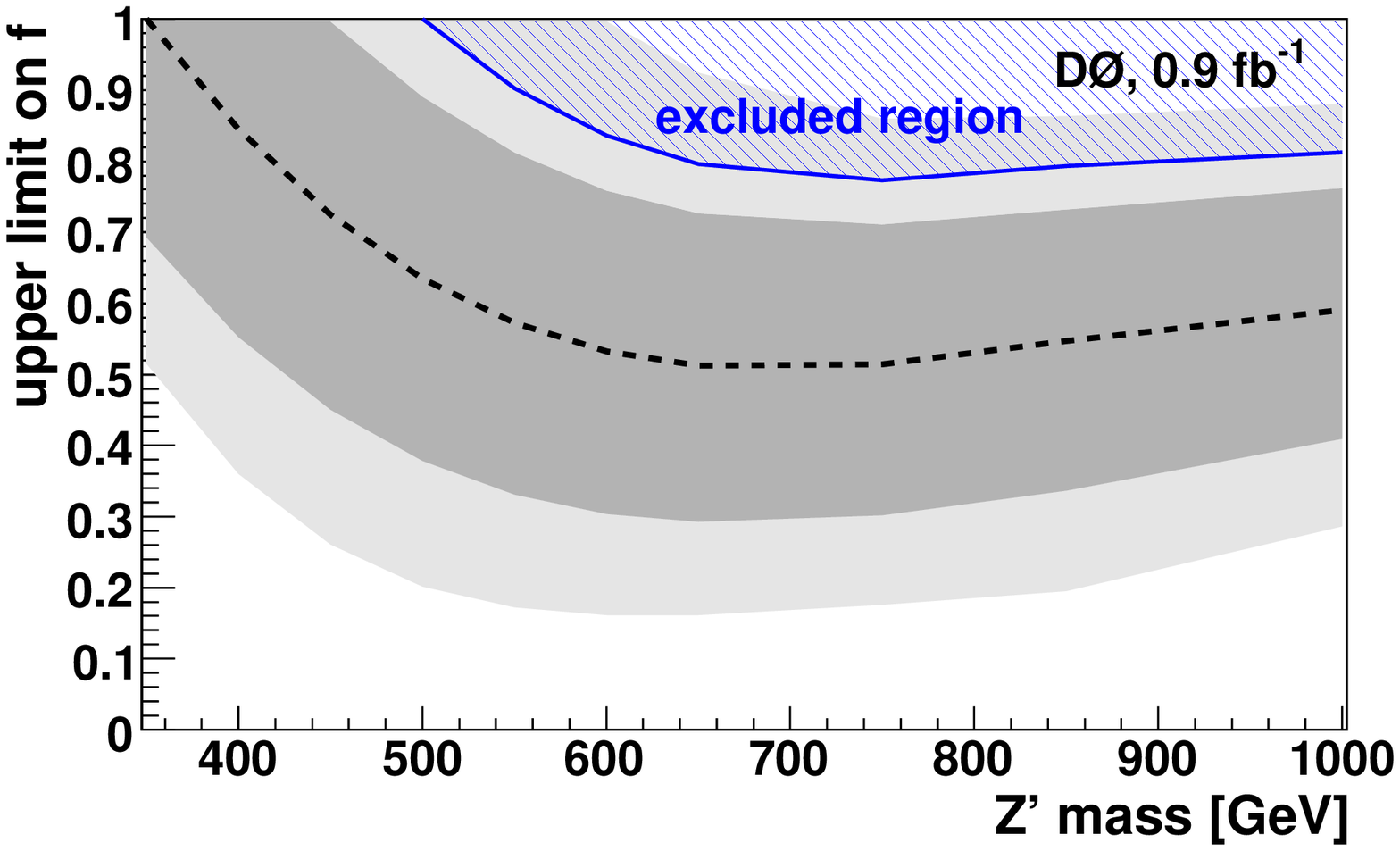}
  \caption{Distribution of the top quark rapidity (left) and rapidity
    difference (middle) as measured by CDF in $1.9\ifb$ compared to the
    SM prediction~\cite{Aaltonen:2008hc}.
    Left: Limits on a possible fraction, $f$, of resonant top pair production
    through a $Z'$ boson obtained from the measurement of the forward backward
  asymmetry in D\O~\cite{d0:2007qb}.}
  \label{fig:d0-afb}
\end{figure}
The obtained distributions depend on the amount of background events and
differ from the true shape due to acceptance and detector effects.  After background subtraction CDF
derives the particle level distributions 
inverting the acceptance efficiencies and migration probability matrices as
derived from  simulation with zero asymmetry using a reduced number of
only four bins. 
%By doing so the unfolding assumes acceptance efficiencies
%generated with \pythia\ to be valid. 
The final unfolded asymmetries obtained by CDF data are~\cite{CdfNote9724,Aaltonen:2008hc}:
\beq
A_{\mathrm{FB}}^{p\bar p} = 0.193 \pm 0.065_{\mathrm{stat}} \pm
0.024_{\mathrm{syst}}\quad (3.2\ifb) \qquad % \nonumber\\
A_{\mathrm{FB}}^{t\bar t} = 0.24 \pm 0.13_{\mathrm{stat}} \pm 0.04_{\mathrm{syst}}\quad(1.9\ifb)\mbox{.}
\eeq
These values are larger than the $0.05$ and $0.08$
expected in the SM at NLO, respectively, by up to two standard deviations.

D\O\ estimates and corrects for the contribution of background event using a
likelihood discriminant insensitive to the asymmetry. 
In $0.9\ifb$ of data D\O\ find a final observed asymmetry~\cite{d0:2007qb} of 
\beq
A_{\mathrm{FB}}^{t\bar t~\mathrm{obs}}=0.12 \pm 0.08_\mathrm{stat} \pm 0.01_\mathrm{syst}\quad\mbox{.}
\eeq
To keep the result model independent and in contrast to the CDF results this
number is not corrected for acceptance and resolution effects.
Instead it needs to be compared to a theory prediction for the phase space
region accepted in this analysis and corrected for dilution effects.  
For NLO QCD and the cuts used in this analysis D\O\ evaluates  $A_{\mathrm{FB}}^{t\bar t}=0.008$.  
Thus as for CDF this  result corresponds to an asymmetry that is slightly higher than
expected in NLO QCD, but not by more than two standard deviations.

In addition to the QCD expectation D\O\  provides a parameterized procedure 
to compute the asymmetry expected in this analysis for an arbitrary model of new
physics. As an example the measurement's sensitivity to top pair production
via a heavy neutral boson, $Z'$, with couplings proportional to that of the
SM $Z$ boson is studied. Limits on the possible fraction of heavy
$Z'$ production are determined as function of the $Z'$ boson mass,
c.f.~\fig{fig:d0-afb}~(right). These can be applied to wide $Z'$ 
resonance by averaging of the corresponding mass range.

\paragraph{Differential Cross-section}
Measurements of the differential cross-sections of top pair production can be
used to verify the production mechanism assumed in the SM. 
Due to the required unfolding these measurements are especially cumbersome.
The CDF collaboration has measured the differential cross section with respect
to the invariant top pair mass 
$
\frac{\mathrm{d}\sigma_{t\bar t}}{\mathrm{d}M_{t\bar t}}(M_{t\bar t})
$ using $2.7\ifb$ of lepton+jets data with one identified $b$
jet~\cite{Aaltonen:2009iz}. 
The invariant mass of the top pairs is reconstructed from the four-momenta of 
the four leading jets in $p_T$, the four momentum of the lepton and the missing
transverse energy. The $z$-component of the neutrino is not
reconstructed  but used as if it was zero.

The expected background is subtracted from the observed distribution  and then 
distortions are unfolded using the
singular value decomposition~\cite{Hocker:1995kb} of the response matrix that
is obtained from simulations.
The differential cross section obtained in $2.7\ifb$ of data using the
semileptonic decay mode is shown in~\fig{fig:cdf-diffxsec}. 
The consistency with the SM expectation 
%using Anderson-Darling statistics~\cite{AndersonDarling:1952}. The observed 
has a $p$-value is 0.28 showing good agreement of with the SM.
\begin{figure}
  \centering
  \includegraphics[width=0.45\textwidth]{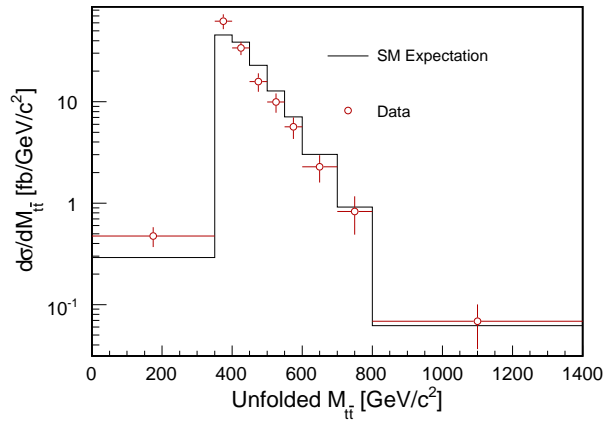}
  \caption{Left: Differential top pair production cross-section measured by CDF in
    $2.7\ifb$ of data using the semileptonic decay mode. Indicated are the
    total uncertainties for each bin, excluding the overall luminosity
    uncertainty of $6\%$~\cite{Aaltonen:2009iz}.}
  \label{fig:cdf-diffxsec}
\end{figure}
The invariant top pair mass was used in further analyses by both CDF and D\O\
to search for new physics. These results are described in Section~\ref{sect:BSM}.

\paragraph{Gluon Production vs. Quark Production}
Production of top quark pairs in the SM occurs through strong force. At the
\tevatron the dominant process is that of $q\bar q$ annihilation, about 15\% are
contributed by gluon gluon fusion. Due to large uncertainties on the gluon PDF
the exact size of the gluon contribution is rather
uncertain~\cite{Kidonakis:2003qe,Cacciari:2003fi,Moch:2008ai}. 

Two properties of the two production processes allow to separate them and to
measure their relative contributions. 
Close to threshold the spin states of the gluon fusion are $J=0$,~$J_z=0$,
while the $q\bar q$ annihilation yields  $J=1$,~$J_z=\pm
1$~\cite{Arens:1992fg}. This yields angular correlations between the 
charged leptons in the dilepton channel.
Alternatively one can exploit the difference in the amount of gluon radiation
from quarks and gluons. The gluon fusion processes are expected to contain
more particles from initial-state radiation.

CDF has used both features to measure the gluon fraction of top-quark pair
production~\cite{CdfNote9432,Abulencia:2008su}.
The dilepton analysis uses the azimuthal angle between the two charged leptons
and compares the distribution to templates of the pure production modes.
In the investigated $2.2\ifb$ of dilepton events CDF obtains a gluon-fusion 
fraction of $0.53\pm0.37$~\cite{CdfNote9432}. The total uncertainty is dominated by
statistical uncertainties and is not yet able to restrict the theoretical
uncertainties on the gluon fusion production.

The lepton+jets analysis uses the average number of  of soft tracks per event, $\left<N_\mathrm{trk}\right>$, with
$0.9\GeV<p_T<2.9\GeV$ in the central detector region $|\eta|\le 1.1$ as a
measure of the amount of initial state radiation. This observable was found to
have a linear relation with the average number of gluons in the hard process. 
For the measurement this relation is calibrated from samples of $W+$0jets
which have low  and dijet events which have a high gluon content. 
In the dataset of $0.96\ifb$ CDF determines  a gluon-fusion fraction of 
top-pair production in semileptonic events of
$%\frac{\sigma(gg\rightarrow t\bar t)}{\sigma(p\bar p\rightarrow t\bar t)}=
0.07 \pm 0.14_\mathrm{stat} \pm 0.07_\mathrm{syst}$~\cite{Abulencia:2008su}. This number corresponds
to an upper limit of $0.33$ at $95\%$~C.L, well in agreement with the SM 
expectations. Also this measurement is statistically limited.

\subsection{Width and Lifetime}
The top-quark width $\Gamma_t$ and its lifetime are related by Heisenberg's uncertainty
principle. In the SM $\Gamma_t$ is expected to be $1.34\GeV$,
corresponding to a very short lifetime of about $5\cdot
10^{-25}s$. Experimentally, these predictions have been
challenged for gross deviations in two different analyses. Lower limits on the
lifetime are determined from the the distribution of reconstructed top mass
values and upper limits on the lifetime from the distribution of lepton track
impact parameters. 

A limit on $\Gamma_t$ was obtained by CDF from the distribution of
reconstructed top mass values in lepton plus jets event using $1\ifb$  of
data~\cite{Aaltonen:2008ir}. The top mass values are reconstructed in each event
using a kinematic fit with constraints on the  $W$ boson mass and the equality
of the top and anti-top masses. The jet parton assignment which yields
the best fit is used in the analysis. CDF claims that the use of
the constraint of the equality of the top quark masses 
does not destroy the sensitivity to $\Gamma_t$.

To find the measured value of $\Gamma_t$ the distribution of top quark
masses reconstructed with the best association in each event is compared to
parameterized templates with varying nominal width. Templates for top pair signal events 
use a nominal top quark mass of $m_t=175\GeV$.  
Including all systematics this CDF analysis of $1\ifb$ yields an upper limit
of the top quark width $\Gamma_t<12.7\GeV$ at the $95\%$ C.L. which corresponds to 
$\tau_t>5.2\cdot10^{-26}\,\mbox{s}$~\cite{Aaltonen:2008ir}. 
These limits would improve if the top quark mass used in the templates was closer to the
current world average of $173.1\GeV$~\cite{:2009ec}.

A limit on the top quark lifetime was obtain by CDF the lepton track impact
parameter distribution  in lepton plus jets events~\cite{CdfNote8104}. The
lepton track impact parameter is  computed with respect to the position of the beam
line in the transverse plane at the reconstructed $z$ position of
the primary vertex.

Templates of the expected distribution are derived from simulations done for
an ideal detector using the resolution of the CDF detector calibrated in
Drell-Yan data. The template with zero life time $c\tau_t=0\um$ describes the
data best. With the  Feldman-Cousins approach a limit of $c\tau_t<52.5\um$ at
the $95\%$~C.L. is set~\cite{CdfNote8104}. 

The measurements of the top quark width and lifetime are fully consistent with
the SM. Only extraordinary large deviations can be excluded with these results.

\subsection{Outlook to LHC}
The measurements of the top quark interaction properties performed at the
\tevatron are in general limited by statistics. The increased cross-section at
the LHC and the relatively reduced SM backgrounds help to improve the
experimental constraints of top quark interaction properties.

ATLAS has investigated the prospects for collisions of $\sqrt{s}=14\TeV$ and a
luminosity of $1\ifb$ for various measurements~\cite{Aad:2009wy}. 
For this reference scenario the expected results surpass the current \tevatron
results significantly. $W$ Helicity reaches uncertainties of 0.03 and 0.05 for $f_+$ and $f_0$,
respectively;  top quark charge measurements  can distinguish  the SM from the
exotic scenario by multiple standard deviations and the branching fractions
of FCNC can be limited to below $10^{-2}$ and $10^{-3}$ for the $tZq$ and the
$t\gamma q$ processes, respectively.
CMS has investigated a scenario of $10\ifb$ and
$\sqrt{s}=14\TeV$~\cite{Ball:2007zza}. At this reference point the top quark
spin correlations are accessible and can be extracted with a precision of $20-30\%$.

%%% Local Variables: 
%%% mode: latex
%%% TeX-master: "TopRevPPNP"
%%% End: 

%% file: new_physics.tex
\section{The Top Quark as a Probe for New Physics}
\label{sec:newPhysics}\label{sect:BSM}
%
% Author: Daniel Wicke
% $Id: new_physics.tex,v 1.15 2009/07/31 13:25:43 wgw Exp $
%

The phenomenology of the top quark may also be altered by particles that are
not expected within the SM. Various such particles are expected by
the many models beyond the SM. They may occur in the top
quark production or its decay, depending on the specific model or its
parameters. Some models also contain new particles with signatures that are very
similar to the SM top quark. The \tevatron experiments have checked for all
these different extensions of the SM.
This section will actually start with a process that is expected in the
SM though at very low rate: associated Higgs production. 
It will continue with new particles occurring in the top quark production, then
move to beyond the SM particles in the top quark decay and finally discuss
production of particles that look like the top quark but aren't.

\subsection{Top Quark Production through New Particles}
\paragraph{Associated Higgs Production}
Top pairs can also take place in association with additional particles. For
parameters where Higgs bosons dominantly decay to bottom quark pairs, this associated
production is a possibility to measure the Yukawa coupling. 
While the corresponding cross-section in the SM is too low to allow a Higgs
discovery in this channel only, it  still contributes to the combination of
the SM Higgs searches. In models some including new physics an enhancement of
$t\bar t H$ production is expected~\cite{Stange:1993td,Feng:2003uv,AguilarSaavedra:2006gw}.

D\O\ performed an analysis searching for associated Higgs production in lepton
plus jets events~\cite{d0note5739}.
The analysis uses the scalar sum of transverse momenta, $H_T$, the number of
jets and the number of jets identified as $b$-jets to discriminate the SM
backgrounds and top pair production containing no Higgs from the signal.

The observed data agree with the background expectations  within statistical
and systematic uncertainties. 
%To compute limits signal and background
%contributions are fitted to the data for a background only assumption 
%and for a signal plus background assumption. 
Limits on $\sigma(t\bar tH)\cdot{\cal B}(H\rightarrow b\bar b)$ are then derived
using the CL$_s$ method for Higgs masses between $105$ and $155\GeV$.
While the limit of about 60 times the SM cross-section for $M_H=115\GeV$
exclude unexpectedly large Higgs production in association with the top quark, 
its contribution to the SM Higgs search remains small.

\paragraph{Resonant Top Quark Pair Production}
Due to the fast decay of the top quark, no  resonant production of top
pairs is expected within the SM. However, unknown heavy resonances decaying to top
pairs may add a resonant part to the SM production mechanism. 
Resonant production is possible for massive $Z$-like bosons in
extended gauge theories \cite{Leike:1998wr}, Kaluza-Klein states of the gluon
or $Z$ boson~\cite{Lillie:2007yh,Rizzo:1999en}, axigluons
\cite{Sehgal:1987wi}, topcolor \cite{Hill:1993hs,Harris:1999ya}, and other
theories beyond the SM.  
Independent of the exact model, such resonant production could
be visible in the reconstructed $t\bar t$ invariant mass. 

CDF has employed several different techniques to search for resonances in the
$t\bar t$ invariant mass distribution~\cite{cdf:2007dia,cdf:2007dz,CdfNote9164}. 
All analyses use lepton plus jets events with
four or more jets. Their main difference is the method to reconstruct the
$t\bar t$ invariant mass distribution.

\begin{figure}
  \centering
  \subfigure[]{
    \includegraphics[height=4.5cm]{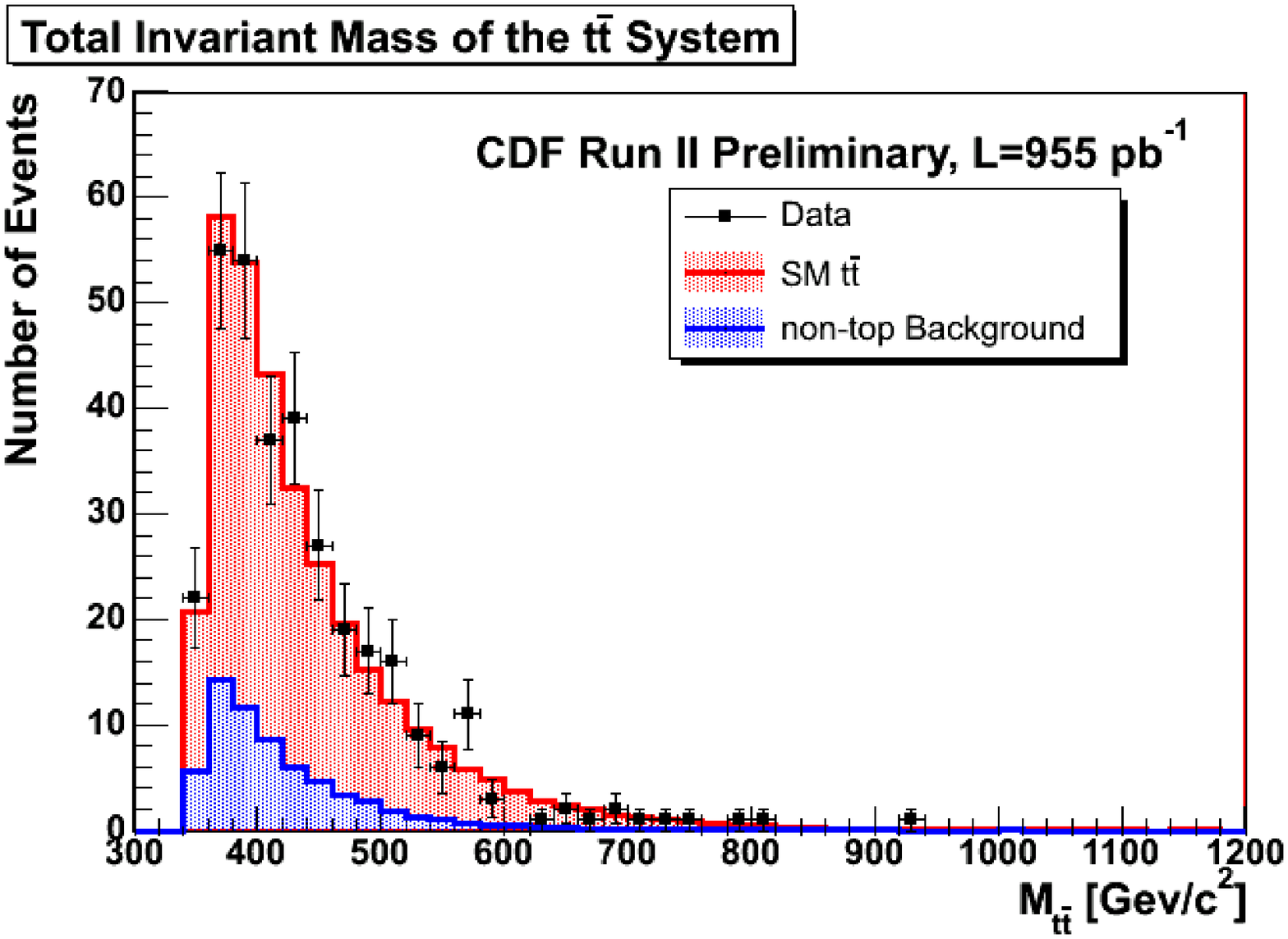}
    \label{fig:CDF_Mtt}
  }
  \hspace{0.15\textwidth}
  \subfigure[]{
    \includegraphics[height=4.5cm]{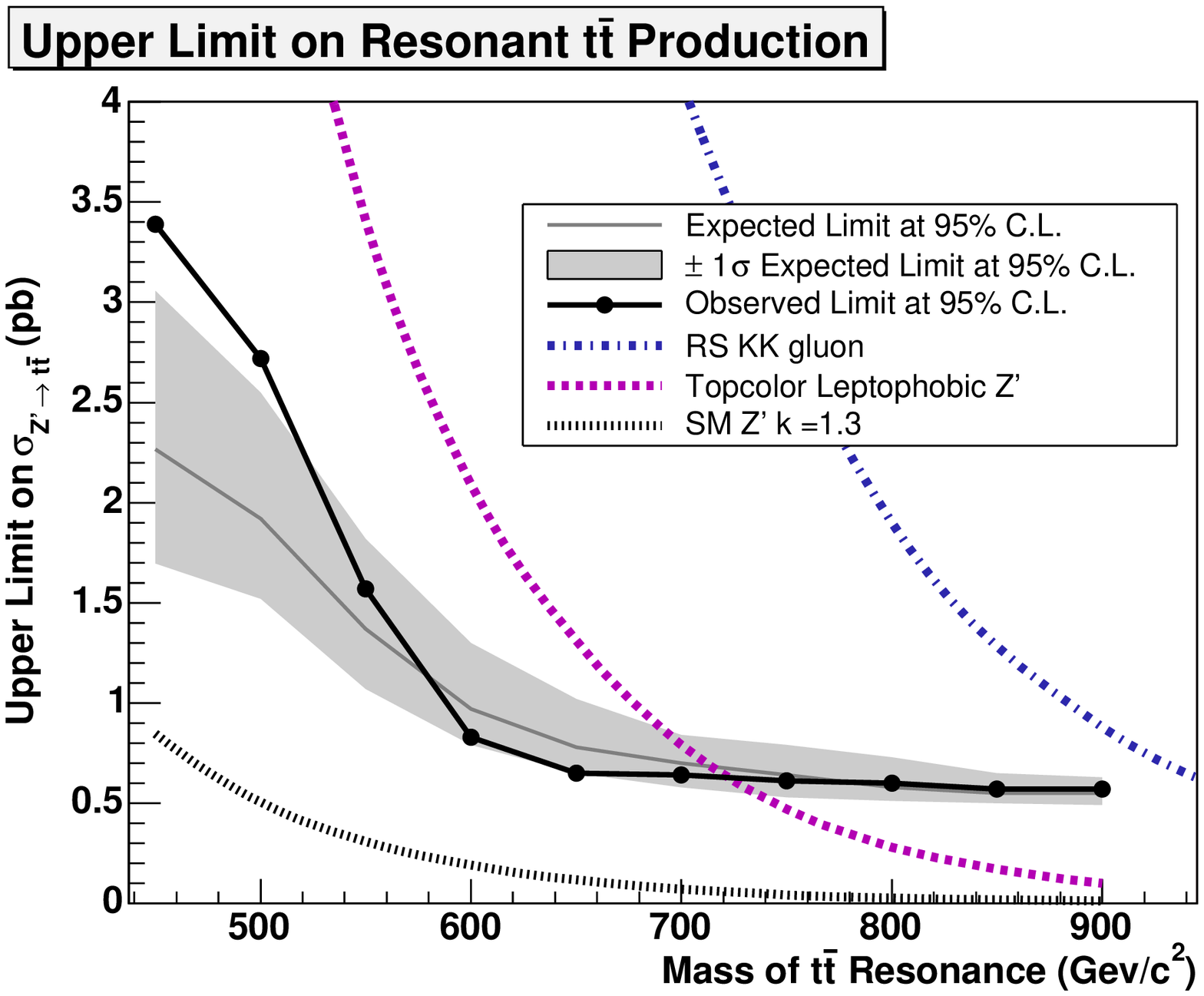}
    \label{fig:CDF_ttbar_Resonance_Limits}
  } \\
  \subfigure[]{ 
    \includegraphics[height=5.5cm]{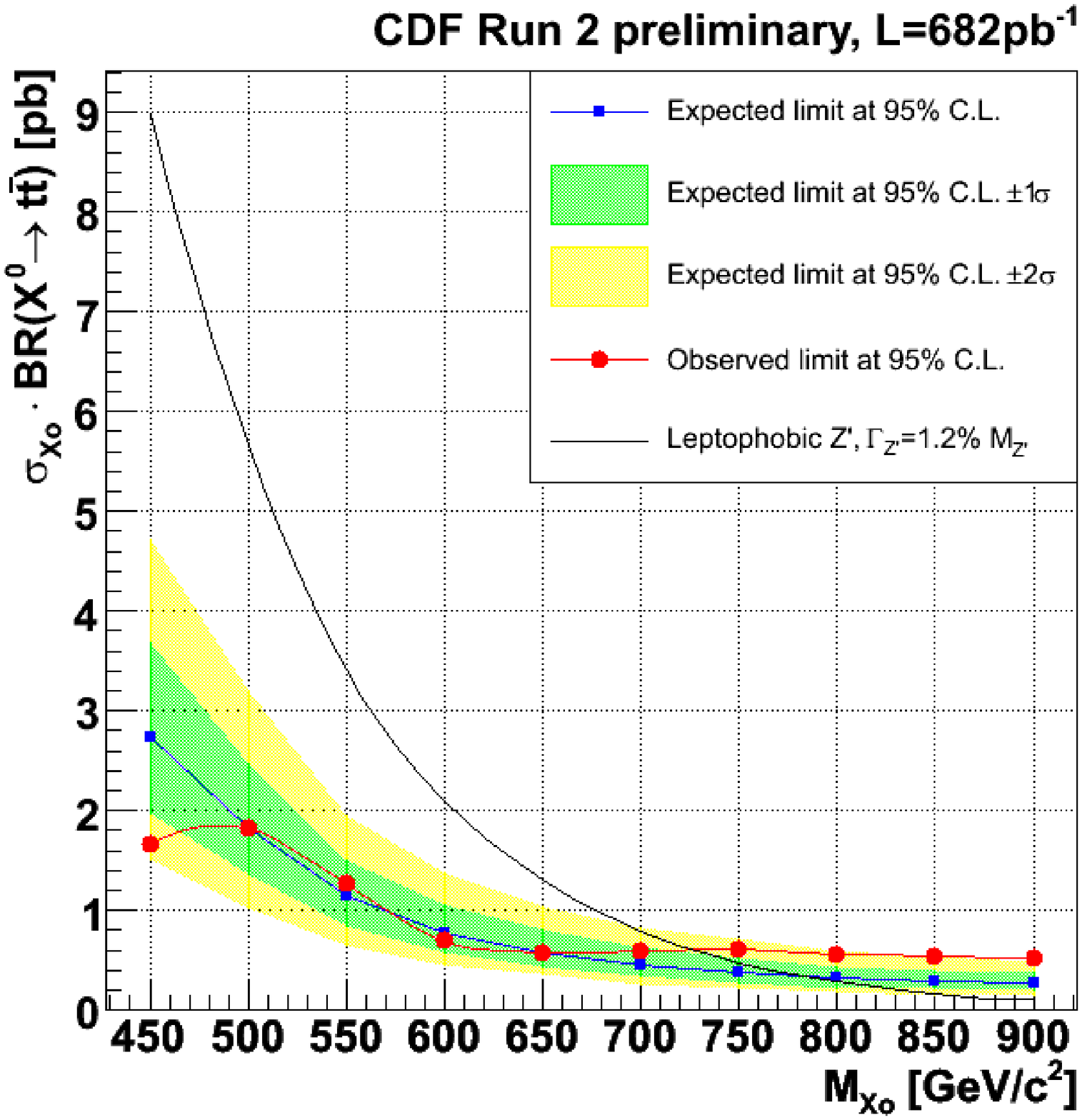}
    \label{fig:CDF_ttbar_Resonanz_ME}
  }
  \hspace{0.15\textwidth}
  \subfigure[]{
    \includegraphics[height=5.5cm]{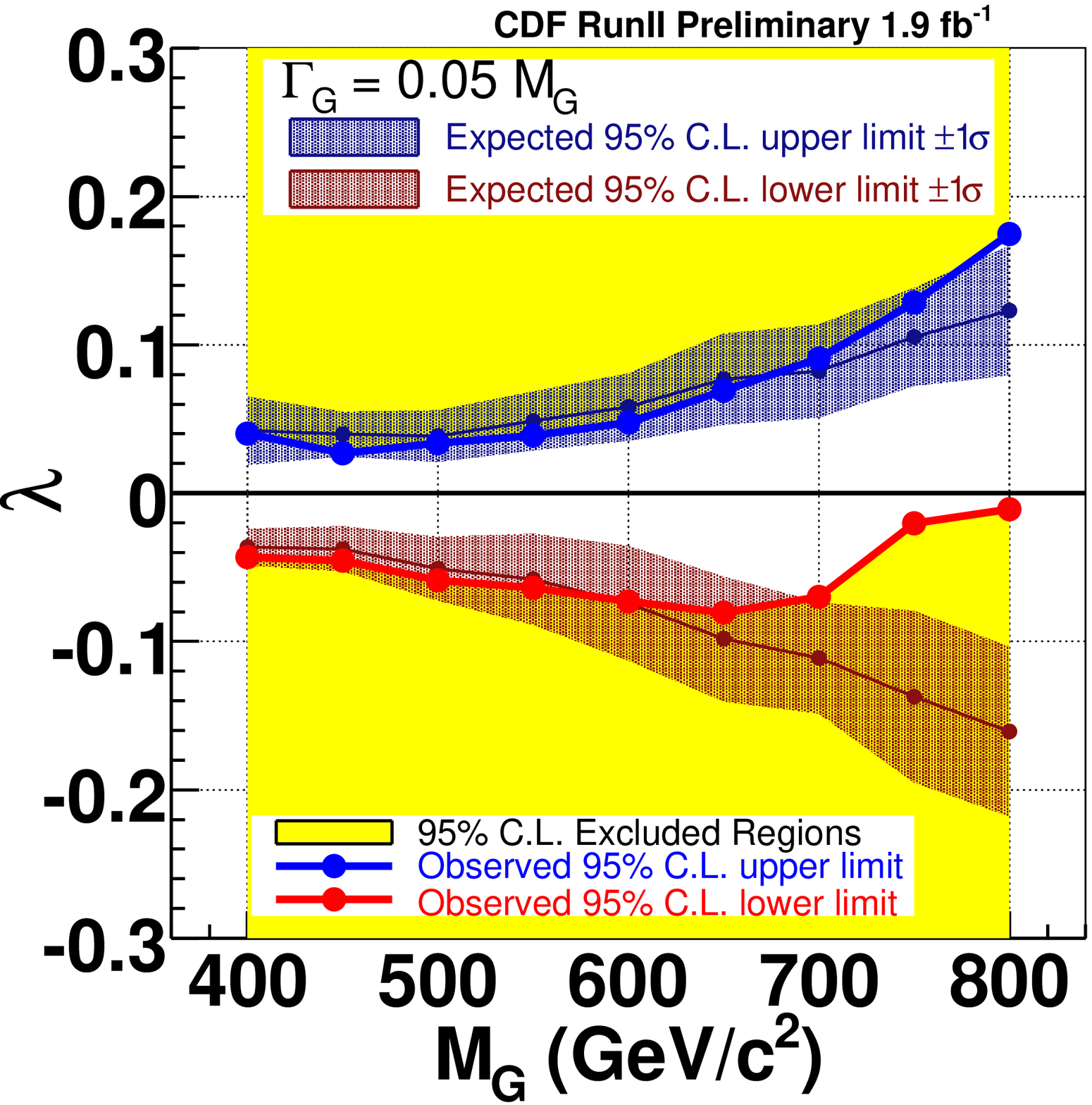}
    \label{fig:CDF_Limits_Massive_Gluon}
  }
  \caption{Distribution of the invariant top pair mass reconstructed with a
    constraint fit in $1\ifb$ of CDF data~\subref{fig:CDF_Mtt} and resulting 
    limits on the
    cross-section times branching fraction for resonant top pair production
    \subref{fig:CDF_ttbar_Resonance_Limits}~\cite{cdf:2007dia}. 
    Corresponding limits obtained 
    the matrix-element technique in $680\ipb$ of CDF 
    data~\subref{fig:CDF_ttbar_Resonanz_ME}~\cite{cdf:2007dz}.  
    Theoretical curves are shown for various model
    and used to set mass limit on the corresponding resonance.  
    Limits on the coupling strength
   of a massive gluon, $G$, contribution deduced for various masses and two
   widths obtained with a Matrix Element technique in 
   $1.9\ifb$~\subref{fig:CDF_Limits_Massive_Gluon}~\cite{CdfNote9164}.
  }
  \label{fig:cdf-ttresonance-1fb-1} 
  \label{fig:cdf-680ipb-mtt.result}  
  \label{fig:mtt-massivegluon}
\end{figure}

The method that uses the least assumptions reconstructs the invariant mass
using a constraint kinematic fit~\cite{cdf:2007dia} and is performed requiring at least
one identified $b$ jet. The fit requires the reconstructed $W$ boson mass to
be consistent with its nominal value and 
the reconstructed top quark mass to be consistent $175\GeV$ within
the corresponding natural widths.
Of the possible jet parton assignments the one with the best fit is used to
compute the top pair invariant mass for each event. The measured distribution
shown in \fig{fig:cdf-ttresonance-1fb-1}\subref{fig:CDF_Mtt}
is compared to templates for SM production, obtained from a combination of
simulation and data, and to templates of resonant production from a narrow
width $Z'$ resonance with
masses between $450$ and $900\GeV$.
The possible cross-section time branching fraction of resonant production,
$\sigma_X{\cal B}(X\rightarrow t\bar t)$, is
computed from a likelihood that includes nuisance parameters to describe the
systematics.
Expected and observed limits are shown 
in~\fig{fig:cdf-ttresonance-1fb-1}\subref{fig:CDF_ttbar_Resonance_Limits}.
At high resonance masses the observed limits excludes a resonant top pair
production with  $\sigma_X{\cal B}>0.55\pb$ at 95\%~C.L. A leptophobic
topcolor production mechanism is excluded for resonance masses up to $720\GeV$.

The resolution of the reconstructed invariant top pair mass can be improved by
assuming additional information. In an analysis of $680\ipb$ CDF employed the
matrix-element technique to reconstruct the invariant mass distribution that is
used to search for resonant production~\cite{cdf:2007dz} following the mass
analysis~\cite{Abulencia:2005pe}. 
For each event a probability distribution,
$P(\{\vec{p}\}|\{\vec{j}\})$, for the momenta of the
top pair decay products (4 quarks, charged lepton and neutrino, $\{\vec{p}\}$) 
is computed given the observed quantities in that event, $\{\vec{j}\}$. This
probability distribution uses parton density functions, the theoretical
matrix-element for SM top pair production and decay and
jet transfer functions that fold in the detector resolution. This probability
density  is then converted to a probability density for the top pair invariant mass:
\beq
P_f(M_{t\bar t}|\{\vec{j}\})=\int\!\mathrm{d}\{\vec{p}\}\,\,P(\{\vec{p}\}|\{\vec{j}\})
\,\delta\!\left(M_{t\bar t}-m(\{\vec{p}\})\right)\quad\mbox{.}
\label{eq:mttFromDLM}
\eeq
The sum of the  probabilities for all possible jet-parton assignments is used 
to compute the reconstructed invariant mass as mean value.
Here the $b$-tagging information is used to reduce the number of allowed jet parton
assignment, but the events are not required to contain $b$-tagged jets in the
event selection.
Again the measured distribution is compared to templates for the SM
expectation and for resonant production through a narrow width $Z'$ boson.
In this analysis Bayesian statistics is employed to compute a possible
contribution from such resonant production. 
As the data show no evidence for resonant
top pair production upper limits are derived for $\sigma_{Z'}\cdot {\cal
  B}(Z'\rightarrow t\bar t)$, 
c.f.~\fig{fig:cdf-680ipb-mtt.result}\subref{fig:CDF_ttbar_Resonanz_ME}.
A comparison to the leptophobic topcolor assisted technicolor model yield a
exclusion of this model for $M_{Z'}<725\GeV$ at 95\%C.L.

A more recent variation of this CDF analysis is used to search for a new color-octet particle, called
a massive gluon, in $1.9\ifb$~\cite{CdfNote9164}. 
In contrast to the previously described analysis
the invariant top pair mass is reconstructed without using the production part
of the Matrix Element to construct the probability densities in
\eq{eq:mttFromDLM} to avoid a bias towards the SM production
mechanism. 
An unbinned likelihood fit based on the production matrix elements with
and without massive gluon contribution  is used to extract the possible
coupling strengths of such a massive gluon contributing to the top pair
production. The likelihood is computed for various the masses and widths 
of the massive gluon.
The observed data agree with the SM expectation within $\sim
1.7\sigma$.
Limits  on the possible coupling strength of a massive gluon, $G$, 
contributing to the top pair production are set at 95\%C.L. for various values of the
width, $\Gamma_G$ as function of the mass, $M_G$. 
\fig{fig:mtt-massivegluon}\subref{fig:CDF_Limits_Massive_Gluon}
shows the expected and observed limits for one choice of the massive gluon
width. 

D\O\ investigated the invariant mass distribution of top pairs in 
up to $3.6\ifb$ of lepton plus jets
events~\cite{Abazov:2008ny,D0Note5600conf,D0Note5882conf}. 
In these analyses the top pair invariant mass
is reconstructed directly from the
reconstructed physics objects. A constraint kinematic fit
is not applied. Instead the momentum of the
neutrino is reconstructed from the transverse missing energy
which is identified with the transverse momentum of the neutrino and by solving
$M_W^2=(p_\ell+p_\nu)^2$ for the $z$-component of the neutrino
momentum. $p_\ell$ and $p_\nu$ are the four-momenta of the lepton and
the neutrino, respectively.
The method uses even fewer assumption first CDF analysis above.
Compared to the constraint fit reconstruction
it gives  better performance at high
resonance masses and in addition allows the inclusion of events with only
three jets.
Templates of the SM expectation and for resonant production through narrow
width $Z'$ are compared to data. 
The cross-sections of resonant production most consistent with the data 
are evaluated using the Bayesian technique for various resonance masses. 
Data agree with pure SM expectation and thus limits
are set on the $\sigma_X\cdot {\cal B}(X\rightarrow t\bar t)$ as function of
the assumed resonance mass. 
The excluded values range from
 about $1\pb$ for low mass resonances to less then $0.2\pb$ for the highest
considered resonance mass of $1\TeV$. 
The benchmark topcolor assisted technicolor model can be excluded for
resonance masses of $M_{Z'}<820\GeV$ at $95\%$CL.

\begin{figure}
  \centering
\includegraphics[width=0.32\textwidth]{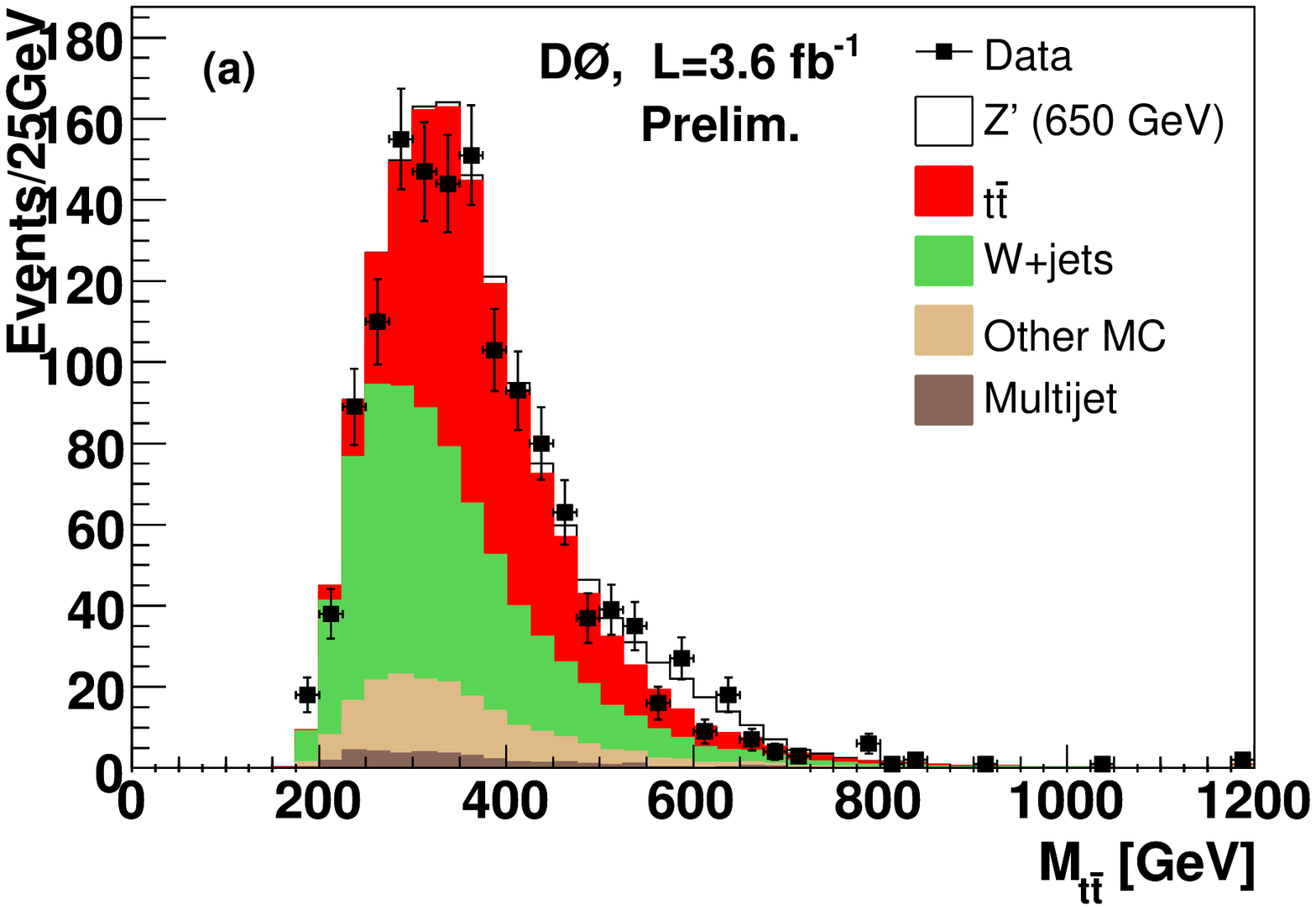}
\includegraphics[width=0.32\textwidth]{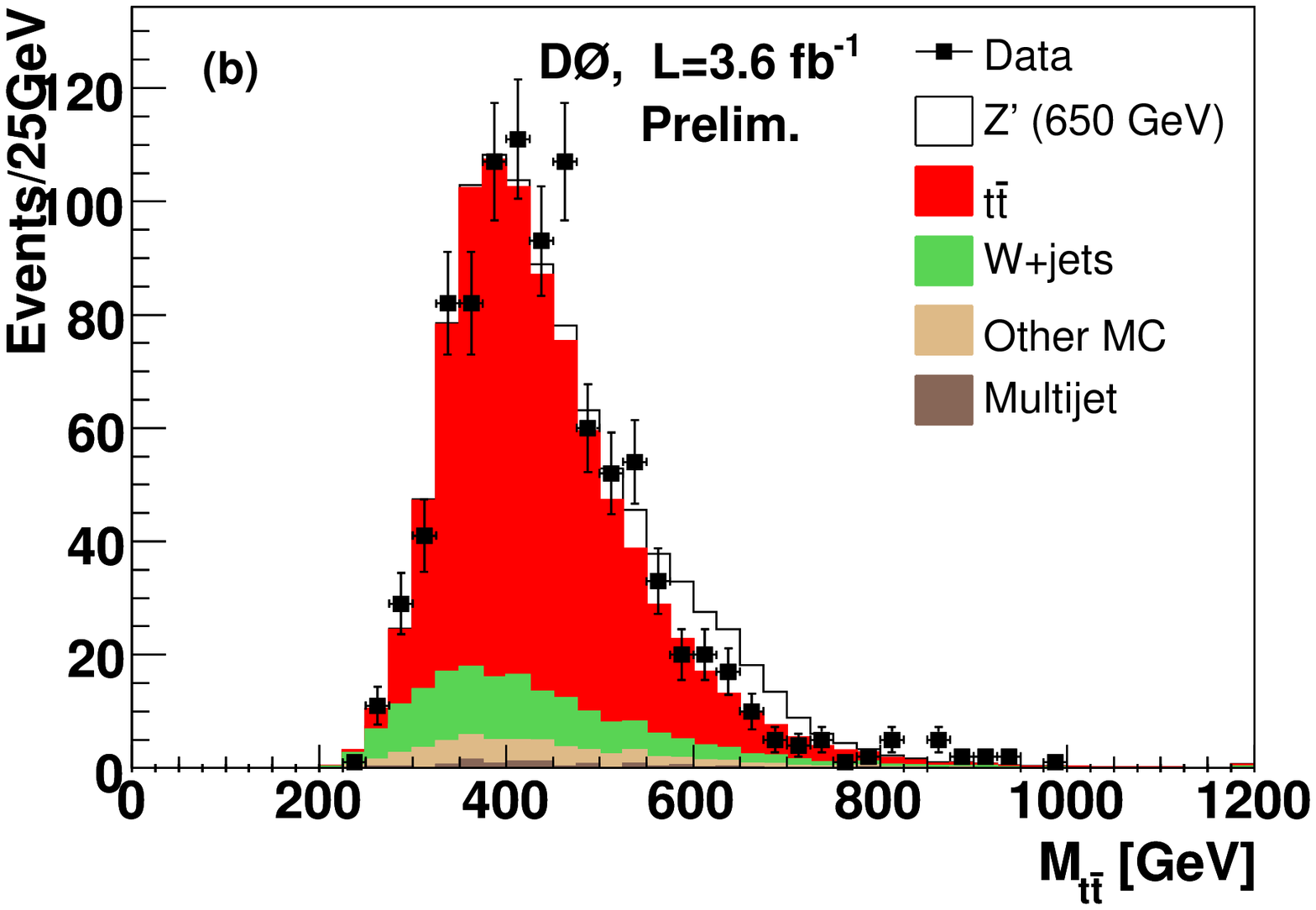}
\unitlength=0.0066\textwidth
\begin{picture}(50,35)
\put(0,0){\includegraphics[width=0.32\textwidth]{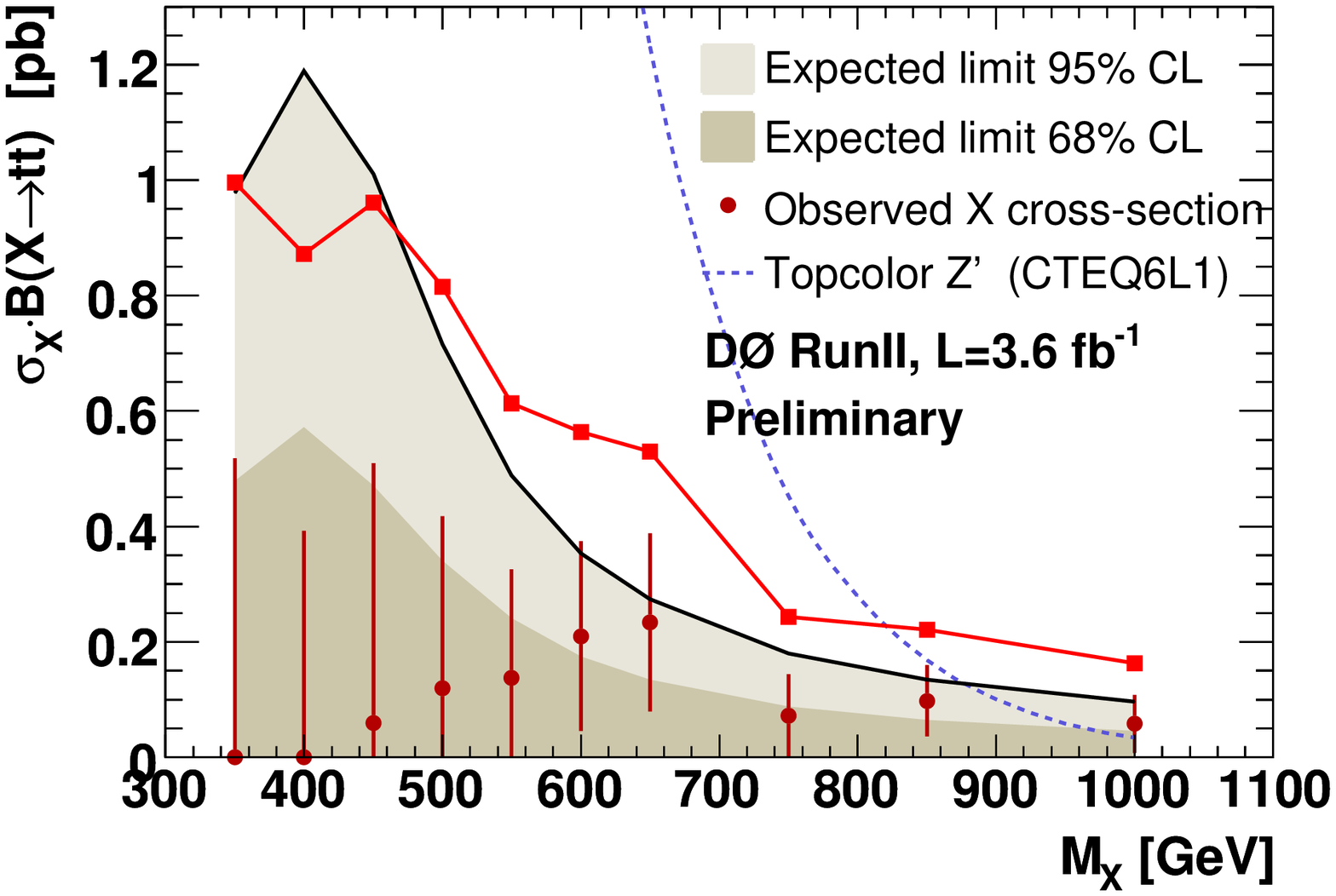}}
\put(25.4,23.2){\includegraphics[height=0.05135\textwidth,clip,trim=104mm 93mm 17mm 10mm]{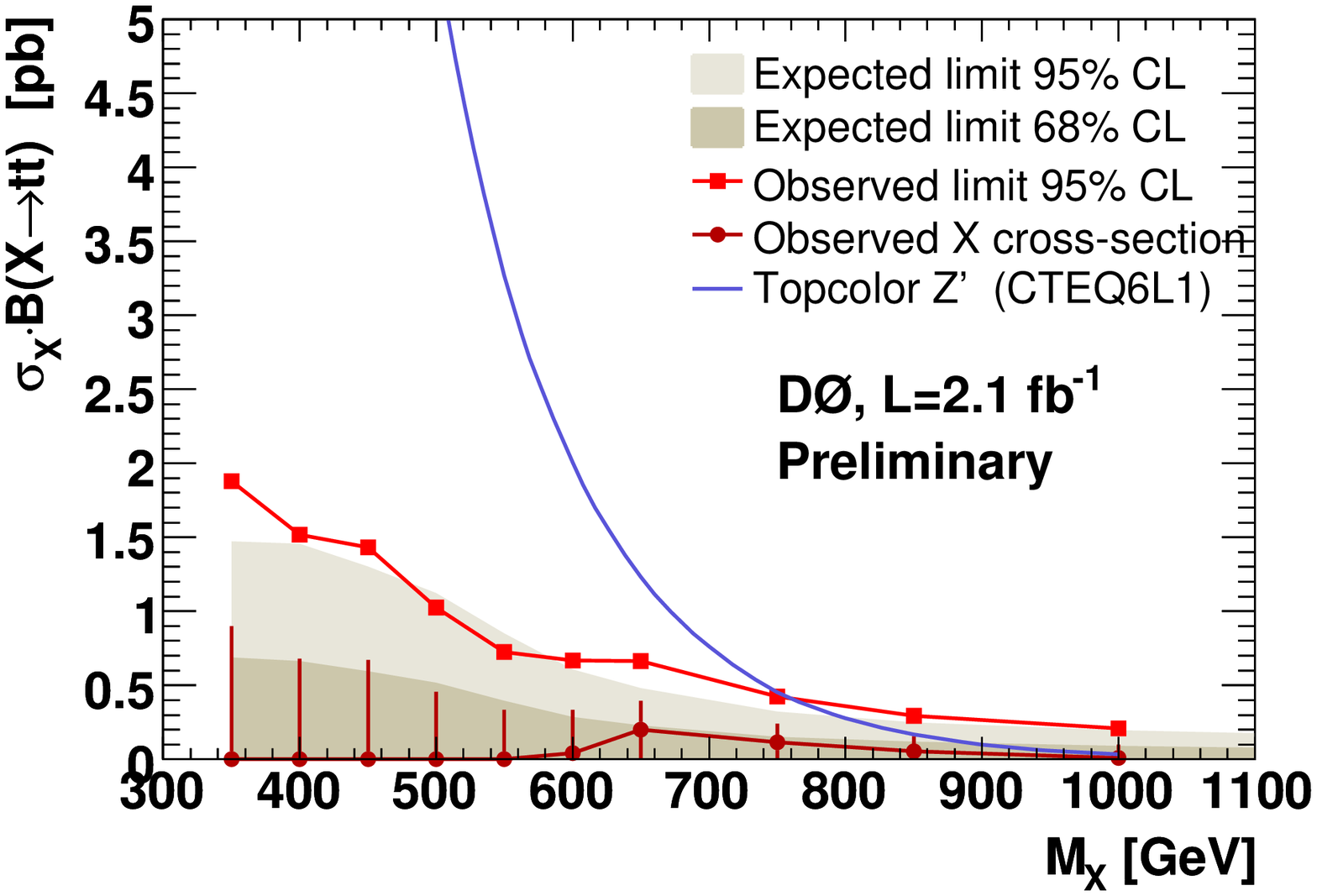}}
\end{picture}
%\null\hfill\includegraphics[height=3.5cm]{fig_wicke/T83F3_lin.eps}%
  \caption{Expected and observed $t\bar t$ invariant mass
  distribution for the combined (left) $\ell+3$~jets and (middle) $\ell+4$ or more jets
  channels, with at least one identified $b$ jet.
  Expected upper limits on
  $\sigma_X\cdot B(X\rightarrow t\bar t)$ (shaded) vs.\ 
the assumed resonance mass compared to the observed
cross-section and exclusion limits at
95\%CL for $3.6\ifb$ of D\O\ data~\cite{D0Note5882conf}.
The prediction of the topcolor assisted technicolor
model used to derive benchmark mass limits. 
  }
  \label{fig:mtt.result} 
\end{figure}

%\subsection{Decays with non-SM particles}

\paragraph{Single Top Production through Charged Higgs}
Charged Higgs bosons appear in many extentions of the SM due to the need 
for an additional Higgs doublet. A charged Higgs can replace the $W$ boson
in single top quark production and in top quark decays. 
Both effects have been searched for at the \tevatron, the latter will be
described in \ref{sec:ChargedHiggs:Decay}.

\label{sec:ChargedHiggs:SingleTop}
The signature for single top quark is identical to that of the SM
$s$-channel, but may have a resonant structure in the
invariant mass distribution of its decay products, the top and the bottom
quark. Following their single top analysis, D\O\ selects events with an isolated
lepton, missing transverse energy and exactly two jets, one of which is
required to be identified as $b$-jet~\cite{Abazov:2008rn}.
SM and charged Higgs production of single tops is separated by reconstructing
the invariant mass of the two jets and the $W$ boson. This distribution shows
good agreement between data and the SM expectation%.
, see \fig{fig:d0:h+singletop}(left)
Bayesian statistics is used to determine the allowed cross-section for
single top production through a charged  Higgs using templates that represent a
narrow width $H^\pm$ boson.
 
Only for the Type I two Higgs doublet model, in which one Higgs gives mass to
all fermions, some region in $\tan\beta$ vs. $m_{H^\pm}$ can be excluded,
c.f.~\fig{fig:d0:h+singletop}(right). A significant fraction of phase space is
not accessible by the analysis in its current form due to the restriction to
narrow $H^\pm$ decay widths.

\begin{figure}
  \centering
  \includegraphics[height=5cm]{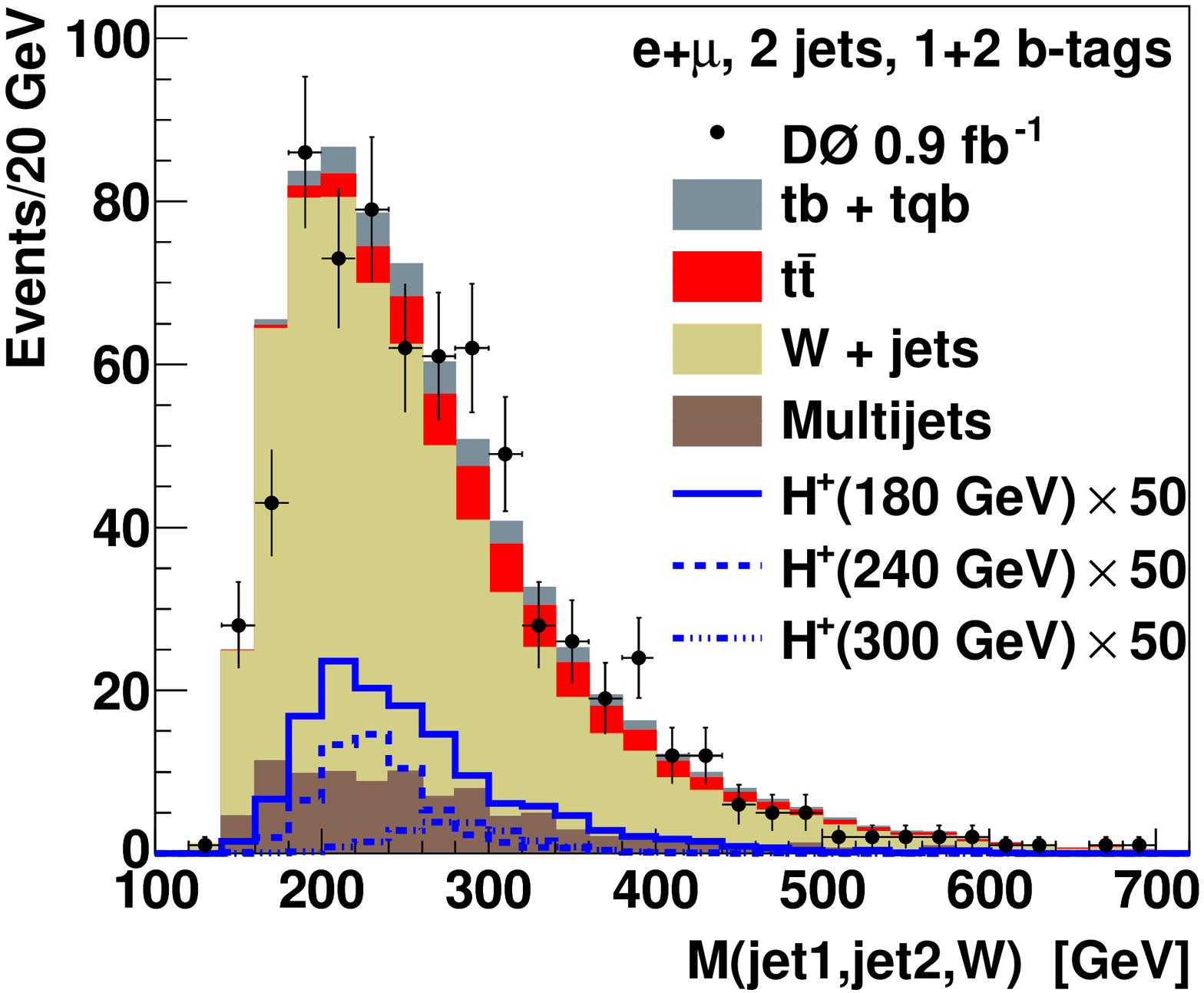}
  \includegraphics[height=5cm,clip,trim=0mm 0mm 0mm 12mm]{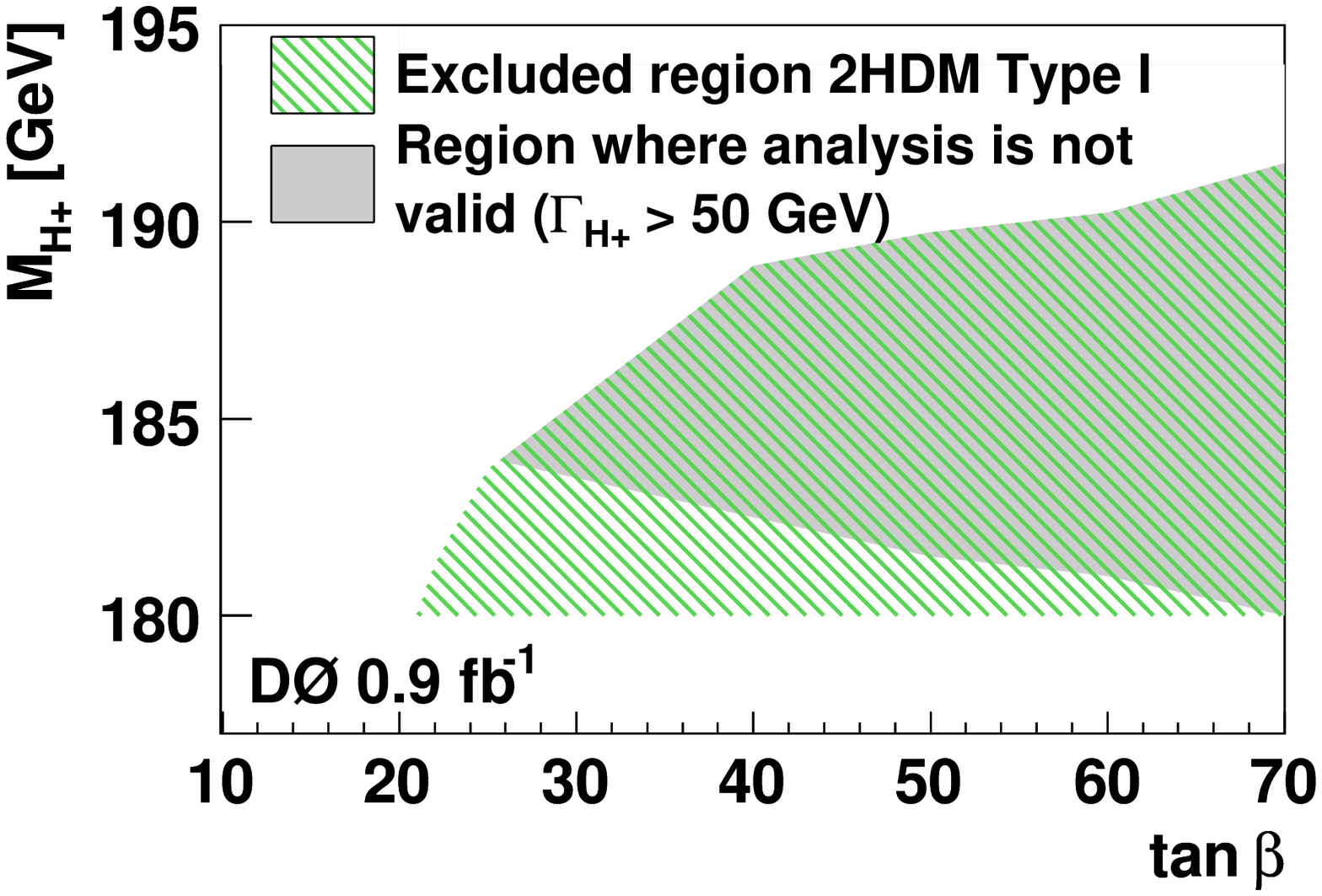}
  \caption{Left: Observed and expected invariant mass of $W$ and the two
    leading jets in $0.9\ifb$ of D\O\ Data. Right: Corresponding exclusion areas
    for Type I two Higgs Doublet Model (2HDM).~\cite{Abazov:2008rn}}
  \label{fig:d0:h+singletop}
\end{figure}

\paragraph{\boldmath Single Top Production through Heavy $W'$ Boson}
Similar to a charged Higgs boson a charged vector boson, commonly called $W'$,
may replace the SM $W$ boson in the single top quark production.
The couplings of such a $W'$ boson 
may be to left-handed fermions like for the SM $W$ boson or
include right-handed fermions.  In general a mixture of these two options is possible.
If the  $W'$ boson has left handed couplings, it will have  a sizeable interference with the SM  $W^\pm$ boson~\cite{Boos:2006xe}.
For purely right-handed couplings, the leptonic decay may only occur when the
right-handed neutrinos are lighter that the $W'$ boson. In this case the decay
to a top and bottom quark is an interesting channel to perform direct searches for such $W'$ bosons.

Both CDF and D0 search for various types of $W'$ bosons decaying to $tb$ pairs
in conjunction with their single top analyses. The main discriminating
observable is the reconstructed invariant mass of the decay products, which
was also utilized to search for a heavy charged Higgs boson, described in the
previous paragraph.

In an investigation of $1.9\ifb$ of data~\cite{Aaltonen:2009qu} CDF selects  
$W+$jets events requiring one lepton ($e$, $\mu$) isolated from
jets, missing transverse energy and two or three energetic jets. At least one
of the jets must be tagged as $b$-jet. 
The unobserved $z$ component of the neutrino is inferred assuming the SM $W$
boson mass. The invariant mass of the lepton, the neutrino
and the two leading jets, $M_{Wjj}$, is then used as a discriminating
observable.
$W'$ signal events are simulated %using Pythia 
for $W'$-masses between $300$
and $950\GeV$ with fermion couplings identical to the $W$ boson. 
When the right-handed $W'$  is heavier than the right-handed neutrinos,
the branching fraction to $\ell\nu$ is corrected according to the additional
decay modes.
Limits on the $W'_{\!R}$ boson production cross-section are set  as a function of $m_{W'_{\!R}}$
assuming the SM coupling strength. These are converted to mass
limits by comparison to the corresponding theoretical expectation and yield
$m_{W'_{\!R}}>800\GeV$ for $W'_{\!R}$ bosons which decay leptonically 
and $m_{W'_{\!R}}>825\GeV$ for  $M_{v_R}>M_{W'_{\!R}}$. 
For the more general case that the $W'_{\!R}$ coupling is a priori unknown
the $W'_{\!R}$ coupling strength, $g'$, 
relative to the SM coupling, $g_W$,  is constrained. Limits
are computed from the above analysis as function of the
assumed $m_{W'}$.
The observed and expected limits derived by CDF for  $m_{W'_{\!R}}$ and
$g'/g_W$  are shown in \fig{fig:CdfWprime}.
\begin{figure}
  \centering
  \includegraphics[height=4cm]{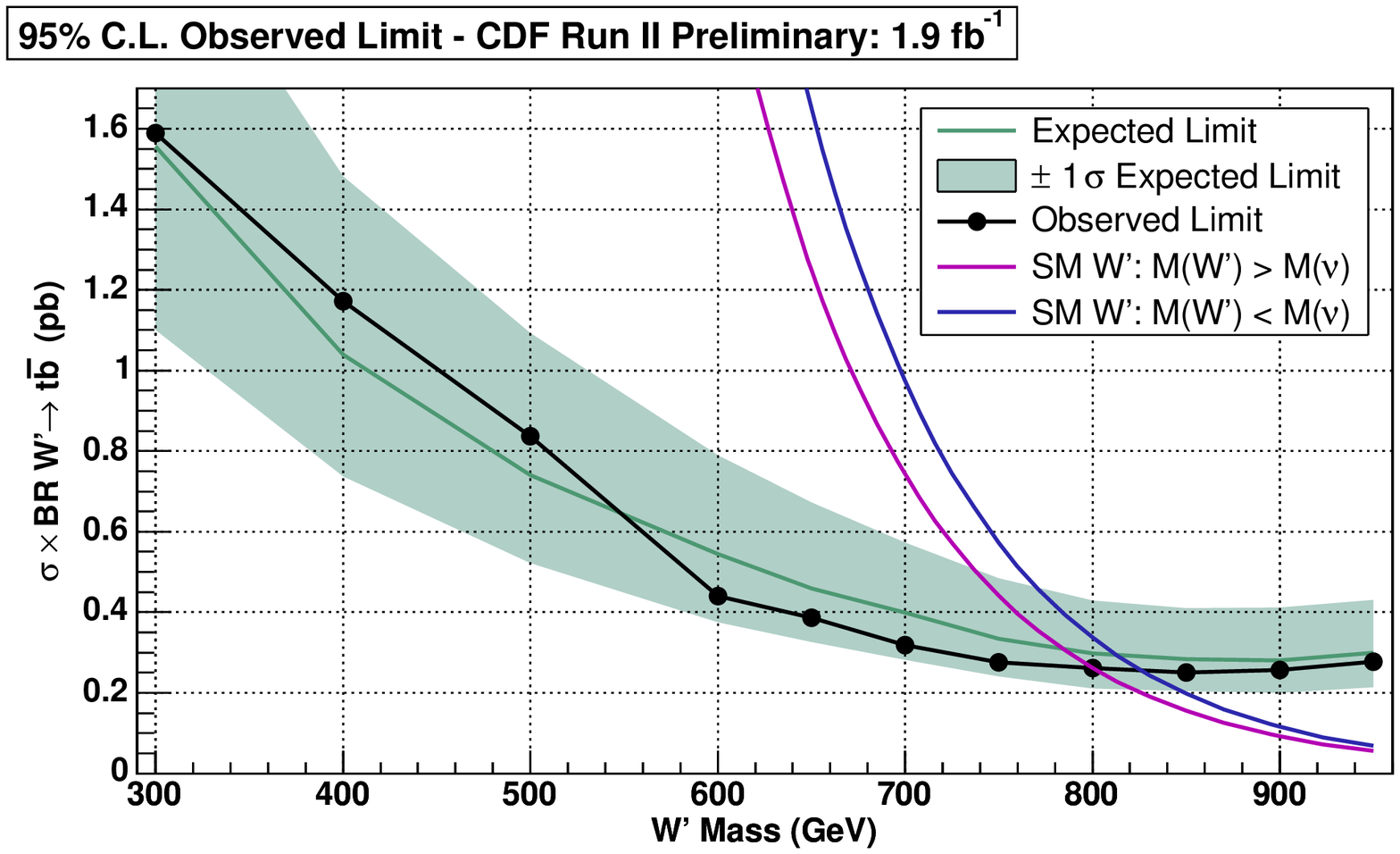}
  \hspace*{0.03\textwidth}
  \includegraphics[height=4cm]{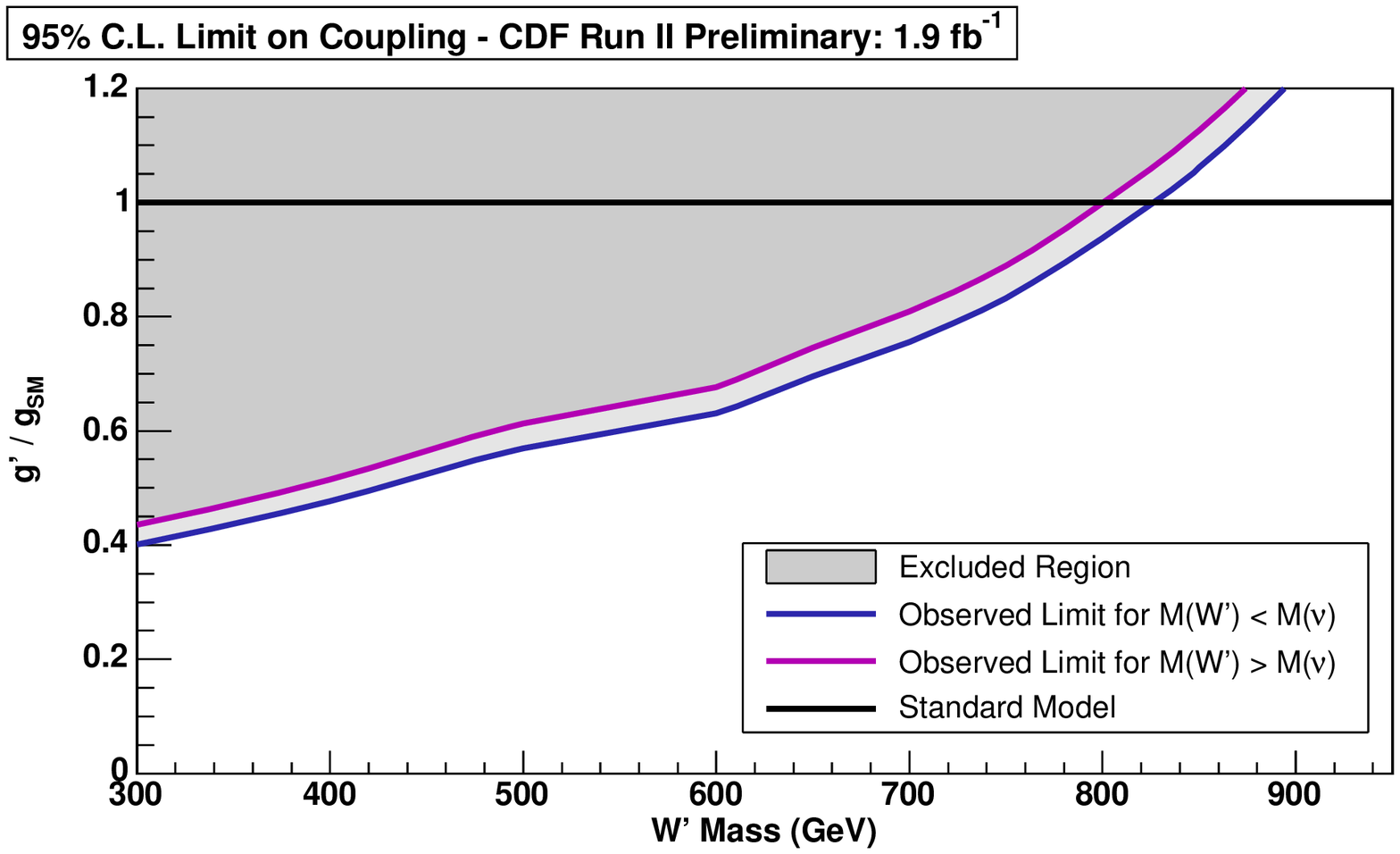}
  \caption{
    CDF results of a search for $W'_{\!R}$  in   $1.9\ifb$ of data~\cite{Aaltonen:2009qu}.
    Left: Limits on $W'_{\!R}$ cross-section times branching
    fraction to $tb$ as function $M_{W'_{\!R}}$ compared to theory.   Right: Limits on $W'_{\!R}$ coupling strength
    relative the SM coupling as function $M_{W_{\!R}'}$.}
  \label{fig:CdfWprime}
\end{figure}

D\O\ searched for a heavy $W'$ boson that decays to a top and a bottom quark using
$0.9\ifb$ of data~\cite{Abazov:2008vj}. Following their single top quark analysis events are required to have an isolated lepton,
missing transverse momentum and two or three jets one of which must be
identified as $b$-jet. The invariant mass of the bottom and
the top decay products is computed from the measured four-momenta of the leading two jets, the charged lepton and the
neutrino. Also D\O\ computes the $z$ component of the neutrino momentum assuming
the SM $W$ boson mass and compares the observed distribution to templates for
the SM expectation. The templates for a production through a $W'$ boson assume
the $W'$ boson to  decay to top and bottom quarks with  couplings as those of
the $W$ boson in the SM, though possibly to right handed fermions.
D\O\ uses the Bayesian approach with a flat non-negative prior on the
cross-section times branching fraction. As the data agree with the expectations
of the SM, limits are derived.
Expected and observed results are shown in \fig{fig:d0Wprime}. 
\begin{figure}
  \centering
  \includegraphics[width=0.32\textwidth]{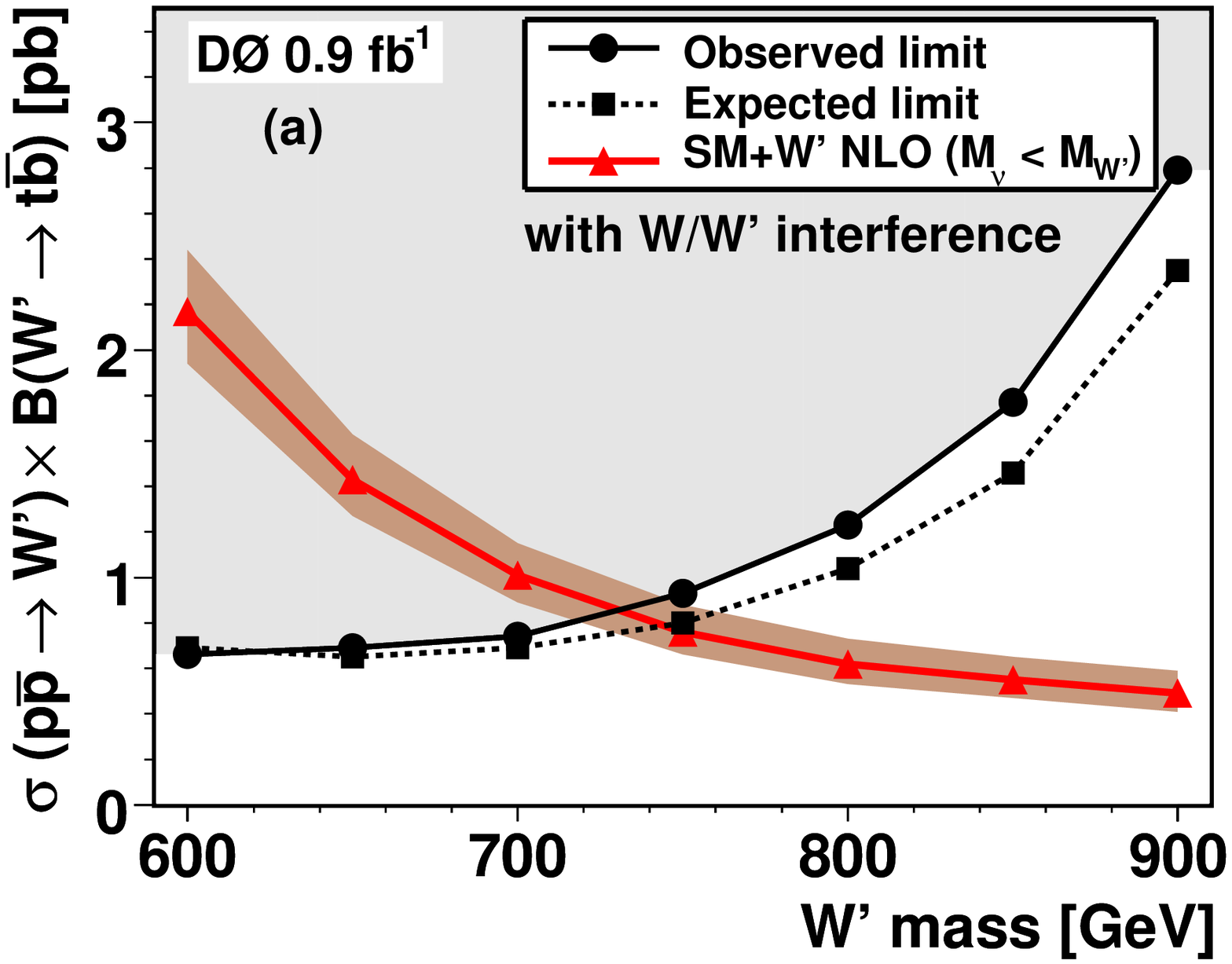}
  \includegraphics[width=0.32\textwidth]{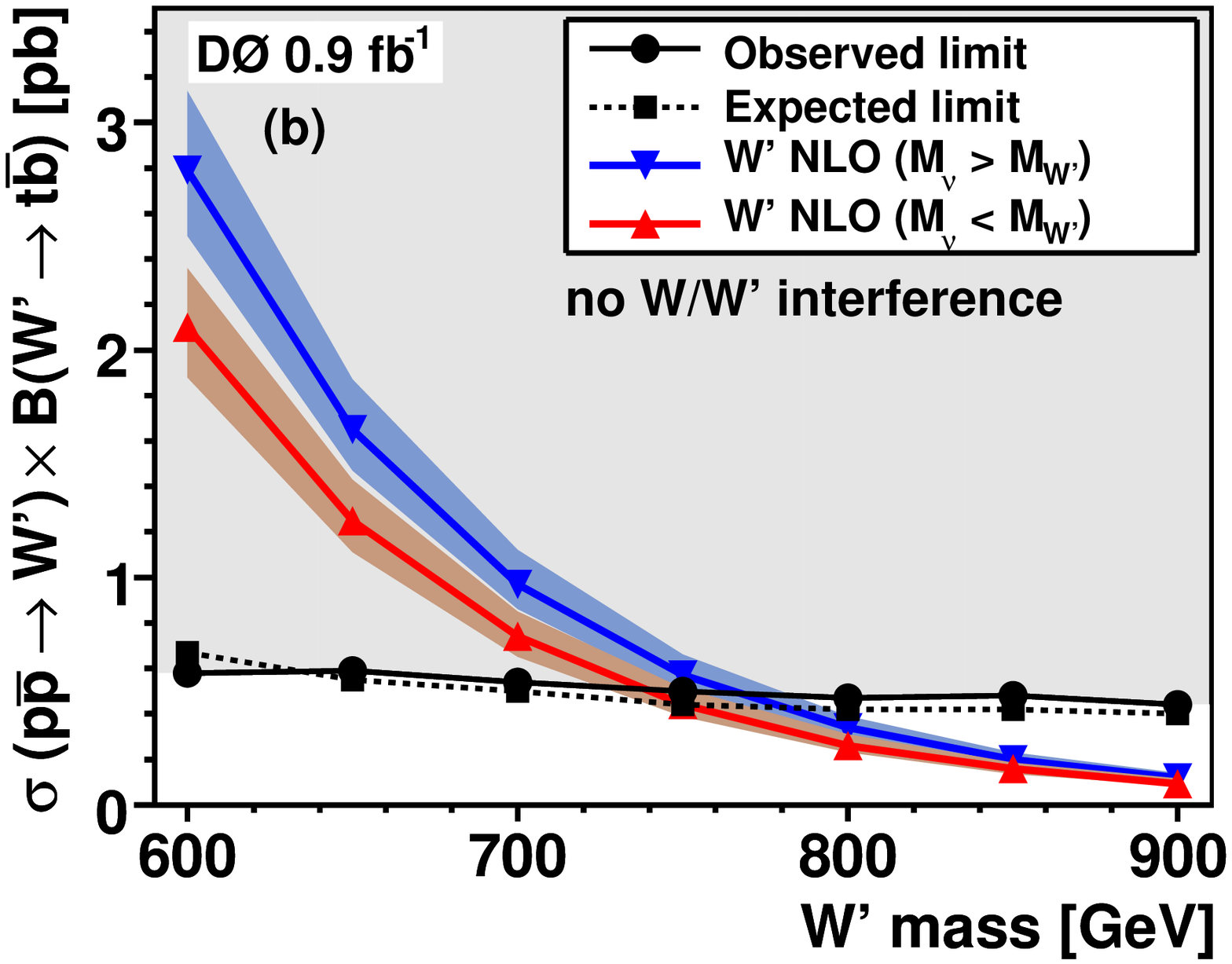}
  \includegraphics[width=0.32\textwidth]{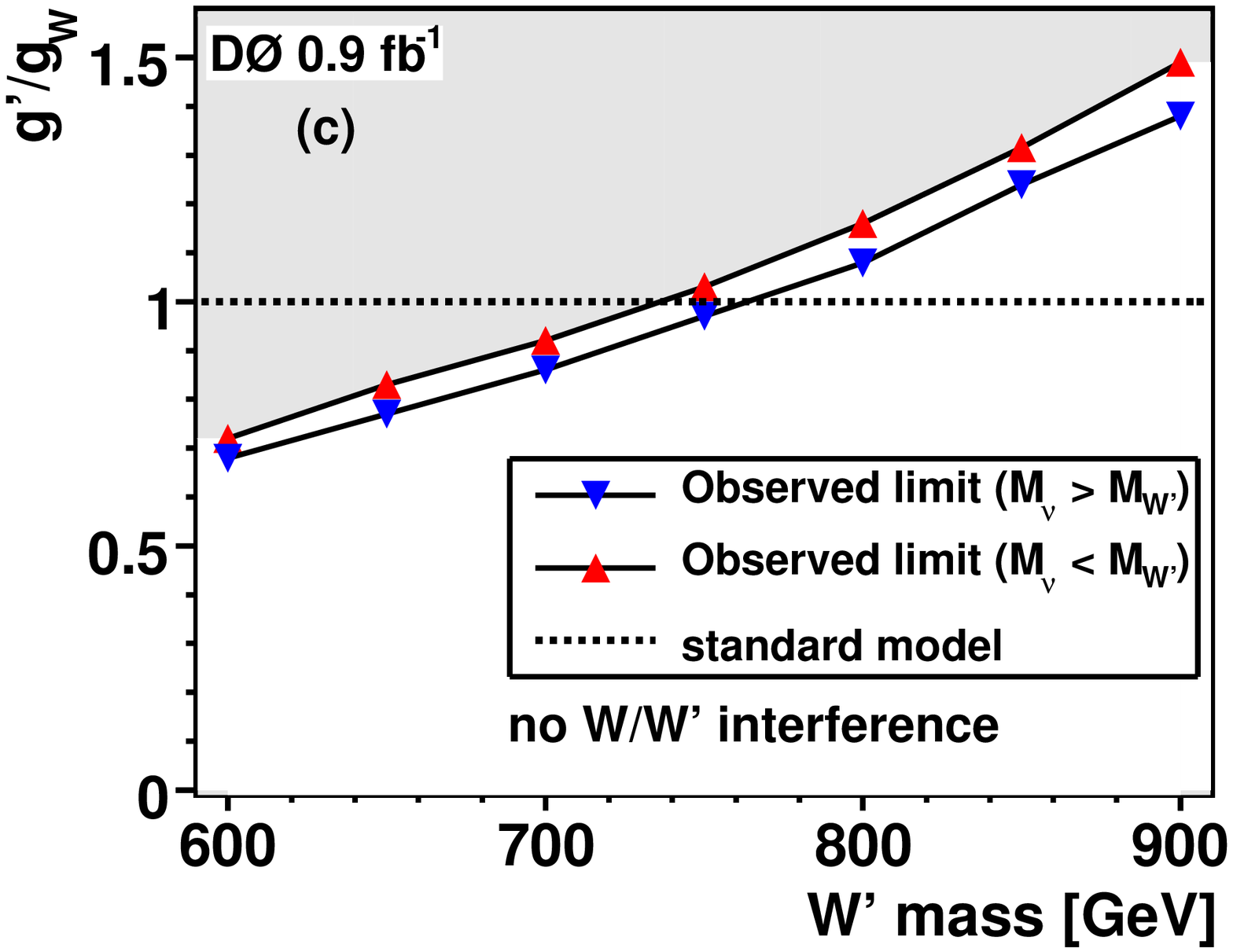}
  \caption{D\O\ results on a search for $W'$ boson decaying to top and
    bottom using $0.9\ifb$ of data. 
    Left: Expected and observed limits on a left-handed $W'$ production cross-section times branching
    fraction to top and bottom as function of $m_{W'}$ compared to the theory
    prediction. Middle: Same but for right-handed  $W'$ production. Right:
    Limits on the $W'$ boson coupling relative to the SM $W$-boson coupling.
  }
  \label{fig:d0Wprime}
\end{figure}
Comparing upper limits on the $W'$ cross-section time branching fraction to top and
bottom to the NLO theory predictions~\cite{Sullivan:2002jt} excludes left handed $W'$ bosons with
$m_{W'_{\!L}}<731\GeV$. If only hadronic decays are allowed the  right handed
$W'_{\!R}$ boson is excluded for $m_{W'_{\!R}}<768\GeV$, when leptonic decays
are also possible the limit is $739\GeV$.
Without assuming the coupling strength the Bayesian approach is used to
determine a limit on the size of this coupling relative to the SM, 
see \fig{fig:d0Wprime} (right). Theses limits assume no interference
between the SM $W$ and the $W'$.

\subsection{Top Quark Decays including new Particles}
New particles in the final state of top pair events may alter the
branching fractions to the various decay channels and modify 
the kinematic properties of the final state. 

\paragraph{Charged Higgs in Top Quark Decays}
\label{sec:ChargedHiggs:Decay}
An obvious candidate for such a particle is a charged Higgs boson. 
Because charged Higgs bosons have different branching fractions than
$W$ bosons this alters the branching fractions to the various top pair decay
channels. If its mass is different from the $m_W$ is also modifies  the
kinematic properties of the top pair final state. 
In the MSSM the decay at low $\tan\beta$ is dominated by hadronic decay to
$c\bar s$  at low Higgs masses and to  $t^*\bar b$ for Higgs masses from
about $130\GeV$. For $\tan\beta$ between $1$ and $2$ a leptonic decay to
$\tau\bar\nu$ dominates.%, c.f.~\fig{figs:ch_brs}.
%The figure also shows the expected branching fraction of 
%$t\rightarrow H^\pm b$ which especially large for very low and very high
%$\tan\beta$ and rather small in the intermediate range

A recast of the cross-section measurements
performed in the $\ell+$jets channel (with
at exactly one $b$-tag or two or more $b$ tags), the dilepton channel and the 
$\tau+\ell$ channel has been performed by both \tevatron
experiments~\cite{Abulencia:2005jd,Abazov:2009ae}.
In addition to the efficiencies for SM top pair decays, efficiencies for
decays including a charged Higgs are computed. 

D\O\ considers $H^+\rightarrow
c\bar{s}$ and $H^+\rightarrow \tau^+\bar{\nu_\tau}$ decays. A likelihood for
the observed number of events in the various channels is build as function of
the branching fraction, ${\cal B}(t\rightarrow H^+b)$. The Feldman-Cousins
procedure is used to find its allowed range. 
The resulting limits exclude branching fraction above
around $20\%$ for the pure leptophobic model, see \fig{fig:d0:h+limits}(left).
\begin{figure}[b]
  \centering
\includegraphics[height=4.5cm]{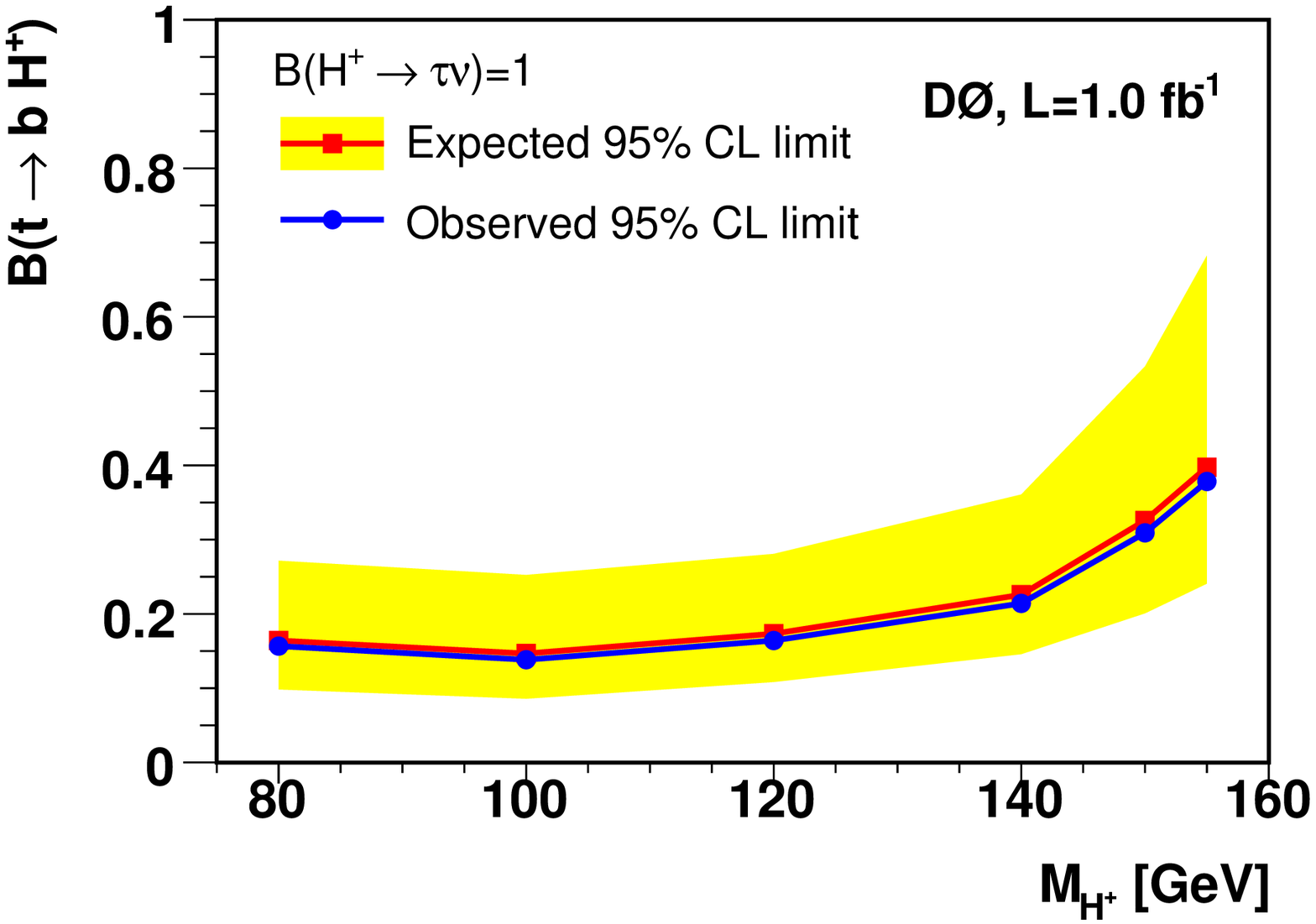}
\includegraphics[height=4.5cm]{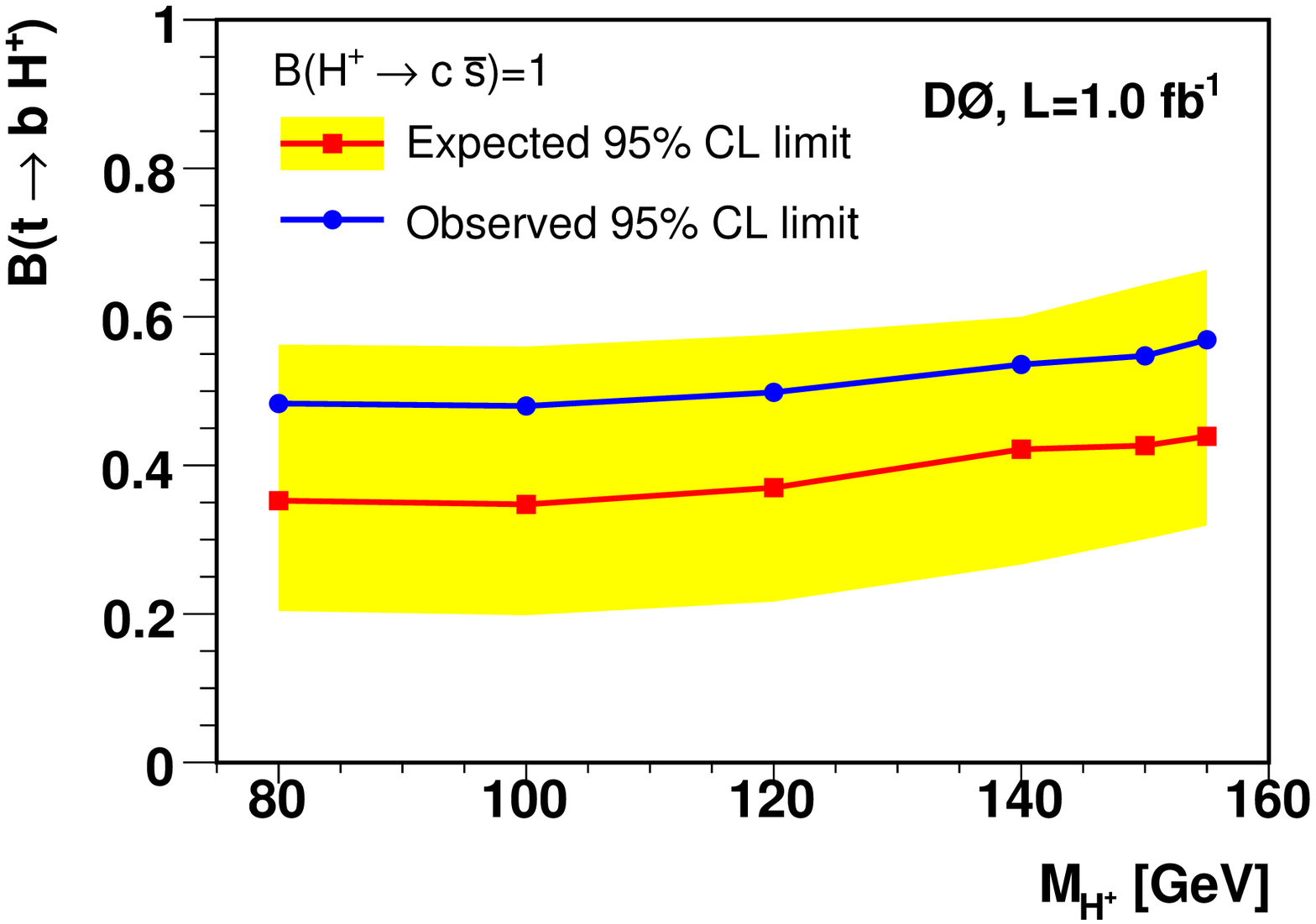}
  \caption{Limits on the contribution of charged Higgs in top decays for a
    leptophobic model (left) and a tauonic model (right) obtained in $1\ifb$
    of D\O\ data~\cite{Abazov:2009ae}.}
  \label{fig:d0:h+limits}
\end{figure}
For a pure tauonic decay of the charged Higgs boson a simultaneous
measurement of the top pair cross-section and the charged Higgs contribution.
In such a two dimensional fit the otherwise dominating systematic
uncertainty due to the assumed top pair cross-section no longer exists  and
the second largest uncertainty, the luminosity uncertainty, is absorbed by the
fit. This results in limits that exclude charged Higgs contributions of more
than $15-25\%$ depending on the Higgs mass (\fig{fig:d0:h+limits}, right).

The CDF analysis includes the $H^+\rightarrow t^*\bar b$ and $H^+\rightarrow
W^+h^0\rightarrow W^+b\bar b$ which are relevant at intermediate values of
$\tan\beta$ in addition to the leptophobic and the tauonic decay.
The event counts observed in data in the four channels are compared to the
expectations in three different ways. For specific benchmarks of the MSSM a Bayesian approach is
used set limits on $\tan\beta$. This analysis uses a flat prior on
$\log\tan\beta$ within the theoretically allowed range. These limits are
computed for various values of the charged Higgs mass and five different
parameter benchmarks. \fig{fig:cdf:h+tanbeta}\subref{fig:MSSM_Exclusion} shows the results
for one specific benchmark.

\begin{figure}
  \centering
  \subfigure[]{
    \includegraphics[height=5.5cm]{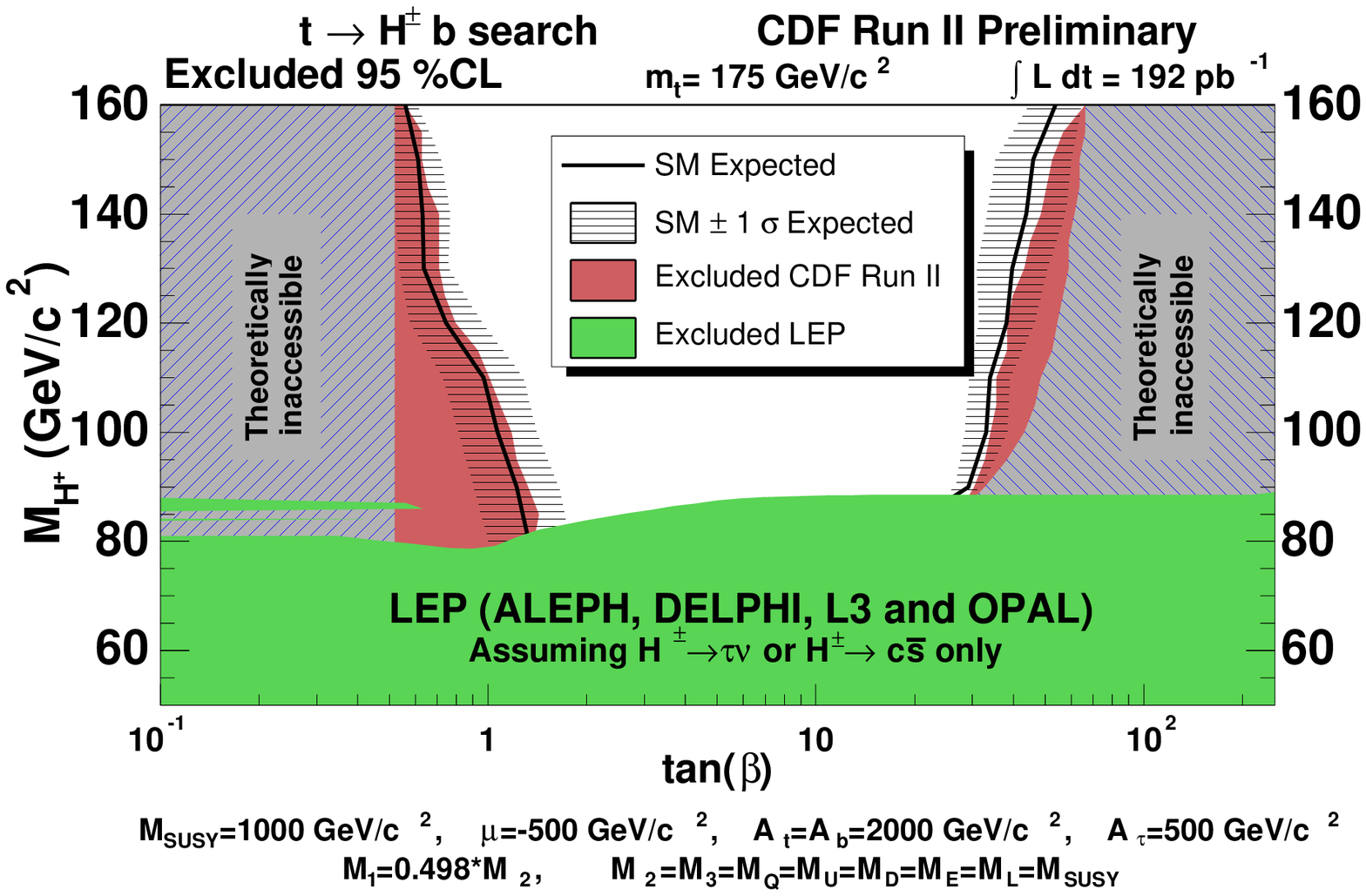}
    \label{fig:MSSM_Exclusion}
  }
  \hspace*{0.1\textwidth}
  \subfigure[]{ 
    \includegraphics[height=5.5cm,trim=0mm -30mm 0mm 0mm]{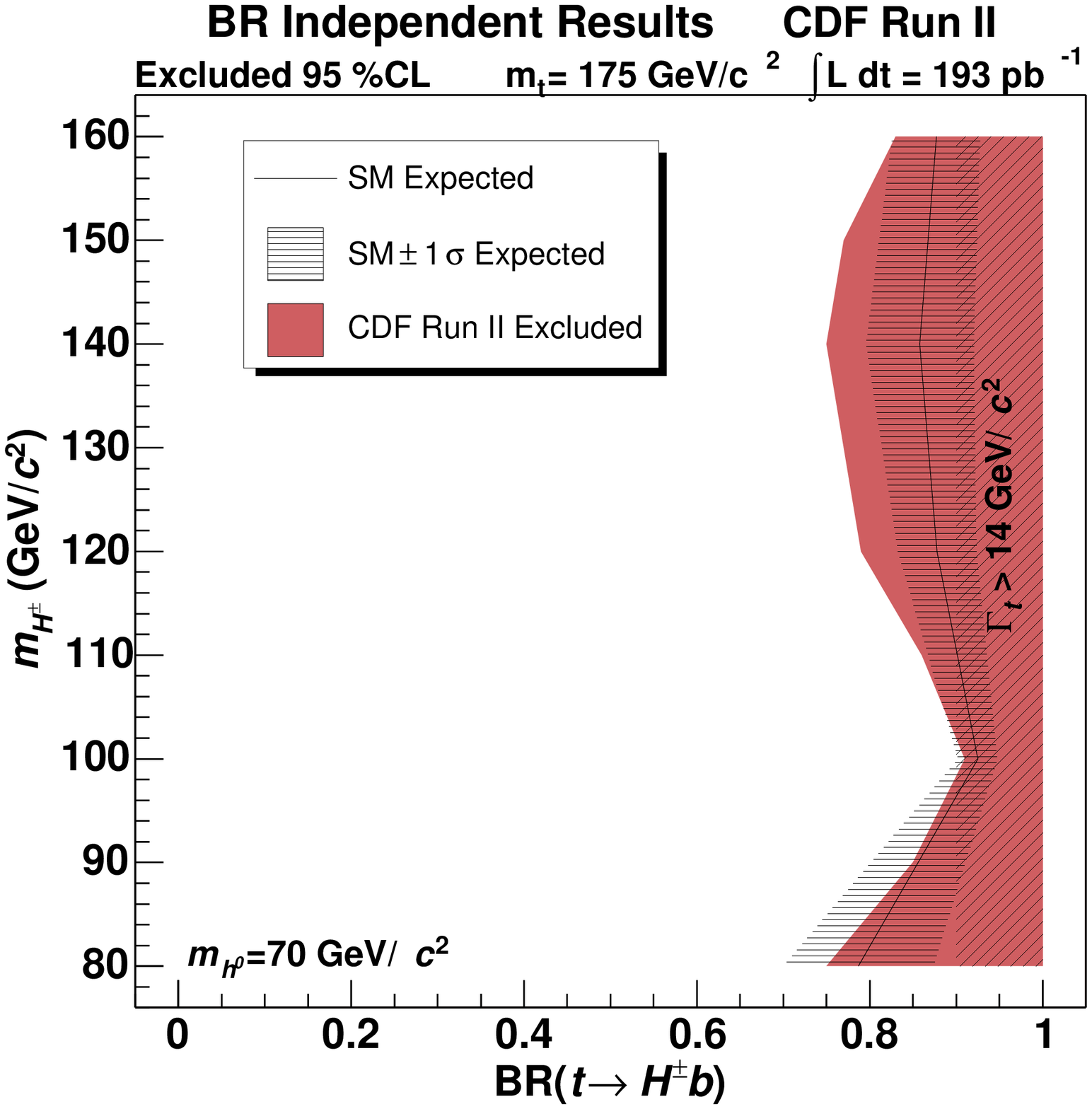}
    \label{fig:CDF_Hplus_Limits}
  }
  \caption{
    Results from recasting the CDF top cross-section
    measurements~\cite{Abulencia:2005jd}. 
    \subref{fig:MSSM_Exclusion}: 
    Exclusion region in the MSSM $m_H$-$\tan\beta$-plane for an example benchmark scenario
    corresponding to the parameters indicated below the plot. 
    \subref{fig:CDF_Hplus_Limits}: Upper limits on ${\cal B}(t\rightarrow H^+b)$ derived without
    assumptions on the charged Higgs branching
    fraction.
       }
  \label{fig:cdf:h+tanbeta}
\end{figure}

\begin{figure}
  \centering
  \subfigure[]{
    \includegraphics[height=6cm]{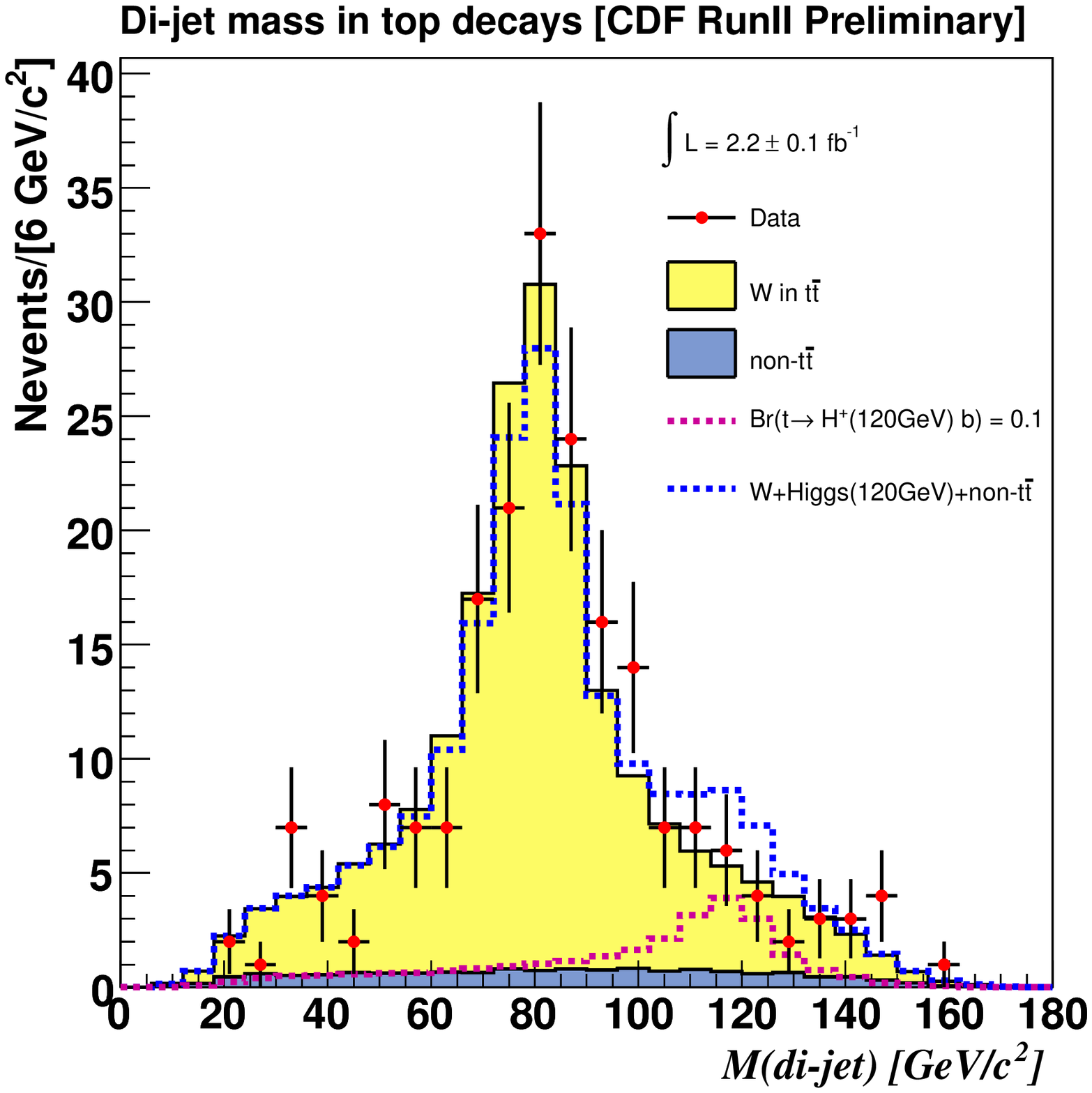}
    \label{fig:CDF_dijet_mass}
  }
  \hspace*{0.1\textwidth}
  \subfigure[]{
    \includegraphics[height=6cm]{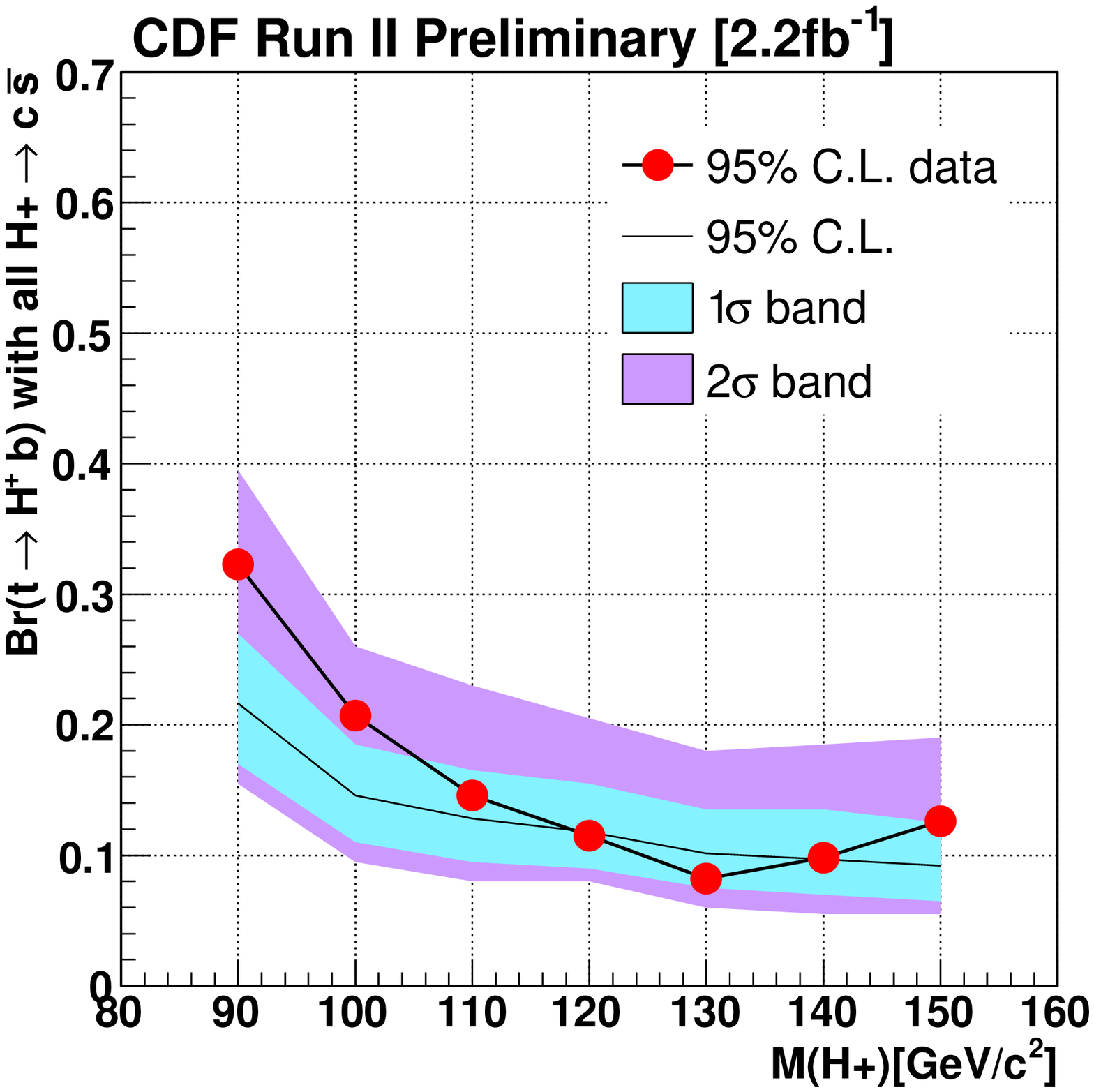}
    \label{fig:CDF_Hplus_CrossSection_Limits}
  }
  \caption{%
      CDF results on search for charged Higgs using kinematic differences~\cite{CdfNote9322}.
      \subref{fig:CDF_dijet_mass}~CDF dijet mass distribution with $120\GeV$ Higgs 
      events assuming 
      ${\cal B}(t\rightarrow H^+b) = 0.1$. 
      The size of Higgs signal corresponds to the expected 
      upper limit branching ratio at $95\%$ C.L. for $120\GeV$.
      \subref{fig:CDF_Hplus_CrossSection_Limits} CDF observed limits on 
      ${\cal B}(t\rightarrow H^+b)$ from $2.2\ifb$ data
      (red dots) compared with the expected limit in the SM (black
      line).
       }
  \label{fig:cdf:h+mjj}
\end{figure}

For the high $\tan\beta$ region $H^+\rightarrow \bar\tau\nu$ dominates in a
large fraction of the MSSM parameter space. Setting the branching fraction of
$H^+\rightarrow \bar\tau\nu$  to $100\%$, limits on the charged Higgs
contribution to top decays are set using Baysian statistics. A flat prior for
${\cal B}(t\rightarrow H^+b)$ between 0 and 1 is used. For  charged Higgs masses
between $80\GeV$ and $160\GeV$ CDF can exclude ${\cal B}(t\rightarrow
H^+b)>0.4$ at 95\%\,C.L.

Finally, a more model independent limit is computed by scanning the full range
of possible charged Higgs decay. For all five $H^\pm$ decay modes considered
the branching fraction is scanned in steps 21 steps, assuring that the sum of
branching fractions adds to one. Limits on ${\cal B}(t\rightarrow H^+b)$ are
computed for each combination. The least restrictive limit is quoted.
Also this analysis is repeated for various charged Higgs masses. The limits
obtained in this more general approach,
shown in \fig{fig:cdf:h+tanbeta}\subref{fig:CDF_Hplus_Limits}, exclude only very high
contribution of charged Higgs to the top decay of above approx. 0.8 to 0.9, depending
on the charge Higgs mass.

At low $\tan\beta$, where the charged Higgs bosons decay also to $c\bar s$, CDF uses
the invariant dijet mass to search for a possible contribution of a charged
Higgs boson in semileptonic top pair events~\cite{CdfNote9322}. A kinematic
fit is used to reconstruct the momenta of the top quark decay products.
Constraints are employed on the consistency of the fitted
lepton and neutrino momenta with the $W$ boson mass and the reconstructed top
masses to be $175\GeV$. 

To determine a possible contribution of charged Higgs in the decay of top pair
production  a binned likelihood fit is performed. The likelihood is
constructed with templates for signal and background with the branching
fraction of top to charged Higgs, the number of top pair events and the number
of background events as parameters. 
%The number of background events is
%constraint  within the uncertainty to the expectation.
The observed dijet mass distribution and he fitted background composition is
shown in \fig{fig:cdf:h+mjj}\subref{fig:CDF_dijet_mass} including a charged 
Higgs contribution of 10\%.
For various assumed Higgs masses 95\%CL limits on the branching fraction are
determined by integrating the likelihood distribution to 95\% of its total.
As shown in~\fig{fig:cdf:h+mjj}\subref{fig:CDF_Hplus_CrossSection_Limits} 
limits between 10\% and 30\% can be
set depending on the mass of the charged Higgs, consistent with the expected
limits for pure SM top decays. This result is less model dependent than the
above CDF limits, but not as strict within the models used above.

\subsection{Non-top quark signatures}
A final very fundamental question that may be asked in the context of top
quark physics beyond the SM is, whether the the events that are considered to be top quarks  actually are
all top quarks or whether some additional unknown new particle is hiding in
the selected data. CDF and D\O\ have followed this question by searching for
two new particles that yield a signature similar to the SM top quark pair
signature, the top quarks supersymmetric  partners, stop $\tilde{t}_{1,2}$,  and 
new heavy quarks, $t'$. 

\paragraph{Admixture of Stop Quarks in Top Pair Events}
In supersymmetry scalar partners of the left and the right handed top quarks are
introduced. The lighter  of the two  supersymmetric  partners of the top
quark, the stop squark $\tilde{t}_{1}$, could be the lightest scalar quark. 
Within the MSSM these would  be preferably produced in pairs by the strong 
interaction.  The decay of the stop squark depends on the details of the
parameters of the MSSM.  Decays through  neutralino and top quark, $\tilde\chi_1^0 t$, or through chargino and
$b$ quark, $\tilde\chi_1^\pm b$, yield neutralino, $b$ quark and $W$ boson,
$\tilde\chi_1^0 b W$, as decay products. These have a signature very similar
to that of semileptonic top quark pairs if the  neutralino
is the lightest supersymmetric particle and is stable due to $R$ parity
conservation. In the decay through,  $\tilde\chi_1^\pm b$, the chargino may
also result in leptonic final states,  $\tilde\chi^0_1\ell\nu(+b)$. If this is
the dominant decay stop quark pairs yield a dilepton signature. 

CDF has searched for a contribution of stop quarks in the dilepton channel
using upto $2.7\ifb$ of data. The reconstructed stop mass is used to distinguish a stop signal from SM
backgrounds including top pair production. The stop mass, $m_{\tilde{t}_1}$,
is determined following an extention of the dilepton neutrino weighting
technique. 
$b$-jets are assigned to their proper lepton  based on jet-lepton 
invariant mass quantities. 
%A correct assignment is reached in $85\%$ to $95\%$ of the cases
%with both $b$-jets being the leading 2 jets.
Neutralino and neutrino are considered as a single, though massive, pseudo
particle. 
For given $\phi$ directions of the pseudo particles the particle momenta are
determined with a fit to the measured quantities using constraints on
the assumed pseudo particle mass, the assumed chargino mass and the equality
of the two stop masses is employed. The reconstructed stop mass is computed as
weighted average of the fitted stop masses, where the average is computed over all
values of $\phi$, with weights according to the fit $\chi^2$ probability.
The combination of reconstructed stop mass templates from the signal and the
various background components is fitted to data. Depending on the dilepton
branching ratio limits are set in the stop vs.\ neutralino mass plane,
c.f. \fig{fig:cdf-stop-limits}. 
\begin{figure}[!tb]
  \centering
  \includegraphics[width=0.35\textwidth]{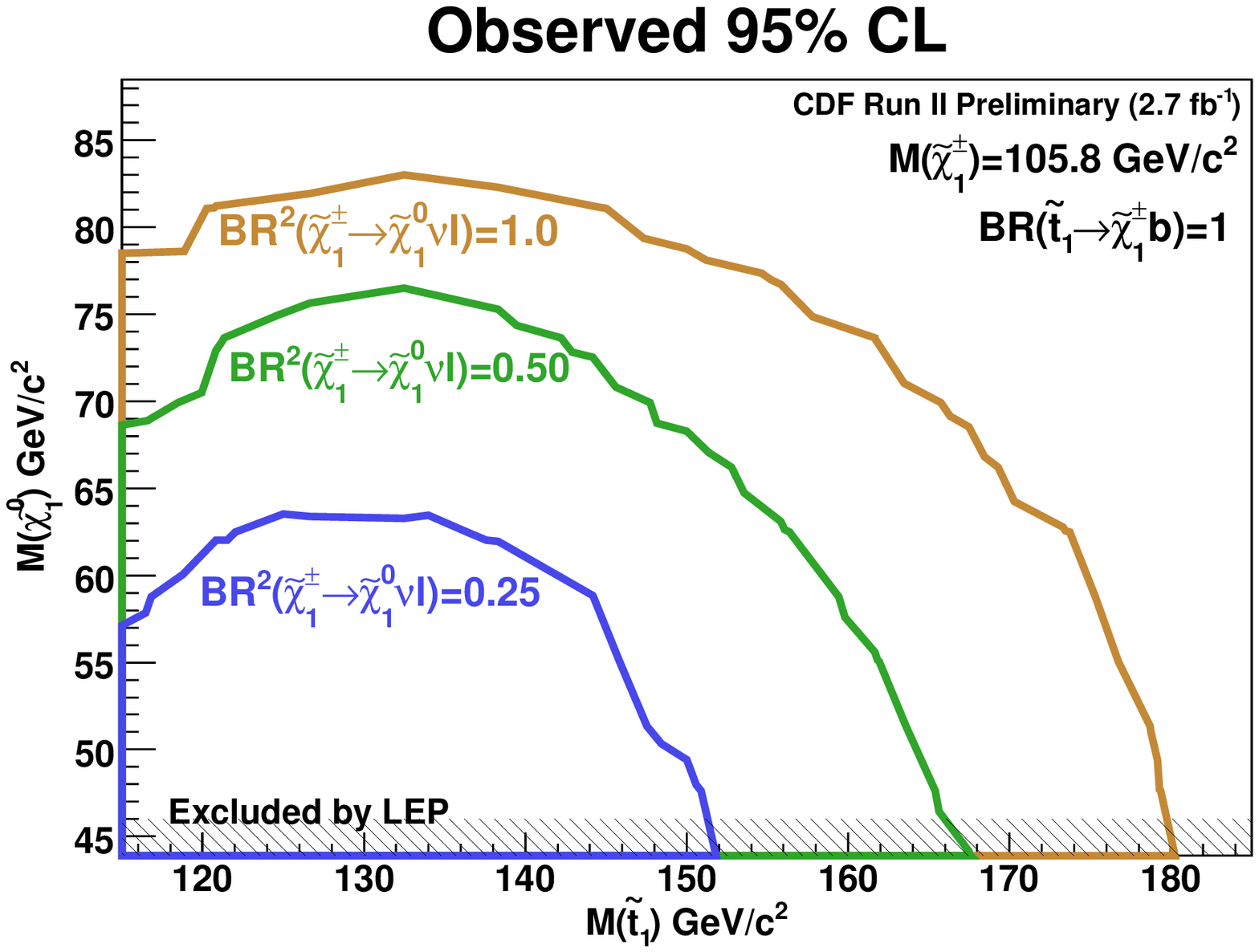}
  \hspace*{0.03\textwidth}
  \includegraphics[width=0.35\textwidth]{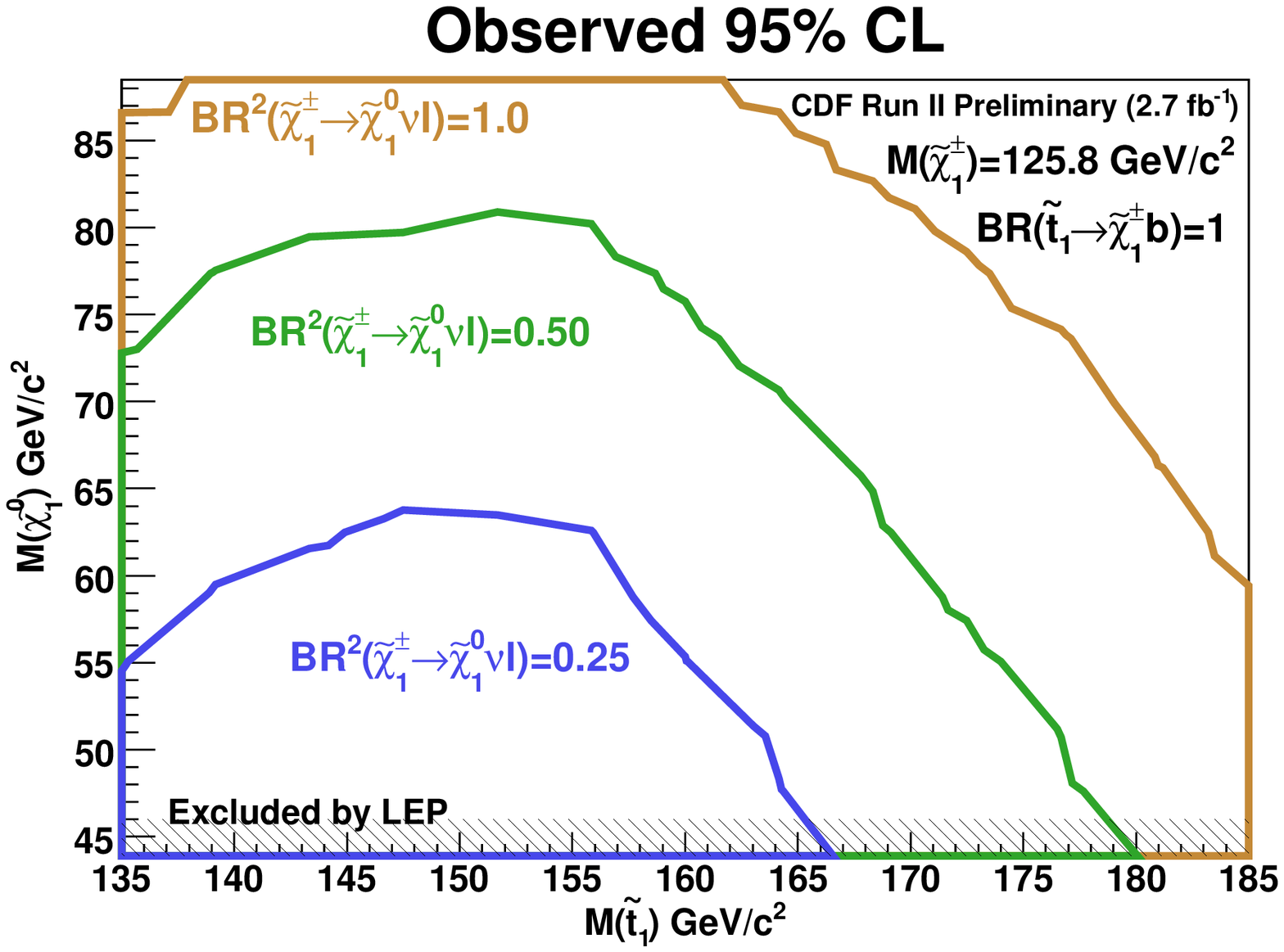}
  \caption{CDF observed 95\%CL limits in the stop mass vs. chargino mass
    plane. The left plot shows limits for a chargino mass of 105.8\GeV, the
    right for 125.8\GeV.}
  \label{fig:cdf-stop-limits}
\end{figure}

D0 has searched for a contribution of such stop pair production in
the semileptonic channel in data with
\mbox{$\sim\!0.9\ifb$}~\cite{Abazov:2009ps}. The analysis uses events with at
least four jets one of which is required to be identified as $b$ jet. 
The event kinematic is reconstructed with a kinematic fitter that assumes SM
top quark pair production and applied constraints on the reconstructed 
$W$ boson masses and the equality on the reconstructed top quark masses. 
Of the possible jet parton assignments only the one with the best
fit quality is considered. 
The differences between stop squark pair events and SM top quark pair production  
is assessed with a likelihood that is based on five kinematic variables. The
the top quark mass reconstructed by the kinematic fitter and the invariant
mass of the second and third non-$b$-tagged jet are found to have the largest
individual separating power among the chosen observables. 

In all cases the theoretically expected
stop signal cross-section in the MSSM is much smaller than the
experimental limits.  Expected and observed limits 
on the stop pair production cross-section are derived in the Bayesian approach.
The limits are compared to the MSSM
prediction for various values of $m_{\tilde{t}_1}$ and $m_{\tilde\chi^\pm_1}$
in Fig.~\ref{fig:stop.limits}. That this point no limits on the MSSM parameter
space can be set from this analysis.
\begin{figure}[t]
  \centering
\includegraphics[width=0.30\textwidth]{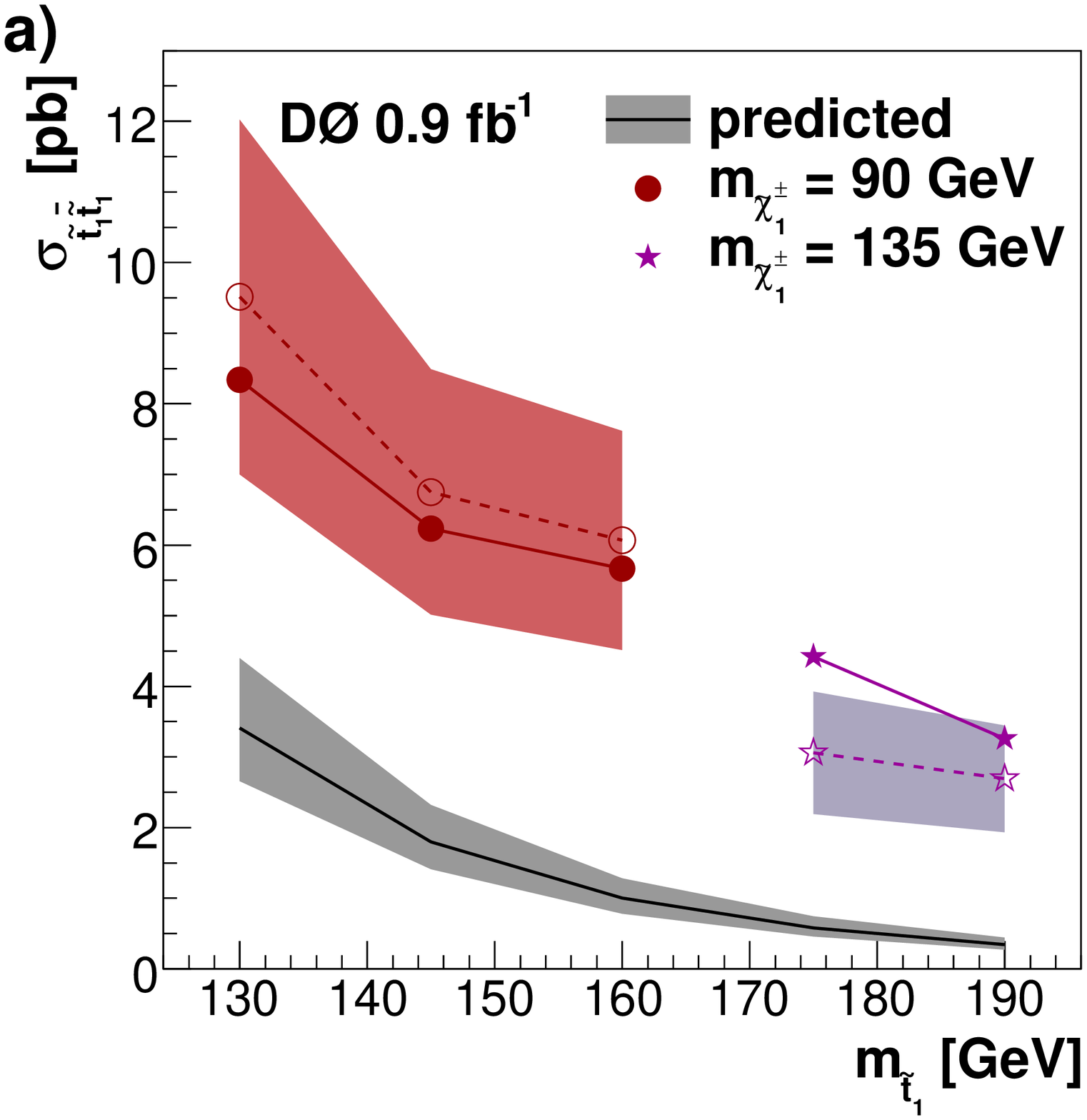}
\hspace*{0.02\textwidth}
\includegraphics[width=0.30\textwidth]{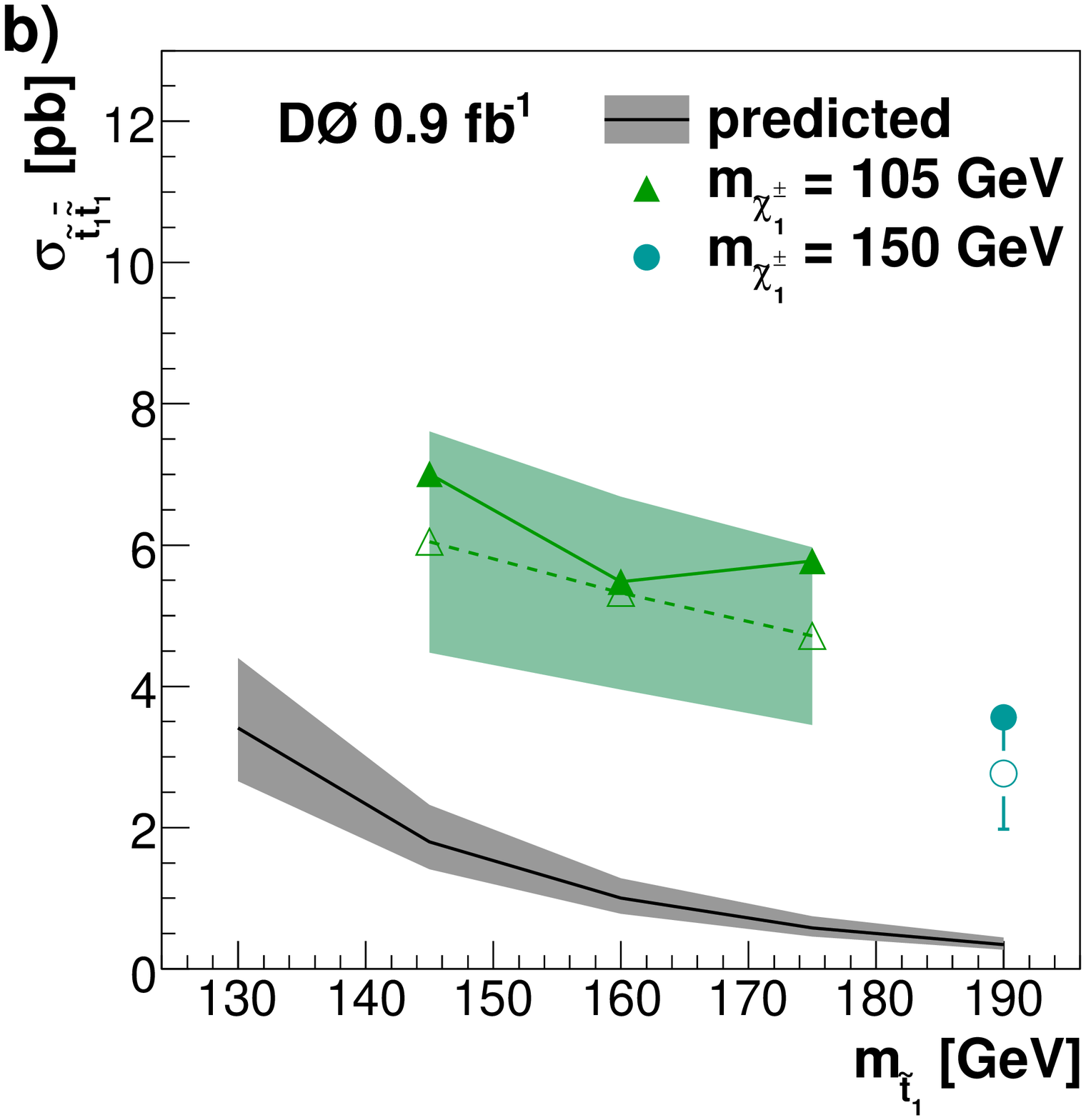}
\hspace*{0.02\textwidth}
\includegraphics[width=0.30\textwidth]{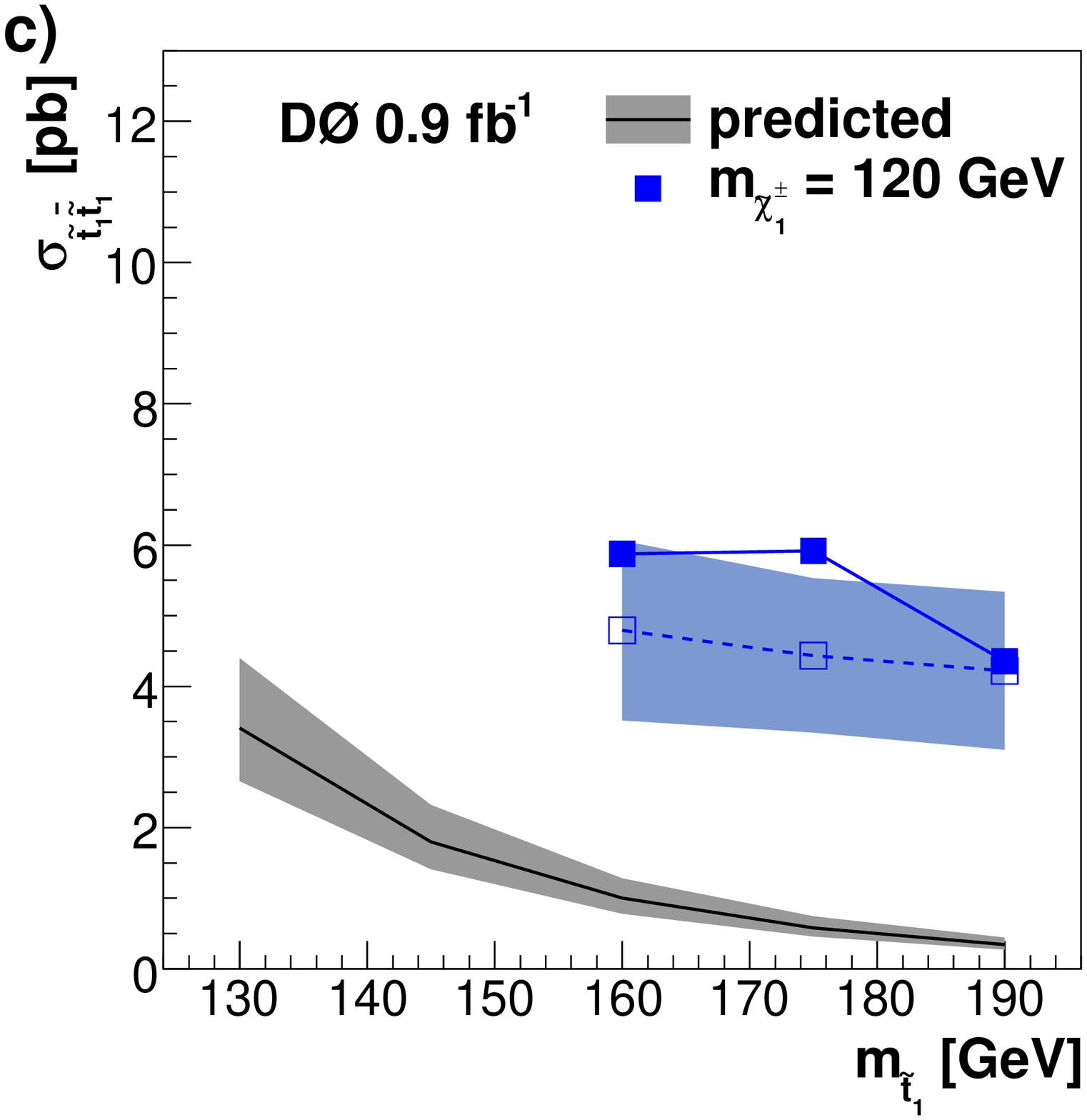}
  \caption{\label{fig:stop.limits}Expected and observed limits on the stop
    pair production cross-section compared to the expectation in the
    various MSSM parameter sets~\cite{Abazov:2009ps}.}
\end{figure}

\paragraph{\boldmath Admixture of additional Heavy Quarks, t'}
Heavy top-like quarks appear in a large number of new physics models: A fourth
generation of fermions~\cite{Kribs:2007nz}, Little Higgs models~\cite{Han:2003gf} and
more named in~\cite{cdf:2008nf}. 
Strong bounds are placed on such models by electroweak precision data, but
for special parameter choices the effects of fourth generation particles
on the electroweak observables compensate each other. Among other settings a small mass
splitting between the fourth $u$-type quark, $t'$, and its isospin partner,
$b'$, is such a case, i.e. $m_{b'}+m_W>m_{t'}$~\cite{Kribs:2007nz}.
Especially, when the new top-like quarks are
very heavy, they should be distinguishable from SM top production
in kinematic observables. So far only CDF has performed a search for such
heavy top-like quarks~\cite{cdf:2008nf,CdfNote9446}.
 As the $t'$ quark decay chain is the same as the top quark decay chain a
constraint fit to the kinematic properties is performed, which puts
constraints on the reconstructed $W$ boson mass and the equality of the
reconstructed  $t^{(\prime)}$ quark masses. The heavy quark signal is
distinguished from SM top quark production using two observables: the
reconstructed  $t^{(\prime)}$ quark mass from the kinematic fit and the scalar
sum of the   transverse energies of the observed
jets, the lepton and the missing transverse energy, $H_T$. The choice
explicitly avoids imposing $b$-quark tagging requirements.
The signal and background shapes in the two dimensional $M_\mathrm{Reco}$,
$H_T$ plane are used to construct a likelihood for the observed data as
function of the assumed $t'$ cross-section, $\sigma_{t'}$. Then Bayesian
statistics is employed to compute expected and observed limits on
$\sigma_{t'}$ that are shown in \fig{fig:cdf-tprime}.
\begin{figure}
  \centering
  \includegraphics[width=0.35\textwidth]{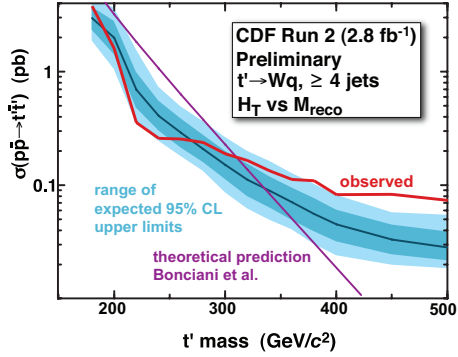}
  \caption{Expected and observed limits on the cross-section of a new top-like
    quark determined from $2.8\ifb$ of CDF data~\cite{CdfNote9446}. The shaded
   bands show the  expected one and two sigma variation on the expected upper limit.}
  \label{fig:cdf-tprime}
\end{figure}
The limits on $t'$ pair production cross-section, $\sigma_{t'}$, determined in 
this search for a new top-like heavy quark are compared to the theoretical
prediction~\cite{Bonciani:1998vc}. Assuming a decay to SM quarks, $t'\rightarrow Wq$,
CDF concludes that  a  $t'$ pair production
can be excluded for $m_{t'}<311\GeV$ at the 95\% C.L. However, for masses of
$m_{t'}\simeq 450\GeV$ the observed limit is worse than the expectation by
more than two standard deviation, which indicates a surplus of data for that range.

\subsection{Outlook to LHC}
Also the \tevatron measurements of top-quark properties including particles beyond the
SM are in general limited by statistics. The increased
cross-section again helps, but not to the same amount for all processes as the
increase for $q\bar q$ annihilation is much lower than that of gluon fusion. 
Many of these analyses profit even more from the increased center-of-mass
energy which open additional phase space for the production of new particles.

An ATLAS study
assuming collisions at $\sqrt{s}=14\TeV$ and a
luminosity of $1\ifb$~\cite{Aad:2009wy} has considered the potential for
discovering resonant top quark pair production through a narrow $Z'$.
A significant  degradation of the selection efficiency is expected at high top
pair invariant masses because the top quark decay products get joined into the
same jet more and more often. At $M_{Z'}=700\GeV$ can be
discovered at $5\sigma$ if its cross-section is $11\pb$ or more.

%%% Local Variables: 
%%% mode: latex
%%% TeX-master: "TopRevPPNP"
%%% End: 

%% file: summary.tex
\section{Summary}
During Run II of the \tevatron top-quark physics has matured to field
providing precision measurements and investigating many different 
aspects of the heaviest elementary particle we know today.
Relatively pure sets of collision events containing a few 
thousand reconstructed top-quark candidates are now available 
for study.

The most important measurement regarding the top quark is the
determination of its mass, since this parameter is essential for 
consistency tests of the SM and the prediction of the Higgs boson mass.
When combining the \tevatron measurements a value of 
$m_t=173.1\pm 1.3\,\mathrm{GeV}/c^2$ is obtained, yielding a relative 
precision of 0.8\%. The mass of the top-quark is thus the most precisely
known quark mass to date.

The most challenging task in top-quark physics was the observation
of single top-quark production via the weak interaction. Using advanced
multivariate techniques CDF and D\O \ were able to discern this rare
process from its huge background, ending a quest which had started 
14 years ago with the discovery of the top quark in pair production
via the strong interaction. The measurements of the
single top-quark cross section indicate that the CKM matrix element
$V_{tb}$ is compatible with one, the value expect in the SM.

Many other production and decay properties of the top quark have been 
tested using the \tevatron data sets. The helicity of $W$ bosons in
top-quark decay, the branching fractions of top quarks, and 
the forward-backward asymmetry in top-quark production were measured and
found to be agreement with the expectations from the SM.
Limits on $\Gamma_t$ and $\tau_t$ were set. Two analyses investigated 
the production mechansim of top quarks, supporting the SM prediction
that $q\bar{q}$ annihilation dominates over $gg$ fusion.

Dedicated analyses searched top-quark data sets for phenomena
beyond the SM. No evidence for right-handed couplings at the $Wtb$
vertex, for FCNC in decay or production, or for heavy resonances
decaying to $t\bar{t}$ pairs was found.

So far all measurements and observations in top-quark physics at the
\tevatron support the picture the SM draws of its heaviest particle.
The \lhc will be a top-quark factory and will allow measurements
that are only limited by systematic uncertainties and may provide 
indications for non-SM physics. A truly encouraging
perspective for a still relatively young field!

%% file: TopRevPPNP.bbl
\begin{thebibliography}{100}\itemsep -2pt

\bibitem{wimpenny_winer}
S.J. Wimpenny and B.L. Winer,
\newblock Annu. Rev. Nucl. Part. Sci. 46 (1995) 149.

\bibitem{campagnari_franklin}
C. Campagnari and M. Franklin,
\newblock Rev. Mod. Phys. 69 (1997) 137.

\bibitem{bhat_prosper_snyder_review}
P.C. Bhat, H.B. Prosper and S.S. Snyder,
\newblock Int. J. Mod. Phys. A 13 (1998) 5113.

\bibitem{tollefson_varnes}
K. Tollefson and E.W. Varnes,
\newblock Annu. Rev. Nucl. Part. Sci. 49 (1999) 435.

\bibitem{chakraborty_konigsberg_rainwater}
D. Chakraborty, J. Konigsberg and D. Rainwater,
\newblock Annu. Rev. Nucl. Part. Sci. 53 (2003) 301.

\bibitem{wagner_2005}
W. Wagner,
\newblock Rept. Prog. Phys. 68 (2005) 2409.

\bibitem{quadt:2006ma}
A. Quadt,
\newblock Eur. Phys. J. C 48 (2006) 835.

\bibitem{bkehoe_2008}
B. Kehoe, M. Narain and A. Kumar,
\newblock Int. J. Mod. Phys.A 23 (2008) 353.

\bibitem{Pleier:2008oc}
M.A. Pleier,
\newblock Int. J. Mod. Phys. A  (2008) 190.

\bibitem{Bernreuther:2008ju}
W. Bernreuther,
\newblock J. Phys. G35 (2008) 083001.

\bibitem{Amsler:2008zzb}
C. Amsler et~al.,
\newblock Phys. Lett. B  667 (2008) 1.

\bibitem{standardmodel0}
S.L. Glashow,
\newblock Nucl. Phys. 22 (1961) 579.

\bibitem{standardmodel1}
S. Weinberg,
\newblock Phys. Rev. Lett. 19 (1967) 1264.

\bibitem{standardmodel2}
A. Salam,
\newblock ed. Nobel Symposium No. 8 (Almqvist \& Wiksell, Stockholm, 1968) .

\bibitem{standardmodel3}
S.L. Glashow, J. Iliopoulos and L. Maiani,
\newblock Phys. Rev. D 2 (1970) 1285.

\bibitem{standardmodel4}
H. Georgi and S.L. Glashow,
\newblock Phys. Rev. Lett. 28 (1972) 1494.

\bibitem{standardmodel5}
H.D. Politzer,
\newblock Phys. Rev. Lett. 30 (1973) 1346.

\bibitem{standardmodel6}
H.D. Politzer,
\newblock Phys. Rept. 14 (1974) 129.

\bibitem{standardmodel7}
D.J. Gross and F. Wilczek,
\newblock Phys. Rev. D 8 (1973) 3633.

\bibitem{standardmodel8}
S. Weinberg,
\newblock Eur. Phys. J. C 34 (2004) 5.

\bibitem{standardmodel9}
G. 't~Hooft,
\newblock Nucl. Phys. B 35 (1971) 167.

\bibitem{standardmodel10}
G. 't~Hooft and M. Veltmann,
\newblock Nucl. Phys. B 44 (1972) 189.

\bibitem{standardmodel11}
G. 't~Hooft and M. Veltmann,
\newblock Nucl. Phys. B 50 (1972) 318.

\bibitem{higgs1}
P.W. Higgs,
\newblock Phys. Lett. 12 (1964) 132.

\bibitem{higgs2}
F. Englert and R. Brout,
\newblock Phys. Rev. Lett. 13 (1964) 321.

\bibitem{higgs3}
G.S. Guralnik, C.R. Hagen and T.W.B. Kibble,
\newblock Phys. Rev. Lett. 13 (1964) 585.

\bibitem{cabibbo}
N. Cabbibo,
\newblock Phys. Rev. Lett. 10 (1963) 531.

\bibitem{kobayashi_maskawa}
M. Kobayashi and T. Maskawa,
\newblock Prog. Theor. Phys. 49 (1973) 652.

\bibitem{herb_1977}
S.W. Herb et~al.,
\newblock Phys. Rev. Lett. 39 (1977) 252.

\bibitem{kane_peskin_1982}
G.L. Kane and M.E. Peskin,
\newblock Nucl. Phys. B 195 (1982) 29.

\bibitem{bean_no_fcnc}
A. Bean et~al.,
\newblock Phys. Rev. D 35 (1987) 3533.

\bibitem{roy_sankar_1990}
D.P. Roy and S.U. Sankar,
\newblock Phys. Lett. B 243 (1990) 296.

\bibitem{argus_1987}
H. Albrecht et~al.,
\newblock Phys. Lett. B 192 (1987) 245.

\bibitem{argus_1987_1}
H. Albrecht et~al.,
\newblock Phys. Lett. B 324 (1994) 249.

\bibitem{cleo_1993}
J. Bartelt et~al.,
\newblock Phys. Rev. Lett. 71 (1993) 1680.

\bibitem{zpole_2005}
S. Schael et~al.,
\newblock Phys. Rept. 427 (2006) 257.

\bibitem{doris_1}
C. Berger et~al.,
\newblock Phys. Lett. B 76 (1978) 243.

\bibitem{doris_2}
J.K. Bienlein et~al.,
\newblock Phys. Lett. B 78 (1978) 360.

\bibitem{doris_3}
W.E. Darden et~al.,
\newblock Phys. Lett. B 76 (1978) 246.

\bibitem{shimonaka_b_weakiso_3}
A. Shimonaka et~al.,
\newblock Phys. Lett. B 268 (1991) 457.

\bibitem{elsen_b_weakiso_1}
E. Elsen et~al.,
\newblock Z. Phys. C 46 (1990) 349.

\bibitem{behrend_b_weakiso_2}
H.J. Behrend et~al.,
\newblock Z. Phys. C 47 (1990) 333.

\bibitem{schaile_zerwas}
D. Schaile and P.M. Zerwas,
\newblock Phys. Rev. D 45 (1992) 3262.

\bibitem{alphaem_bejing}
J.Z. Bai et~al.,
\newblock Phys. Rev. Lett. 88 (2002) 101802.

\bibitem{muon_lifetime}
T. van Ritbergen and R.G. Stuart,
\newblock Phys. Rev. Lett. 82 (1999) 488.

\bibitem{Abe:1995hr}
F. Abe et~al.,
\newblock Phys. Rev. Lett. 74 (1995) 2626.

\bibitem{Abachi:1995iq}
S. Abachi et~al.,
\newblock Phys. Rev. Lett. 74 (1995) 2632.

\bibitem{lepew_2008}
S. Schael et~al.,
\newblock CERN-PH-EP/2008-020  (2008), hep-ex/08114682.

\bibitem{lepew_2009}
. {The LEP Electroweak Working Group},
\newblock {Updated for 2009 winter conferences: http://www.cern.ch/LEPEWWG},
  2009 .

\bibitem{:2009ec}
T. Aaltonen et~al.,
\newblock arXiv: 0903.2503 [hep-ex] (2009).

\bibitem{lep_sm_higgs}
R. Barate et~al.,
\newblock Phys. Lett. B 565 (2003) 61.

\bibitem{tevatron_sm_higgs}
T. Aaltonen et~al.,
\newblock arXiv: 0903.4001 [hep-ex] (2009).

\bibitem{nutev_sin2th}
G.P. Zeller et~al.,
\newblock Phys. Rev. Lett. 88 (2002) 091802.

\bibitem{priv_comm_cquigg}
C. Quigg,
\newblock Phys. Today 50N5 (1997) 20.

\bibitem{Smith97}
M. Smith and S. Willenbrock,
\newblock Phys. Rev. Lett. 79 (1997) 3825.

\bibitem{martinez_miquel}
M. Martinez and R. Miquel,
\newblock Eur. Phys. J. C 27 (2003) 49.

\bibitem{hepph_0001286}
A.H. Hoang et~al.,
\newblock Eur. Phys. J. direct C 2 (2000) 1.

\bibitem{hepph_0011254}
A.H. Hoang et~al.,
\newblock Phys. Rev. Lett. 86 (2001) 1951.

\bibitem{Chang:1998pt}
D. Chang, W.F. Chang and E. Ma,
\newblock Phys. Rev. D  59 (1999) 091503.

\bibitem{Chang:1999zc}
D. Chang, W.F. Chang and E. Ma,
\newblock Phys. Rev. D  61 (2000) 037301.

\bibitem{Choudhury:2001hs}
D. Choudhury, T.M.P. Tait and C.E.M. Wagner,
\newblock Phys. Rev. D  65 (2002) 053002.

\bibitem{bauer_orr_01}
U. Baur, M. Buice and L.H. Orr,
\newblock Phys. Rev. D 64 (2001) 094019.

\bibitem{tesla_physics_tdr}
J.A. Aguilar-Saavedra et~al.,
\newblock SLAC-REPRINT-2001-002, DESY-2001-011, ECFA-2001-209, 2001
  (unpublished) , hep-ph/0106315.

\bibitem{nlc_snowmass01}
T. Abe et~al.,
\newblock SLAC-570, 2001 (unpublished) , hep-ex/0106055, hep-ex/0106056,
  hep-ex/0106057, hep-ex/0106058.

\bibitem{bigi_1986}
I.I.Y. Bigi et~al.,
\newblock Phys. Lett. B 181 (1986) 157.

\bibitem{tspin_kuehn}
J.H. K{\"u}hn,
\newblock Nucl. Phys. B 237 (1984) 77.

\bibitem{czarnecki:1990pe}
A. Czarnecki, M. Je{\.z}abek and J.H. {K\"uhn},
\newblock Nucl. Phys. B 351 (1991) 70.

\bibitem{Bernreuther:1994cx}
W. Bernreuther and A. Brandenburg,
\newblock Phys. Rev. D 49 (1994) 4481.

\bibitem{Bernreuther:1995cx}
W. Bernreuther, A. Brandenburg and P. Uwer,
\newblock Phys. Lett. B 368 (1996) 153.

\bibitem{Dharmaratna:1996xd}
W.G.D. Dharmaratna and G.R. Goldstein,
\newblock Phys. Rev. D 53 (1996) 1073.

\bibitem{Mahlon:1995zn}
G. Mahlon and S. Parke,
\newblock Phys. Rev. D  53 (1996) 4886.

\bibitem{stelzer_willenbrock}
T. Stelzer and S. Willenbrock,
\newblock Phys. Lett. B 374 (1996) 169.

\bibitem{Bernreuther:2005is}
W. Bernreuther, M. {F\"ucker} and Z.G. Si,
\newblock Phys. Lett. B 633 (2006) 54.

\bibitem{kuehn_rodrigo_prl}
J.H. K{\"u}hn and G. Rodrigo,
\newblock Phys. Rev. Lett. 81 (1998) 49.

\bibitem{Kuhn:1998kw}
J.H. K{\"u}hn and G. Rodrigo,
\newblock Phys. Rev. D  59 (1999) 054017.

\bibitem{bowsen_ellis_rainwater}
M.T. Bowen, S.D. Ellis and D. Rainwater,
\newblock Phys. Rev. D 73 (2006) 014008.

\bibitem{willenbrock_0211067}
S. Willenbrock,
\newblock Lectures presented at the 12th Advanced Study Institute on Techniques
  and Concepts of High Energy Physics, St. Croix, U.S. Virgin Islands, 2002 ,
  hep-ph/0211067.

\bibitem{ATLAS_TDR2}
. ATLAS,
\newblock CERN-LHCC-99-15 (1999).

\bibitem{Ball:2007zza}
G.L. Bayatian et~al.,
\newblock J. Phys. G  34 (2007) 995.

\bibitem{Aad:2008zza}
G. Aad et~al.,
\newblock CERN-OPEN-2008-20  (2008) 1.

\bibitem{Bodek:1979rx}
A. Bodek et~al.,
\newblock Phys. Rev. D  20 (1979) 1471.

\bibitem{Adloff:1997mf}
C. Adloff et~al.,
\newblock Nucl. Phys. B  497 (1997) 3.

\bibitem{Chekanov:2005nn}
S. Chekanov et~al.,
\newblock Eur. Phys. J. C  42 (2005) 1.

\bibitem{Pumplin:2002vw}
J. Pumplin et~al.,
\newblock JHEP 07 (2002) 012.

\bibitem{Martin:2007bv}
A.D. Martin et~al.,
\newblock Phys. Lett. B  652 (2007) 292.

\bibitem{Alekhin:2006zm}
S. Alekhin, K. Melnikov and F. Petriello,
\newblock Phys. Rev. D  74 (2006) 054033.

\bibitem{Blumlein:2006be}
J. Blumlein, H. Bottcher and A. Guffanti,
\newblock Nucl. Phys. B  774 (2007) 182.

\bibitem{Moch:2008qy}
S. Moch and P. Uwer,
\newblock Phys. Rev. D  78 (2008) 034003.

\bibitem{bargerPhillips}
V.D. Barger and R.J.N. Phillips,
\newblock Collider Physics,Frontiers in Physics (Addison-Wesley, Reading,
  1987).

\bibitem{nason1988}
P. Nason, S. Dawson and R.K. Ellis,
\newblock Nucl. Phys. B  303 (1988) 607.

\bibitem{beenakker1989}
W. Beenakker et~al.,
\newblock Phys. Rev. D  40 (1989) 54.

\bibitem{Bardeen:1978yd}
W.A. Bardeen et~al.,
\newblock Phys. Rev. D  18 (1978) 3998.

\bibitem{Bernreuther:2001rq}
W. Bernreuther et~al.,
\newblock Phys. Rev. Lett. 87 (2001) 242002.

\bibitem{Czakon:2008ii}
M. Czakon and A. Mitov,
\newblock (2008).

\bibitem{Cacciari:2003fi}
M. Cacciari et~al.,
\newblock JHEP 0404 (2004) 068.

\bibitem{laenen1994}
E. Laenen, J. Smith and W.L. van Neerven,
\newblock Phys. Lett. B  321 (1994) 254.

\bibitem{laenen1992}
E. Laenen, J. Smith and W.L. van Neerven,
\newblock Nucl. Phys. B  369 (1992) 543.

\bibitem{Cacciari:2008zb}
M. Cacciari et~al.,
\newblock JHEP 09 (2008) 127.

\bibitem{Kidonakis:2008mu}
N. Kidonakis and R. Vogt,
\newblock Phys. Rev. D  78 (2008) 074005.

\bibitem{Moch:2008ai}
S. Moch and P. Uwer,
\newblock Nucl. Phys. Proc. Suppl. 183 (2008) 75.

\bibitem{cdfnote9448}
T. Aaltonen et~al.,
\newblock CDF conf. note 9448, 2008.

\bibitem{Abazov:2008gc}
V.M. Abazov et~al.,
\newblock Phys. Rev. Lett. 100 (2008) 192004.

\bibitem{heinson1997}
A.P. Heinson, A.S. Belyaev and E.E. Boos,
\newblock Phys. Rev. D  56 (1997) 3114.

\bibitem{WillenbrockDicus1986}
S.S.D. Willenbrock and D.A. Dicus,
\newblock Phys. Rev. D  34 (1986) 155.

\bibitem{DawsonWillenbrock1987}
S. Dawson and S.S.D. Willenbrock,
\newblock Nucl. Phys. B  284 (1987) 449.

\bibitem{Yuan1990}
C.P. Yuan,
\newblock Phys. Rev. D  41 (1990) 42.

\bibitem{stelzerSullivan1997}
T. Stelzer, Z. Sullivan and S.S.D. Willenbrock,
\newblock Phys. Rev. D  56 (1997) 5919.

\bibitem{harris2002}
B.W. Harris et~al.,
\newblock Phys. Rev. D  66 (2002) 054024.

\bibitem{sullivan2004}
Z. Sullivan,
\newblock Phys. Rev. D  70 (2004) 114012.

\bibitem{Campbell:2004ch}
J. Campbell, R.K. Ellis and F. Tramontano,
\newblock Phys. Rev. D  70 (2004) 094012.

\bibitem{Kidonakis:2006bu}
N. Kidonakis,
\newblock Phys. Rev. D  74 (2006) 114012.

\bibitem{Kidonakis:2007ej}
N. Kidonakis,
\newblock Phys. Rev. D  75 (2007) 071501.

\bibitem{carlson1993}
D. Carlson and C.P. Yuan,
\newblock Phys. Lett. B  306 (1993) 386.

\bibitem{mahlon1997}
G. Mahlon and S. Parke,
\newblock Phys. Rev. D  55 (1997) 7249.

\bibitem{stelzer1998}
T. Stelzer, Z. Sullivan and S.S.D. Willenbrock,
\newblock Phys. Rev. D  58 (1998) 094021.

\bibitem{Jezabek:1994qs}
M. Jezabek,
\newblock Nucl. Phys. Proc. Suppl. 37B (1994) 197.

\bibitem{Tait:1999cf}
T.M.P. Tait,
\newblock Phys. Rev. D  61 (1999) 34001.

\bibitem{Belyaev:2000me}
A. Belyaev and E. Boos,
\newblock Phys. Rev. D  63 (2001) 034012.

\bibitem{Aad:2009wy}
G. Aad et~al.,
\newblock arXiv: 0901.0512 [hep-ex] (2009).

\bibitem{Kuhn:1996ug}
J.H. K\"uhn,
\newblock arXiv: hep-ph/9707321 (1997).

\bibitem{jezabekKuehn1989}
M. Jezabek and J.H. K\"uhn,
\newblock Nucl. Phys. B  314 (1989) 1.

\bibitem{jezabekKuehn1988}
M. Jezabek and J.H. K\"uhn,
\newblock Phys. Lett. B  207 (1988) 91.

\bibitem{mrenna92}
S. Mrenna and C.P. Yuan,
\newblock Phys. Rev. D  46 (1992) 1007.

\bibitem{denner1991}
A. Denner and T. Sack,
\newblock Nucl. Phys. B  358 (1991) 46.

\bibitem{migneron1991}
R. Migneron et~al.,
\newblock Phys. Rev. Lett. 66 (1991) 3105.

\bibitem{jezabekKuehn1993}
M. Jezabek and J.H. K\"uhn,
\newblock Phys. Rev. D  48 (1993) R1910,
\newblock Erratum: Phys. Rev. D 49 (1994) 4970.

\bibitem{kuehn1981}
J.H. K\"uhn,
\newblock Act. Phys. Pol. B  12 (1981) 347.

\bibitem{Fischer:2000kx}
M. Fischer et~al.,
\newblock Phys. Rev. D  63 (2001) 031501.

\bibitem{Fischer:2001gp}
M. Fischer et~al.,
\newblock Phys. Rev. D  65 (2002) 054036.

\bibitem{Do:2002ky}
H.S. Do et~al.,
\newblock Phys. Rev. D  67 (2003) 091501.

\bibitem{jezabekKuehn1994}
M. Jezabek and J.H. K\"uhn,
\newblock Phys. Lett. B  329 (1994) 317.

\bibitem{Mahlon:1998uv}
G. Mahlon,
\newblock arXiv: hep-ph/9811219 (1998).

\bibitem{Jacob:1984vc}
M. Jacob and K. Johnsen,
\newblock Invited talk given at last Meeting of the ISR Committee, Geneva,
  Switzerland, Jan 27, 1984.

\bibitem{Arnison:1983rp}
G. Arnison et~al.,
\newblock Phys. Lett. B122 (1983) 103.

\bibitem{Banner:1983jy}
M. Banner et~al.,
\newblock Phys. Lett. B122 (1983) 476.

\bibitem{:2008ut}
T. Aaltonen et~al.,
\newblock arXiv: 0808.0147 [hep-ex] (2008).

\bibitem{Incandela:2007zz}
J.R. Incandela,
\newblock Nucl. Instrum. Meth. A  579 (2007) 712.

\bibitem{Amidei:1994iq}
D.E. Amidei et~al.,
\newblock Nucl. Instrum. Meth. A  350 (1994) 73.

\bibitem{Cihangir:1994gx}
S. Cihangir et~al.,
\newblock Nucl. Instrum. Meth. A  360 (1995) 137.

\bibitem{Blair:1996kx}
R. Blair et~al.,
\newblock  FERMILAB-Pub-96/390-E (1996).

\bibitem{Sill:2000zz}
A. Sill et~al.,
\newblock Nucl. Instr. and Meth. A  447 (2000) 1.

\bibitem{Bardi:2001uv}
A. Bardi et~al.,
\newblock Nucl. Instrum. Meth. A  485 (2002) 178.

\bibitem{Hill:2004qb}
C.S. Hill,
\newblock Nucl. Instrum. Meth. A  530 (2004) 1.

\bibitem{Abulencia:2006ze}
A. Abulencia et~al.,
\newblock Phys. Rev. Lett. 97 (2006) 242003.

\bibitem{Affolder:2000tj}
A. Affolder et~al.,
\newblock Nucl. Instr. and Meth. A  453 (2000) 84.

\bibitem{Affolder:2003ep}
T. Affolder et~al.,
\newblock Nucl. Instr. and Meth. A  526 (2004) 249.

\bibitem{Acosta:2004kc}
D. Acosta et~al.,
\newblock Nucl. Instr. and Meth. A  518 (2004) 605.

\bibitem{deBarbaro:1998sc}
P. de~Barbaro,
\newblock AIP Conf. Proc. 450 (1998) 405.

\bibitem{Abazov:2005pn}
V.M. Abazov et~al.,
\newblock Nucl. Instrum. Meth. A  565 (2006) 463.

\bibitem{Cooper:2008zzb}
W.E. Cooper,
\newblock Nucl. Instrum. Meth. A598 (2009) 41.

\bibitem{Abazov:2006ka}
V.M. Abazov et~al.,
\newblock Phys. Rev. D  74 (2006) 112004.

\bibitem{Evans:2008zz}
L. Evans and P. Bryant,
\newblock JINST 3 (2008) S08001.

\bibitem{:2008zzm}
G. Aad et~al.,
\newblock JINST 3 (2008) S08003.

\bibitem{Lindstrom:2001ww}
G. Lindstrom et~al.,
\newblock Nucl. Instrum. Meth. A  466 (2001) 308.

\bibitem{:2008zzk}
R. Adolphi et~al.,
\newblock JINST 0803 (2008) S08004.

\bibitem{kalman}
R. Kalman,
\newblock Transactions of the ASME-Journal of Basic Engineering, Series D 82
  (1960) 35.

\bibitem{Abazov:2007kg}
V.M. Abazov et~al.,
\newblock Phys. Rev. D  76 (2007) 092007.

\bibitem{Abulencia:2006kv}
A. Abulencia et~al.,
\newblock Phys. Rev. D  74 (2006) 072006.

\bibitem{Bayatian:2006zz}
G.L. Bayatian et~al.,
\newblock {CMS physics: Technical design report},
\newblock CERN-LHCC-2006-001.

\bibitem{Fruhwirth:2004ma}
R. Fruhwirth and T. Speer,
\newblock Nucl. Instrum. Meth. A  534 (2004) 217.

\bibitem{Bethe:1934za}
H. Bethe and W. Heitler,
\newblock Proc. Roy. Soc. Lond. A  146 (1934) 83.

\bibitem{Seymour:2006vv}
M.H. Seymour and C. Tevlin,
\newblock JHEP 11 (2006) 052.

\bibitem{Aaltonen:2007qf}
T. Aaltonen et~al.,
\newblock Phys. Rev. D  76 (2007) 072009.

\bibitem{Abulencia:2006in}
A. Abulencia et~al.,
\newblock Phys. Rev. Lett. 97 (2006) 082004.

\bibitem{Buskulic:1993ka}
D. Buskulic et~al.,
\newblock Phys. Lett. B  313 (1993) 535.

\bibitem{Abe:1994st}
F. Abe et~al.,
\newblock Phys. Rev. D  50 (1994) 2966.

\bibitem{Acosta:2005zd}
D. Acosta et~al.,
\newblock Phys. Rev. D  72 (2005) 032002.

\bibitem{Acosta:2004hw}
D. Acosta et~al.,
\newblock Phys. Rev. D  71 (2005) 052003.

\bibitem{Abazov:2005ey}
V.M. Abazov et~al.,
\newblock Phys. Lett. B  626 (2005) 35.

\bibitem{Shary:2008pr}
V. Shary,
\newblock Nuovo Cim. 123B (2008) 1053.

\bibitem{Schmaltz:2005ky}
M. Schmaltz and D. Tucker-Smith,
\newblock Ann. Rev. Nucl. Part. Sci. 55 (2005) 229.

\bibitem{Hill:1993hs}
C.T. Hill and S.J. Parke,
\newblock Phys. Rev. D  49 (1994) 4454.

\bibitem{Nilles:1983ge}
H.P. Nilles,
\newblock Phys. Rept. 110 (1984) 1.

\bibitem{Mangano:2002ea}
M.L. Mangano et~al.,
\newblock JHEP 07 (2003) 001.

\bibitem{Maltoni:2002qb}
F. Maltoni and T. Stelzer,
\newblock JHEP 02 (2003) 027.

\bibitem{Sjostrand:2000wi}
T. Sj\"ostrand,
\newblock Comp. Phys. Commun. 135 (2001) 238.

\bibitem{Corcella:2000bw}
G. Corcella et~al.,
\newblock JHEP 01 (2001) 010.

\bibitem{Lange:2001uf}
D.J. Lange,
\newblock Nucl. Instrum. Meth. A  462 (2001) 152.

\bibitem{Agostinelli:2002hh}
S. Agostinelli et~al.,
\newblock Nucl. Instrum. Meth. A  506 (2003) 250.

\bibitem{Allison:2006ve}
J. Allison et~al.,
\newblock IEEE Trans. Nucl. Sci. 53 (2006) 270.

\bibitem{cdfnote9399}
T. Aaltonen et~al.,
\newblock CDF Conf. Note 9399, 2008.

\bibitem{Acosta:2004uw}
D. Acosta et~al.,
\newblock Phys. Rev. Lett. 93 (2004) 142001.

\bibitem{Aaltonen:2009ve}
T. Aaltonen et~al.,
\newblock Phys. Rev. D  79 (2009) 112007.

\bibitem{Abazov:2007bu}
V.M. Abazov et~al.,
\newblock Phys. Rev. D  76 (2007) 052006.

\bibitem{Abazov:2009si}
V.M. Abazov et~al.,
\newblock arXiv: 0901.2137 [hep-ex] (2009).

\bibitem{Abulencia:2006mf}
A. Abulencia et~al.,
\newblock Phys. Rev. D  78 (2008) 012003.

\bibitem{Abulencia:2006yk}
A. Abulencia et~al.,
\newblock Phys. Rev. Lett. 96 (2006) 202002.

\bibitem{:2009ax}
T. Aaltonen et~al.,
\newblock Phys. Rev. D  79 (2009) 052007.

\bibitem{cdfnote9474}
T. Aaltonen et~al.,
\newblock CDF Conf. Note 9474, 2008.

\bibitem{cdfnote9616}
T. Aaltonen et~al.,
\newblock CDF Conf. Note 9616, 2008.

\bibitem{Abe:1997rh}
F. Abe et~al.,
\newblock Phys. Rev. Lett. 79 (1997) 1992.

\bibitem{Abbott:1999mr}
B. Abbott et~al.,
\newblock Phys. Rev. Lett. 83 (1999) 1908.

\bibitem{Abazov:2009ae}
V.M. Abazov et~al.,
\newblock arXiv: 0903.5525 [hep-ex] (2009).

\bibitem{Lyons:1989gh}
L. Lyons, A.J. Martin and D.H. Saxon,
\newblock Phys. Rev. D  41 (1990) 982.

\bibitem{Mangano:2008ag}
M.L. Mangano,
\newblock Int. J. Mod. Phys. A  23 (2008) 3833.

\bibitem{CDFsingleTopRun1}
D. Acosta et~al.,
\newblock Phys. Rev. D  65 (2002) 091102.

\bibitem{CDFsingleTopANNRun1}
D. Acosta et~al.,
\newblock Phys. Rev. D  69 (2004) 052003.

\bibitem{CDFsingleTop2005}
D. Acosta et~al.,
\newblock Phys. Rev. D  71 (2005) 012005.

\bibitem{d0SingleTopCutsRun1}
B. Abbott et~al.,
\newblock Phys. Rev. D  63 (2000) 031101.

\bibitem{d0SingleTopNNRun1}
V.M. Abazov et~al.,
\newblock Phys. Lett. B  517 (2001) 282.

\bibitem{Abazov:2005zz}
V.M. Abazov et~al.,
\newblock Phys. Lett. B  622 (2005) 265.

\bibitem{Abazov:2006uq}
V.M. Abazov et~al.,
\newblock Phys. Rev. D  75 (2007) 092007.

\bibitem{Abazov:2006gd}
V.M. Abazov et~al.,
\newblock Phys. Rev. Lett. 98 (2007) 181802.

\bibitem{Abazov:2008kt}
V.M. Abazov et~al.,
\newblock Phys. Rev. D  78 (2008) 012005.

\bibitem{Aaltonen:2008sy}
T. Aaltonen et~al.,
\newblock Phys. Rev. Lett. 101 (2008) 252001.

\bibitem{Aaltonen:2009jj}
T. Aaltonen et~al.,
\newblock arXiv: 0903.0885 [hep-ex] (2009).

\bibitem{Abazov:2009ii}
V.M. Abazov et~al.,
\newblock arXiv: 0903.0850 [hep-ex] (2009).

\bibitem{Bobrowski:2009ng}
M. Bobrowski et~al.,
\newblock arXiv: 0902.4883 [hep-ph] (2009).

\bibitem{Alwall:2006bx}
J. Alwall et~al.,
\newblock Eur. Phys. J. C  49 (2007) 791.

\bibitem{Bhatti:2005ai}
A. Bhatti et~al.,
\newblock Nucl. Instrum. Meth. A  566 (2006) 375.

\bibitem{Scanlon:2006wc}
T. Scanlon,
\newblock FERMILAB-THESIS-2006-43.

\bibitem{Sjostrand:2006za}
T. Sjostrand, S. Mrenna and P. Skands,
\newblock JHEP 05 (2006) 026.

\bibitem{Alwall:2007st}
J. Alwall et~al.,
\newblock JHEP 09 (2007) 028.

\bibitem{Lueck:2006hz}
J. Lueck,
\newblock FERMILAB-MASTERS-2006-01.

\bibitem{Boos:2006af}
E.E. Boos et~al.,
\newblock Phys. Atom. Nucl. 69 (2006) 1317.

\bibitem{Boos:2004kh}
E. Boos et~al.,
\newblock Nucl. Instrum. Meth. A  534 (2004) 250.

\bibitem{tait2001}
T.M.P. Tait and C.P. Yuan,
\newblock Phys. Rev. D  63 (2001) 014018.

\bibitem{Simmons:1998my}
E.H. Simmons,
\newblock arXiv: hep-ph/9908511 (1998).

\bibitem{abe_prd50_1994}
F. Abe et~al.,
\newblock Phys. Rev. D 50 (1994) 2966.

\bibitem{cdf_9679}
T. Aaltonen et~al.,
\newblock \cdf note 9679, 2009 (unpublished) .

\bibitem{dlm1}
K. Kondo,
\newblock J. Phys. Soc. Jpn. 57 (1988) 4126.

\bibitem{dlm2}
K. Kondo,
\newblock J. Phys. Soc. Jpn. 60 (1991) 836.

\bibitem{dlm3}
K. Kondo, T. Chikamatsu and S.H. Kim,
\newblock J. Phys. Soc. Jpn. 62 (1993) 1177.

\bibitem{dlm5}
K. Kondo,
\newblock arXiv: hep-ex/0508035 (2005).

\bibitem{dalitz_goldstein_1}
R.H. Dalitz and G.R. Goldstein,
\newblock Phys. Rev. D 45 (1992) 1531.

\bibitem{dalitz_goldstein_2}
R.H. Dalitz and G.R. Goldstein,
\newblock Phys. Lett. B 287 (1992) 225.

\bibitem{dalitz_goldstein_3}
R.H. Dalitz and G.R. Goldstein,
\newblock Proc. R. Soc. Lond. A 445 (1999) 2803.

\bibitem{d0_5877}
V.M. Abazov et~al.,
\newblock \dzero\ note 5877, 2009 (unpublished) .

\bibitem{Abazov:2007rk}
V.M. Abazov et~al.,
\newblock Phys. Rev. D  75 (2007) 092001.

\bibitem{delphi_ideogram}
P. Abreu et~al.,
\newblock Eur. Phys. J. C 2 (1998) 581.

\bibitem{mulders_ideogram}
M. Mulders,
\newblock Int. J. Mod. Phys. A 16S1A (2001) 284.

\bibitem{Fiedler:2007tf}
F. Fiedler,
\newblock Eur. Phys. J. C  53 (2008) 41.

\bibitem{Hill:2005zy}
C.S. Hill, J.R. Incandela and J.M. Lamb,
\newblock Phys. Rev. D  71 (2005) 054029.

\bibitem{Abulencia:2006rz}
A. Abulencia et~al.,
\newblock Phys. Rev. D  75 (2007) 071102.

\bibitem{cdf_9414}
T. Aaltonen et~al.,
\newblock \cdf note 9414, 2008 (unpublished) .

\bibitem{Sonnenschein:2006ud}
L. Sonnenschein,
\newblock Phys. Rev. D  73 (2006) 054015.

\bibitem{Abbott:1997fv}
B. Abbott et~al.,
\newblock Phys. Rev. Lett. 80 (1998) 2063.

\bibitem{Abbott:1998dn}
B. Abbott et~al.,
\newblock Phys. Rev. D  60 (1999) 052001.

\bibitem{Abe:1998bf}
F. Abe et~al.,
\newblock Phys. Rev. Lett. 82 (1999) 271.

\bibitem{Abulencia:2005uq}
A. Abulencia et~al.,
\newblock Phys. Rev. Lett. 96 (2006) 152002.

\bibitem{cdf_7797}
D. Acosta et~al.,
\newblock CDF conf. note 7797, 2005 (unpublished) .

\bibitem{d0_5746}
V.M. Abazov et~al.,
\newblock \dzero\ note 5746, 2009 (unpublished) .

\bibitem{d0_5463}
V.M. Abazov et~al.,
\newblock \dzero\ note 5463, 2007 (unpublished) .

\bibitem{Aaltonen:2008bd}
T. Aaltonen et~al.,
\newblock Phys. Rev. Lett. 102 (2009) 152001.

\bibitem{cdf_8959}
D. Acosta et~al.,
\newblock CDF conf. note 8959, 2007 (unpublished) .

\bibitem{d0_5459}
V.M. Abazov et~al.,
\newblock \dzero\ note 5459, 2009 (unpublished) .

\bibitem{cdf_9694}
T. Aaltonen et~al.,
\newblock \cdf note 9694, 2009 (unpublished) .

\bibitem{Abbott:1998dc}
B. Abbott et~al.,
\newblock Phys. Rev. D  58 (1998) 052001.

\bibitem{Abachi:1997jv}
S. Abachi et~al.,
\newblock Phys. Rev. Lett. 79 (1997) 1197.

\bibitem{Abazov:2004cs}
V.M. Abazov et~al.,
\newblock Nature 429 (2004) 638.

\bibitem{d0_0407005}
V.M. Abazov et~al.,
\newblock arXiv: hep-ex/0407005 (2004).

\bibitem{d01_mt_allj_plb}
V.M. Abazov et~al.,
\newblock Phys. Lett. B 606 (2005) 25.

\bibitem{d0_5907}
V.M. Abazov et~al.,
\newblock \dzero\ note 5907, 2009 (unpublished) .

\bibitem{Abazov:2008ds}
V.M. Abazov et~al.,
\newblock Phys. Rev. Lett. 101 (2008) 182001.

\bibitem{d0_5897}
V.M. Abazov et~al.,
\newblock \dzero\ note 5897, 2009 (unpublished) .

\bibitem{d0_09012137}
V.M. Abazov et~al.,
\newblock arXiv: hep-ex/09012137 (2009).

\bibitem{d0_5900}
V.M. Abazov et~al.,
\newblock \dzero\ note 5900, 2009 (unpublished) .

\bibitem{Abe:1997vq}
F. Abe et~al.,
\newblock Phys. Rev. Lett. 80 (1998) 2767.

\bibitem{Affolder:2000vy}
T. Affolder et~al.,
\newblock Phys. Rev. D  63 (2001) 032003.

\bibitem{cdf1_alljets_prl}
F. Abe et~al.,
\newblock Phys. Rev. Lett. 79 (1997) 1992.

\bibitem{cdf_9518}
A. Abulencia et~al.,
\newblock \cdf note 9518, 2009 (unpublished) .

\bibitem{Abulencia:2007br}
A. Abulencia et~al.,
\newblock Phys. Rev. Lett. 99 (2007) 182002, hep-ex/0703045.

\bibitem{cdf_9692}
T. Aaltonen et~al.,
\newblock \cdf note 9692, 2009 (unpublished) .

\bibitem{cdf_8669}
A. Abulencia et~al.,
\newblock \cdf note 8669, 2007 (unpublished) .

\bibitem{cdf_9683}
T. Aaltonen et~al.,
\newblock \cdf note 9683, 2009 (unpublished) .

\bibitem{cdf_9135}
A. Abulencia et~al.,
\newblock \cdf note 9135, 2007 (unpublished) .

\bibitem{cdf_8951}
A. Abulencia et~al.,
\newblock \cdf note 8951, 2009 (unpublished) .

\bibitem{Abulencia:2006ry}
A. Abulencia et~al.,
\newblock Phys. Rev. D  75 (2007) 031105.

\bibitem{cdf_8955}
A. Abulencia et~al.,
\newblock \cdf note 8955, 2007 (unpublished) .

\bibitem{cdf_09013773}
T. Aaltonen et~al.,
\newblock arXiv: 0901.3773 [hep-ex].

\bibitem{:2007jw}
T. Aaltonen et~al.,
\newblock Phys. Rev. Lett. 100 (2008) 062005.

\bibitem{cdf_9265}
T. Aaltonen et~al.,
\newblock \cdf note 9265, 2008 (unpublished) .

\bibitem{cdf_08111062}
T. Aaltonen et~al.,
\newblock arXiv: 0811.1062 [hep-ex].

\bibitem{cdf_9714}
A. Abulencia et~al.,
\newblock \cdf note 9714, 2009 (unpublished) .

\bibitem{tevewwg_1}
P. Azzi et~al.,
\newblock arXiv: hep-ex/0404010 (2004).

\bibitem{Hoang:2008xm}
A.H. Hoang and I.W. Stewart,
\newblock Nucl. Phys. Proc. Suppl. 185 (2008) 220.

\bibitem{nlc_physics_tdr}
S. Kuhlman et~al.,
\newblock FERMILAB-PUB-96-112 (unpublished) , arXiv: hep-ex/9605011 (1996).

\bibitem{Abulencia:2006ei}
A. Abulencia et~al.,
\newblock Phys. Rev. D  75 (2007) 052001.

\bibitem{Aaltonen:2008ei}
T. Aaltonen et~al.,
\newblock Phys. Lett. B  674 (2009) 160.

\bibitem{Abazov:2006hb}
V.M. Abazov et~al.,
\newblock Phys. Rev. D  75 (2007) 031102.

\bibitem{Abazov:2007ve}
V.M. Abazov et~al.,
\newblock Phys. Rev. Lett. 100 (2008) 062004.

\bibitem{Abulencia:2006iy}
A. Abulencia et~al.,
\newblock Phys. Rev. Lett. 98 (2007) 072001.

\bibitem{Lyons:1988rp}
L. Lyons, D. Gibaut and P. Clifford,
\newblock Nucl. Instrum. Meth. A  270 (1988) 110.

\bibitem{CdfNote9144}
T. Aaltonen et~al.,
\newblock CDF conf. note 9144  (2007).

\bibitem{Amsler:2008zz}
C. Amsler et~al.,
\newblock Phys. Lett. B  667 (2008) 1.

\bibitem{Acosta:2005hr}
D. Acosta et~al.,
\newblock Phys. Rev. Lett. 95 (2005) 102002.

\bibitem{Abazov:2008yn}
V.M. Abazov et~al.,
\newblock Phys. Rev. Lett. 100 (2008) 192003.

\bibitem{Fritzsch:1989qd}
H. Fritzsch,
\newblock Phys. Lett. B  224 (1989) 423.

\bibitem{AguilarSaavedra:2004wm}
J.A. Aguilar-Saavedra,
\newblock Acta Phys. Polon. B  35 (2004) 2695.

\bibitem{Aaltonen:2008qr}
T. Aaltonen et~al.,
\newblock Phys. Rev. Lett. 102 (2009) 151801.

\bibitem{Abazov:2007ev}
V.M. Abazov et~al.,
\newblock Phys. Rev. Lett. 99 (2007) 191802.

\bibitem{Abe:1997fz}
F. Abe et~al.,
\newblock Phys. Rev. Lett. 80 (1998) 2525.

\bibitem{Aaltonen:2008aaa}
T. Aaltonen et~al.,
\newblock Phys. Rev. Lett. 101 (2008) 192002.

\bibitem{Heister:2002xv}
A. Heister et~al.,
\newblock Phys. Lett. B  543 (2002) 173.

\bibitem{Abdallah:2003wf}
J. Abdallah et~al.,
\newblock Phys. Lett. B  590 (2004) 21.

\bibitem{Achard:2002vv}
P. Achard et~al.,
\newblock Phys. Lett. B  549 (2002) 290.

\bibitem{Abbiendi:2001wk}
G. Abbiendi et~al.,
\newblock Phys. Lett. B  521 (2001) 181.

\bibitem{Chekanov:2003yt}
S. Chekanov et~al.,
\newblock Phys. Lett. B  559 (2003) 153.

\bibitem{Aktas:2003yd}
A. Aktas et~al.,
\newblock Eur. Phys. J. C  33 (2004) 9.

\bibitem{H1:FCNC2007}
H1,
\newblock {Search for single top quark production in e p collisions at HERA},
\newblock Contributed paper to EPS2007, abstract 776, H1prelim-07-163, 2007.

\bibitem{Liu:2005dp}
J.J. Liu et~al.,
\newblock Phys. Rev. D  72 (2005) 074018.

\bibitem{Yang:2006gs}
L.L. Yang et~al.,
\newblock Phys. Rev. D  73 (2006) 074017.

\bibitem{Zhang:2008yn}
J.J. Zhang et~al.,
\newblock Phys. Rev. Lett. 102 (2009) 072001.

\bibitem{Aaltonen:2008hc}
T. Aaltonen et~al.,
\newblock Phys. Rev. Lett. 101 (2008) 202001.

\bibitem{CdfNote9724}
T. Aaltonen et~al.,
\newblock CDF conf. note 9724  (2009).

\bibitem{d0:2007qb}
V.M. Abazov et~al.,
\newblock Phys. Rev. Lett. 100 (2008) 142002.

\bibitem{Antunano:2007da}
O. Antunano, J.H. Kuhn and G. Rodrigo,
\newblock Phys. Rev. D  77 (2008) 014003.

\bibitem{Dittmaier:2007wz}
S. Dittmaier, P. Uwer and S. Weinzierl,
\newblock Phys. Rev. Lett. 98 (2007) 262002.

\bibitem{Aaltonen:2009iz}
T. Aaltonen et~al.,
\newblock Phys. Rev. Lett. 102 (2009) 222003.

\bibitem{Hocker:1995kb}
A. Hocker and V. Kartvelishvili,
\newblock Nucl. Instrum. Meth. A  372 (1996) 469.

\bibitem{Kidonakis:2003qe}
N. Kidonakis and R. Vogt,
\newblock Phys. Rev. D  68 (2003) 114014.

\bibitem{Arens:1992fg}
T. Arens and L.M. Sehgal,
\newblock Phys. Lett. B  302 (1993) 501.

\bibitem{CdfNote9432}
T. Aaltonen et~al.,
\newblock CDF conf. note 9432  (2008).

\bibitem{Abulencia:2008su}
T. Aaltonen et~al.,
\newblock Phys. Rev. D  79 (2009) 031101.

\bibitem{Aaltonen:2008ir}
T. Aaltonen et~al.,
\newblock Phys. Rev. Lett. 102 (2009) 042001.

\bibitem{CdfNote8104}
T. Aaltonen et~al.,
\newblock CDF conf. note 8104  (2006).

\bibitem{Stange:1993td}
A. Stange and S. Willenbrock,
\newblock Phys. Rev. D  48 (1993) 2054.

\bibitem{Feng:2003uv}
T.F. Feng, X.Q. Li and J. Maalampi,
\newblock Phys. Rev. D  69 (2004) 115007.

\bibitem{AguilarSaavedra:2006gw}
J.A. Aguilar-Saavedra,
\newblock JHEP 12 (2006) 033.

\bibitem{d0note5739}
V.M. Abazov et~al.,
\newblock D\O\ conf. note 5739  (2008).

\bibitem{Leike:1998wr}
A. Leike,
\newblock Phys. Rept. 317 (1999) 143.

\bibitem{Lillie:2007yh}
B. Lillie, L. Randall and L.T. Wang,
\newblock JHEP 09 (2007) 074.

\bibitem{Rizzo:1999en}
T.G. Rizzo,
\newblock Phys. Rev. D 61 (2000) 055005.

\bibitem{Sehgal:1987wi}
L.M. Sehgal and M. Wanninger,
\newblock Phys. Lett. B 200 (1988) 211.

\bibitem{Harris:1999ya}
R.M. Harris, C.T. Hill and S.J. Parke,
\newblock arXiv: hep-ph/9911288  (1999).

\bibitem{cdf:2007dia}
T. Aaltonen et~al.,
\newblock Phys. Rev. D  77 (2008) 051102.

\bibitem{cdf:2007dz}
T. Aaltonen et~al.,
\newblock Phys. Rev. Lett. 100 (2008) 231801.

\bibitem{CdfNote9164}
T. Aaltonen et~al.,
\newblock CDF conf. note 9164  (2008).

\bibitem{Abulencia:2005pe}
A. Abulencia et~al.,
\newblock Phys. Rev. D  73 (2006) 092002.

\bibitem{Abazov:2008ny}
V.M. Abazov et~al.,
\newblock Phys. Lett. B  668 (2008) 98.

\bibitem{D0Note5600conf}
V.M. Abazov et~al.,
\newblock {{D\O \ }~conf. note~5600}  (2008).

\bibitem{D0Note5882conf}
V.M. Abazov et~al.,
\newblock {{D\O \ }~conf.~note~5882}  (2009).

\bibitem{Abazov:2008rn}
V.M. Abazov et~al.,
\newblock Phys. Rev. Lett. 102 (2009) 191802.

\bibitem{Boos:2006xe}
E. Boos et~al.,
\newblock Phys. Lett. B  655 (2007) 245.

\bibitem{Aaltonen:2009qu}
T. Aaltonen et~al.,
\newblock arXiv: 0902.3276 [hep-ex]  (2009).

\bibitem{Abazov:2008vj}
V.M. Abazov et~al.,
\newblock Phys. Rev. Lett. 100 (2008) 211803.

\bibitem{Sullivan:2002jt}
Z. Sullivan,
\newblock Phys. Rev. D  66 (2002) 075011.

\bibitem{Abulencia:2005jd}
A. Abulencia et~al.,
\newblock Phys. Rev. Lett. 96 (2006) 042003.

\bibitem{CdfNote9322}
T. Aaltonen et~al.,
\newblock CDF conf. note 9322  (2008).

\bibitem{Abazov:2009ps}
V.M. Abazov et~al.,
\newblock Phys. Lett. B  674 (2009) 4.

\bibitem{Kribs:2007nz}
G.D. Kribs et~al.,
\newblock Phys. Rev. D  76 (2007) 075016.

\bibitem{Han:2003gf}
T. Han et~al.,
\newblock Phys. Lett. B  563 (2003) 191.

\bibitem{cdf:2008nf}
T. Aaltonen et~al.,
\newblock Phys. Rev. Lett. 100 (2008) 161803.

\bibitem{CdfNote9446}
T. Aaltonen et~al.,
\newblock CDF conf. note 9446  (2008).

\bibitem{Bonciani:1998vc}
R. Bonciani et~al.,
\newblock Nucl. Phys. B  529 (1998) 424.

\end{thebibliography}
